
\input epsf

\magnification\magstep1



\overfullrule=0pt  
\hbadness=10000      
\vbadness=10000

\font\eightit=cmti8  
\font\eightrm=cmr8 \font\eighti=cmmi8                 
\font\eightsy=cmsy8 
\font\sixrm=cmr6


\def\eightpoint{\normalbaselineskip=10pt 
\def\rm{\eightrm\fam0} \let\it\eightit
\textfont0=\eightrm \scriptfont0=\sixrm 
\textfont1=\eighti \scriptfont1=\seveni
\textfont2=\eightsy \scriptfont2=\sevensy 
\normalbaselines \eightrm
\parindent=1em}



\def\eq#1{{\noexpand\rm(#1)}}          
\newcount\eqcounter                    
\eqcounter=0                           
\def\numeq{\global\advance\eqcounter by 1\eq{\the\eqcounter}}           
\def\relativeq#1{{\advance\eqcounter by #1\eq{\the\eqcounter}}}
\def\lasteq{\relativeq0}               
\def\beforelasteq{\relativeq{-1}}      

\def\nexttonexteq{\relativeq2}         

\def\namelasteq#1{\global\edef#1{{\eq{\the\eqcounter}}}}  

\def\A{{\rm A}}
\def\Ac{{\cal A}}                      
\def\a{\alpha}                        
\def\aprime{{\alpha'}}
\def\arrowsim#1{{
  \buildrel\leftrightarrow\over{#1} }} 
\def\aS{\a_{\rm S}}              
\def\b{\beta}                          
\def\bG{{\bf G}}
\def\BM{{\rm BM}}                      
\def\bS{\b_{\rm S}}              

\def\bmu{{\bar\mu}}
\def\bnu{{\bar\nu}}
\def\brho{{\bar\rho}}    
\def\cite#1{{\rm[#1]}}                 
\def\BonneauGraph#1{\overline          
 {\bigl\langle                              
	{#1}\; 
	\N[\check\Delta](q\!=\!\!-\!{\textstyle\sum} p_i) 
	\bigr\rangle}^\irr_{K=0}}
\def\Graph#1{ \bigl\langle                             
	{#1}\; 
	\N[\check\Delta](q\!=\!\!-\!{\textstyle\sum} p_i) 
	\bigr\rangle^{\irr}_{K=0}}
\def\BonneauGraphZero#1{\overline          
 {\bigl\langle                              
	{#1}\;                                    
	\N[\check\Delta](q=0) 
	\bigr\rangle}^\irr_{K=0}}

\def\coefZero#1{ \times                  
	\hbox{coef. in r.s.p. $\BonneauGraphZero{#1}$ of $\,$} }
\def\CA{C_\A}
\def\Cno{{\rm C}\!\!\!{\rm I}\;}
\def\CL{C_\L}
\def\CR{C_\R}
\def\CLpR{C_{\L(\R)}}
\def\clr{\rm cl}
\def\dL{d_\L}
\def\dR{d_\R}
\def\dLpR{d_{\L(\R)} }
\def\dLR{d_{\L-\R} }
\def\dRL{d_{\R-\L} }
\def\DL{{\cal D}_\L}
\def\DR{{\cal D}_\R}
\def\DLR{{\cal D}_{\L-\R} }

\def\Dslash{{ {\rm D}\mkern-11mu/}}    
\def\eps{\varepsilon}                  
\def\F{{\rm F}\,}
\def\FF{ \Tr\, F_{\mu\nu} 
		  F^{\mu\nu} }                   

\def\ga{\gamma}                     
\def\ghat{\hat g}                      
\def\gtah{\check g}                      
\def\gbar{\bar g}                      
\def\gamhat{\hat\gamma}                
\def\gambar{\bar\gamma}                
\def\gam5{\gamma_5}                    
\def\GR{\Gamma_{\rm ren}}                
\def\GRpuno{{\tilde\Gamma}^{\rm R\,(1)}} 

\def\Gapuno{{\tilde\Gamma}^{(1)}}      
\def\Gapcero{{\tilde\Gamma}^{(0)}}     

\def\intd{\int\! {\rm d}}              
\def\intdk{\mu^\eps\!
	\int\! {{\rm d}^dk\over (2\pi)^d}\;} 
\def\irr{{\rm 1PI}}                    
\def\kslash{{k\mkern-9mu/}}            
\def\L{{\rm L}}                        
\def\LmR{{\L-\R}}
\def\LpR{{\L(\R)}}
\def\RpL{{\R(\L)}}
\def\la{\lambda}                       
\def\M{{\cal M}}                       
\def\muGR{\mu{\pr\GR\over\pr\mu}}
\def\N{{\rm N}}                        
\def\NN{{\cal N}}                      
\def\NS{{N_{\!\rm \SS}}}          
\def\NA{{N_{\!\rm \Ac}}}
\def\n{{\rm naive}}                    
\def\nocorr{\kern0pt}                  
\def\om{\omega}                        
\def\Op{{\cal O}}                      
\def\Ouno{ {}^{(1)} }
\def\ouno{ {(1)} }
\def\pp{{(\prime)}}                    
\def\pr{\partial}                      
\def\PL{{P_\L}}                        
\def\PR{{P_\R}}                        
\def\PLpR{{P_\LpR}}                    
\def\PRpL{{P_\RpL}}                    
\def\PT{{\rm P}_{\rm T} }              
\def\psipsi{\bar\psi\arrowsim\prslash
				 \psi}                     
\def\psipsiprime{\bar\psi' 
		\arrowsim\prslash \psi'}          
\def\psipsicov{\bar\psi\arrowsim
		\Dslash_\L\psi}                  
\def\psipsiprimecov{\bar\psi' 
		\arrowsim\Dslash_\R \psi'}       
\def\pslash{{p\mkern-8mu/}{\!}}        
\def\prslash{{\partial\mkern-9mu/}}    
\def\R{{\rm R}}                        
\def\rsp{{\rm r.s.p.}\;}               
\def\sa#1{a_{\! #1 \!}}                
\def\sm#1{\mu_{\! #1 \!}}              
\def\sepeq{\noalign{\vskip10pt}}        
\def\square{\vbox{\hrule\hbox{\vrule height5.2pt \hskip 5.2pt
  \vrule}\hrule}}                      
\def\Sfct{S_{\rm fct} }                
\def\SS{{\cal S}}     
\def\T{{\rm T}}
\def\Tr{{\rm Tr}}                      
\def\TA{T_\A}
\def\TL{T_\L}
\def\TR{T_\R}
\def\TLpR{T_{\rm L(R)}}
\def\tLpR{t_{\rm L(R)}}
\def\tL{t_\L}                          
\def\tR{t_\R}                          
\def\TLR{T_{\L+\R}}

\def\TRmL{T_{\R-\L}}
\def\ucpc{{1 \over (4 \pi)^2}}         
\def\hcpc{{\hbar^1 \over (4 \pi)^2}}   
\def\icpc{{i\over (4\pi)^2}}           
\def\un{{\rm I}\!{\rm I}}              
\def\V{{\rm V}}
\def\W{{\cal W}}
\def\bW{{\bar \W}}
\def\hW{{\hat \W}}




\newif\ifstartsec                   

\outer\def\section#1{\vskip 0pt plus .15\vsize \penalty -250
\vskip 0pt plus -.15\vsize \bigskip \startsectrue
\message{#1}\centerline{\bf#1}\nobreak\noindent}

\outer\def\doublesection#1#2{\vskip 0pt plus .15\vsize \penalty -250
\vskip 0pt plus -.15\vsize \bigskip \startsectrue
\message{#1}\centerline{\bf#1}\par
\centerline{\bf#2}\nobreak\noindent}

\def\subsection#1{\ifstartsec\medskip\else\bigskip\fi \startsecfalse
\noindent{\it#1}\penalty100\medskip}

\def\refno#1. #2\par{\smallskip\item{\rm\lbrack#1\rbrack}#2\par}

\hyphenation{geo-me-try}


\def\Ash{1}
\def\HV{2}
\def\BMa{3}
\def\BMb{4}
\def\BMc{5}
\def\BonneauA{6}
\def\BonneauB{7}
\def\BonneauC{8}
\def\Leib{9}
\def\Collins{10}
\def\Min{11}
\def\Tech{12}
\def\Multi{13}
\def\BonneauRemarks{14}
\def\Aky{15}
\def\Kreimer{16}
\def\Colnor{17}
\def\PiguetSorella{18}
\def\Brandt{19}
\def\Hepp{20}
\def\BonneauReview{21}
\def\naive{22}
\def\Alwit{23}
\def\Wilson{24}
\def\PiguetRouet{25}
\def\Korner{26}
\def\Ferrari{27}
\def\Barroso{28}
\def\Gottlieb{29}
\def\Ovrut{30}
\def\IZ{31}
\def\Bare{32}


\rightline{FT/UCM--20--99}

\vskip 1cm

\centerline{\bf  Action principles, restoration of BRS symmetry 
           and the    }
 \centerline{\bf renormalization group equation for chiral non-Abelian gauge 
theories}
\centerline{\bf in dimensional renormalization with a non-anticommuting 
$\gam5$}

\bigskip

\centerline{\rm C. P. Mart{\'\i}n* and D. S\'anchez-Ruiz\dag}
\medskip
\centerline{\eightit Departamento de F{\'\i}sica Te\'orica I,			Universidad Complutense, 28040 Madrid, Spain}
\vfootnote*{email: {\tt carmelo@elbereth.fis.ucm.es}}
\vfootnote\dag{email: {\tt domingos@eucmos.sim.ucm.es}}

\bigskip\bigskip

\begingroup\narrower\narrower
\eightpoint
The one-loop renormalization of a general chiral gauge theory without scalar
and Majorana fields  is fully worked out within Breitenlohner and Maison 
dimensional renormalization scheme. The coefficients of the anomalous terms 
introduced in the Slavnov-Taylor equations by the minimal subtraction 
algorithm are calculated and the asymmetric counterterms needed to restore the 
BRS symmetry, if the anomaly cancellation conditions are met, are
computed. The renormalization group equation and its coefficients
are worked out in the anomaly free case. The computations draw heavily from
the existence of  action principles and  BRS cohomology theory.
\par
\endgroup 

\bigskip

\section{1. Introduction}

Dimensional Regularization \cite{\Ash,\HV } is the standard regularization 
method in four dimensional perturbative quantum field theory as applied to
particle physics. Both its axiomatics and properties were rigorously 
established long ago 
\cite{\BMa,\BMb,\BMc,\BonneauA,\BonneauB,\BonneauC,\Leib,\Collins } 
and quite a number of computational techniques based on the method
have been developed over the years \cite{\Min,\Tech }. Involved multiloop 
computations of the parameters of the Standard Model, a must due to the 
availability of high precision tests of the model in particle accelerators,
have been carried out in the simplest possible setting thanks to dimensional 
regularization \cite{\Multi}. The success of Dimensional Regularization
as a practical regularization method stems from the fact that it preserves
the BRS symmetry of vector-like non-supersymmetric  gauge theories without
distorting neither the shape of the integrand of a regularized Feynman diagram
nor the properties of the algebraic objects involved (no chiral objects being
present). The minimal subtraction scheme, the famous MS scheme \cite{\Min},
then leads \cite{\HV } to a renormalized BRS-invariant theory 
\cite{\BMa,\BonneauA }. Counterterms are generated by multiplicative 
renormalization of the tree-level Lagrangian, so that $\beta$ functions and
anomalous dimensions can be computed easily \cite{\Collins}.

It is an empirical fact that electroweak interactions are chiral and hence
vector-like gauge theories does fail to account for them. It turn thus out that
chiral gauge theories are of key importance to understanding Nature. 
Unfortunately, Dimensional Regularization loses its smartness when applied
to chiral gauge theories. The algebraic properties that the matrix $\gamma_5$
has in four dimensions cannot be maintained without introducing algebraic 
inconsistencies as we move away from four dimensions \cite{\BonneauRemarks }. 
Hence, within the framework of Dimensional Regularization, the definition 
of the object $\gamma_5$ in ``d complex dimensions'' demands a new 
set of algebraic identities. There exists such a
set of identities, they were introduced in ref.~\cite{\BMa}, following 
refs.~\cite{\HV,\Aky}, and they entail that the object $\gamma_5$ anticommutes
no longer with the object $\gamma_{\mu}$. The axiomatics of dimensional 
regularization so established is the only one which has been shown to be 
thoroughly consistent at any order in perturbation theory if cyclicity of the
trace is not given up (see ref.~\cite{\Kreimer} for the  non-cyclic trace
alternative). Minimal subtraction \cite{\Min} of the singular 
part of the dimensionally regularized Feynman diagrams, a subtraction 
procedure known as minimal dimensional renormalization
\cite{\BMa,\BMb,\BMc,\BonneauA}, leads to a renormalized quantum field 
theory that satisfies Hepp axioms \cite{\Hepp} of renormalization theory.
Field equations, the action principles and Zimmerman-Bonneau identities hold
in dimensional renormalization. This has been rigorously shown in refs.
\cite{\BMa,\BMb,\BMc,\BonneauA,\BonneauB,\Colnor} and it constitutes one
of the key ingredients of the modern approach to the quantization of gauge
theories by using BRS methods \cite{\PiguetSorella,\Brandt}.

That Dimensional Regularization can be used along with algebraic BRS 
techniques is of the utmost importance, for this regularization method and
the minimal subtraction algorithm that comes with it break, 
generally speaking, chiral gauge symmetries. This breaking gives rise 
to both physical and non-physical anomalies 
\cite{\BonneauB,\BonneauReview }. Physical 
anomalies in the currents of the Lagrangian impose anomaly cancellation 
conditions \cite{\Alwit}, lest the theory ceases to make sense as a quantum
theory. These anomalies belong to the cohomology of the Slavnov-Taylor 
operator that governs the chiral gauge symmetry at the quantum level 
and they are not artifacts of the regularization method nor due to the 
renormalization scheme employed. Non-physical anomalies correspond to trivial
objects in the cohomology of the BRS operator and, therefore, they can be set
to zero by an appropriate choice of finite counterterms \cite{\BonneauReview}.
Non-physical anomalies are artifacts either of the regularization method, the
renormalization method or both.

 Notwithstanding the unique status that as a thoroughly  consistent framework 
the Breitenlohner and Maison scheme enjoys among the dimensional 
renormalization prescriptions, there is, up to the best of our knowledge,
no complete one-loop study of a general non-Abelian chiral gauge theory, 
let alone the Standard Model. Most of the implementations of dimensional
regularization in quantum field theories  which involve the 
matrix $\gam5$ and are anomally-free, {\it e.g.} the Standard 
Model, take  a fully anticommuting $\gam5$ in ``$d$  complex dimensions''. 
This is the so-called ``naive'' prescription for the object  
$\gam5$ \cite{\naive}. The purpose of this paper is  to remedy this 
situation by carrying out such  one-loop study for a general  chiral gauge 
theory with neither scalar nor Majorana fields. The inclusion of scalar
fields will be discussed elsewhere.

The layout of this paper is as follow. In Section 2 we give a quick account
of the dimensional renormalization algorithm  
{\it \'a la Breitenlohner and Maison}, the  Regularized and Quantum 
Action Principles and Bonneau identities. 
Section 3 is devoted to a thorough study of the
one-loop dimensional renormalization of a general chiral gauge theory
for a simple gauge group. In this section the BRS anomaly is computed
for the first time in the Breitenlohner and Maison framework. Of course,
the standard anomaly cancellation conditions are thus obtained. The
beta function and the anomalous dimensions of the theory are computed in
this section as well. Again, the one-loop renormalization group equation
for the model at hand had never been work out as yet in Breitenlohner and
Maison scheme. In section 4 we adapt the results obtained in the
previous section to the case of a non-simple  gauge group. We also include
three appendices. In Appendices A and  B we show how to use the  action 
principles to obtain the breaking of the Slavnov-Taylor identities. In
Appendix C we compare, for a simple gauge group, the one-loop renormalized 
chiral gauge theory obtained by means of the Breitenlohner 
and Maison scheme with the corresponding one-loop renormalized theory 
obtained with the help of an fully anticommuting $\gam5$ in 
``$d$  complex dimensions''. 
  
\medskip

\section{2. Dimensional renormalization: notation and general results}

As explained in the Introduction, we will use the formulation of dimensional
renormalization given by Breitenlohner and Maison \cite{\BMa,\BMb,\BMc}, 
because it is a systematic and consistent procedure (valid in the sense 
of Hepp \cite{\Hepp} to all orders in perturbation theory) in which tools 
like field equations,  action principles and Ward identities 
can be rigorously implemented. This applies specially to the treatment 
of $\gam5$.

We will also compare  Breitenlohner and Maison formulation with the 
``naive'' prescription which sets the object $\gam5$ to anticommute with
the object $\gamma_{\mu}$\cite{\naive}.

\smallskip

\subsection{2.1. The Breitenlohner and Maison ``$d$-dimensional covariants''}

Breitenlohner and Maison define the usual $d$-dimensional 
``Lorentz Covariants'' ($g_{\mu\nu}$, $p_\mu$, $\gamma_\mu$, {\it etc.}.) 
to be formal objects obeying the standard algebraic identities that they 
would satisfy in spaces of integral dimension \cite{\BMa} 
($d$ is a {\it complex\/} number, indices do 
not take any value and questions like of ``Lorentz invariance'' are  
meaningless !). 

Besides the ``$d$-dimensional'' metrics
$g_{\mu\nu}$, they introduce a new one, $\ghat_{\mu\nu}$, which can be 
considered as a ``$(d-4)$-dimensional covariant''. Moreover 
the $\epsilon$ tensor is  considered to be 
a ``4 dimensional covariant'' object (because of  axiom in 
eq.~\nexttonexteq\ below).

The symbols are required to obey (apart from obvious rules concerning 
contractions of indices, addition, multiplication by numbers, commutation
of some symbols, {\it etc.}):
$$
\eqalignno{
&g^{\mu\nu} p_\nu=p^\mu,\qquad \ghat^{\mu\nu} p_\nu=\hat p^\mu, \qquad
	g^{\mu\nu} \gamma_\nu=\gamma^\mu,\qquad \ghat^{\mu\nu} 
	\gamma_\nu=\gamhat^\mu,                                       \cr
&g_{\mu\nu} g^\nu{}_\rho = g_{\mu\rho},\qquad
	\ghat_{\mu\nu} g^\nu{}_\rho = \ghat_{\mu\nu} \ghat^\nu{}_\rho =
		\ghat_{\mu\rho},    &\numeq \cr \namelasteq{\EqSymbolsAxioms}
&g^\mu{}_\mu=d  ,\qquad \hbox{Tr}\,\un=4, \qquad
	\{\gamma^\mu,\gamma^\nu\}=2g^{\mu\nu} \,\un;                \cr
&{}\cr
&\epsilon_{\mu_1\ldots\mu_4} \epsilon_{\nu_1\ldots\nu_4} =
	-\sum_{\pi\in\, S_4} \hbox{sign}\, \pi \prod_{i=1}^4 \,
(g_{\mu_i\nu_{\pi(i)}}- \ghat_{\mu_i\nu_{\pi(i)}}),     &\numeq \cr
} \namelasteq{\EqTwoEpsilons}
$$
where $S_4$ denotes the permutation group of 4 objects, and $\un$ is
the unit of the symbolic algebra of gammas.

With the definitions
$$
\eqalignno{
&\gbar^{\mu\nu} = g^{\mu\nu} - \ghat^{\mu\nu},\qquad                     
 \gbar^{\mu\nu} p_\nu=\bar p^\mu,\qquad
	\gbar^{\mu\nu} \gamma_\nu=\gambar^\mu,            \cr
&{}\cr
&\gam5={i \over 4!} \epsilon_{\mu_1\ldots\mu_4} 
	\gamma^{\mu_1}\cdots\gamma^{\mu_4}         &\numeq \cr
}\namelasteq{\Gammafive}
$$
and the {\it assumption of cyclicity\/} of the symbol Tr, 
the following properties can be proved algebraically \cite{\BMa}
$$
\eqalignno{
&\ghat_{\mu\nu} \gbar^\nu{}_\rho = 0,\quad \hat p \cdot \bar q =0, \qquad
	g_{\mu\nu} \gbar^\nu{}_\rho =\gbar_{\mu\rho},\qquad   
	\ghat^\mu{}_\mu=d-4,\quad \gbar^\mu{}_\mu=4, \cr
&\{\gambar^\mu,\gambar^\nu\}=2\gbar^{\mu\nu} \,\un, \qquad 
	\{\gamhat^\mu,\gamhat^\nu\}=2\ghat^{\mu\nu}\,\un,       \cr
&\epsilon_{\mu_1\ldots\mu_4}=\hbox{sign}\,\pi\,\,
	\epsilon_{\mu_{\pi(1)}\ldots\mu_{\pi(4)}},                 \cr
&\epsilon_{\mu_1\ldots\mu_4} \ghat^{\mu_i\nu_i}=0,\qquad
 \epsilon_{\mu_1\ldots\mu_4} g^{\mu_i\nu_i}=\epsilon_{\mu_1\ldots\mu_4} 
	\gbar^{\mu_i\nu_i};                                \cr
&{}\cr
&\Tr\,\gamma^\mu = \Tr\,\gam5=0,\qquad\ 
	\gam5^2=\un,                                            \cr
&\Tr\,[\gamma^\mu\gamma^\nu\gam5]=\Tr\,[\gambar^\mu\gambar^\nu\gam5]=
	\Tr\,[\gamhat^\mu\gamhat^\nu\gam5]=\Tr\,[\gambar^\mu\gamhat^\nu\gam5]=
	0,                     &\numeq\cr     \namelasteq{\EqGammaTraces}
&\Tr\,[\gamma^{\mu_1}\cdots\gamma^{\mu_4}\gam5]=
 \Tr\,[\gambar^{\mu_1}\cdots\gambar^{\mu_4}\gam5]=
 i\, \hbox{Tr}\,\un \,\,\epsilon_{\mu_1\ldots\mu_4};               \cr 
&{}\cr
&\{\gam5,\gamma^\mu\}=\{\gam5,\gamhat^\mu\}=2\gam5 \gamhat^\mu=\,
	2\gamhat^\mu \gam5,                                        \cr
&[\gam5,\gamma^\mu]=[\gam5,\gambar^\mu]=2\gam5 \gambar^\mu=
	-2\gambar^\mu \gam5,                               &\numeq\cr
&\{\gam5,\gambar^\mu\}=[\gam5,\gamhat^\mu]=0.                         \cr
} \namelasteq{\EqGammaCommutators}
$$

$\gbar^{\mu\nu}$ can be thought of as a projector over the ``4-dimensional
space'' and $\ghat^{\mu\nu}$ as a projector over the ``$(d-4)$-dimensional''
one.

All usual formulae (not involving $\gam5$) used for computing 
traces of strings of gammas in terms of combinations of the metric 
remain valid, even when the gammas are hatted or barred: 
in these cases one has only to put hats or bars over the 
corresponding metrics. Also the trace of an
odd number of (normal, hatted or barred) $\gamma$s vanish.

Strings of gammas with contracted indices are simplified with the aid of
formulae like:
$$
\eqalignno{
\gamma^\mu \gamma^\nu \gamma_\mu=(2-d)\gamma^\nu&,          \cr
\gambar^\mu \gambar^\nu \gambar_\mu=-2\,\gambar^\nu&,\qquad
   \gambar^\mu \gamhat^\nu \gambar_\mu=-4\,\gamhat^\nu,      \cr
\gamhat^\mu \gamhat^\nu \gamhat_\mu=(6-d)\gamhat^\nu&,\qquad
	\gamhat^\mu \gambar^\nu \gamhat_\mu=(4-d)\gambar^\nu.    \cr
}
$$

Of course, when $\gam5$ of the $\epsilon$ tensor
appear the situation is very different. In the ``naive'' prescription, one 
assumes the {\it anticommutativity\/} of $\gam5$:
$\{\gamma_\mu,\gam5\}=0$, and the {\it cyclicity\/} of the trace. From these 
assumptions one obtains \cite{\BMa,\Collins,\BonneauRemarks,\BonneauReview}
$$
\eqalignno{
2d \,\Tr[\gam5]&=0,                                               \cr 
2(d-2)\,\Tr[\gamma_{\mu_1}\gamma_{\mu_2}\gam5]&=0 \qquad
		\hbox{if $d\ne0$},                           &\numeq\cr
2(d-4)\,\Tr[\gamma_{\mu_1}\ldots\gamma_{\mu_4}\gam5]&=0 \qquad
		\hbox{if $d\ne0$, 2};                             \cr
}  \namelasteq\EqNaiveInconsistency
$$ 
which has the consequence that $\Tr[\gamma_{\mu_1}\ldots\gamma_{\mu_4}\gam5]$
is identically 0 in the dimensionally regularized theory ($d\ne4$). 
So, it also vanishes in the dimensionally renormalized theory, 
which is incompatible with the true property (see eq.~\EqGammaTraces) in four 
dimensions. We could introduce the symbol $\epsilon$ with the requirement 
that it satisfies eq.~\EqGammaTraces, but then the symbol 
$\epsilon$ would be identically 0. Despite this well known fact, 
people use the ``naive'' prescription in {\it e.g.} the Standard Model, 
but they do not set $\epsilon$ equal to 0 and 
at the end of calculations the symbol $\epsilon$ is supposed to have 
its usual meaning as the Levi-Civita tensor. This is clearly a mathematical 
inconsistency (the axioms are not compatible) and when mathematical 
inconsistencies are present, results obtained from the axioms are ambiguous 
and cannot be trusted. 

 Eq.~\lasteq\ is obtained by computing
$\Tr\,[\gamma_{\mu_1}\ldots\gamma_{\mu_m}\gamma^\a\gamma_\a
\gam5]$ in two different ways: first, one contracts the index $\a$ without
moving around the $\gam5$ and, second, one anticommutes 
the  $\gamma_\a$ which is next to the $\gam5$ with this very 
$\gam5$ and puts it on the left thanks to 
the cyclicity of the trace; then, one anticommutes it to the right 
through the rest of $\gamma$s until the other $\gamma^\a$ is met,  
contraction of the index $\a$ is now carried out. So, at least the trace of 
six $\gamma$s, having two indices contracted, and a $\gam5$ are needed to 
obtain this inconsistency.

Here, the ambiguity in the results is a consequence of the choice of the
position of the $\gam5$. For example, if the $\gam5$ is not moved in any 
case,
$$
\eqalignno{
\Tr\,[\gamma^{\mu_1}\ldots\gamma^{\mu_4}\gamma^\a\gamma_\a
\gam5]&=d\,\Tr\,[\gamma^{\mu_1}\ldots\gamma^{\mu_4}\gam5],         \cr
\ne -\Tr\,[\gamma_\a\gamma^{\mu_1}\ldots\gamma^{\mu_4}\gamma^\a\gam5]
	&=(8-d)\,\Tr\,[\gamma^{\mu_1}\ldots\gamma^{\mu_4}\gam5],      
					&\numeq\cr
}  \namelasteq\EqNaiveAmbiguity
$$
unless the trace of four $\gamma$s and one $\gam5$ (or the $\epsilon$ 
tensor) is 0.

Notice that the ambiguity will be always proportional to $(d-4)$ or a
$(d-4)$ object like $\gamhat_\mu$, but there are poles in $(d-4)$ in
the divergent diagrams that
will make finite this ambiguity in the dimensionally renormalized theory, 
even in one loop calculations.

Then, why do people go on using the ``naive'' prescription in calculations of, 
{\it e.g.} Standard Model? The key is that the ambiguities are claimed to
be always proportional to the coefficient of the (chiral gauge) anomaly. So,
it would appear that one could freely use the ``naive'' prescription in 
theories with cancellation of anomalies (like the Standard Model). 
Calculations to low orders in perturbation theory 
in some models support this idea  but there is not a rigorous 
proof of it  valid to all orders in 
perturbation theory. (Also, it seems difficult 
to build a consistent theory with inconsistent elements).

We shall close this subsection with a few words regarding the construction
of Lorentz covariants due to Wilson and Collins 
\cite{\Wilson,\Collins}. It is worth mentioning it since it furnishes 
explicit expressions for the Lorentz covariants satisfying 
eqs.~\EqSymbolsAxioms-\EqTwoEpsilons, thus showing that no 
inconsistencies arise. In this construction gamma ``matrices''  
are represented as infinite dimensional objects. The ``matrix'' 
$\gam5$ being defined by eq.~\Gammafive, with 
$\epsilon_{\mu_1\ldots\mu_4}$ as defined by eq.~\EqTwoEpsilons.  
However, we will not use this construction but the algebraic approach 
of abstract symbols by Breitenlohner and Maison. 

\subsection{2.2. Minimal dimensional renormalization and the renormalized 
theory}

Dimensional renormalization {\it \'a la} Breitenlohner and Maison is carried
out by applying the pole subtraction algorithm 
\cite{\BMa,\BMb,\BMc,\BonneauA} as given by the forest formula to each
dimensionally regularized 1PI Feynman diagram. This algorithm can also be 
implemented by adding to the ``$d$-dimensional'' Lagrangian (see below) 
counterterms whose coefficients have poles at $d=4$. Both the set of
dimensionally regularized Feynman diagrams, which conforms what is referred to 
as the regularized theory, and the renormalized 1PI functional obtained from 
them are best established in the following way:

\item{\it i)} Write a classical, {\it i.e.} order zero in $\hbar$, action, 
$S_0$, in $d$ space-time dimensions ($d$ being a generic positive integer) 
which yield the four dimensional action of the theory when the formal
limit $d\rightarrow 4$ is taken. The ``$d$-dimensional'' covariants in 
$S_0$ are  defined as in the previous subsection. We shall assume as in
ref.~\cite{\BMa} that each free term, $\int d^d x\,{1\over 2}\phi D\phi$, in 
$S_0$ is such that $D^{-1}$ is the same algebraic expression as in four 
dimensions, although it is expressed in terms of ``$d$-dimensional''
 covariants.
$\phi$ denotes a collection of bosonic or fermionic fields. $i\hbar D^{-1}$
gives the free propagator. $S_0$ will be referred to as the Dimensional 
Regularization classical action. We shall call the ``$d$-dimensional'' 
space-time used above to set the Dimensional Regularization 
classical action, $S_0$, the  ``$d$-dimensional'' space-time of 
Dimensional Regularization. 

\item{\it ii)} Use the Dimensional Regularization classical action, $S_0$, 
along with standard path integrals textbook techniques, 
formally applied, to obtain the set of ``$d$-dimensional'' 
Feynman rules stemming from it. These Feynman
rules will lead to the collection of Feynman diagrams in the $d$-dimensional 
space-time of Dimensional Regularization,
$d$ being a generic positive integer, which is to be turned into
the dimensionally regularized Feynman diagrams by using the algorithms in
refs.~\cite{\BMa,\BMb,\BMc,\Collins }. This set of dimensionally 
regularized Feynman diagrams defines  the regularized theory.
 
\item{\it iii)} Introduce the Dimensional Regularization generating functional
$Z_{\rm DReg}[J;K;\lambda]$:
$$
Z_{\rm DReg}[J;K;\lambda]\,=\,\int\;{\cal D}\phi\; 
{\rm exp}\{{i\over \hbar}
(\,S_0[\phi;K;\lambda]+\int d^d x\,J_i(x)\phi_i(x)\,)\},
\eqno\numeq \namelasteq\EqZdefinition
$$
where the right hand side of eq.~\EqZdefinition\ is defined as the formal 
power  expansion in $\hbar$, $K$ and $J$ given by the Feynman diagrams 
obtained by using the ``$d$-dimensional'' Feynman rules previously established.
The symbols $K$ and $\lambda$ denote, respectively, any external field,
usually coupled linearly to composite operators, and any parameter occurring 
in the action. Let us stress that eq.~\EqZdefinition\ is merely a symbol 
which denotes the whole set of Feynman diagrams in the ``$d$-dimensional'' 
space-time of Dimensional Regularization,
$d$ being a generic positive integer, which are converted into dimensionally
regularized Feynman diagrams by analytic continuation in $d$. However, this
symbol is very useful due to the Regularized Action Principle 
\cite{\BMa,\BMb,\BMc} discussed below. Indeed, the standard formal 
manipulations of the path integral, {\it e.g} variations of fields and 
differentiation with respect to parameters, that in four dimensions lead to
equations of motion, Ward identities and so on, can be formally performed in
the path integral in eq.~\EqZdefinition ; thus leading to formal functional
equations involving $Z_{\rm DReg}$. These formal manipulations are 
mathematically well defined if expressed in terms of the dimensionally 
regularized Feynman diagrams arising from the right hand side of 
eq.~\EqZdefinition , and the formal functional equations they lead to are
also mathematically sound if they are considered as symbols denoting in a
condensed manner the equations verified by the corresponding dimensionally
regularized Feynman diagrams. These equations are the dimensionally
regularized counterparts of the four dimensional equations of motion, Ward
identities, {\it etc.} It is also useful to introduce 
the Dimensional Regularization 1PI functional, 
$\Gamma_{\rm DReg}[\phi;K;\lambda]$, which can be obtained
from eq.~\EqZdefinition\ through formal Legendre transform; a procedure well
defined in terms of regularized Feynman diagrams. Again, the Regularized
Action Principle guarantees that the formal functional equations verified
by  $\Gamma_{\rm DReg}[\phi;K;\lambda]$ make sense mathematically when
expressed   in terms of dimensionally regularized Feynman diagrams.
We shall refer to the formal equations satisfied by $Z_{\rm DReg}$ and
$\Gamma_{\rm DReg}$ as dimensionally regularized equations, keeping always
in mind the previous discussion.   

\item{\it iv)} If there are classical symmetries in the four dimensional 
classical theory that should hold in the quantum theory, 
one generalizes next the corresponding field transformations to 
the $d$-dimensional  space-time of Dimensional Regularization. 
Objects in this $d$ dimensional space-time should be defined 
according to the algebra of ``$d$-dimensional'' covariants 
given in the previous subsection. The field transformations in 
the $d$-dimensional space-time thus obtained should yield,
in the limit $d\rightarrow4$, the classical four dimensional transformations.
Generally speaking, the transformations in the $d$ dimensional space-time 
of Dimensional Regularization do not leave invariant 
the Dimensional Regularization classical action $S_0$. 
This lack of invariance will make the Dimensional Regularization
generating functionals, $Z_{\rm DReg}$ and  $\Gamma_{\rm DReg}$,  satisfy 
anomalous Ward identities, which can be derived (see Appendix A) by
performing formal manipulations of the path integral in eq.~\EqZdefinition .
Again, the Regularized Action Principle guarantees that both these formal
manipulations and the identities they lead to have a well defined
mathematical meaning when expressed in terms of dimensionally regularized
Feynman diagrams. The dimensionally regularized identities so obtained 
are of enormous help in the computation of the anomalous breaking 
terms of the dimensionally renormalized four dimensional 
theory (see next subsection), which otherwise
would have to be computed by evaluating the complete dimensionally 
renormalized 1PI functional.

\item{\it v)} The minimal subtraction algorithm of references 
\cite{\BMa,\BonneauA} is applied next to every dimensionally regularized
Feynman diagram coming   from the   1PI functional  
$\Gamma_{\rm DReg}[\phi;K;\lambda]$. The minimally renormalized
1PI functional $\Gamma_{\rm minren}[\phi;K;\lambda;\mu]$ ($\mu$ stands for
the Dimensional Regularization scale) is obtained by 
taking the limit $d\rightarrow 4$, first, and then setting every hatted 
object to zero in every subtracted 1PI diagram.
The minimal subtraction algorithm amounts to subtracting the pole at $d=4$ 
from every diagram once subdivergences have been taken care of, 
and it can be formulated in terms of singular, at $d=4$,
counterterms \cite{\BMa} added to the Dimensional Regularization
classical action $S_0$. These singular counterterms, which
can be read from the forest formula \cite{\BMa,\Collins}, are local 
polynomials of the fields and their derivatives in the $d$-dimensional,  
with $d\neq 4$, space-time of Dimensional Regularization. The 
coefficients of these polynomials are the principal part at $d=4$ 
of a certain meromorphic function of complex $d$. The singular counterterms in
question give rise to new ``$d$-dimensional'' vertices, which in turn yield
new Feynman diagrams that cancel, after dimensionally regularized, the
singular behaviour of the dimensionally regularized diagrams provided by 
$S_0$.  
\item{} According to ref.~\cite{\BonneauA} the dimensional 
regularization scale, $\mu$, is introduced by replacing every loop 
momentum measure ${{d^d p}\over {(2\pi)}^d}$ with 
$\mu^{4-d}{{d^d p}\over {(2\pi)}^d}$ before applying the subtraction 
algorithm. If one follows the procedure of singular counterterms, 
the previous replacement should be made regardless the
diagram involves singular counterterms. The introduction of the scale $\mu$ 
will render dimensionally homogeneous the Laurent expansion around 
$d=4$ of a given  dimensionally regularized diagram.
\item{} Unfortunately, if there is a symmetry we wanted preserved at the
quantum level, the renormalized functional 
$\Gamma_{\rm minren}[\phi;K;\lambda;\mu]$ would not do, since it would not 
define, in general, a quantum theory having such a symmetry: the regularization
and also the subtraction process may break the symmetry. However, the 
situation is not hopeless. Algebraic BRS renormalization along with the
Quantum Action Principle \cite{\PiguetSorella}  comes in our aid: if the
anomaly cancellation conditions \cite{\Alwit} are met, the anomalous
breaking terms, called spurious, can be canceled at each order in $\hbar$
by adding appropriate counterterms to the four dimensional classical 
action of the theory.

\noindent
Suppose that $n$ is the order in the $\hbar$ expansion at which a non-anomalous
symmetry is broken for the first time in the dimensionally renormalized theory.
Then,

\item{\it vi)} Compute the breaking, which will be a local four dimensional
functional, with the help of the action principles and the Bonneau
identities \cite{\BonneauA,\BonneauB,\BonneauC}. This can be easily done
(see next subsection) by taking as action, not the classical one, $S_0$, 
but a new action $S^{(n)}=S_0 + S^{(n)}_{\rm sct}$, where 
$S^{(n)}_{\rm sct}$ denotes the set of singular counterterms 
needed to minimally renormalize the theory at order $\hbar^{n}$, 
and then redo steps {\it i)} to {\it v)}. Indeed, if $d\neq 4$, 
the Regularized Action Principle still holds
for the Dimensional Regularization generating functionals,
$Z_{\rm DReg}[J;K;\lambda]$ and $\Gamma_{\rm DReg}[\phi;K;\lambda]$, 
defined for $S^{(n)}$, and, hence, we can take advantage of the dimensionally
regularized equations, in the sense explained in {\it ii)}, to obtain the
renormalized equations verified by $\Gamma_{\rm minren}[\phi;K;\lambda;\mu]$.
\item{} Next, extract from the breaking the four dimensional finite 
counterterms needed to restore the symmetry, generalize them to 
$d$ space-time dimensions with the help of the algebra of $d$-dimensional
covariants (we shall denote this generalization by $S^{(n)}_{\rm fct}$), and
add them to the action $S^{(n)}$ to obtain yet another new action 
$S^{(n)}_{\rm DReg}=S_0+S^{(n)}_{\rm sct}+S^{(n)}_{\rm fct}$. 
$S^{(n)}_{\rm DReg}$ will be called the Dimensional Regularization action.
Furnished with this new action $S^{(n)}_{\rm DReg}$, 
establish then  new perturbation theory: redo steps {\it i)} to {\it v)} 
upon replacing $S_0$ with $S^{(n)}_{\rm DReg}$. 
The new renormalized 1PI functional, $\Gamma_{\rm ren}[\phi;K;\lambda;\mu]$, 
will satisfy up to order $\hbar^n$ the Ward identities that 
govern the given symmetry at the quantum level. Move on to {\it vi)} 
and compute the breaking at order $\hbar^{n+1}$, 
and so on and so forth.

\item{\it vii)} Use the Bonneau \cite{\BonneauA,\BonneauB}
identities to obtain the renormalization group equation 
at every order in the perturbative expansion for the theory being analyzed. 
Thus the beta functions and anomalous dimensions of the theory 
will be evaluated.

\noindent
Let us close this subsection by making some remarks on the generalization
to $d$ space-time dimensions, {\it i.e.} to the $d$-dimensional space-time of
Dimensional Regularization, of the four dimensional classical action, the 
four dimensional counterterms needed to restore non-anomalous symmetries and
the four dimensional symmetry transformations. If the classical action of the
theory involves objects whose properties depend on the dimension of space-time
 ({\it e.g.} $\gamma_5$, the Levi-Civita symbol,...) there is no canonical
Dimensional Regularization classical action, $S_0$, 
in $d$ space-time dimensions. Any local functional which 
formally go to the four dimensional classical action 
as $d\rightarrow 4$ would do, provided the free terms of the 
former lead to the propagators described in {\it i)}. 
For instance, let $\kappa_1$ and $\kappa_2$ be a couple of
positive real numbers, then, the four dimensional metric and gamma matrices
in the interaction part of the four dimensional classical action can be
replaced with $\gbar_{\mu\nu}+\kappa_1 \ghat_{\mu\nu}$ and 
$\gambar_{\mu}+\kappa_2 \gamhat_{\mu}$, respectively, to obtain an 
admissible biparametric family of Dimensional Regularization classical actions.
Each member of this family yields a particular  regularization of the theory, 
this situation is somewhat reminiscent of the lattice regularization method.
Generally speaking the difference between two admissible ``$d$-dimensional''
lagrangians will always be a local ``evanescent'' operator\dag~vanishing 
\vfootnote\dag{ An ``evanescent'' operator is an operator whose tree level
contribution vanishes as $d\rightarrow 4.$}
as the coupling constants go to zero. Analogously, in 
the type of theories under scrutiny, namely, chiral gauge theories , 
there is no canonical generalization to $d$-dimensional 
space-time of Dimensional Regularization of the four 
dimensional finite counterterms needed to restore a 
non-anomalous broken symmetry. Again, two such 
generalizations of a given four dimensional finite counterterm which agree
up to order $\hbar^{n-1}$ will differ in a ``$d$-dimensional'' integrated 
``evanescent'' operator of order $\hbar^n$. However, the dependence 
on the choice of generalization will show in the renormalized theory at order 
$\hbar^{n+1}$, never at order $\hbar^{n}$. Last, but not least, from the
point of view of renormalization, there is no canonical generalization to
$d$ space-time dimensions of the symmetry transformations of the fields for
theories which are not vector-like. Again, two such generalizations differ in
an ``evanescent'' operator. However, this arbitrariness in the choice
of Dimensional Regularization classical  action, ``$d$-dimensional'' finite
counterterms and symmetry transformations is useful since one can play around
with it so as to simplify the form of the symmetry breaking contributions.
Finally, for vector-like theories such as QCD, there is, of course, a
canonical choice of Dimensional Regularization classical action and vector-like
transformations: the usual one \cite{\Collins}.

\subsection{2.3. Regularized and Quantum Action Principles }
As we said in the previous subsection, one of the main characteristics of the 
dimensional renormalization procedure is that action principles 
are precisely stated and proved \cite{\BMa-\BMc}.
These principles are most efficiently expressed in terms of the 
Dimensional Regularization  generating 
functional, $Z_{\rm DReg}[J;K;\lambda]$, 
for the Greens function given by the Gell-Mann-Low series and the dimensional
renormalization 1PI functional, $\Gamma_{\rm ren}[\phi;K;\lambda;\mu]$,  
obtained from it (see previous subsection).

Let us take as Dimensional Regularization action (see {\it vi)} in the previous
subsection) 
$$
S^{(n)}_{\rm DReg}=
S_{\rm free}[\phi;\lambda]+S_{\rm int} [\phi;K;\lambda],
\eqno\numeq \namelasteq\Actiondef
$$
where $\phi$ denotes a collection of commuting or anticommuting quantum
fields,  $K$ set of commuting or anti-commuting external fields, 
$\lambda $ is a generic  parameter, and
$$
\eqalign{&S_{\rm free}[\phi;\lambda]=S_{0\rm free}[\phi;\lambda],\cr
&S_{\rm int} [\phi;K;\lambda]=S_{0 \rm int}[\phi;K;\lambda]+
S^{(n)}_{\rm sct}[\phi;K;\lambda]+S^{(n)}_{\rm fct}[\phi;K;\lambda].\cr}
$$

We next introduce
$$S_{\rm INT}[\phi,J,K,\lambda]=
	S_{\rm int}[\phi,K,\lambda]+ \intd^d x\, J_i(x) \phi_i(x)      
$$
and define, following eq.~\EqZdefinition , the functional $Z_{\rm DReg}$
$$
Z_{\rm DReg}[J;K;\lambda]=\int{\cal D}\phi\;
{\rm exp}\{{i\over \hbar}
(S_{\rm DReg}^{(n)}+\int  J_i\phi_i)\}\equiv
\bigl\langle \exp\Bigl\{ {i\over\hbar}
	S_{\rm INT}[\phi;J;K;\lambda]\Bigr\}
		\bigr\rangle_0.    \eqno\numeq \namelasteq\EqZDefinition
$$
The symbol exp in the previous equation stands for its formal power series and
the symbol $\langle \cdots \rangle_0$ denotes the usual vacuum expectation
value 
$$
\langle \cdots \rangle_0 = \int\,{\cal D}\phi\; \cdots\;
{\rm exp}\{{i\over \hbar}S_{0\rm free}[\phi;\lambda]\},
$$
defined by gaussian integration, which gives a formal power series in $\hbar$
and the external fields $J$ and $K$. Each coefficient of this series is a
sum of ``$d$-dimensional'' Feynman diagrams. Every ``$d$-dimensional'' Feynman 
diagram can be converted into a meromorphic function of $d$ by 
promoting $d$ to a complex variable and  understanding the ``$d$-dimensional''
covariants as in subsection {\it 2.1.}

Now, the {\it Regularized Action Principle} states that the following
three functional equations hold in the dimensionally regularized theory:

\item{}1) Arbitrary polynomial variations of the quantized fields $\phi$,
$\delta\phi(x)=\delta\theta(x)\,P(\phi(x))$ leave $Z_{\rm DReg}$ 
invariant and
$$
\delta Z_{\rm DReg}[J,K]\equiv
\bigl\langle \delta(S_{\rm free}+S_{\rm INT}) \exp\Bigl\{ {i\over\hbar}
	S_{\rm INT}[\phi,J,K,\lambda]\Bigr\}
		\bigr\rangle_0 =0,            
           \eqno\numeq\namelasteq\Varfield
$$
where the variations are the linear parts in $\delta\theta$.

\item{}2) variations of external fields $E(x)\equiv (K(x), J(x))$ give
$$
\bigl\langle {\delta S_{\rm INT}\over \delta E(x)} \exp\Bigl\{ {i\over\hbar}
	S_{\rm INT}[\phi,J,K,\lambda]\Bigr\}
		\bigr\rangle_0 =
 -i\hbar{\delta Z_{\rm DReg}[J,K,\lambda] \over \delta E(x)}. 
\eqno\numeq\namelasteq\Varexternal
$$

\item{}3) variations of parameters give
$$
\bigl\langle {\pr(S_{\rm free} + S_{\rm INT})\over\pr\lambda} 
	\exp\Bigl\{ {i\over\hbar}
	S_{\rm INT}[\phi,J,K,\lambda]\Bigr\}
		\bigr\rangle_0 =
		-i\hbar{\pr Z_{\rm DReg}[J,K,\lambda] \over\pr\lambda}. 
   \eqno\numeq\namelasteq\Varparameter
$$

 Let us explain in what sense are the functional equations 
\Varfield , \Varexternal\ and  \Varparameter\ mathematically meaningful. First,
these equations are to be understood diagramatically only. The formal
expansion in powers of $\hbar$, $K$ and $J$ of both sides of each equation
leads to an infinite set of equations involving only ``$d$-dimensional''
Feynman diagrams. It is this infinite set of diagramatic equations in the 
$d$-dimensional space-time of Dimensional Regularization what is taken 
as the definition of the corresponding
functional equation. The {\it Regularized Action Principle}  
\cite{\BMa,\BonneauA} states that this very set of diagramatic 
equations still holds if every formal 
``$d$-dimensional'' Feynman diagram in them is replaced with its dimensionally
regularized counterpart, with the scale $\mu$ introduced as explained in 
{\it 2.2.v)}.\dag
\vfootnote\dag{ The scale $\mu$ is introduced so as to 
render dimensionally homogeneous the Laurent expansions around $d=4$. } 
A mathematically sound set of equations is thus obtained.
Of course, it is easier to handle the functional equations above than their
dimensionally regularized diagramatic definition. Furthermore, this 
functional equations  can be deduced by formal path integral manipulations,
which in turn can be given a diagramatic definition valid within the
Dimensional Regularization method {\it \'a la } Breitenlohner and Maison. 
Let us finally note that  the limit $d\rightarrow 4$ of dimensionally 
regularized the Green functions coming from the right hand side 
of eqs.~\Varexternal\ and \Varparameter\ does exist 
at each order in the $\hbar$ expansion up to order $n$. Notice that  we have
included in the action $S_{\rm DReg}^{(n)}$, defined in 
eq.~\Actiondef , the appropriate singular
counterterms up to order $\hbar^n$. Hence, eqs.~\Varexternal\ and 
\Varparameter\ have a well-defined $d\rightarrow 4$ limit up to order
$\hbar^n$ and can be used to compute the variations with $E(x)$ and $\lambda$
of the renormalized functional $Z_{\rm ren}[J;K;\lambda]$ up to order 
$\hbar^n$. 
 
The {\it Regularized Action Principle} given  by eqs.~\Varfield ,
\Varexternal\ and \Varparameter\ leads 
\cite{\BMa,\BonneauA,\BonneauB,\Collins } to the 
{\it Quantum Action Principle} (see ref.~\cite{\PiguetRouet} and 
references therein) which states that the
following functional equations hold for the dimensionally renormalized theory:

\item{} QAP1)
$$
{\pr \Gamma_{\rm ren}\over \pr \lambda} = 
{\rm N}[{\pr (S_0+S^{(n)}_{\rm fct})\over \pr\lambda}]\cdot\Gamma_{\rm ren}.
\eqno\numeq\namelasteq\QAPone
$$ 

\item{} QAP2)
$$
{\delta \Gamma_{\rm ren}\over \delta E(x)} = 
{\rm N}[{\delta (S_0+S^{(n)}_{\rm fct})
\over \delta E(x)}]\cdot\Gamma_{\rm ren}.
\eqno\numeq\namelasteq\QAPtwo
$$
 
\item{} QAP3) Let $\phi'(x)= P(\phi(x))$ denote a linear transformation of 
$\phi(x)$ whose coefficients do not depend explicitly on $d$, then
$$
\phi'(x){\delta \Gamma_{\rm ren}\over \delta \phi(x)} = 
{\rm N}[\phi'(x){\delta (S_0+S^{(n)}_{\rm fct})\over \delta\phi(x)}]
\cdot\Gamma_{\rm ren}.
\eqno\numeq\namelasteq\QAPthree
$$

\item{} QAP4) If $\delta \phi(x)$ is a non-linear field
transformation in the four dimensional space-time and $K(x)$ is the
external field coupled to it, then
$$
{\delta \Gamma_{\rm ren}\over \delta K(x)}
{\delta \Gamma_{\rm ren}\over \delta \phi(x)} = 
{\rm N}[{\cal O}(x)]\cdot\Gamma_{\rm ren},
\eqno\numeq\namelasteq\QAPfour 
$$ 
where the symbols ${\rm N}[\cdots]$
and ${\rm N}[\cdots]\cdot\Gamma_{\rm ren}$
denote normal product as defined in
refs.~\cite{\BonneauA,\BonneauB,\Collins}
and its insertion in the renormalized
1PI functional, respectively.  ${\rm N}[{\cal O}(x)]$ is a local normal
product of  ultraviolet dimension $4-{\rm dim}(\phi)+{\rm dim }(\delta \phi)$. 
$S_{0}$ and $S_{\rm fct}^{(n)}$ have been defined 
in {\it i)} and {\it v)} of  subsection {\it 2.2.} 

Eqs.~\QAPone -\QAPfour have been shown to hold by using the forest 
formula \cite{\BMa,\BMb,\BonneauA,\BonneauB}  and the singular 
counterterms algorithm \cite{\Collins}, respectively. Of course, both proofs
are equivalent.   

The {\it Quantum Action Principle } is of the utmost importance to the 
renormalization of BRS symmetries \cite{\PiguetSorella}, for it guarantees
(see QAP1 and QAP2) that the breaking of such symmetry is given by the 
insertion into the dimensionally renormalized 1PI $\Gamma_{\rm ren}$ 
of a certain integrated normal operator, 
${\rm N}[{\cal O}]=\int d^4 x\,{\rm N}[{\cal O} (x)]$, where
${\rm N}[{\cal O} (x)]$ is a local normal operator of ultraviolet dimension 
$4-{\rm dim}(\Phi)+{\rm dim}(s\Phi)$ and
ghost number 1. Indeed, the following equation holds for 
$\Gamma_{\rm ren}[\varphi,\Phi;K_\Phi]$
$$
\SS(\Gamma_{\rm ren})\equiv\intd^4 x\,\,
(s\varphi){\delta\Gamma_{\rm ren}\over\delta\varphi} +
{\delta\Gamma_{\rm ren}\over\delta K_\Phi} 
{\delta\Gamma_{\rm ren}\over\delta\Phi}=\Delta_{\rm breaking}.
\eqno\numeq\namelasteq\EqRenSTBreaking
$$
with $\Delta_{\rm breaking}={\rm N}[{\cal O}]\cdot\Gamma_{\rm ren}$.
In the previous equation $\varphi$ and $\Phi$ stand for fields which undergo
linear and non-linear BRS transformations, respectively. As it is customary
$s\varphi$ denotes a BRS transformation. $K_\Phi$ stands for the  external
field coupled to the non-linear BRS transformation $s\Phi$ in the 
four-dimensional classical action \cite{\PiguetSorella}.

Now, if finite counterterms have been added to the four-dimensional classical
action so that the BRS symmetry is preserved up to order $\hbar^{n-1}$ 
({\it i.e.} the first non-vanishing contribution to the right hand side 
of eq.~\EqRenSTBreaking\ is order $\hbar^n$), eq.~\EqRenSTBreaking\ reads 
$$
 \SS(\Gamma_{\rm ren})={\rm N}[{\cal O}]^{(n)}+ {\rm O}(\hbar^{n+1}).
$$
Where ${\rm N}[{\cal O}]^{(n)}$ is the contribution to ${\rm N}[{\cal O}]$ of 
order $\hbar^n$. We have taken into account that 
${\rm N}[{\cal O}]\cdot\Gamma_{\rm ren}= {\rm N}[{\cal O}]^{(n)}+ 
{\rm O}(\hbar\,{\rm N}[{\cal O}]^{(n)})$, since by assumption 
${\rm N}[{\cal O}]\cdot\Gamma_{\rm ren}={\rm O}(\hbar^n)$.

Next, if ${\rm N}[{\cal O}]^{(n)}= -\hbar^n\, b\, S_{{\rm fct}, n}$ 
({\it i.e.} ${\rm N}[{\cal O}]^{(n)}$ is $b$-exact) for some four-dimensional 
integrated local functional of the fields, the BRS 
symmetry can be restored, up to order $\hbar^{n}$, by adding 
to the four-dimensional classical action the finite counterterm 
$S_{{\rm fct}, n}$ \cite{\PiguetSorella}. Here $b$ denotes the linearized
BRS operator:
$$b=s+\intd^4 x\,\Bigl\{
{\delta {\Gamma^{(0)}_{\rm ren}}\over\delta\Phi(x)} 
{\delta\over\delta K_\Phi(x)}\Bigr\},
$$
and $\Gamma^{(0)}_{\rm ren}$ is the order
$\hbar^0$ contribution to $\Gamma_{\rm ren}$,
{\it i.e.} the BRS invariant classical four dimensional action.
Then, we set $S_{\rm fct}^{(n)}$ $=$ $\sum_{m=1}^n$
$\hbar^m S_{{\rm fct}, m}$.
Of course, if  ${\rm N}[{\cal O}]^{(n)}$ is not $b$-exact the 
theory is anomalous at order $\hbar^{n}$. 

The computation of the right hand side of eq.~\EqRenSTBreaking ,
$\Delta_{\rm breaking}$, is the issue we will discuss now. 
One can always compute    the complete 1PI functional 
$\Gamma_{\rm ren}$ at each order in $\hbar$, insert it in the
left hand side of equation \EqRenSTBreaking , work out the functional 
derivatives and thus obtain the right hand side. This method is, of course,
very impractical. The {\it Regularized Action principle}, as given in 
eqs.~\Varfield\ and \Varexternal , provides us with
a more efficient way to compute the symmetry breaking term of the BRS symmetry.
Indeed, in Appendix A we show that the following functional equation
hold, in the same sense as eqs.~\Varfield\ and \Varexternal , for the
1PI formal functional $\Gamma_{\rm DReg}$
$$
\eqalignno{
\SS_{d}(\Gamma_{\rm DReg})\equiv&\intd^d x\,\,
(s_d\varphi){\delta\Gamma_{\rm DReg}\over\delta\varphi} +
{\delta\Gamma_{\rm DReg}\over\delta K_\Phi} 
{\delta\Gamma_{\rm DReg}\over\delta\Phi}=     \cr
=&\,\Delta\cdot\Gamma_{\rm DReg} + 
\Delta_{\rm ct}\cdot\Gamma_{\rm DReg} +
\intd^dx\,\bigl[{\delta S_{\rm ct}^{(n)}\over \delta K_\Phi(x)}
 \cdot\Gamma_{\rm DReg}\bigr]
      {\delta\Gamma_{\rm DReg}\over\delta\Phi(x)},    &\numeq\cr
} \namelasteq\EqRegSTBreakingct
$$
were $\Gamma_{\rm DReg}$ is the 1PI functional computed with the 
Dimensional Regularization action, $S^{(n)}_{\rm DReg}$, 
defined at the beginning of the current subsection. 
The symbol $s_d$ denotes the generalization to 
$d$ dimensions of the four-dimensional BRS transformations 
(see {\it 2.2.iv)}). The operators $\Delta$ and $\Delta_{\rm ct}$ are
given by $\Delta =s_d S_0$ and $\Delta_{\rm ct}=s_d S_{\rm ct}^{(n)}$, 
respectively. Here, $S_{0}=S_{0free}+S_{0int}$ and
$S_{\rm ct}^{(n)}=S_{\rm fct}^{(n)}+S_{\rm sct}^{(n)}$ (see eq.~\Actiondef\
and the paragraph below it for definitions).

Let us recall that $\Gamma_{\rm DReg}$ is  defined as a formal series 
expansion in powers of $\hbar$ and the fields $\varphi$, $\Phi$ and $K_{\phi}$.
The coefficients of this power series are 1PI Green functions of the
theory at a given order in $\hbar$. Taking into account that 
$\Gamma_{\rm DReg}$ contains the contributions coming from the singular 
counterterms up to order $\hbar^{n}$, denoted by $S_{\rm sct}^{(n)}$, 
one concludes that the dimensionally regularized 
1PI Green functions obtained from $\Gamma_{\rm DReg}$ 
are finite in the limit $d\rightarrow 4$ up to order
$\hbar^n$. These 1PI Green functions are the dimensionally renormalized
1PI Green functions upon introduction of the Dimensional Regularization scale 
$\mu$ (see subsection {\it 2.2.}). Hence, up to order $\hbar^n$, one may  
formally write
$$
{\rm LIM}_{d\rightarrow 4}\;\Gamma_{\rm DReg}[\varphi,\Phi;K_{\Phi};\mu]=
\Gamma_{\rm ren}[\varphi,\Phi;K_{\Phi};\mu]
\eqno\numeq\namelasteq\Limit
$$
In the previous equation,  ${\rm LIM}_{d\rightarrow 4}$ is defined by
the following process: one  takes first the limit $d\rightarrow 4$ of 
the dimensionally regularized 1PI Green functions and then sets to zero 
every hatted object ({\it i.e.} $(d-4)$-dimensional covariant). 
Recall that the Dimensional Regularization scale, $\mu$, is introduced 
by hand to render dimensionally homogeneous the regularized Green 
functions upon Laurent expansion around $d=4$ \cite{\BonneauA}.

Now, by taking ${\rm LIM}_{d\rightarrow 4}$ of the left hand side of  
eq.~\EqRegSTBreakingct\ after discarding any contribution of order 
greater than $\hbar^n$, and bearing in mind eq.~\Limit , one obtains
the left hand side of eq.~\EqRenSTBreaking\ up to order $\hbar^n$.  
Hence, $\Delta_{\rm breaking}$ in the latter equation is given by 
$$
\Delta_{\rm breaking}=
{\rm LIM}_{d\rightarrow 4}\Big\lbrace
\Delta\cdot\Gamma_{\rm DReg} + 
\Delta_{\rm ct}\cdot\Gamma_{\rm DReg} +
\intd^dx\,\bigl[{\delta S_{\rm ct}^{(n)}\over \delta K_\Phi(x)}
 \cdot\Gamma_{\rm DReg}\bigr]
		{\delta\Gamma_{\rm DReg}\over\delta\Phi(x)}\Big\rbrace.
\eqno\numeq\namelasteq\Breaking
$$
Of course, the previous equation only makes sense up to order $\hbar^n$.
The order $\hbar$ contribution to eq.~\Breaking\ reads particularly simple 
since $s_d S_0$ is an evanescent operator (there is no $O(\hbar^0)$ 
contribution to ${\rm N}[\Delta]$ thereby)  and the lowest order contribution
to $S_{\rm ct}$ is order $\hbar$:
$$
\Delta_{\rm breaking}^{(1)}=
{\rm LIM}_{d\rightarrow 4}\Big\lbrace
\Bigl[\Delta\cdot\Gamma_{\rm DReg}\Bigr]^{(1)}_{\rm singular} +
b_d\,S^{(1)}_{\rm sct}\Big\rbrace 
+\Bigl[{\rm N}[\Delta]\cdot\Gamma_{\rm ren}\Bigr]^{(1)}+b\,S_{\rm fct}^{(1)}, 
\eqno\numeq\namelasteq\Oneloopbreaking
$$
where the superscript $(1)$ denotes contributions of order $\hbar$ and $\Delta$
is equal to the evanescent operator $s_d S_0$. In eq.~\Oneloopbreaking, 
$b_d$ stands for  the linearized  BRS operator in $d$ dimensions:
$$
b_d=s_d+\intd^d x\,\Bigl\{
	{\delta S_0\over\delta\Phi(x)} {\delta\over\delta K_\Phi(x)}\Bigr\}.
$$
The symbol $\Bigl[\Delta\cdot\Gamma_{\rm DReg}\Bigr]^{(1)}_{\rm singular}$ 
denotes the expression, singular at $d=4$, that it is subtracted from 
$\Delta\cdot\Gamma_{\rm DReg}$ to obtain 
$\Bigl[{\rm N}[\Delta]\cdot\Gamma_{\rm ren}\Bigr]^{(1)}$ according to 
the forest formula \cite{\BMa,\BonneauA}. Hence,
$$
\Bigl[\Delta\cdot\Gamma_{\rm DReg}\Bigr]^{(1)}=
\Bigl[\Delta\cdot\Gamma_{\rm DReg}\Bigr]^{(1)}_{\rm singular}+ 
\Bigl[{\rm N}[\Delta]\cdot\Gamma_{\rm ren}\Bigr]^{(1)}+
{\rm vanishing\;\; terms}.
\eqno\numeq\namelasteq\SingInsert
$$
By ``vanishing terms'' we mean contributions that vanish as $d$ is send to
$4$ first and then every hatted object is set to zero.
 
There are some comments regarding eq.~\Oneloopbreaking\ which we would 
like to make. First, if the theory is not anomalous, the finite 
counterterms in $S_{\rm fct}^{(1)}$ can be chosen so 
that $\Delta_{\rm breaking}^{(1)}=0$ \cite{\PiguetSorella}.
Second, there are two possible sources of symmetry breaking
in eq.~\Oneloopbreaking . One is the
regularization method; in particular, the Dimensional Regularization classical
action, $S_0$, one began with. The other is the minimal subtraction algorithm
({\it i.e.} the renormalization algorithm) we are employing. That the 
regularization method breaks the symmetry shows itself in the contributions
$\bigl[\Delta\cdot\Gamma_{\rm DReg}\bigr]^{(1)}_{\rm singular}$ and
$\bigl[{\rm N}[\Delta]\cdot\Gamma_{\rm ren}\bigr]^{(1)}$. The subtraction
algorithm gives rise to the contribution $b_d\,S^{(1)}_{\rm sct}$, which 
need not be zero as we shall show later on. Third, the ``limit'' 
$$
{\rm LIM}_{d\rightarrow 4}\Big\lbrace
\Bigl[\Delta\cdot\Gamma_{\rm DReg}\Bigr]^{(1)}_{\rm singular}+
b_d\,S^{(1)}_{\rm sct}\Big\rbrace
$$
need not be zero either, as we shall  discuss in section {\it 3}. Of course,
this ``limit'' yields a finite local four-dimensional functional of the fields
and their derivatives. Fourth, the operator $\Delta=s_d S_0$ being evanescent
ushers in the techniques introduced by Bonneau \cite {\BonneauA,\BonneauB} to 
express normal products of evanescent operators, also called {\it anomalous}
normal products, in terms of a basis of standard 
(non-evenescent operators) normal products. We shall
address this subject in the next subsection. Suffice it to say that expansion
of the normal product of an evanescent operator in terms of ``standard''
normal products has coefficients that are series expansions in $\hbar$ with
no ${\rm O}(\hbar)$ term. Hence, there is no breaking of the BRS symmetry
at order $\hbar^0$ and 
$\bigl[{\rm N}[\Delta]\cdot\Gamma_{\rm ren}\bigr]^{(1)}$ is, in agreement
 with the {\it Quantum Action Principle}, a sum of standard local operators.

We shall finally recall that the equation of motion holds in dimensional
renormalization \cite{\BMa,\BonneauA,\Collins}:
$$
{\delta \Gamma_{\rm ren}\over \delta \phi(x)} = 
{\rm N}[{\delta (S_0+S^{(n)}_{\rm fct})\over \delta\phi(x)}]
\cdot\Gamma_{\rm ren}={\rm LIM}_{d\rightarrow 4}\Big\lbrace
{\delta S^{(\infty, n)}_{\rm DReg}\over\delta\phi(x)}\cdot
 \Gamma_{\rm DReg}\Big\rbrace,
\eqno\numeq\namelasteq\EOM
$$
where $S^{(n)}_{\rm fct}$ and $S^{(\infty,n)}_{\rm DReg}$ are defined
in subsection {\it 2.2.} and eq.~\Actiondef . The superscript $\infty$
tell us that the  singular counterterms needed to minimally renormalize
the theory at any order in $\hbar$ have been included; however, one
has added finite counterterms only up to order $\hbar^n$. Eq.~\EOM\ 
can obtained from QAP3 (eq.~\QAPthree  ) by setting $\phi'(x)$ to $1$. 
Alternatively, eq.~\Varfield, for $P(\phi(x))=1$, leads to eq.~\EOM . 

The use of both the equation of motion of the ghost field and the equation of
motion of the auxiliary field coupled to  the gauge fixing condition  
simplifies the renormalization of theories with BRS symmetries.

\subsection{2.4. Expression of anomalous terms with the aid of Bonneau 
	identities}

The computation of the insertions of normal products of evanescent operators,
also called anomalous normal products, are, in dimensional renormalization, 
of key importance to the  calculation of the symmetry breaking terms in  
Ward identities \cite{\BonneauA,\Collins}. We shall see in the next section
that  $\big[{\rm N}[\Delta]\cdot\Gamma_{\rm ren}\bigr]^{(1)}$ 
(see eq.~\Oneloopbreaking ) yields both
the essential and the spurious BRS anomalies for the regularized theory 
we will consider. The purpose of the current subsection is to summarily
recall that the normal product of an evanescent operator 
can be expressed as linear combinations of normal products of 
standard (non-evanescent) operators. Further details can be found in 
refs.~\cite{\BonneauA,\BonneauB,\BonneauC}.  

In renormalized perturbation theory, any set of insertions of quantum 
operators such that their classical approximations form a basis in the linear 
space of classical operators of dimensions bounded by $D$ is also a basis 
in the linear space of insertions, with dimension bounded by the same $D$, of 
quantum operators \cite{\PiguetSorella}. So, in dimensional
renormalization, if $\{\Op^i\}$ form such a basis in the  space of classical
operators, then $\{\N[\Op^i]\cdot\Gamma_{\rm ren}\}$, where $\Op^i$ is any of 
the possible generalizations to $d$ dimensions  of the corresponding classical 
four-dimensional operator, is a basis in the space of quantum insertions. 
We will call
standard  operators the operators in the ``$d$-dimensional'' space-time of
Dimensional Regularization which are generalizations of the
 classical operators in four dimensions. These standard operators are obtained
with the help of the algebra ``$d$-dimensional'' covariants and they are
non-evanescent. Hence, $\N[\Op^i]\cdot\GR$ is an insertion
of an standard normal product.

It is thus expected that  anomalous normal products, {\it i.e.} normal 
products of non-evanescent operators minimally subtracted should be 
decomposable in terms of some basis of standard normal products, 
minimally subtracted  again. This was shown by Bonneau \cite{\BonneauC} who, 
remarkably, proved a linear system of Zimmerman-like identities which 
expresses a decomposition of any anomalous normal product on all 
possible normal products of monomials of operators, 
including standard and anomalous ones.

If there is only a scalar field, the Bonneau identities have the form
$$
\eqalignno{
&\N[\ghat_{\mu\rho}\Op^{\mu\rho}](x)\cdot\GR = -
	\sum_n^4 \;\; \sum_r^{4-n} 
\sum_{\scriptscriptstyle\{i_1\,\ldots\, i_r\} \atop
		\scriptscriptstyle 1\le i_j\le n  }                   \cr
&\qquad\quad\left\{\rsp {(-i)^r\over r!} 
	{\pr^r\over\pr p_{i_1}^{\mu_1}\cdots\pr p_{i_r}^{\mu_r} }
	\overline{\bigl\langle\,
		\tilde\phi(p_1)\ldots\tilde\phi(p_n)\;
		\N[\gtah_{\mu\rho}\Op^{\mu\rho}]\, \bigr\rangle}{}^\irr 
	\Big|_{p_i\equiv\gtah\equiv0}  \right\}                        \cr
&\qquad\quad\times\;\; \N\Bigl[ 1/n! \prod_{k=1}^n 
   \Bigl\{\Bigl(\prod_{\{\a/i_\a=k\}}\!
     \pr_{\mu_\a}\Bigr)\phi\Bigr\}\Bigr] (x)
		\cdot\GR,                                      &\numeq\cr
}  \namelasteq\EqBonneauIdentities
$$
where the tilde indicates Fourier transformed fields, the bar means
that only the minimal subtraction of the subdivergences has been done,
r.s.p.~stands for ``residue of the simple pole in $\eps=4-d$'' (this is the 
reason why the global sign $-1$ occurs: $\ghat^\mu{}_\mu=-\eps\,$) 
and the $\times$ symbol means that the tensorial structure 
$\{\ldots\}$ really appears {\it inside} the normal product. Of course, 
any colour or pure number factor can be taken out of the normal product, 
but note that 
$g_{\mu\nu} \N[\Op^{\mu\nu}]=\N[\gbar_{\mu\nu}\Op^{\mu\nu}]$
$\ne\N[g_{\mu\nu} \Op^{\mu\nu}]$, the difference being the anomalous normal 
product $\N[\ghat_{\mu\nu}\Op^{\mu\nu}]$, which need not vanish.

Notice that the sum in $n$ is a sum in the number of fields in the 
monomial and the sum in $r$, a sum in the number of derivatives 
in the mononial.

The tensor $\gtah$ is a new one, which has been introduced in order to
simplify the calculations. Its properties are:
$$
\eqalignno{
&\gtah_{\mu\nu}=\gtah_{\mu\nu},\qquad \gtah_{\mu}{}^{\mu}=1,             \cr
&\gtah_{\mu\rho} g^\rho{}_\nu=
      \gtah_{\mu\rho} \ghat^\rho{}_\nu=\gtah_{\mu\nu}   \cr
&\gtah_{\mu\rho} \gbar^{\rho\mu}=0,
\qquad \N[\gtah_{\mu\nu}\Op^{\mu\nu}]=\gtah_{\mu\nu} \N[\Op^{\mu\nu}]. \cr
}
$$

 Eq.~\EqBonneauIdentities\ is easily generalized when there are 
several fields, not  necessarily scalar, involved.  One just sums over 
all kind of fields, taking special care of the symmetry factors and 
the fermionic signs. We shall give below the generalization of 
eq.~\EqBonneauIdentities\ that is relevant to our computations in 
subsection {\it 3.5}. 

Due to the fact that all subdivergences have been previously subtracted,
the barred 1PI function of eq.~\EqBonneauIdentities\ has a polynomial 
singular part. We are interested in the coefficients of the simple pole. 
These coefficients will be in general combinations of metrics, constants 
and colour factors. In vector-like gauge theories with the usual 
regularizations, its tensorial structure will involve only usual 
metrics, which combined with the tensor indices of the normal product 
will get the expansions of the anomalous normal product in terms of a 
basis of standard monomial normal products.
However, if in the regularized theory or in the calculations 
some hatted objects appear, then the Bonneau identities 
(eq.~\EqBonneauIdentities ) would  express any anomalous normal product
in terms of a collection of standard ($\M^i$) and also evanescent ($\hat\M^j$) 
monomial normal products.
That is                    
$$
\N[\ghat_{\mu\rho}\Op^{\mu\rho}](x)\cdot\GR=
	\sum_i \a_i\; \N[\M^i](x)\cdot\GR +
   \sum_j \hat\a_j\; \N[\hat\M^j](x)\cdot\GR.                  
$$
Therefore, the rest of independent anomalous monomials should be also
expanded by a similar formula, then getting a system of identities.

But notice that the left hand side of the Bonneau identities and also the 
coefficients given by the r.s.p.~of the right 
hand side are both  of order $\hbar^1$, in the least. Therefore, 
the Bonneau identities are not a trivial system but a 
linear system whose unique solution give the desired expansion of any
anomalous operator in terms of a quantum basis of standard insertions:
$$
\N[\ghat_{\mu\rho}\Op^{\mu\rho}](x)\cdot\GR=
   \sum_i q_i\; \N[\M^i](x)\cdot\GR.                          
$$
Of course, the coefficients $q_i$ are formal series in $\hbar$.

Clearly, at lowest order the linear system is decoupled and has an easy
interpretation: the loops with the anomalous insertion are replaced
with sums of tree diagrams corresponding to insertion of 
standard operators, but with coefficients of order $\hbar^1$. This will be
most relevant to our computations below. 
In general, for the calculation of $q_i$ at order $\hbar^m$, it is needed 
the coefficients $\a_i$ up to order $\hbar^m$ and the coefficients 
$\hat\a_i$ up to order $\hbar^{(m-1)}$.

\bigskip

\section{3. Chiral non-Abelian Yang-Mills theories: simple gauge groups}

This section is devoted to the study of the one-loop dimensional 
renormalization of a general Chiral non-Abelian gauge theory with neither
Majorana nor scalar fields. We shall assume for the time being that the
gauge group is a compact simple Lie group and leave for section
{\it 4.} the case of a general compact group.

Let us give first some definitions and display some properties
$$
\eqalignno{
&\PL={1-\gam5 \over 2},\quad \PR={1+\gam5 \over 2},\quad             
	\PL^2=P_L,\quad \PR^2=\PR,\quad \PL\PR=\PR\PL=0         \cr
&\psi_\L=P_L\psi,\qquad \psi_\R=\PR\psi,
	\qquad \bar\psi_\L=\bar\psi \PR,
   \qquad \bar\psi_\R=\bar\psi \PL,                            \cr
}
$$
and also the special properties of the symbols in the algebra of  covariants:
$$
\eqalignno{
&\PR\gamma^\mu \PR=\PR\gamhat^\mu \PR=\gamhat^\mu \PR\ne0,           \cr
&\PL\gamma^\mu \PL=\PL\gamhat^\mu \PL=\gamhat^\mu \PL\ne0,           \cr
&\PL\gamma^\mu \PR=\PL\gambar^\mu \PR=\gambar^\mu \PR=
		\PL\gambar^\mu,                             &\numeq\cr
&\PR\gamma^\mu \PL=\PR\gambar^\mu \PL=\gambar^\mu \PL=
		\PR\gambar^\mu,                                    \cr
&\PL\gambar^\mu \PL=\PR\gambar^\mu \PR=0=
	\PL \gamhat^\mu \PR = \PR \gamhat^\mu \PL.                 \cr
}  \namelasteq\EqRegProjectors
$$

\medskip

\subsection{3.1. The Classical Action}

The BRS invariant classical  four dimensional action is
$$
S_{\rm cl} = S_{\rm inv} + S_{\rm gf} + S_{\rm ext},
   \eqno\numeq\namelasteq\Classicalact
$$

with
$$
\eqalignno{
S_{\rm inv} &= \intd^4x \;
	-{1\over 4 g^2} \Tr\, F_{\mu\nu} F^{\mu\nu} + 
	{i\over2} \psipsicov + {i\over2}\psipsiprimecov,                  \cr
S_{\rm gf} &= \intd^4x \; \Tr\,{\a\over2}\,B^2+ \Tr\,B(\pr_\mu A^\mu)-
              \Tr\, \bar\om \,\pr^\mu \nabla_\mu\om,      \cr
S_{\rm ext} &= \intd^4x \; \Tr\,\rho^\mu sA_\mu + \Tr\, \zeta s\om+
	\bar L s\psi + L s\bar\psi + \bar R s\psi' + R s\bar\psi',        \cr
}
$$
where $A_\mu=A_\mu^a \tau^a, \om, \bar\om, B, \rho$ and $\zeta$ take values on 
the Lie algebra of a compact simple Lie group $G$.  
$\tau^a$ are the generators of $G$ in a given finite dimensional 
representation, normalized so that  $\Tr\,[\tau^a\tau^b]=\delta^{ab}$. We thus 
have  $[\tau^a,\tau^b]=i c^{abc}\,\tau^c$,  
$c^{abc}$ being completely antisymmetric,
which defines the adjoint representation $(\TA^a)_{ij}=-i\,c^{aij}$ with a
certain normalization $\Tr\,[\TA^a\TA^b]=\TA\,\delta^{ab}$, 
$(\TA^e\,\TA^e)_{ij}=\CA\,\delta_{ij}$, $\TA=\CA$.
The following definitions will be used in the sequel
$$
\eqalignno{
F_{\mu\nu} &=F_{\mu\nu}^a \tau^a= 
	\pr_\mu A_\nu-\pr_\nu A_\mu - i[A_\mu,A_\nu],              \cr
\nabla_\mu\phi^a &=\pr_\mu\phi^a+c^{abc}\,A_\mu^b\phi^c,
\quad \phi\;\hbox{being a Lie algebra valued object}                 \cr
D_{\L\mu} \psi &= (\pr_\mu - i A_\mu^a \TL^a \PL) \psi,     \cr
D_{\R\mu} \psi' &= (\pr_\mu - i A_\mu^a \TR^a \PR) \psi'.           \cr
}
$$
$\psi$ ($\psi'$) represents a collection of left-handed (right-handed) 
fermionic multiplets carrying finite representations, $\TL^a$ 
($\TR^a$) of the group generators. The following equations hold 
$$
\eqalignno{
[\TL^a,\TL^b]&=ic^{abc}\,\TL^c,\qquad [\TR^a,\TR^b]=ic^{abc}\,\TR^c,     \cr
\Tr\,[\TL^a\TL^b]&=\TL\delta^{ab},\qquad 
   \Tr\,[\TR^a\TR^b]=\TR\delta^{ab},                            \cr
\TL^e\,\TL^e&=\CL,\quad \TR^e\,\TR^e=\CR.
}
$$
We also introduce the  shorthand notation:
$$
\eqalignno{
&\Tr\,[\TL^{a_1}\cdots\TL^{a_n}]\equiv\TL^{a_1\ldots a_n},\qquad
\Tr\,[\TR^{a_1}\cdots\TR^{a_n}]\equiv\TR^{a_1\ldots a_n},     \cr
&\TL^{a_1\ldots a_n}+\TR^{a_1\ldots a_n}\equiv\TLR^{a_1\ldots a_n},\qquad
	\TL^{a_1\ldots a_n}-
      \TR^{a_1\ldots a_n}\equiv T_\LmR^{a_1\ldots a_n}, \cr
&\Tr_{\rm L}\,[\phi_1\cdots\phi_n]\equiv
	\phi_1^{a_1}\ldots\phi_n^{a_n}\;\Tr\,[\TL^{a_1}\cdots\TL^{a_n}], 
\;
\Tr_{\rm R}\,[\phi_1\cdots\phi_n]\equiv
	\phi_1^{a_1}\ldots\phi_n^{a_n}\;\Tr\,[\TR^{a_1}\cdots\TR^{a_n}].\cr
}
$$

The appropriate index labeling different fermions will be understood
throughout this paper:
every left handed multiplet can yield a different, say, $\TL,\,\CL \,
\ldots$.

For the  action we have chosen, the free boson propagator 
is (in momentum space):
$$
g^2\,{-i\over k^2 + i\epsilon}\,\delta^{ab}\, 
	\Bigl[ g^{\mu\nu}-\bigl(1-{\a\over g^2}\bigr)\,{k^\mu k^\nu \over k^2}
			\Bigr]=
g^2\,{-i\over k^2 + i\epsilon}\,\delta^{ab}\, 
	\Bigl[ \bigl(g^{\mu\nu}-{k^\mu k^\nu \over k^2}\bigr)+
      {\a\over g^2}\,{k^\mu k^\nu \over k^2}\Bigr].      
$$
Because in perturbative calculations the combination $\a /g^2$ will
often appear, we denote it with  $\a'\;$. Therefore, the Feynman gauge here is 
defined by $\a=g^2$ or $\a'=1$ \dag.%
\vfootnote\dag{If the fields 
are rescaled by $g$: $A\rightarrow g A'$, $\om\rightarrow g \om'$, 
	$\bar\om\rightarrow g^{-1} \bar\om'$; and $\a\rightarrow g^2 \a'$, 
	then the action adopt the other usual form with 
$g$ being interpreted as the ``coupling'' constant. But let us remember 
that the loop expansion is an expansion in $\hbar$ rather than in $g$.}

$S_{\rm cl}$ is left invariant in four dimensions by the BRS
transformations:
$$
\eqalignno{
&s \psi= i \om^a \TL^a\PL\psi,\qquad 
	s\bar\psi=i\bar\psi\TL^a\PR\om^a,                  \cr
&s \psi'= i \om^a\TR^a \PR\psi',\quad 
		s\bar\psi'=i\bar\psi'\TR^a\PL\om^a,         &\numeq\cr
&s A_\mu = \nabla_\mu\om,\quad s\om=i\om^2, \quad
			s \bar\om=B, \quad s B=0,   \cr
&s \rho^\mu = 0,\quad s \zeta = 0,\quad s L = s\bar L = s R = s\bar R =0. \cr
} \namelasteq\BRStrans
$$

This is due to the anticommutativity of
$\gam5$ in four dimensions, which allow us to write $S_{\rm inv}$ in the 
gauge invariant form
$$
\eqalignno{
S_{\rm inv} = \intd^4x \;
	&-{1\over 4 g^2} F_{\mu\nu} F^{\mu\nu} + 
	{i\over2}\bar\psi_\L\arrowsim\Dslash_\L\psi_\L +
		{i\over2}\bar\psi'_\R\arrowsim\Dslash_\R \psi'_\R +  
	 {i\over2}\bar\psi_\R\arrowsim\prslash\psi_\R +
      {i\over2}\bar\psi'_\L\arrowsim\prslash\psi'_\L.     \cr
}
$$

The action preserves the ghost number. Its value together with the dimensions
and the commutativity for the different fields are shown in Table 1.

\begingroup
\parindent=0pt \leftskip=1cm \rightskip=1cm \baselineskip=11pt
\topinsert     

\centerline{\vbox{\offinterlineskip
  \def\tabrule{\noalign{\hrule}}
  \def\hpt{height3pt&\omit&&\omit&&\omit&&\omit&&\omit&&\omit&&
			\omit&&\omit&&\omit&&\omit&}
 \halign{\vrule#&\strut\quad\hfil#\hfil\quad&\vrule width 1.2 pt#&
			&\strut\quad\hfil#\hfil\quad&\vrule#\cr
 \tabrule\hpt\cr
 &&& $s$ && $\!A_\mu\!$ && $\!\!\psi,\psi'\!\!\!$ && $\om$ &
   & $\bar\om$ && $B$ && $\rho_\mu\!$ && $\zeta\!$ && $\!\!L,R\!\!$ &\cr
 \hpt\cr
 \noalign{\hrule height 1.2pt}
 \hpt\cr
 & Ghost numb.$\!\!$  && 1  && 0  && 0  
&& 1  && -1 && 0 && -1 && -2 && -1 &\cr
 \hpt\cr \tabrule \hpt\cr
 & Dimension && 0  && 1  && $\!$3/2$\!$&& 0 
 && 2 && 2 && 3  && 4  && $\!$5/2$\!$&\cr
 \hpt\cr \tabrule \hpt\cr
 & Commutat.$\!$  && -1 && $\!$+1$\!$ && -1 && -1 && -1$\!$ && $\!$+1$\!$
          && -1$\!$ && +1$\!$ && +1 &\cr
 \hpt\cr \tabrule}}}
\vskip 12pt
{\bf Table 1:}
{\eightrm Ghost number and dimension of the fields. 
In the last row, +1 (-1) means 
that the symbol commutes (anticommutes)}
\vskip 0.4cm
\endinsert     
\endgroup

\smallskip

The four-dimensional BRS linear operator $b$ is:

$$
\eqalignno{
b\equiv s\,+\,
\intd^4 x\,\,\Big\{
&\Tr\,{\delta S_{\rm cl}\over\delta A_\mu}{\delta\over\delta\rho^\mu} +
   \Tr\,{\delta S_{\rm cl}\over\delta\om}{\delta\over\delta\zeta} + \cr
&+ {\delta S_{\rm cl}\over\delta\psi}{\delta\over\delta\bar L} +
   {\delta S_{\rm cl}\over\delta\psi'}{\delta\over\delta\bar R} +
   {\delta S_{\rm cl}\over\delta\bar\psi}{\delta\over\delta L} +
   {\delta S_{\rm cl}\over\delta\bar\psi'}{\delta\over\delta R}
   \Big\}\,, &\numeq\cr
}              \namelasteq\EqOperatorb
$$
which satisfies $b^2=0$ because is the linearization of the BRS operator
and the classical action satisfy the BRS identities \cite{\PiguetSorella}.
The symbol $s$ has been defined in eq.~\BRStrans .

The classical $b$-invariant combinations are:
$$
\eqalignno{
L_g&=g{\pr S_{\rm cl}\over\pr g}={1\over2g^2}\,\intd^4x\;\FF,\cr
L_A&=\,b\,\cdot\intd^4x\; \tilde\rho_\mu^a\,A^{a\mu},\qquad
   L_\om=\,-b\,\cdot\intd^4x\;\zeta^a\,\om^a,       &\numeq\cr
L_{\psi}^{\rm L}&=-\, b\,\cdot\intd^4x\;\bar L\PL\psi + \bar\psi\PR L=
   2\,\intd^4x\;{i\over2}\,\bar\psi\arrowsim\prslash\PL\psi+
	\bar\psi\ga^\mu\PL\TL^a\psi A_\mu^a,    \cr
L_{\psi^\prime}^{\rm R}&=-\, b\,\cdot\intd^4x\;\bar R\PR\psi' +
    \bar\psi'\PL R=
   2\,\intd^4x \; {i\over2}\,\bar\psi'\arrowsim\prslash\PR\psi'+
	\bar\psi'\ga^\mu\PR\TR^a\psi' A_\mu^a,    \cr
L_{\psi}^{\rm R}&=-\, b\,\cdot\intd^4x\;\bar L\PR\psi + \bar\psi\PL L=
   2\,\intd^4x\;{i\over2}\,\bar\psi\arrowsim\prslash\PR\psi,    \cr
L_{\psi^\prime}^{\rm L}&=-\, b\,\cdot\intd^4x\;\bar R\PL\psi' + \
     \bar\psi'\PR R=
   2\,\intd^4x \; {i\over2}\,\bar\psi'\arrowsim\prslash\PL\psi'\,;    \cr
}             \namelasteq\EqbInvariants
$$
where $\tilde\rho^\mu=\rho^\mu+\pr^\mu\bar\om$.
$L_{\psi}^{\rm R}$ and $L_{\psi^\prime}^{\rm L}$ will prove not to be
useful due to our choice of the regularized chiral vertex.

It will be shown later that the non-trivial $L_g$ is
associated with the finite renormalizations of the coupling constant $g$, 
whereas the $b$-exact terms, $L_\psi$, $L_{\psi^\prime}$, $L_A$ and
$L_\om$ give rise to finite renormalizations of the corresponding 
wave functions in the space of field functionals in four dimensional 
space-time.

\medskip
\vfill\eject

\subsection{3.2.The Dimensional Regularization  Action}

We shall now follow {\it i)} of subsection {\it 2.2.} and set the
Dimensional Regularization classical action $S_0$. The kinetic terms are
uniquely defined, not so the interaction terms. Notice, for instance,
that the Dirac matrix part of the left-handed-fermion-gauge-boson vertex has 
the following equivalent forms in 4 dimensions: $\gamma^\mu\PL$ 
$=\PR\gamma^\mu$ $=\PR\gamma^\mu\PL$. All these forms
are different in the $d$-dimensional space-time of Dimensional Regularization
 because of the non-anticommutativity of $\gam5$. 
Of course, the generalization of the interaction to the 
Dimensional Regularization space is not unique, and any choice is {\it equally 
correct}, although some choices will be more convenient than others.

We choose the following generalization of $S_{\rm inv}$: 
$$
\eqalignno{
 & \intd^dx \;
	-{1\over 4 g^2} \Tr\, F_{\mu\nu} F^{\mu\nu} + 
	{i\over2} \psipsi + {i\over2}\psipsiprime  +
	(\bar\psi \PR \gamma^\mu\PL \TL^a\psi +
	 \bar\psi'\PL \gamma^\mu\PR \TR^a\psi') A_\mu^a                   \cr
 &=\intd^dx \;
	-{1\over 4 g^2} \Tr\, F_{\mu\nu} F^{\mu\nu} + 
	{i\over2} \psipsi + {i\over2}\psipsiprime  +
	(\bar\psi \gambar^\mu\PL\TL^a\psi + 
         \bar\psi'\gambar^\mu \PR\TR^a\psi') A_\mu^a.           \cr
}
$$

We next generalize, in the obvious way, to the $d$-dimensional
space-time of Dimensional Regularization the  BRS variations 
(which we will denote by $s_d$ and call Dimensional Regularization 
BRS variations) and the gauge-fixing terms. The Dimensional
Regularization BRS transformations read
$$
\eqalignno{
&s_d  \psi= i \om^a \TL^a\PL\psi,\qquad 
s_d \bar\psi=i\bar\psi\TL^a\PR\om^a,                       \cr
&s_d  \psi'= i \om^a\TR^a \PR\psi',\quad 
s_d \bar\psi'=i\bar\psi'\TR^a\PL\om^a,                      &\numeq\cr
&s_d A_\mu = \nabla_\mu\om,\quad s_d \om=i\om^2, \quad
			s_d  \bar\om=B, \quad s_d  B=0,   \cr
&s_d  \rho^\mu = 0,\quad s_d  \zeta = 0,\quad s_d  L = s_d \bar L = 
s_d  R = s_d \bar R =0. \cr
} \namelasteq\DBRStrans
$$

Notice that we have chosen the Dimensional Regularization BRS transformations
so that no explicit evanescent operator occurs in them. This will certainly
simplify the computation of the anomalous breaking term in  BRS identities for
the minimally renormalized theory. We refer the reader to Appendix B for
the discussion of the computation of the BRS anomalous breaking term when
the Dimensional Regularization BRS transformations involve coefficients that
vanish as $d\rightarrow 4$.

In summary,  our Dimensional Regularization classical action, $S_0$, has 
the following expression:
$$
\eqalignno{
S_0 = &\intd^dx \;
	-{1\over 4 g^2} \Tr\, F_{\mu\nu} F^{\mu\nu} + 
	+{i\over2} \psipsi + {i\over2}\psipsiprime  +
	(\bar\psi \gambar^\mu\PL\TL^a\psi + 
			\bar\psi'\gambar^\mu \PR\TR^a\psi') A_\mu^a +\cr
& \intd^dx \; \Tr\,{\a\over2}\,B^2+ \Tr\,B(\pr_\mu A^\mu)-
	\Tr\, \bar\om \,\pr^\mu \nabla_\mu\om\quad+ &\numeq\cr
& \intd^dx \; \Tr\,\rho^\mu s_d A_\mu + \Tr\, \zeta s_d\om+
\bar L s_d\psi + L s_d\bar\psi + \bar R s_d\psi' + R s_d\bar\psi'.    \cr
} \namelasteq\DRegclass
$$

The fermionic part of the Dimensional Regularization classical 
Lagrangian has the following non-gauge invariant form:
$$   
\eqalignno{
&{i\over2}\bar\psi_\L\arrowsim{\bar\Dslash}_\L\psi_\L +
		{i\over2}\bar\psi'_\R\arrowsim{\bar\Dslash}_\R \psi'_\R + 
	{i\over2}\bar\psi_\R\arrowsim{\bar\prslash}\psi_\R +
		{i\over2}\bar\psi'_\L\arrowsim{\bar\prslash}\psi'_\L       \cr
&\qquad\qquad   +
	{i\over2}\bar\psi_\R\arrowsim{\hat\prslash} \psi_\R +             
{i\over2}\bar\psi_\L\arrowsim{\hat\prslash} \psi_\L +             
{i\over2}\bar\psi'_\L\arrowsim{\hat\prslash}\psi'_\L +    
{i\over2}\bar\psi'_\R\arrowsim{\hat\prslash}\psi'_\R.     &\numeq\cr
}\namelasteq\nongaugeinvariant
$$
Hence, $S_0$ is not BRS invariant, the breaking, $s_d S_0$,   coming from the
the last four terms of eq.~\nongaugeinvariant:
$$
\eqalignno{
s_d S_0 &= s_d \intd^dx  \;
	{i\over2}\bar\psi\gamhat^\mu\PL\arrowsim\pr\!_\mu\psi + 
	{i\over2}\bar\psi\PR\gamhat^\mu\arrowsim\pr\!_\mu\psi + 
	{i\over2}\bar\psi'\gamhat^\mu\PR\arrowsim\pr\!_\mu \psi'+
	{i\over2}\bar\psi'\PL\gamhat^\mu\arrowsim\pr\!_\mu \psi   \cr
&=\;\;\intd^dx \; {1\over2} \om^a\Big\{                                 
	\bar\psi\gamhat^\mu\gam5\TL^a\arrowsim\pr\!_\mu\psi +
		\pr_\mu (\bar\psi\gamhat^\mu\TL^a\psi) -
	\bar\psi'\gamhat^\mu\gam5\TR^a\arrowsim\pr\!_\mu\psi' +
		\pr_\mu (\bar\psi'\gamhat^\mu\TR^a\psi') \Big\}         \cr
&\equiv \hat\Delta\equiv \intd^dx\;\hat\Delta(x).            &\numeq\cr
}  \namelasteq\EqBreaking   
$$ 

The Feynman rule of 
the insertion of this anomalous breaking is given by fig.~1.
We denote hereafter the line of the fermion which interact 
left(right)-handedly as L(R). 

\midinsert
\def\graphwidth{1.5in}        
{\eightpoint
$$
\eqalign{\epsfxsize=\graphwidth\epsffile{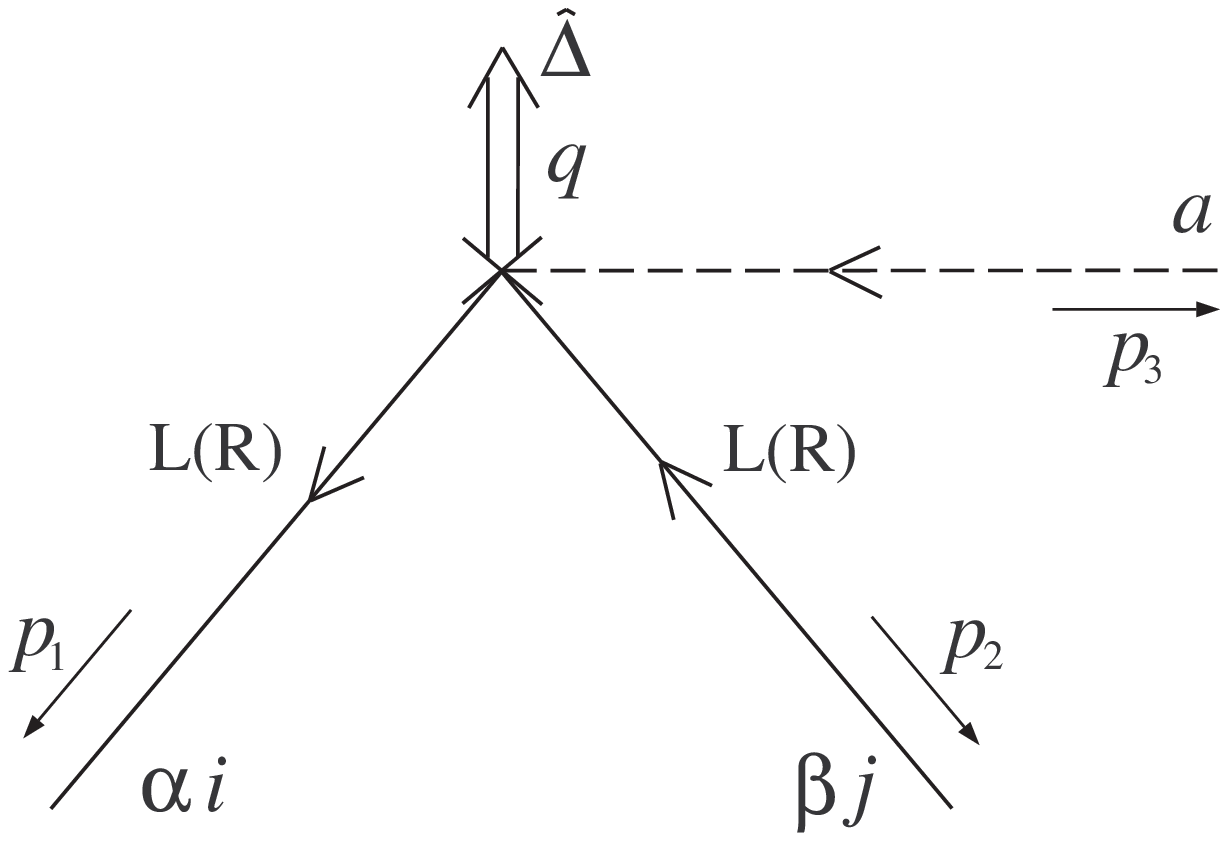}}
\quad\
\eqalign{
\equiv&\quad
   \Gapcero_{\psi^\pp\bar\psi^\pp\om;\hat\Delta}
		{\,}^{\b\a,a}_{ji} (p_1,p_2;q)\quad=    \cr
=&\quad
{i\over2}\,[\TLpR^a]_{ij}\,[+(-) (\hat\pslash_2-\hat\pslash_1)\gam5+
	(\hat\pslash_1+\hat\pslash_2)]{\big|}_{\a\b}
}
$$
}
\vskip -0.3cm
\narrower\noindent {\bf Figure 1.}
{\eightrm Feynman rule of the insertion of the non-integrated breaking
	eq.~\EqBreaking}
\vskip 0.4cm
\endinsert

Notice that if the fermion representation(s) were 
compatible with a CP invariance of the classical action, then the 
Dimensional Regularization action with this choice of chiral vertex 
would also be formally CP invariant, and the breaking  eq.~\EqBreaking\  
would have a definite CP value ($+1$ if we assign $(\om_\LpR)'=-(\om_\LpR)^t$,
$-1$ if we assign $(\om_\LpR)'=(\om_\LpR)^t$, where 
$\om_\LpR\equiv\om^a \TLpR$ and ${}^t$ stands for transposition of the colour
matrices).

The breaking
is an (implicit) ``$(d-4)$-object'', {\it i.e.} an evanescent operator,
 and, clearly, this would be also true for any other choice Dimensional
Regularization classical action. For example, if we had chosen as 
the regularized interaction 
$$
\bar\psi\gamma^\mu\PL\TL^a\psi A_\mu^a + 
   \bar\psi'\gamma^\mu\PR\TR^a\psi' A_\mu^a,                  
$$
the breaking would have been
$$
\eqalignno{
&\intd^dx \; {1\over2} \om^a\Big\{                                 
	\bar\psi\gamhat^\mu\gam5\TL^a\arrowsim{\prslash}_\mu\psi +
		\pr_\mu (\bar\psi\gamhat^\mu\gam5\TL^a\psi) -
	\bar\psi'\gamhat^\mu\gam5\TR^a\arrowsim{\prslash}_\mu\psi' -
		\pr_\mu (\bar\psi'\gamhat^\mu\gam5\TR^a\psi')\Big\}   \cr
&\qquad -
	c^{abc}\,\om^b
		(\bar\psi\gamhat^\mu\PL\TL^a\psi+
	\bar\psi'\gamhat^\mu\PR\TR^a\psi') A_\mu^c           \cr
&\qquad +
	i\om^b (\bar\psi\gamhat^\mu\PL\TL^c\TL^b\psi+
            \bar\psi'\gamhat^\mu\PR\TR^c\TL^b\psi') A_\mu^c,  \cr
}
$$
which is certainly more involved. (And with this vertex, the formal
 CP invariance of the Dimensional Regularization classical  action 
for CP invariant classical actions would be lost).

Remember that in the Breitenlohner and Maison  prescription 
$\gamhat^\mu$ anticommutes whereas $\gambar^\mu$ commutes with 
$\gam5$; so, in general the algebraic manipulations of the strings of $\ga$s 
would be a bit more tedious with the second kind of vertex because 
$\ga^\mu=\gambar^\mu+\gamhat^\mu$.

Obviously, due to the fact that these two vertices are different, the
results obtained by using minimal subtraction would be different ---they
would be {\it two different renormalization schemes}---and they should be
related by finite terms. Moreover, in general, it should be expected that
both results are different from the one obtained with the usual ``naive''
prescription of an anticommuting $\gam5$ every time a fermionic  
loop occurs. But the ``hermitian vertex'' of ref.~\cite{\Korner} 
is not the only correct one as claimed there, but also any other 
one differing from it by a evanescent 
operator of dimension 4 as correctly remarked in \cite{\Ferrari}.

The fact that in general the results obtained by minimal subtraction 
{\it \'a la} Breitenlohner and Maison 
in theories with cancellation of anomalies {\it do not satisfy} 
the Ward identities is a drawback from the practical point
of view but not a reason to cast doubt on Breitenlohner and Maison 
schemes \cite{\Barroso}. Indeed, we should always remember that we have 
the freedom to add {\it any} finite counterterm to restore the identities. 
And {\it in theories with anomalies the mathematical rigour of the 
regularization scheme we are considering has not been surpassed by any other 
dimensional regularization prescription }.

Furnished with the action $S_0$ given in eq.~\DRegclass\ one next develops
a dimensionally regularized perturbation theory by following steps
{\it i)} to {\it iv)} in subsection {\it 2.2.} This regularized theory is
not invariant under the Dimensional Regularization BRS transformations in
eq.~\DBRStrans. Indeed, the Dimensional Regularization 1PI functional
$\Gamma_{0}$ obtained from $S_0$ satisfies  anomalous BRS identities,
the symmetry breaking terms being given by the insertion in $\Gamma_0$
of the operator $\hat\Delta$ in eq.~\EqBreaking .
 
Since our ultimate goal is to obtain a BRS invariant theory (if the anomaly 
cancellation conditions  \cite {\Alwit} are met) we shall further proceed 
and develop a dimensionally regularized perturbation theory by using the 
following action
$$
S_{\rm DReg}^{(n)}=S_0+S_{\rm ct}^{(n)},\quad
S_{\rm ct}^{(n)}=S_{\rm sct}^{(n)}+S_{\rm fct}^{(n)},
\eqno\numeq \namelasteq\Dchiralact
$$
instead of the Dimensional Regularization classical action $S_0$. We 
have named, in subsection {\it 2.2.}, $S_{\rm Dreg}^{(n)}$ the 
{\it Dimensional Regularization action }(at order $\hbar^n$). 
It is computed order by order in the perturbative expansion
in $\hbar$ by proceeding as in steps {\it v)} to {\it vii)} of subsection
{\it 2.2.} Let us recall that $S_{\rm sct}^{(n)}$ denotes the singular 
(at $d=4$) counterterms needed to minimally renormalize the theory 
dimensionally up to order $\hbar^n$, whereas $S_{\rm fct}^{(n)}$ stands 
for the finite (at $d=4$) counterterms needed to turn the minimally 
renormalized theory into a BRS invariant theory up to order $\hbar^n$.
In this paper we shall be concerned only with the renormalized
theory at order $\hbar$. Hence, we will just compute $S_{\rm sct}^{(1)}$
and $S_{\rm fct}^{(1)}$. 

Note that we have denoted by $\Gamma_0$, instead of $\Gamma_{\rm DReg}$,
the Dimensional Regularization 1PI functional for $S_0$ to avoid confusion
with the Dimensional Regularization 1PI functional for $S_{\rm Dreg}^{(n)}$, 
which shall denote by $\Gamma_{\rm DReg}$.
\medskip

\bigskip

\subsection{3.3. The one-loop singular counterterms: $S_{\rm sct}^{(1)}$}

Let $\Gamma^{\ouno}_{0\,{\rm sing}}$ be the one-loop singular contribution at
$d=4$ to $\Gamma_0$, the latter being the Dimensional Regularization 1PI 
functional obtained from $S_0$. $S_0$ is given in eq.~\DRegclass . Then, by
definition 
$$
S_{\rm sct}^{(1)}=-\Gamma^{\ouno}_{0\,{\rm sing}}.
\eqno\numeq\namelasteq\Singcount
$$
We show  the 1PI functions contributing to $S_{\rm sct}^{(1)}$ 
in figs.~2 and 3.

\topinsert

{\settabs 4\columns \def\graphwidth{1.1in} 
\eightpoint

\+\hfil$\vcenter{\epsfxsize=\graphwidth\epsffile{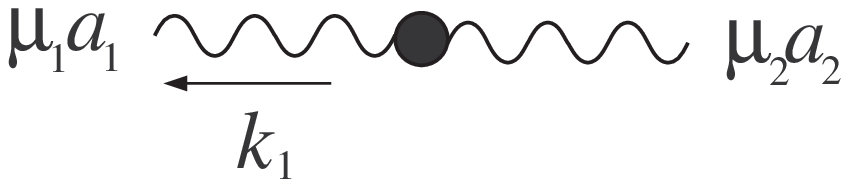}}$\hfil
      & $\vcenter{
\hbox{$-\,\tilde\Gamma_{AA\;\,0\,{\rm sing}}^{\ouno\,\sm1\sm2,\sa1\sa2}
 (k_1)=            $}
      \hbox{$\qquad -
        \icpc\CA\delta^{\sa1\sa2}\big[{5\over3}+{(1-\aprime)\over2}\big]
            \,{2\over\eps}\,(k_{1\!}{}^2 g^{\sm1\sm2}-
                  k_{1\!}{}^{\sm1} k_{1\!}{}^{\sm2})
            $}
      \hbox{$\qquad-
        \icpc\TLR\delta^{\sa1\sa2}{2\over3}\,{2\over\eps}
            \big[\,(\bar k_{1\!}{}^{\sm1} \bar k_{1\!}{}^{\sm2} -
               \bar k_{1\!}{}^2 \gbar^{\sm1\sm2})-
               {1\over2}\hat k_{1\!}{}^2 \gbar^{\sm1\sm2}\big]
            $}
                  }$\cr
\bigskip
\medskip

\+\hfil$\vcenter{\epsfxsize=\graphwidth\epsffile{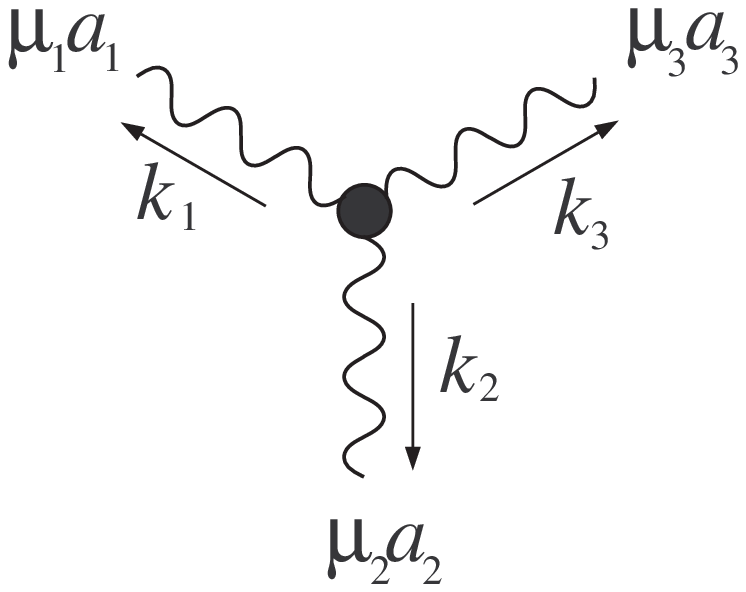}}$\hfil
      & $\vcenter{
 \hbox{$-\,\tilde\Gamma_{AAA\;\;0\,{\rm sing}}^{\ouno\,
\sm1\sm2\sm3,\sa1\sa2\sa3}
                  (k_1,k_2,k_3\!=\!-\!k_1\!-\!k_2)=
            $}
      \vskip 2pt
      \hbox{$\qquad -
        \ucpc\CA c^{\sa1\sa2\sa3}\big[{2\over3}\!+
                     \!{3\over4}(1\!-\!\aprime)\big]
            {2\over\eps}\,
            \Bigg[\vcenter{\hbox{$ g^{\sm1\sm2}(k_1-k_2)^{\sm3} +$}
                           \hbox{$ g^{\sm1\sm3}(k_3-k_1)^{\sm2} +$}
                           \hbox{$ g^{\sm2\sm3}(k_2-k_3)^{\sm1}$} }
                 \Bigg]
            $}
      \vskip 2pt
      \hbox{$\qquad+
        \ucpc\,{\TLR\over2}\,c^{\sa1\sa2\sa3}\,{4\over3}\;{2\over\eps}
            \Bigg[\vcenter{
               \hbox{$\gbar^{\sm1\sm2}(\bar k_1-\bar k_2)^{\sm3}+$}
               \hbox{$\gbar^{\sm1\sm3}(\bar k_3-\bar k_1)^{\sm2}+$}
               \hbox{$\gbar^{\sm2\sm3}(\bar k_2-\bar k_3)^{\sm1}$} }
               \Bigg]
            $}
                  }$\cr
\bigskip
\medskip
\+\hfil$\vcenter{\epsfxsize=\graphwidth\epsffile{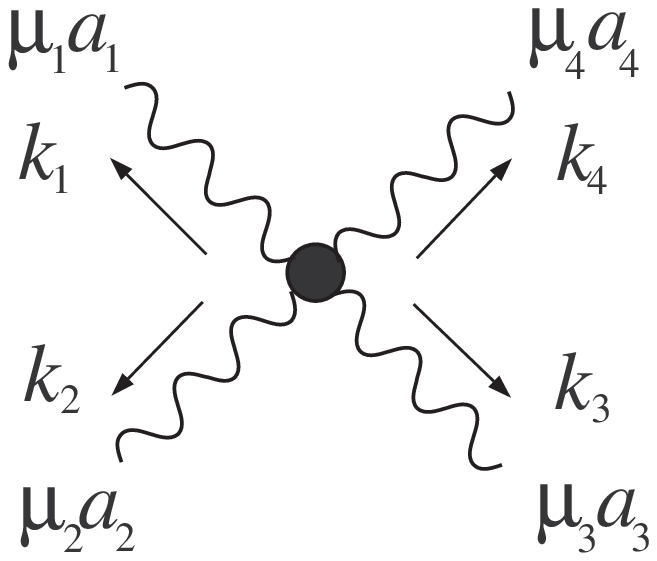}}$\hfil
      & $\vcenter{
\hbox{$-\,\tilde\Gamma_{AAAA\;\;\,0\,{\rm sing}}^{\ouno\,\sm1\sm2\sm3\sm4,
             \sa1\sa2\sa3\sa4}(k_1,k_2,k_3,k_4=-k_1-k_2-k_3)=
            $}
      \vskip 2pt
      \hbox{$\qquad +
          \icpc\CA \big[\!-\!{1\over3}+(1\!-\!\aprime)\big]\,{2\over\eps}\,
              \Bigg[\vcenter{
                 \hbox{$c^{e\sa1\sa2}c^{\sa3\sa4 e}
                    (g^{\sm2\sm3}g^{\sm1\sm4}- g^{\sm2\sm4}g^{\sm1\sm3})+$}
                 \hbox{$c^{e\sa1\sa3}c^{\sa4\sa2 e}
                    (g^{\sm3\sm4}g^{\sm1\sm2}- g^{\sm3\sm2}g^{\sm1\sm4})+$}
                 \hbox{$c^{e\sa1\sa4}c^{\sa2\sa3 e}
                    (g^{\sm1\sm3}g^{\sm2\sm4}-g^{\sm1\sm2}g^{\sm4\sm3}) $}
              }\Bigg]
              $}
      \vskip 2pt
      \hbox{$\qquad -
        \icpc\,{\TLR\over2}\,{4\over3}  \,{2\over\eps}\,
            \Bigg[\vcenter{
               \hbox{$c^{e\sa1\sa2}c^{\sa3\sa4 e}
                  (\gbar^{\sm2\sm3}\gbar^{\sm1\sm4}-
                   \gbar^{\sm2\sm4}\gbar^{\sm1\sm3})+$}
               \hbox{$c^{e\sa1\sa3}c^{\sa4\sa2 e}
                  (\gbar^{\sm3\sm4}\gbar^{\sm1\sm2}-
                        \gbar^{\sm3\sm2}\gbar^{\sm1\sm4})+$}
               \hbox{$c^{e\sa1\sa4}c^{\sa2\sa3 e}
                  (\gbar^{\sm1\sm3}\gbar^{\sm2\sm4}-
                        \gbar^{\sm1\sm2}\gbar^{\sm4\sm3}) $}
            }\Bigg]
            $}
                   }$\cr

\bigskip
\medskip

\+\hfil$\vcenter{\epsfxsize=\graphwidth\epsffile{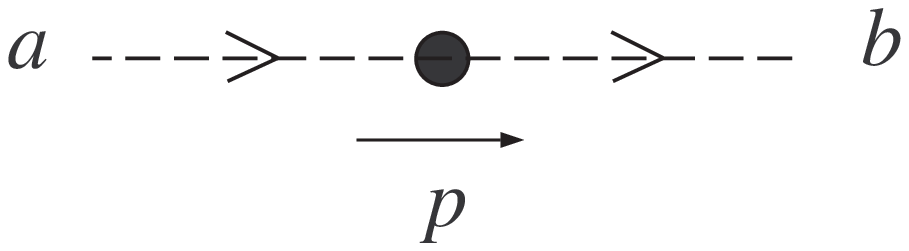}}$\hfil
      & $\vcenter{
      \hbox{$-\,\tilde\Gamma_{\om\bar\om\;0\,{\rm sing}}^{\ouno\,ab}(p)=
            $}
      \hbox{$\qquad +
         \icpc\CA\delta^{ab}\,g^2 {1\over2}\big[1+{(1-\aprime)\over2}\big]
            \,{2\over\eps}\,p^2
            $}
                  }$\cr
\bigskip
\medskip

\+\hfil$\vcenter{\epsfxsize=\graphwidth\epsffile{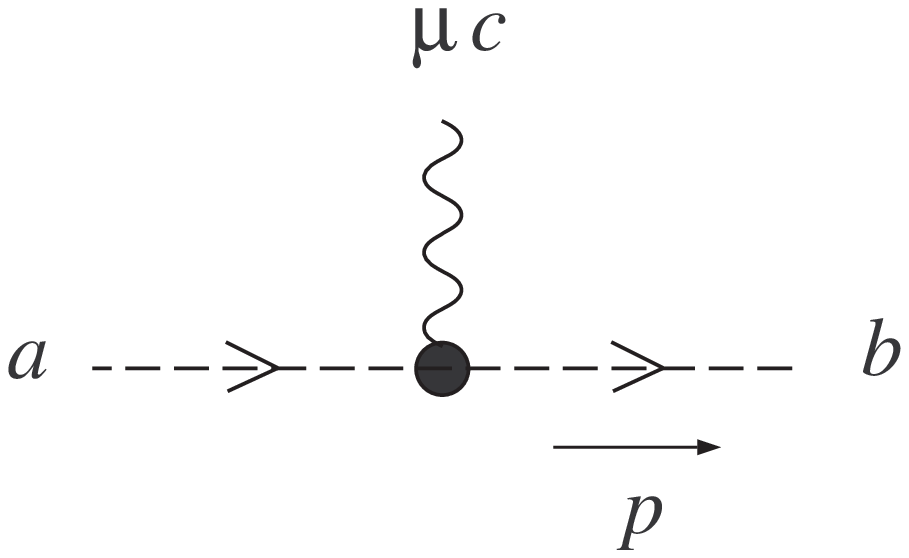}}$\hfil
      & $\vcenter{
    \hbox{$-\,\tilde\Gamma_{\om\bar\om A\;0\,
{\rm sing}}^{\ouno\,abc \,\mu} (p,q)
           =
            $}
      \hbox{$\qquad +
          \ucpc\CA\,g^2 \, c^{abc} {\aprime\over2} \,{2\over\eps}\,p^{\mu}
            $}
                  }$\cr

\bigskip
\medskip

\+\hfil$\vcenter{\epsfxsize=\graphwidth\epsffile{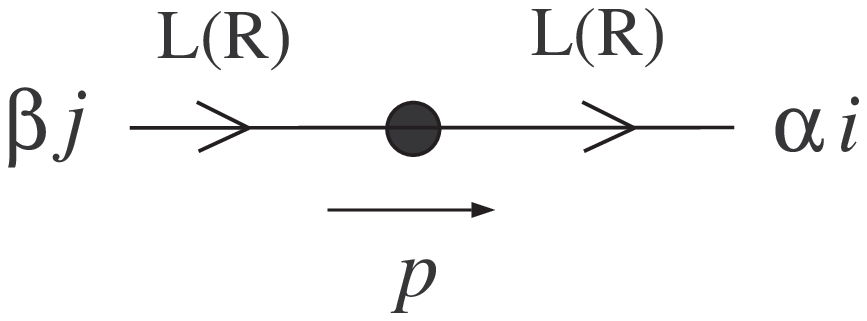}}$\hfil
      & $\vcenter{
      \hbox{$-\,\tilde\Gamma_{\psi^\pp\bar
         \psi^\pp\;\,0\,{\rm sing}}^{\ouno\,\b\a,ji} (p)=
            $}
      \hbox{$\qquad-
        \icpc\,g^2 \delta_{ji}\,\CLpR \,\aprime\,{2\over\eps}\,\bar\pslash\,
                          \, [\PLpR]^{\a\b}
            $}
                  }$\cr
\bigskip
\medskip

\+\hfil$\vcenter{\epsfxsize=\graphwidth\epsffile{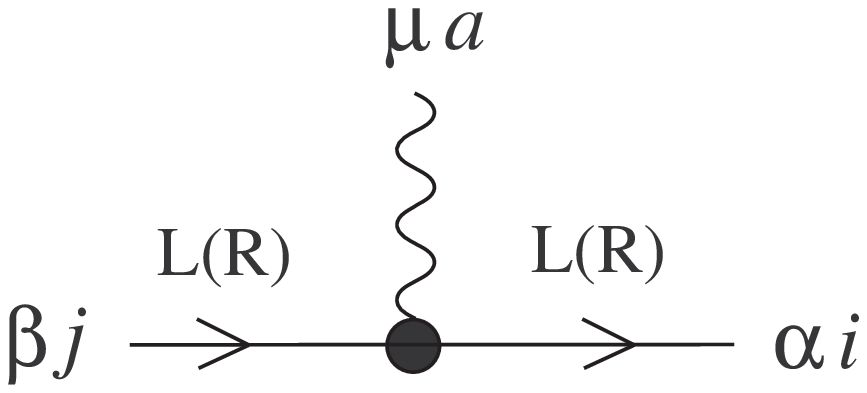}}$\hfil
      & $\vcenter{
      \hbox{$-\,\tilde\Gamma_{\psi^\pp\bar\psi^\pp A
         \;\;0\,{\rm sing}}^{\ouno\,\b\a\,a\mu, ji} (p,q)=
            $}
      \hbox{$\qquad -
        \icpc\,g^2  \big[(1-{1-\aprime\over4})\CA +\aprime\CLpR
        \big] \;{2\over\eps}\; [\TLpR^a]_{ij} \,\gambar^\mu
                        \,   [\PLpR]^{\a\b}
            $}
                  }$\cr

}
\vskip 12pt
\narrower\noindent {\bf Figure 2:}
{\eightrm Feynman rules for the order $\hbar$ singular counterterms.
Here, only functions with no external fields are shown}

\vskip 0.1cm
\endinsert

\topinsert

{\settabs 3\columns \def\graphwidth{1.1in} 
\eightpoint

\+\hfil$\vcenter{\epsfxsize=\graphwidth\epsffile{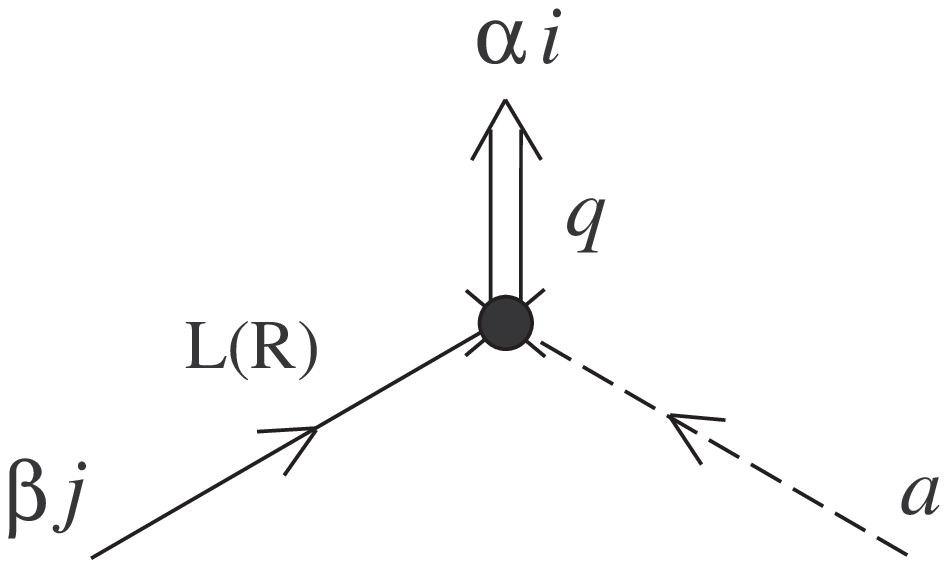}}$\hfil
      & $\vcenter{
      \hbox{$-\,\tilde\Gamma_{\psi^\pp\om;s\psi\;0\,
           {\rm sing}}^{\ouno\,\b a;\a i} 
		(p;q)=
            \, -
        \icpc g^2 \CA {\aprime\over2} \,{2\over\eps}\,
            [T^a_{\LpR}]^{ij}\, [\PLpR]^{\a\b}
            $}
                  }$\cr
\bigskip
\medskip

\+\hfil$\vcenter{\epsfxsize=\graphwidth\epsffile{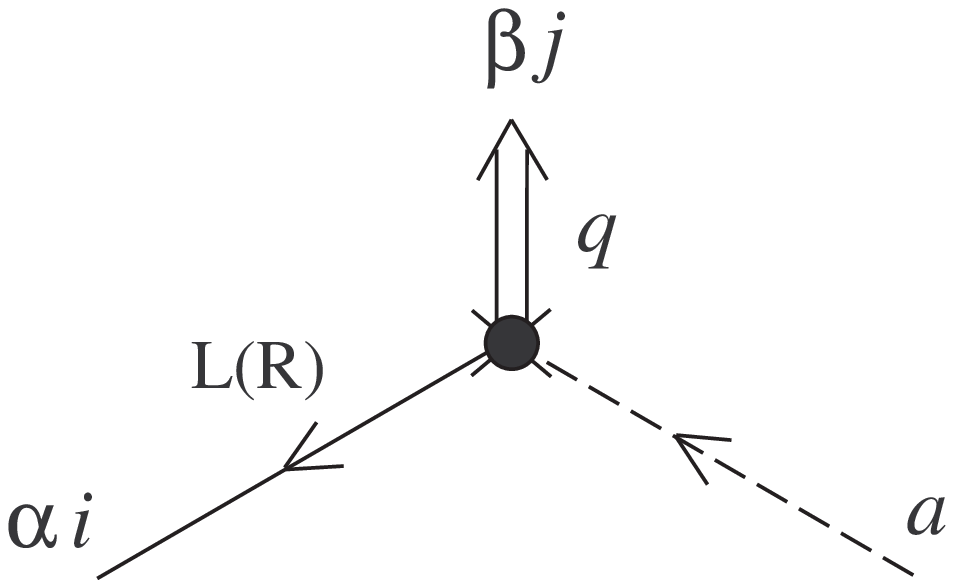}}$\hfil
      & $\vcenter{
      \hbox{$-\,\tilde\Gamma_{\bar\psi^\pp\om;s\bar\psi\;
          0\,{\rm sing}}^{\ouno\,\b a;\a i}(p;q)=
        \icpc g^2 \CA {\aprime\over2} \,{2\over\eps}\,
            [T^a_{\LpR}]^{ij}\, [\PRpL]^{\a\b}
            $}
                  }$\cr
\bigskip
\medskip

\+\hfil$\vcenter{\epsfxsize=\graphwidth\epsffile{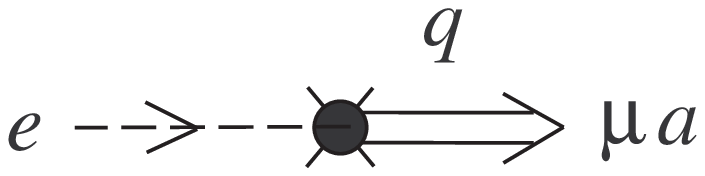}}$\hfil
      & $\vcenter{
      \hbox{$-\,\tilde\Gamma_{\om;s\A\;0\,{\rm sing}}^{\ouno\,e;a\,\mu}(;q)=
        \icpc g^2 \CA \big(1+{1-\aprime\over2}\big) \,{1\over\eps}\,
               \delta^{ae}\,q^\mu
            $}
                  }$\cr
\bigskip
\medskip

\+\hfil$\vcenter{\epsfxsize=\graphwidth\epsffile{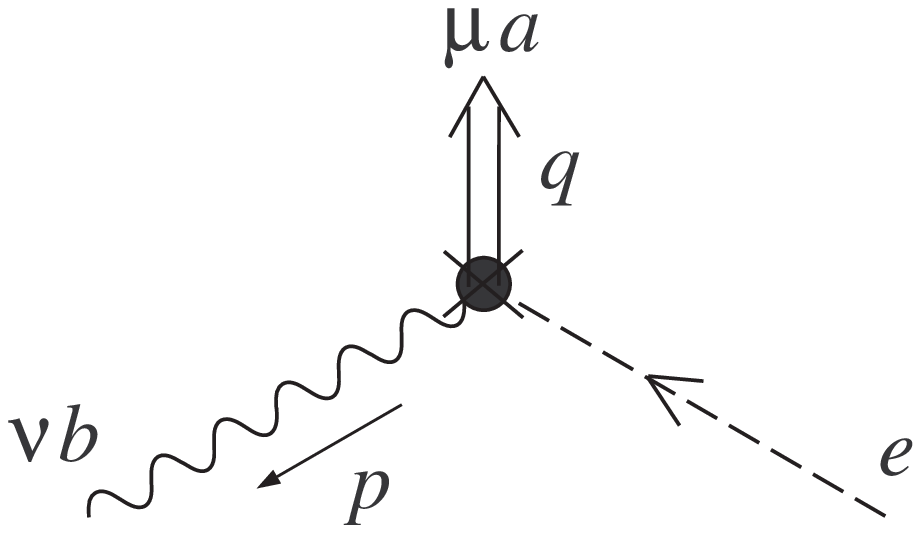}}$\hfil
      & $\vcenter{
      \hbox{$-\,\tilde\Gamma_{\om A;s\A\;0\,{\rm sing}}^{\ouno\,e b\nu;a\,\mu}
                  (p;q)=
        -\ucpc g^2 \CA \aprime \,{1\over\eps}\, c^{abe}
            $}
                  }$\cr
\bigskip
\medskip

\+\hfil$\vcenter{\epsfxsize=\graphwidth\epsffile{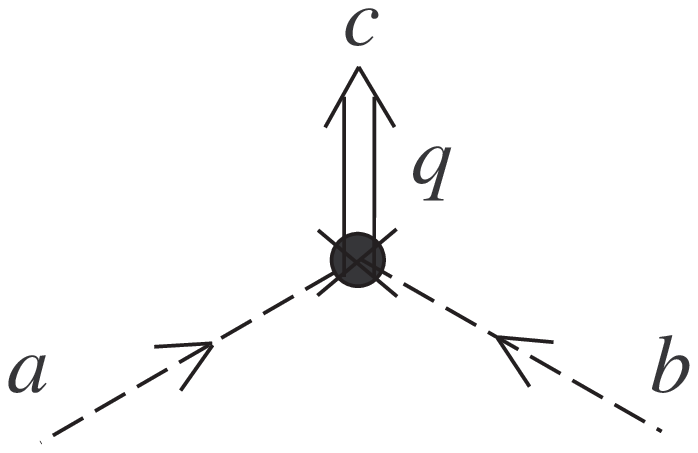}}$\hfil
      & $\vcenter{
  \hbox{$-\,\tilde\Gamma_{\om\om;s\om\;0\,{\rm sing}}^{\ouno\,a b\;c}(p;q)=
    \ucpc\,g^2\,\CA\, \aprime \,{1\over\eps}\, c^{abc}
            $}
                  }$\cr
\bigskip
\medskip
}
\vskip 12pt
\narrower\noindent {\bf Figure 3:}
{\eightrm The same as for fig.~2 but with functions involving external fields}
\vskip 0.4cm
\endinsert

\topinsert

{\settabs 6\columns \def\graphwidth{1.1in} 
\eightpoint

\+&\hfil$\vcenter{\epsfxsize=\graphwidth\epsffile{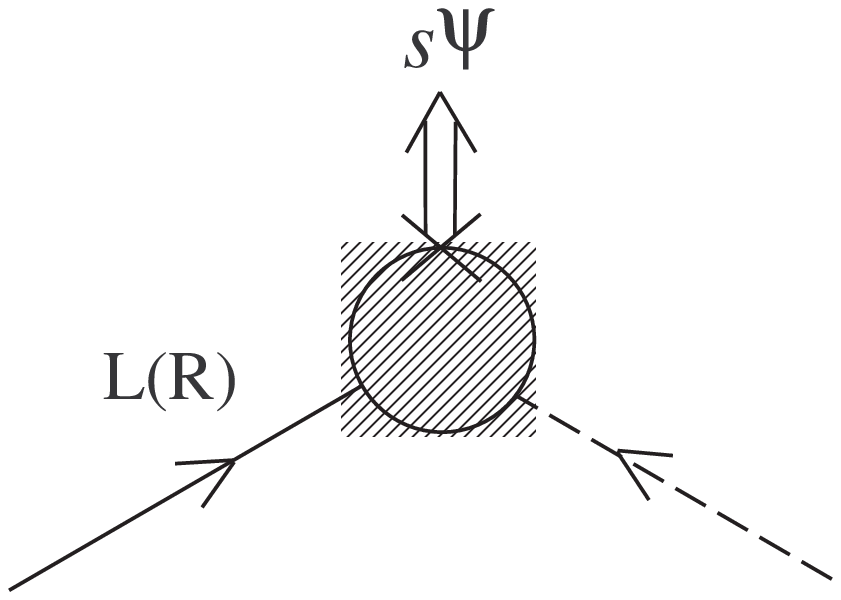}}$\hfil
      & $\qquad\vcenter{=}$
      & \hfil$\vcenter{\epsfxsize=\graphwidth\epsffile{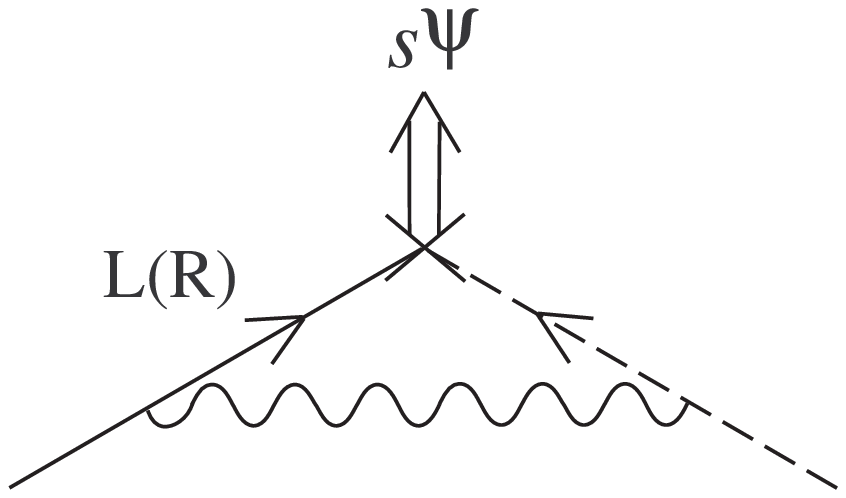}}$\hfil
                     \cr

\+&\hfil$\vcenter{\epsfxsize=\graphwidth\epsffile{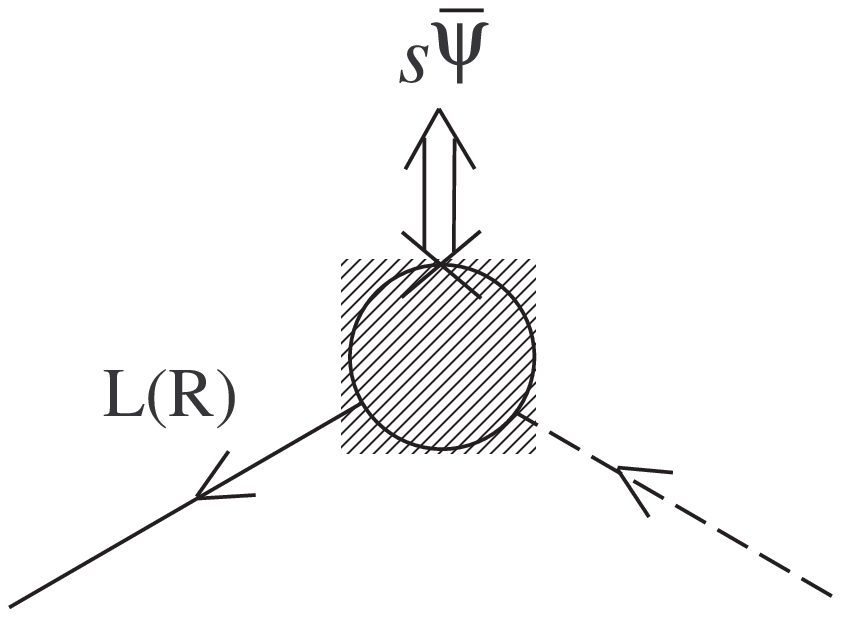}}$\hfil
      & $\qquad\vcenter{=}$
      & \hfil$\vcenter{\epsfxsize=\graphwidth\epsffile{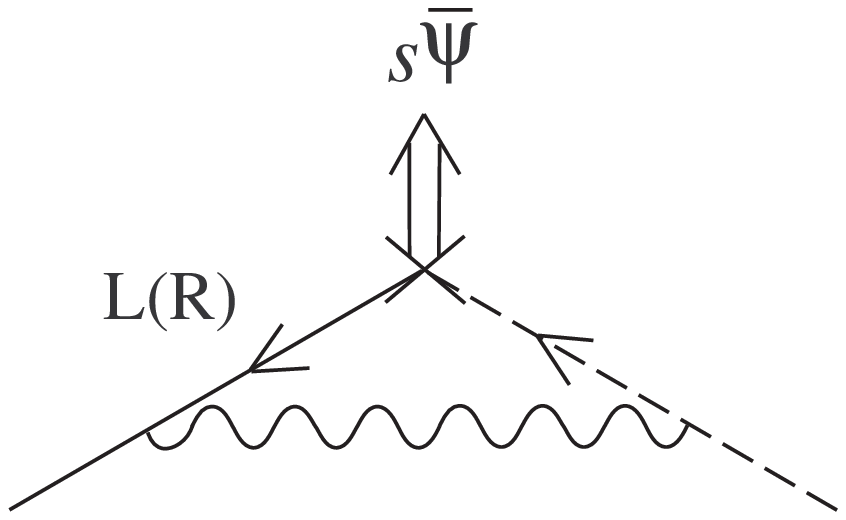}}$\hfil
                     \cr

\+&\hfil$\vcenter{$\,$\epsfxsize=0.9in\epsffile{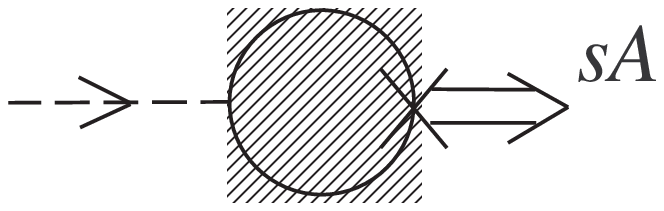}}$\hfil
      & $\qquad\vcenter{=}$
      & \hfil$\vcenter{\epsfxsize=\graphwidth\epsffile{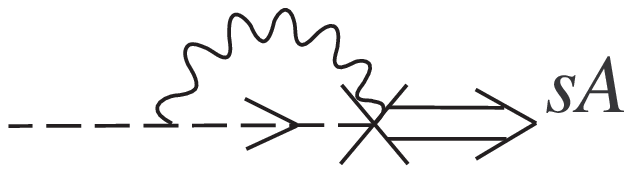}}$\hfil
                     \cr

\+&\hfil$\vcenter{\epsfxsize=\graphwidth\epsffile{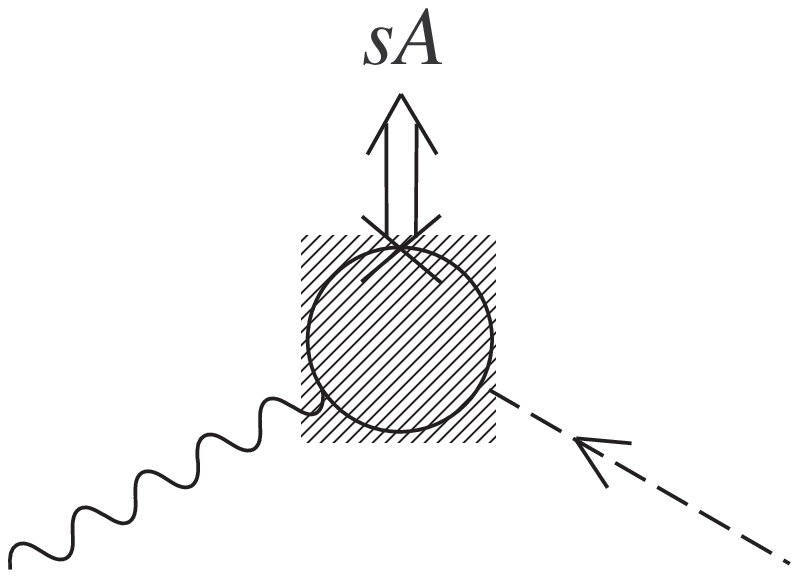}}$\hfil
      & $\qquad\vcenter{=}$
      & \hfil$\vcenter{\epsfxsize=0.6in\epsffile{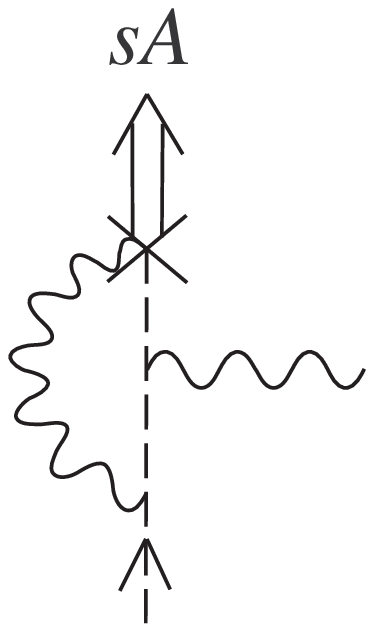}}$\hfil
      & $+
         \quad
         \vcenter{\epsfxsize=\graphwidth\epsffile{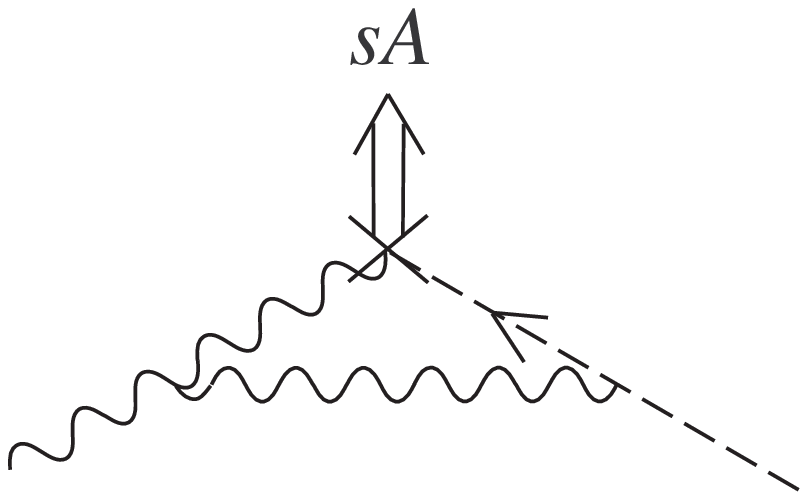}}$
                     \cr

\+&\hfil$\vcenter{$\,$\epsfxsize=0.9in\epsffile{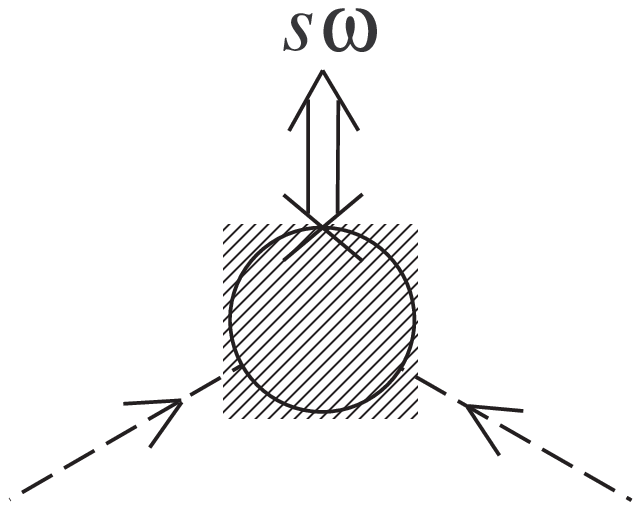}}$\hfil
      & $\qquad\vcenter{=}$
      & \hfil$\vcenter{\epsfxsize=\graphwidth\epsffile{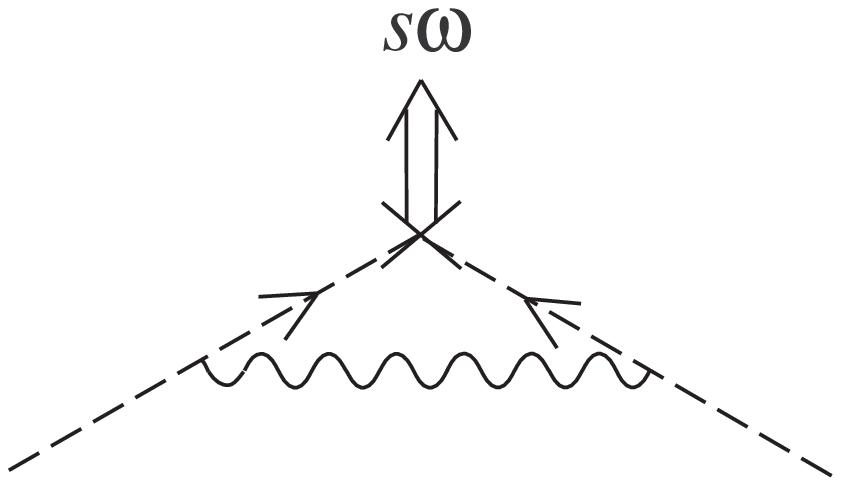}}$\hfil
                     \cr
}
\vskip 12pt
\narrower\noindent {\bf Figure 4:}
{\eightrm One-loop Feynman graphs  contributing to the 1PI functions
involving external fields}

\vskip 0.4cm
\endinsert

Let $\eps=4-d$. Then, the functional $\Gamma^{\ouno}_{0\,{\rm sing}}$
is given by 
 
$$
\eqalignno{
\Gamma^{\ouno}_{0\,{\rm sing}} =& {\hbar\over\eps}\;\ucpc\,g^2\,\bigg\{
 -C_A \left[{10\over3}+(1-\a')\right]\,S_{0\,AA}\; +\;
 {8\over3}{\TLR\over2}\,\bar S_{0\,AA}  \cr
&\qquad -C_A \left[{4\over3} + {3\over2}(1-\a')\right]\,S_{0\,AAA}\; +\;
 {8\over3}{\TLR\over2}\,\bar S_{0\,AAA} \cr
&\qquad -C_A \left[-{2\over3} + 2(1-\a')\right]\,S_{0\,AAAA}\;+\;
 {8\over3}{\TLR\over2}\,\bar S_{0\,AAAA}\cr
&\qquad +\CL 2\,\a'\;\bar S_{0\,\bar\psi\psi}
      +\CR 2\,\a'\;\bar S_{0\,\bar\psi'\psi'} \cr
&\qquad +\left[\left(2-{(1-\a')\over2}\right)C_A + 2\,\a'\CL\right]
         \;\bar S_{0\,\bar\psi\psi A}    \cr
&\qquad +\left[\left(2-{(1-\a')\over2}\right)C_A + 2\,\a'\CR\right]
         \;\bar S_{0\,\bar\psi'\psi' A} \cr
&\qquad -\left(1+{(1-\a')\over2}\right)C_A\;S_{0\,\bar\om\om}
   + \a'\,C_A\;S_{0\,\bar\om\om A} \cr    
&\qquad +\,\a'\,C_A\;
	\left[S_{0\, \bar L s\psi}+S_{0\, \bar R s\psi'}+
		S_{0\,L s\bar\psi}+ S_{0\,R s\bar\psi'}\right]\cr    
&\qquad -\left(1+{(1-\a')\over2}\right)C_A\;S_{0\,\rho\om} 
      +a'\,C_A\; S_{0\,\rho\om A} \cr
&\qquad +\a'\,C_A\; S_{0\,\zeta\om\om}\;\bigg\} \cr
& +{\hbar\over\eps}\;\ucpc\; {\TLR\over 2}\,{4\over3}\;
      \intd^dx\;{1\over2}\,\bar A_\mu\hat{\square}\bar A^\mu,
&\numeq\cr
}   \namelasteq\OnePIpole
$$
where the $S_{0\,X}$ terms are the corresponding ($\a$-independent) terms of
the Dimensional Regularization classical  action $S_0$. The bar above some 
of them means that all the index in the term are barred.

Let us define the Dimensional Regularization linearized BRS operator
$b_d$ as follows
$$
\eqalignno{
b_d\equiv s_d\,+\,
\intd^d x\,\,\Big\{
&\Tr\,{\delta S_0\over\delta A_\mu}{\delta\over\delta\rho^\mu} +
   \Tr\,{\delta S_0\over\delta\om}{\delta\over\delta\zeta} + \cr
&+ {\delta S_0\over\delta\psi}{\delta\over\delta\bar L} +
   {\delta S_0\over\delta\psi'}{\delta\over\delta\bar R} +
   {\delta  S_0\over\delta\bar\psi}{\delta\over\delta L} +
   {\delta S_0 \over\delta\bar\psi'}{\delta\over\delta R}
   \Big\}\,, \cr
}
$$
Notice that $b_d^2\ne0$ since $s_d S_0\ne0$. The action of $s_d$ on the fields
has been defined in eq.~\DBRStrans .
If we define in the $d$-dimensional space-time of Dimensional Regularization
the following integrated field polynomials
$$
\eqalignno{
L_g&\equiv{1\over2g^2}\,\intd^dx\;\FF=g{\pr\over\pr g} S_0,
   \qquad \bar L_g={1\over2g^2}\,\intd^dx\;\Tr\, \bar F_{\mu\nu} 
\bar F^{\mu\nu},                \cr
L_{\psi^\pp}^\LpR\!&\equiv- b_d\,\cdot\intd^dx\;\bigl\{
    \bar L (\bar R)\PLpR\psi^\pp +\bar\psi^\pp\PRpL L(R)\bigr\}=
   \left( N_{\psi}^\LpR-N_{L(R)} \right)\,S_0= \cr
 &=2\,\intd^dx\;\bigl\{{i\over2}\,\bar\psi^\pp\arrowsim\prslash\PLpR\psi^\pp+
   \bar\psi^\pp\gambar^\mu\!\PLpR\TLpR^a\psi^\pp A_\mu^a\bigr\},    \cr
\bar L_{\psi^\pp}^\LpR\!&\!\equiv\!L_{\psi^\pp}\!-\! \!
   \intd^dx\,i\,\bar\psi^\pp\!\hat{\arrowsim\prslash}\PLpR\psi^\pp\!=\!
   2\!\intd^dx\,\!\bigl\{\!{i\over2}\,\bar\psi^\pp\!\bar{\arrowsim\prslash}
   \PLpR\psi^\pp\!\!+\!
   \bar\psi^\pp\!\gambar^\mu\!\PLpR\TLpR^a\psi^\pp \!A_\mu^a\!\bigr\},\cr
L_A&\equiv\,b_d\,\cdot\intd^dx\;\tilde\rho_\mu^a\,A^{a\mu}=
    \left({\rm Tr}\,[N_A-N_{\rho}-N_{B}-N_{\bar\omega}]+2\alpha{\partial\over
\partial \alpha}\right)\,S_0,\cr
L_\om&\equiv\,-b_d\,\cdot\intd^dx\; \zeta^a\,\om^a=\,
            {\rm Tr}\,[N_{\omega}-N_{\zeta}]\,S_0,     \cr
}
$$
where
{\settabs 2 \columns
\openup1\jot
\+$\qquad{\displaystyle
   N_\phi\equiv\intd^dx\;\phi(x)\,{\delta\over\delta\phi(x)}}$,
	&${\displaystyle
   \phi=A_{\mu},\rho_{\mu}, B,{\bar\omega},\omega\,\hbox{and}\;\zeta}$,  \cr
\+$\qquad{\displaystyle N_{\psi}^\LpR\equiv\intd^dx\;(\PLpR)\psi)_{\b}\,
	{\delta\over\delta\psi_\b}}$,
	&${\displaystyle N_{\psi'}^\RpL\equiv\intd^dx\;(\PRpL)\psi')_{\b}\,
			{\delta\over\delta\psi'_\b}}$,  \cr
\+$\qquad{\displaystyle N_{\bar\psi}^\RpL\equiv\intd^dx\;(\bar\psi\PRpL)_{\b}\,
	{\delta\over\delta\bar\psi_\b}}$,
		&${\displaystyle N_{\bar\psi'}^\LpR\equiv
			\intd^dx\;(\bar\psi'\PLpR)_{\b}\,
 {\delta\over\delta\bar\psi'_\b}}$;  \hfill\numeq\namelasteq\Fieldcounting&\cr
}
\medskip
\noindent
then, the functional $\Gamma^{\ouno}_{0\,{\rm sing}}$ in eq.~\OnePIpole\
can be cast into the form:
$$
\eqalignno{
\Gamma^{\ouno}_{0\,{\rm sing}}=&\ucpc g^2\,{\hbar\over\eps}\,
         \left[ {11\over3}\CA\,L_g+(2-{1-\aprime\over2})\CA\,L_A+
            \aprime\CA\,L_\om\right]\cr
&+ \ucpc g^2\,{\hbar\over\eps}\,
          \left[-{4\over3}{\TLR\over2}\bar L_g +
            \aprime\CL \bar L_\psi + \aprime\CR \bar L_{\psi^\prime}
          \right] \cr
&+ \ucpc \,{\hbar\over\eps}\,{\TLR\over2}\,{4\over3}\,
      \intd^dx\;{1\over2}\,\bar A_\mu \hat{\square}\bar A^{\mu};&\numeq\cr
} \namelasteq\EqGammaSing
$$
{\it i.e.} $b_d\,\Gamma^{\ouno}_{0\,{\rm sing}}=O(\ghat_{\mu\nu})$.

The second and third lines of the previous equation show explicitely that
the standard  ``infinite'' multiplicative renormalization of the 
coupling constant $g$ and of the fields of $S_0$ is already lost at the 
one-loop level.

\bigskip

\subsection{3.4. The BRS identity and the anomalous symmetry breaking term}

The {\it Regularized Action Principle}, explained in subsection {\it 2.3.} 
and Appendix A, leads to the following BRS identity
$$
\eqalignno{
{\cal S}_d(\Gamma_{\!{\rm DReg}})\equiv
&\intd^d x\,\,\Bigl\{
\Tr\,{\delta\Gamma_{\!{\rm DReg}}\over\delta\rho^\mu}
 {\delta\Gamma_{\rm DReg}\over\delta A_\mu}+
\Tr\,{\delta\Gamma_{\!{\rm DReg}}\over\delta\zeta}
 {\delta\Gamma_{\rm DReg}\over\delta\om}+
\Tr\; B {\delta\Gamma_{\!{\rm DReg}}\over\delta\bar\om} \cr
&+ {\delta\Gamma_{\!{\rm DReg}}\over\delta\bar L} 
{\delta\Gamma_{\!{\rm DReg}}\over\delta\psi}+
{\delta\Gamma_{\!{\rm DReg}}\over\delta\bar R}
 {\delta\Gamma_{\!{\rm DReg}}\over\delta\psi'}+
{\delta\Gamma_{\!{\rm DReg}}\over\delta L}
 {\delta\Gamma_{\!{\rm DReg}}\over\delta\bar\psi}+
{\delta\Gamma_{\!{\rm DReg}}\over\delta R} 
{\delta\Gamma_{\!{\rm DReg}}\over\delta\bar\psi'}\Bigr\}=\cr
=&\,\hat\Delta\cdot\Gamma_{\!{\rm DReg}} + 
\Delta_{\rm ct}\cdot\Gamma_{\!{\rm DReg}} + 
\intd^dx\,\sum_{\Phi}\Bigl\{
\bigl[{\delta S_{\rm ct}^{(n)}\over \delta K_\Phi(x)}
\cdot\Gamma_{\!{\rm DReg}}\bigr]
      {\delta\Gamma_{\!{\rm DReg}}\over\delta\Phi(x)}\Bigr\},
	 &\numeq\cr
} \namelasteq\EqRegYMSTBreaking             
$$
when applied to the Dimensional Regularization 1PI functional,
$\Gamma_{\rm DReg}$, obtained from the Dimensional Regularization 
action $S^{(n)}_{\rm DReg}$, the latter being given in eq.~\Dchiralact . 
The symbols $\Phi$ and $K_{\Phi}$ in the previous equation stand, 
respectively, for any field that undergoes non-linear 
BRS transformation, {\it i.e} 
$A_\mu$, $\om$, $\psi$, $\psi'$, $\bar\psi$ and $\bar\psi'$ and the
external field which couple to the corresponding BRS variation, 
{\it i.e.} $\rho_\mu$, $\zeta$, $\bar L$, $\bar R$, $L$, and  $R$.
The operator $\hat\Delta$ is given in eq.~\EqBreaking.
The symbol $\Delta_{\rm ct}$ is
equal to $s_d S^{(n)}_{\rm ct}$, where $S^{(n)}_{\rm ct}$ and $s_d$ are  
defined in eqs.~\Dchiralact\ and \DBRStrans , respectively. 
Eq.~\EqRegYMSTBreaking\ is a particular
instance of eq.~\EqRegSTBreakingct , and it is obtained by considering the
concrete realization of the Dimensional Regularization BRS transformations 
given in eq.~\DBRStrans . This equation is a rigorous equation, valid to 
all orders in the expansion in powers of $\hbar$ and fields.

Notice that $\Gamma_{\rm DReg}$ is not BRS invariant since 
${\cal S}_d(\Gamma_{\rm DReg})$ does not vanish. Indeed, 
the far right hand side of eq.~\EqRegYMSTBreaking ,
$$
\hat\Delta\cdot\Gamma_{\rm DReg} + 
\Delta_{\rm ct}\cdot\Gamma_{\rm DReg} + 
\intd^dx\,\sum_{\Phi}\Bigl\{
\bigl[{\delta S_{\rm ct}^{(n)}\over \delta K_\Phi(x)}
\cdot\Gamma_{\rm DReg}\bigr]
		{\delta\Gamma_{\rm DReg}\over\delta\Phi(x)}\Bigr\},
$$
is not zero. This term is the symmetry breaking term of the dimensionally
regularized theory defined by the Dimensional Regularization action 
$S^{(n)}_{\rm DReg}$. 

 We have to ask now about the renormalized counterpart of 
eq.~\EqRegYMSTBreaking . Since $S^{(n)}_{\rm DReg}$ contains the singular
counterterms needed to render non-singular at $d=4$ the 1PI functions of
the theory up to order $\hbar^n$, eq.~\Limit\  defines the
renormalized 1PI functional $\Gamma_{\rm ren}$ up to order $\hbar^n$:
$$
{\rm LIM}_{d\rightarrow 4}\;\Gamma_{\rm DReg}[\varphi,\Phi;K_{\Phi};\mu]=
\Gamma_{\rm ren}[\varphi,\Phi;K_{\Phi};\mu].
\eqno\numeq\namelasteq\LimitYM 
$$
Let us recall that the limiting process denoted by 
${\rm LIM}_{d\rightarrow 4}$ is acomplished by taking the ordinary limit 
$d\rightarrow 4$ first and then replacing with zero every hatted object.
The Dimensional Regularization scale has been introduced by hand as explained
in {\it v)} of subsection {\it 2.2.}
 
Now, by particularizing  eqs.~\EqRenSTBreaking --\Limit\
to the field theory under study, one concludes that the following
equation, which is the renormalized counterpart of eq.~\EqRegYMSTBreaking ,
 holds up to order $\hbar^n$:
$$
\eqalignno{
{\cal S}(\Gamma_{\rm ren})&\equiv
\intd^4 x\,\,\Bigl\{
\Tr\,{\delta\Gamma_{\rm ren}\over\delta\rho^\mu}
 {\delta\Gamma_{\rm ren}\over\delta A_\mu}+
\Tr\,{\delta\Gamma_{\rm ren}\over\delta\zeta}
 {\delta\Gamma_{\rm ren}\over\delta\om}+
\Tr\; B {\delta\Gamma_{\rm ren}\over\delta\bar\om} \cr
&+ {\delta\Gamma_{\rm ren}\over\delta\bar L} 
{\delta\Gamma_{\rm ren}\over\delta\psi}+
{\delta\Gamma_{\rm ren}\over\delta\bar R}
 {\delta\Gamma_{\rm ren}\over\delta\psi'}+
{\delta\Gamma_{\rm ren}\over\delta L}
 {\delta\Gamma_{\rm ren}\over\delta\bar\psi}+
{\delta\Gamma_{\rm ren}\over\delta R} 
{\delta\Gamma_{\rm ren}\over\delta\bar\psi'}\Bigr\}=
\Delta_{\rm breaking},&\numeq\cr
}
$$\namelasteq\EqRenYMSTBreaking
where $\Gamma_{\rm ren}$ is the dimensionally renormalized 1PI functional as
defined by eq.~\LimitYM . The BRS symmetry breaking term 
$\Delta_{\rm breaking}$ is given up to order $\hbar^n$ by the following 
``limit''
$$
\Delta_{\rm breaking}={\rm LIM}_{d\rightarrow 4}\Bigl\{
\hat\Delta\cdot\Gamma_{\rm DReg}+ 
\Delta_{\rm ct}\cdot\Gamma_{\rm DReg}+\! 
\intd^dx\,\sum_{\Phi}\Bigl\{
\bigl[{\delta S_{\rm ct}^{(n)}\over \delta K_\Phi(x)}
\cdot\Gamma_{\rm DReg}\bigr]
		{\delta\Gamma_{\rm DReg}\over\delta\Phi(x)}\Bigr\}\Bigr\}.
\eqno\numeq\namelasteq\YMBreaking              
$$

The fact \cite{\PiguetSorella} that if the anomaly cancellation conditions are
met \cite{\Alwit} the cohomology of the linearized BRS operator $b$ 
in eq.~\EqOperatorb\ is trivial on the space of local 
polynomials of ghost number one, guarantees that $S^{(n)}_{\rm fct}$ can be
chosen so that $\Delta_{\rm breaking}$ above vanishes whatever the value of
$n$. Hence, if the BRS symmetry is non-anomalous, non-symmetric counterterms 
can be added to the Dimensional Regularization classical action so that
the resulting renormalized action is BRS invariant: 
${\cal S}(\Gamma_{\rm ren})=0$. In this paper we shall compute 
$S^{(1)}_{\rm fct}$ explicitly (see subsection {\it 3.6.}). 

Let us close this subsection and express in terms of local operators in
four dimensions the order $\hbar$ contribution,
$\Delta^{(1)}_{\rm breaking}$,
to the symmetry breaking term given in eq.~\YMBreaking . Adapting
eqs.~\Breaking\ and \SingInsert\ to the regularized theory at hand
we conclude that
$$
\eqalignno{\Delta_{\rm breaking}^{(1)}=&
{\rm LIM}_{d\rightarrow 4}\Big\lbrace
\Bigl[\hat{\Delta}\cdot\Gamma_{\rm DReg}\Bigr]^{(1)}_{\rm singular} +
b_d\,S^{(1)}_{\rm sct}\Big\rbrace 
+\Bigl[{\rm N}[\hat\Delta]\cdot\Gamma_{\rm ren}\Bigr]^{(1)}+
b\,S_{\rm fct}^{(1)}\cr
=&\Bigl[{\rm N}[\hat\Delta]\cdot\Gamma_{\rm ren}\Bigr]^{(1)}+
b\,S_{\rm fct}^{(1)}.&\numeq\cr
}
$$\namelasteq\YMOneloopbreaking
where $\hat\Delta$ and $b_d$ are defined in eqs.~\EqBreaking\  and 
\EqOperatorb , respectively. Indeed, we shall show below that 
$$
\Bigl[\hat{\Delta}\cdot\Gamma_{\rm DReg}\Bigr]^{(1)}_{\rm singular}\!\! =
\ucpc {\TLR\over 2}\,{4\over3}\,{\hbar\over\eps}\,
      \intd^dx\,-\om^a\,\left\{\hat{\square}\,\bar\pr_\mu\bar A^{\mu\,a}\! +
      c^{abc}\,(\hat{\square}\bar A_\mu^a) \bar A^{\mu\,b}\right\}=
-b_d\,S^{(1)}_{\rm sct}.
$$

 The order $\hbar$ singular (pole)  contribution, 
$\Bigl[\hat{\Delta}\cdot\Gamma_{\rm DReg}\Bigr]^{(1)}_{\rm singular}$, to 
$\Delta\cdot\Gamma_{\rm DReg}$ only comes from the diagrams depicted in
figs.~5 and 6, after replacing  $\check \Delta$ with $\hat\Delta$. These
contributions read
$$
\eqalignno{
\Bigl[\Bigl[\hat{\Delta}\cdot\Gamma_{\rm DReg}\Bigr]^{\mu\, b a}_{Aw}
\Bigr]^{(1)}_{\rm singular}(p)&=
   \icpc \TLR \delta^{ab} {2\over3}\,{1\over\eps}\,\bar p^\mu
            {\hat p}^2 ;\cr
\Bigl[\Bigl[\hat{\Delta}\cdot\Gamma_{\rm DReg}\Bigr]^{\mu\nu\, bca}_{AAw}
\Bigr]^{(1)}_{\rm singular}(p_1,p_2)&=
   \icpc i c^{abc} {4\over3}\,{1\over\eps}\,
            ({\hat p}_{\!1}^2-{\hat p}_{\!2}^2)\,\gbar^{\mu\nu}.
                          &\numeq\cr
} \namelasteq\Poleone
$$

Now, $b_d\, L_g=b_d\,L_A=b_d\,L_\om=b_d\,\bar L_g=
b_d\,\bar L_{\psi^\pp}=0$, and therefore the only non- vanisning 
$b_d$-variation of $\Gamma^{(1)}_{0 {\rm sing}}$ comes from the third 
line of eq.~\EqGammaSing, which in turns is equal to eq.~\Poleone . The fact 
that  $S^{(1)}_{\rm sct}=-\Gamma^{(1)}_{0\, {\rm sing}}$ concludes the proof.

The next task to face is the computation of the order $\hbar$ contribution
to the anomalous insertion ${\rm N}[\hat\Delta]\cdot\Gamma_{\rm ren}$. 
We shall carry out this
calculation in the next section. The computation is somewhat simplyfied
if we bare in mind that the field $B$, enforcing the gauge-fixing condition,
has no dynamics (no interaction vertices) unless forcibly introduced  by
appropriate choice  of $S^{(n)}_{\rm fct}$ (finite countertems). Hence,
it will be very advisable to imposse that the finite counterms be independent
of $B$, so that the only contribution to $\Gamma_{\rm DReg}$ involving $B$ is 
order $\hbar^0$ and given by $S_0$ in eq.~\DRegclass.  It is thus plain
that the so-called gauge-fixing equation
$$ 
{\cal B} (\Gamma_{\rm ren})\equiv{\delta \Gamma_{\rm ren}
\over\delta B}-\pr_\mu A^\mu-\a\,B=0\,, 
			\eqno\numeq
$$ \namelasteq\EqGaugeFixingCondition
holds for the renormalized theory. Notice that eq.~\EqGaugeFixingCondition\ 
does not clash with restoring the BRS symmetry since $\Delta_{\rm breaking}$
defined in eq.~\YMBreaking\ does not depend on $B$.  Of course,
eq.~\EqGaugeFixingCondition\ is the equation of motion of $B$.

Finally, the ghost equation is another interesting equation. It is just the
equation of motion  of $\bar\om$:
$$
{\cal G}\,\Gamma_{\rm ren}\equiv\left\{ {\delta\over\delta\bar\om} + 
	\pr_\mu {\delta\over\delta\rho_\mu}\right\}\;\Gamma_{\rm ren} =0, 
 \eqno\numeq
$$ \namelasteq\EqGhostEquation
and is a consequence of eq.~\EqRenYMSTBreaking\ and the gauge-fixing 
equation. Some functional differentiation and the fact that 
$\Delta_{\rm breaking}$ does not depend on $B$ lead to eq.~\EqGhostEquation .
A furher simplication, the ghost equation implies that the functional 
which satisfy it depends on $\rho_\mu$ and $\bar\om$ through the combination 
$\tilde\rho_\mu=\rho_\mu+\pr_\mu\bar\om$. Finally, it is not difficult to 
come to  the conclusion that this very simplification applies to each term
on the right hand side of eq.~\YMBreaking\  
and $S^{(n)}_{\rm ct}$ in eq.~\Dchiralact .

\bigskip

\subsection{3.5. Expansion of the anomalous insertion}

The anomalous insertion  ${\rm N}[\hat\Delta]\cdot\Gamma_{\rm ren}$ has 
ghost number $+1$ and it is the insertion into the 1PI functional 
$\Gamma_{\rm ren}$ of an evanescent integrated  
polynomial operator of ultraviolet dimension $4$ with neither 
free Lorentz indices nor free  indices for 
the gauge group. This anomalous insertion is  a functional of 
$\psi$, $\psi'$, $A_\mu$, $\bar\om$ (only through $\tilde\rho_\mu$), 
$\om$, but not of $B$, 
and {\it also\/} of the external fields $L$, $\bar L$, $R$, $\bar R$, 
$\rho_\mu$ (only through $\tilde\rho_\mu$) and $\zeta$. It is thus compulsory 
to generalize first eq.~\EqBonneauIdentities\ so that 
it  includes also Feynman diagrams with the external fields as vertices.

Due to the explicit power-counting in our model, 
only diagrams with either no or one external fields can be divergent. 
Consider a diagram with the vertex  $K_{\Phi} s\Phi$ 
($\Phi$ denotes any field which undergoes non-linear BRS 
transformations). The Feynman rule in  momentum space of this vertex 
is an integration over a momentum $q'$, 
which is absorbed in the definition of the Fourier transform of the 
function with an insertion of an operator,
and the factors ${i\over\hbar}\,\tilde K_{\Phi}(q')$ times the Feynman rule of 
the operator insertion $s\Phi$ at momentum $q'$. The singular part is also
a polynomial in $q'$, therefore it can be expanded, together with the rest
of momenta, in a finite Taylor series in $q'$. 
A factor $q'{}^{\mu_1}\cdots q'{}^{\mu_s}$ together with $\tilde K(q')$ will 
lead to a $(-i)^s\;\pr_{\mu_1}\cdots\pr_{\mu_s}\,K_{\Phi}(x)$ which multiply 
the insertion of the monomial operator obtained with the rest of momenta.
Therefore the Bonneau identities (eq.~\EqBonneauIdentities ) are 
generalized to:
$$
\eqalignno{
&\N[\hat\Delta](x)\cdot\Gamma_{\rm ren} = -
\sum_{n=0}^4 \;\; \sum_{\{j_1\,\ldots\, j_n\}}
 \bigg[             
 \sum_{r=0}^{\delta(J)}\sum_{\scriptscriptstyle\{i_1\,\ldots\, i_r\} \atop
         \scriptscriptstyle 1\le i_j\le n  }  \cr
&\qquad\Big\{ {(-i)^r\over r!}
	{\pr^r\over\pr p_{i_1}^{\mu_1}\cdots\pr p_{i_{r}}^{\mu_r} }
\,(-i\hbar)\,\rsp\,
\BonneauGraph{\tilde\phi_{j_1}(p_1)\ldots\tilde\phi_{j_n}(p_n)}
	\Big|_{p_i=0,\;\gtah=0}  \Big\}                        \cr
&\qquad\quad\times\;\; \N\Bigl[ {1\over n!} \prod_{k=n}^1 
\Bigl\{\Bigl(\prod_{\{\a/i_\a=k\}}
 \!\!\pr_{\mu_\a}\Bigr)\phi_{j_k}\Bigr\}\Bigr] (x)
		\cdot\Gamma_{\rm ren}                                          
\;+\,\cr
&\quad+\;\sum_{\Phi}\;
\sum_{s,t\atop 0\le s+t\le\delta(J;\Phi)}
\sum_{\scriptscriptstyle\{i_1\,\ldots\, i_s\} \atop
			\scriptscriptstyle 1\le i_j\le n  }
	\Big\{ 
{(-i)^{s+t}\over (s+t)!}
 {\pr^{s+t}\over\pr p_{i_1}^{\mu_1}\cdots\pr p_{i_s}^{\mu_s}
      \pr p_{n+1}^{\nu_1}\cdots\pr p_{n+1}^{\nu_t}} \cr
&\qquad\rsp
\BonneauGraph{\tilde\phi_{j_1}(p_1)\ldots\tilde\phi_{j_n}(p_n);
			\N[s\Phi](p_{n+1})}
	\Big|_{p_i=0,\;\gtah\equiv0}  \Big\}                        \cr
&\qquad\quad\times\;\; (\pr_{\nu_1}\ldots\pr_{\nu_t}\,K_{\Phi})(x)\;
	\N\Bigl[ {1\over n!} \prod_{k=n}^1 
   \Bigl\{\Bigl(\prod_{\{\a/i_\a=k\}}\!\!\pr_{\mu_\a}\Bigr)\phi_{j_k}
                                                   \Bigr\}\Bigr] (x)
      \cdot\Gamma_{\rm ren} \quad  \bigg]\; ,                &\numeq\cr
} \namelasteq\EqBonneauGeneralization 
$$
where $J\equiv\{j_1,\cdots,j_n\}$ and the indexes $j_k,\;k=1,\cdots,n$ label 
the different types of quantum fields, with the proviso that
fields having different values of gauge group indices are taken as different.
The symbols $\delta(J)$, $\delta(J;\Phi)$ denote,  
respectively, the overall degrees of ultraviolet divergence of the 
1PI functions $\Graph{\tilde\phi_{j_1}(p_1)\ldots\tilde\phi_{j_n}(p_n)}$ and
$\Graph{\tilde\phi_{j_1}(p_1)\ldots\tilde\phi_{j_n}(p_n);\N[s\Phi](p_{n+1})}$.
Notice that since our propagators are massless the only non-vanishing 
contributions to the left hand side of eq.~\EqBonneauGeneralization\ come from
$r=\delta(J)$ and $s+t=\delta(J;\Phi)$. $\prod_{k=n}^1$ means that fields in
the product are ordered from left to right according to decreasing values of 
$k$.

\medskip
The previous formula gives rise to an expansion of 
${\rm N}[\hat\Delta](x)\cdot\Gamma_{\rm ren}$ in terms of both standard and 
evanescent monomials \cite{\BonneauA,\BonneauB} (see section {\it 2.4.}).
If we denote the monomials generically by the letter $\M$, the expansion  
will have the form 

$$
\eqalignno{
\N[\hat\Delta]\cdot\Gamma_{\rm ren}&=
	\intd^4x\;\N[\hat\Delta](x)\cdot\Gamma_{\rm ren}\cr
&=
\sum_i \b_i^{abc\ldots}\; \N[\M_i^{abc\ldots}]\cdot\Gamma_{\rm ren} +
\sum_j \hat\b_j^{abc\ldots}\; \N[\hat\M_j^{abc\ldots}]
\cdot\Gamma_{\rm ren}  \cr
&=\sum_i {\bar \beta}_i^{abc\ldots}\; \N[{\bar\M}_i^{abc\ldots}]
\cdot\Gamma_{\rm ren} +
	\sum_j \hat\b_j^{\prime abc\ldots}\; 
\N[\hat\M_j^{abc\ldots}]\cdot\Gamma_{\rm ren},       &\numeq\cr
} 
$$\namelasteq\EqBonneauBetas
with each coefficient being a formal series in $\hbar$ (starting at order
$\hbar^1$). The latin letters $a bc $ stand for the gauge group indices. 
We shall assume that repeated gauge group indices are summed over. The
monomials in eq.~\EqBonneauBetas\ have no free Lorentz index. The 
objects $\{{\bar\M}_i^{abc\ldots}\}$, called ``barred'' nonomials, 
are  monomials where all Lorentz contractions are carried out with  
${\bar g}^{\mu\nu}$. Notice that there is a one-to-one 
correspondence between $\{\M_i^{abc\ldots}\}$ and 
$\{{\bar\M}_i^{abc\ldots}\}$. The objects $\{{\hat\M}_i^{abc\ldots}\}$, 
called ``hatted'' nonomials, are evanescent operators.

Notice that in our case the monomials will be intregated local functionals
of the fields and its derivatives with ghost number 
one and ultraviolet dimension $4$.

Expanding in the same way all the normal insertions of  the 
evanescent monomials ${\hat\M}_i^{abc\ldots}$ on the right hand side of 
eq.~\EqBonneauBetas\ and solving the system of Bonneau identities 
we would get finally the true  expansion of the anomalous
insertions in terms of  the basis of standard (or barred) normal 
products (also called standard quantum insertions):
$$
\eqalignno{
\N[\hat\Delta]\cdot\Gamma_{\rm ren}&= 
   \sum_i k_{i}^{abc\ldots}\; \N[\M_i^{abc\ldots}]\cdot\Gamma_{\rm ren}   \cr
&=
	\sum_i {\bar k}_{i}^{abc\ldots}\; 
            \N[{\bar\M}_i^{abc\ldots}]\cdot\Gamma_{\rm ren};     \cr
}
$$
where the barred monomials  simply means again that all the Lorentz 
contractions are done with the $\gbar$ tensor.

Let us shorthand $\intd^dx$ to $\int$. The list of integrated monomials of
ghost number $1$, ultraviolet dimension $4$ and no free of Lorentz 
indices which is relevant to our computation is the following:

\noindent $*$ a) {\sl Monomials with 1 $\om$ and 1 $A$}

{\settabs 3 \columns
\openup1\jot
\+$\M_{1}^{ab}\equiv\intd^dx\;\om^a\,\square\,\pr_\mu A^{b\mu}$. \cr
}
\bigskip

\noindent $*$ b) {\sl Monomials with 1 $\om$ and 2 $A$'s and no
	$\eps_{\mu\nu\a\b}$ tensor}

{\settabs 2 \columns
\openup1\jot
\+$\M_{2}^{abc}\equiv\int\!\om^a\,(\square\,A_\mu^b)A^{c\mu}$,
	&$\M_{3}^{abc}\equiv\int\!\om^a\,(\pr_\mu A_\nu^b)(\pr^\mu A^{c\nu})=
		\M_{3}^{acb}$,        \cr
\+$\M_{4}^{abc}\equiv\int\!\om^a\,(\pr_\mu\pr_\nu A^{b\mu}) A^{c\nu}$,
	&$\M_{5}^{abc}\equiv\int\!\om^a\,(\pr_\mu A^{b\mu})(\pr_\nu A^{c\nu})=
		\M_{5}^{acb}$,        \cr
\+$\M_{6}^{abc}\equiv\int\!\om^a\,(\pr_\mu A_\nu^b)(\pr^\nu A^{c\mu})=
		\M_{6}^{acb}$.          \cr
}
\bigskip

\noindent $*$ c) {\sl Monomials with 1 $\om$ and 3 $A$'s and no
	$\eps_{\mu\nu\a\b}$ tensor}

{\settabs 2 \columns
\openup1\jot
\+$\M_{7}^{abcd}\equiv\int\!\om^a\,(\pr_\mu A^{b\mu}) A_\nu^c A^{d\nu}=
	\M_{7}^{abdc}$,
	&$\M_{8}^{abcd}\equiv\int\!\om^a\,(\pr_\mu A_\nu^b)
				A^{c\mu} A^{d\nu}$.        \cr
}
\bigskip

\noindent $*$ d) {\sl Monomials with 1 $\om$ and 4 $A$'s and no
	$\eps_{\mu\nu\a\b}$ tensor}
	
{\settabs 2 \columns
\openup1\jot
\+$\M_9^{abcde}\equiv\int\!\om^a\,A_\mu^b A^{c\mu} A_\nu^d A^{e\nu}=
	\M_9^{acbde}= \M_9^{abced}=\M_9^{adebc}$.                     \cr
}
\bigskip

Since the formulae with fermions $\psi$ and $\psi'$ are very
similar we collect them in a way obvious to interpret. In general, we will
denote with a roman letter L or R the fermionic lines or loops in Feynman
diagrams, the monomials, their coefficients,~$\ldots$ corresponding to the
fermions $\psi$, whose interaction is left-handed, or to the fermions
fermions $\psi'$, whose interaction is right-handed.
\medskip
\noindent $*$ e) {\sl Monomials with 1 $\om$ and $\psi,\bar\psi$
		($\psi',\bar\psi'$)}
	
{\settabs 2 \columns
\openup1\jot
\+$\M_{10\LpR}^{a,ij}\equiv\int\!\om^a\,\bar\psi^\pp_i\gambar^\mu\PLpR
		\pr_\mu\psi^\pp_j$,
	&$\M_{11\LpR}^{a,ij}\equiv\int\!\om^a\,(\pr_\mu\bar\psi^\pp_i)
		\gambar^\mu\PLpR\pr_\mu\psi^\pp_j$,         \cr
\+$i$, $j$ denote the group indices of the corresponding fermion fields.  \cr
}
\bigskip

\noindent $*$ f) {\sl Monomials with 1 $\om$, 1 $A$ and  
	$\psi,\bar\psi$ ($\psi',\bar\psi'$)}
	
{\settabs 2 \columns
\openup1\jot
\+$\M_{12\LpR}^{ab,ij}\equiv\int\!
	\om^a\,\bar\psi^\pp_i\gambar^\mu\PLpR\psi^\pp_j A_\mu^b$.
	\cr
}

\bigskip

\noindent $*$ g) {\sl Monomials with fermions and external fields}
	
{\settabs 2 \columns
\openup1\jot
\+$\M_{14\LpR}^{ab,ij}\equiv\int\!\om^a\om^b\,\bar L_i(R_i)\PLpR\psi^\pp_j$,
	&$\M_{15\LpR}^{ab,ij}\equiv\int\!
		\om^a\om^b\,\bar\psi_i^{\pp}\PRpL L_j(R_j)$     \cr
}

\bigskip

We have not considered operators like 
$\om^a\,\bar\psi_i\gambar^\mu\PR\pr_\mu\psi_j$
as  admissible monomials since it  will not be generated by the 
Bonneau identities due to the form of our choice for fermion 
vertex at order $\hbar^0$.

Will shall adopt the notation that the {$\,\{\cdots\}\,$ enclosing 
indices denotes symmetrization and $\,[\cdots]\,$ antisymmetrization.

\bigskip

\noindent $*$ h) {\sl Monomials with $\eps_{\mu\nu\a\b}$}

{\settabs 2 \columns
\openup1\jot
\+$\M_{50}^{abc}\equiv\int\!
	\eps_{\mu\nu\a\b}\,\om^a\,(\pr^\a A^{b\mu})(\pr^\b A^{c\nu})=
	\M_{50}^{acb}$,
 &$\M_{51}^{abcd}\!\equiv\!\int\!
\eps_{\mu\nu\rho\a}\,\om^a(\pr^\a A^{b\mu}) A^{c\nu} A^{d\rho}\!=
		\!-\M_{51}^{abdc}\!\! ,$  \cr
\+$\M_{52}^{abc}\equiv\int\!
\eps_{\mu\nu\rho\la}\,\om^a\,A^{b\mu} A^{c\nu}A^{d\rho} A^{e\la}=
		\M_{52}^{a[bcde]}$;    \cr
}                                                              
\bigskip

\noindent $*$ i) {\sl Monomials with $\tilde\rho$ or $\zeta$}

Notice that it is possible to construct other monomials with ghost number $+1$
and dimension 4:
$\tilde\rho_\mu^a$ $\om^b\om^c$ $A^{d\mu}$, 
$\tilde\rho_\mu^a (\pr^\mu\om^b) \om^c$,
$\zeta^a\om^b\om^c\om^d$. The formulae for their coefficients in 
the Bonneau expansion are also easily obtained, but we do not write 
them because in the one loop computation all this coefficients turn out 
to be zero. This is clear because all the 1PI
functions we would have to compute involve at least two loops. 

\noindent $*$ {\sl Anomalous monomials, {\it i.e.} with some $\ghat$}

Take all the monomials written above and write all monomials obtained
by adding hats to the Lorentz indices in all possible ways. No  
hatted index should appear contracted with the $\eps_{\mu\nu\a\b}$,
because this contraction vanish.
Notice that with the  fermionic vertices 
$\bar\psi^{\pp}\gambar^\mu P_{\rm L(R)} T_{\rm L(R)}^a\psi^{\pp} A_\mu^a=
\bar\psi^{\pp}P_{\rm R(L)}\gambar^\mu P_{\rm L(R)} 
T_{\rm L(R)}^a\psi^{\pp} A_\mu^a$
in the Dimensional Regularization classical action $S_0$, neither
$\om^a\,\bar\psi_i\PL\gamhat^\mu\PL\pr_\mu\psi_j$ 
nor 
$\om^a\,\bar\psi^{\pp}_i\PR\gamhat^\mu\PR\pr_\mu\psi^{\pp}_j$ occur
in the Bonneau expasion we are considering; however, with the vertex
$\bar\psi^{\pp}\gamma^\mu P_{\rm L(R)}T_{\rm L(R)}^a\psi^{\pp} 
A_\mu^a=$ $A_\mu^a$
$(\bar\psi^{\pp}P_{\rm R(L)}\gambar^\mu P_{\rm L(R)} 
T_{\rm L(R)}^a\psi^{\pp}$ $+$
$\bar\psi^{\pp} P_{\rm L(R)}\gamhat^\mu T_{\rm L(R)}^a
 P_{\rm L(R)}\psi^{\pp})$, both do.

\bigskip

It is convenient to use the expansion in terms of barred and
hatted monomials (so the 1PI functions with $\check\Delta$ inserted have 
to be expressed in terms of $\gbar$ and $\ghat$). 
Then, the formulae of the coefficients for all orders and 
the result for the first order read (see figs.~7-12):

\medskip
\noindent $*$ a) {\sl From 1PI functions with one $\om$ and one $A$}
   (fig.~5)
$$
\eqalignno{
\b_{1}^{ab}&=-(-i)^3 \coefZero{\om^a(p_2\equiv -p_1)A_\mu^b(p_1)}
				\bar p_1{}^2 \;\bar p_{1\mu}     \cr
&=-{1\over48} 
	{(-i)^3\pr^3\over\pr\bar p_1^\nu\pr\bar p_{1\nu}\pr\bar p_{1\mu}}
\,\rsp\BonneauGraphZero{\om^a(-p_1)A_\mu^b(p_1)}\Big|_{p_i\equiv0}  \cr
&=-\ucpc {\TL+\TR\over3}\delta^{ab}\, \hbar^1 + O(\hbar^2),   &\numeq\cr
}
$$

\midinsert
\def\graphwidth{1.5in}        
$$
\eqalign{\epsfxsize=\graphwidth\epsffile{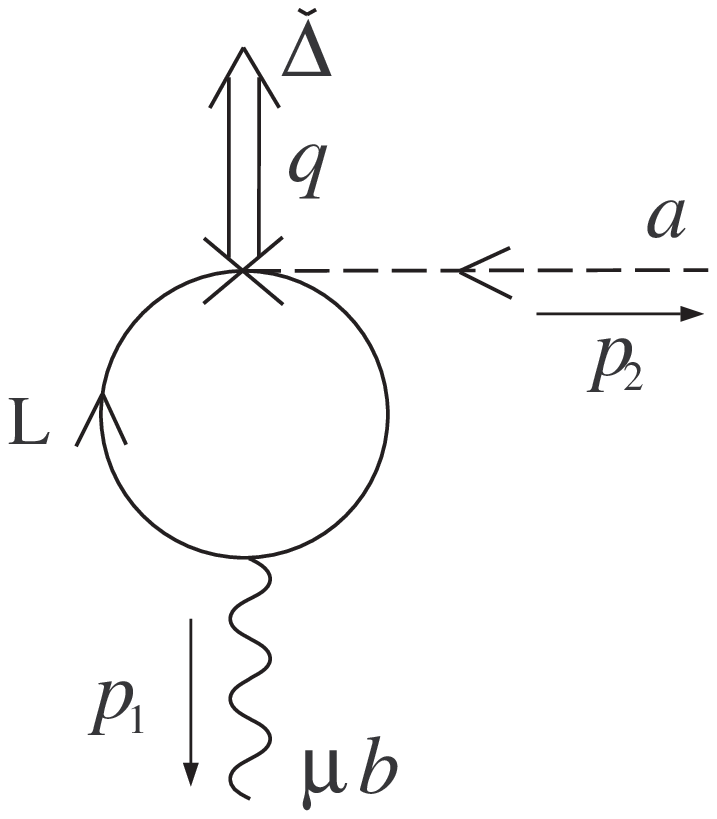}}
  \qquad\eqalign{+\qquad \hbox{idem with R fermions}}
$$
\vskip -0.3cm
\narrower\noindent {\bf Figure 5:}
{\eightrm 1PI Feynman diagrams needed to compute eq.~\lasteq}
\vskip 0.4cm
\endinsert

\noindent $*$ b) {\sl From 1PI functions with one $\om$ and two $A$s,  
   and no $\eps_{\mu\nu\a\b}$ tensor} (fig.~6)

$$
\eqalignno{
\{\b_{2}^{abc}&, \,\b_{4}^{abc}\}=  \cr
&=-(-i)^2 \coefZero{\om^a(p_3\equiv -p_1-p_2)A_\mu^b(p_1)A_\nu^c(p_2)}   \cr
&\qquad  \{\bar p_1{}^2 \;\gbar_{\mu\nu},
			\,\bar p_{1\mu} \;\bar p_{1\nu}\}\quad{\rm respectively}    \cr
&=\ucpc (\TL+\TR)\,c^{abc}\,\{-{1\over3}, {2\over3}\}\,\hbar^1 
+ O(\hbar^2), &\numeq\cr
\{\b_{3}^{abc}&=\b_{3}^{acb}, \,\b_{5}^{abc}=\b_{5}^{acb}, 
				\,\b_{6}^{abc}=\b_{6}^{acb}\}=  \cr
&=-{(-i)^2 \over2} \coefZero{\om^a(p_3\equiv -p_1-p_2)%
	A_\mu^b(p_1)A_\nu^c(p_2)}          \cr
&\qquad  \{\bar p_1\cdot \bar p_2 \;\gbar_{\mu\nu},
			\,\bar p_{1\mu} \;\bar p_{2\nu},
	\,\bar p_{1\nu} \;\bar p_{2\mu}\}\quad{\rm resp.}  \cr
&= 0\,\hbar^1 + O(\hbar^2),       &\numeq\cr
}
$$
\medskip

\midinsert
\def\graphwidth{1.5in}
$$
\eqalign{\epsfxsize=\graphwidth\epsffile{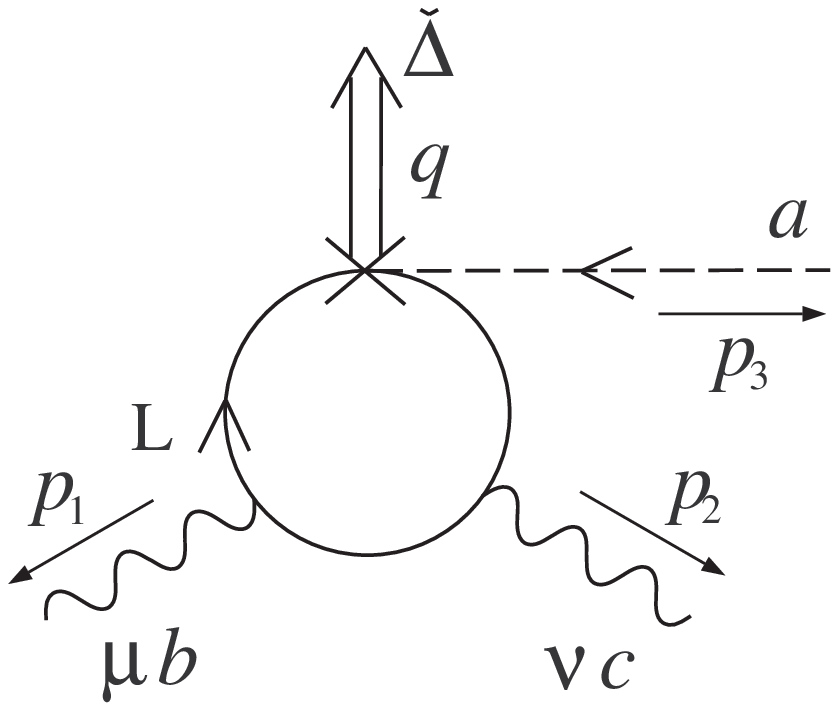}}
  \quad\eqalign{+}\quad
		\eqalign{\matrix{\hbox{permutations} \cr
				\hbox{of}\cr
				\hbox{bosonic legs}\cr
								}
					}
	\quad\eqalign{+}\quad
  \eqalign{\hbox{idem with R fermions}}
$$
\vskip -0.3cm
\narrower\noindent {\bf Figure 6:}
{\eightrm 1PI Feynman diagrams needed to compute eqs.~\beforelasteq,  
\lasteq\ and~(61)}
\vskip 0.4cm
\endinsert
\noindent $*$ c) {\sl From 1PI functions with one $\om$ and three $A$s, 
   and no $\eps_{\mu\nu\a\b}$ tensor} (fig.~7)

$$
\eqalignno{
\b_{7}^{abcd}&=\b_{7}^{abdc}=    \cr
&=-{(-i)\over2} \coefZero{\om^a(-\sum_i^3 p_i)%
		A_\mu^b(p_1)A_\nu^c(p_2)A_\rho^d(p_3)} \cr
&\qquad  \bar p_{1\mu} \;\gbar_{\nu\rho} \cr
&=\ucpc {1\over6} [\,\TLR^{abcd}+\TLR^{acdb}+\TLR^{abdc}+ 
				\TLR^{adcb}-\TLR^{acbd}-\TLR^{adbc}\,]
		\,\hbar^1 + O(\hbar^2),                &\numeq\cr
\b_{8}^{abcd}&=
-(-i) \coefZero{\om^a(-\sum_i^3 p_i)%
	A_\mu^b(p_1)A_\nu^c(p_2)A_\rho^d(p_3)} \cr
&\qquad  (\bar p_{1\nu} \;\gbar_{\mu\rho} + 
	 \bar p_{1\rho} \;\gbar_{\mu\nu}) \cr
&=\ucpc {1\over3} [\,\TLR^{abdc}+\TLR^{acbd}+\TLR^{acdb}+ 
				\TLR^{adbc}-\TLR^{abcd}-\TLR^{adcb}\,]
		\,\hbar^1 + O(\hbar^2),                &\numeq\cr
}
$$
\medskip
\midinsert
\def\graphwidth{2in}
$$
\eqalign{\epsfxsize=\graphwidth\epsffile{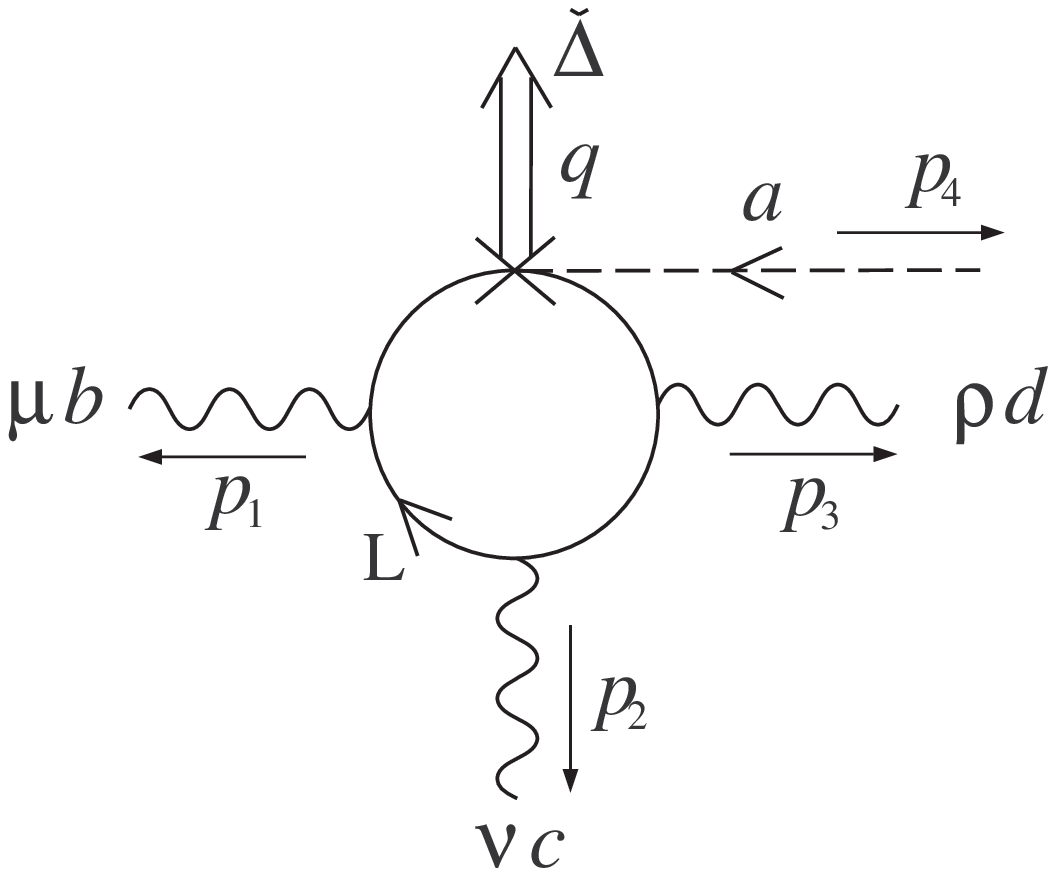}}
  \quad\eqalign{+}\quad
		\eqalign{\matrix{\hbox{permutations} \cr
				\hbox{of}\cr
				\hbox{bosonic legs}\cr
								}
					}
	\quad\eqalign{+}\quad
  \eqalign{\hbox{idem with R fermions}}
$$
\vskip -0.3cm
\narrower\noindent {\bf Figure 7:}
{\eightrm 1PI Feynman diagrams needed to compute 
eqs.~\beforelasteq, \lasteq\ and~(62)}
\vskip 0.4cm
\endinsert
\medskip
\midinsert
\def\graphwidth{2in}
$$
\eqalign{\epsfxsize=\graphwidth\epsffile{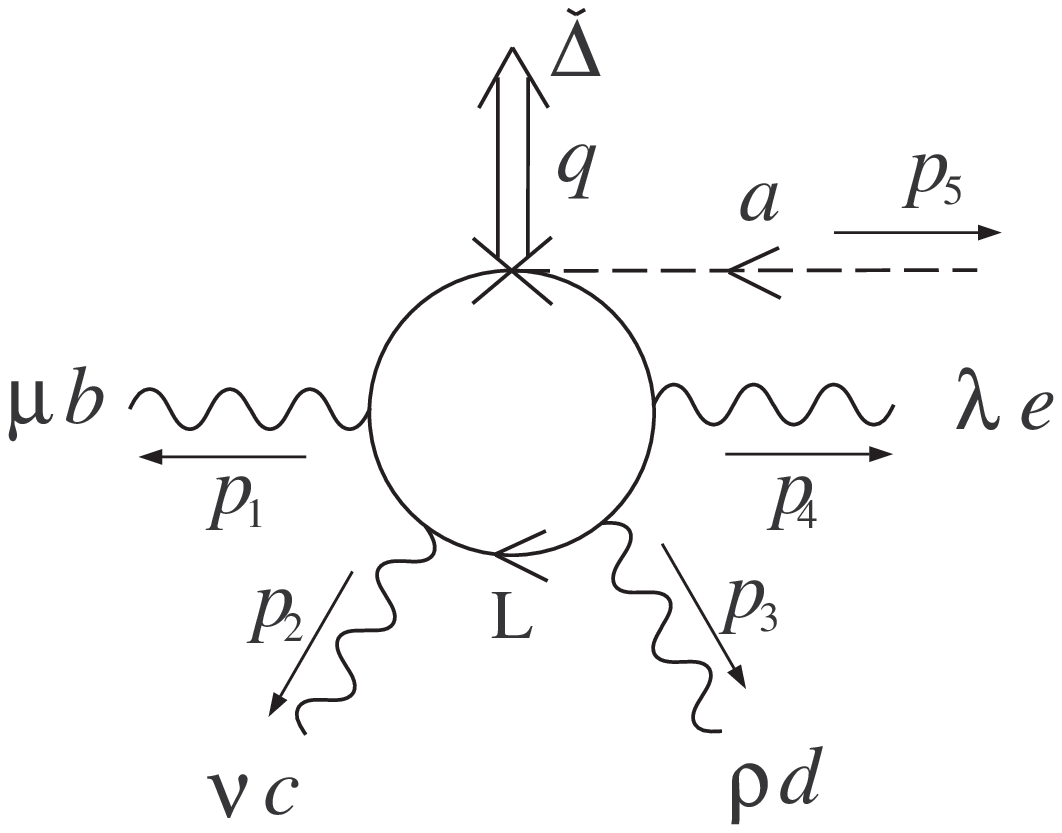}}
  \quad\eqalign{+}\quad
		\eqalign{\matrix{\hbox{permutations} \cr
				\hbox{of}\cr
			\hbox{bosonic legs}\cr
								}
					}
	\quad\eqalign{+}\quad
  \eqalign{\hbox{idem with R fermions}}
$$
\vskip -0.3cm
\narrower\noindent {\bf Figure 8:}
{\eightrm 1PI Feynman diagrams needed to compute eqs.~\relativeq1\ and~
(63)}
\vskip 0.4cm
\endinsert
\noindent $*$ d) {\sl From 1PI functions with one $\om$ and four $A$s, 
					and no $\eps_{\mu\nu\a\b}$ tensor} 
               (fig.~8) 

$$
\eqalignno{
\b_{9}^{abcde}&=\b_{9}^{a\{bc\}\{de\}}=\b_{9}^{a\{de\}\{bc\}}=     \cr
&=-{1\over8} \coefZero{\om^a A_\mu^b A_\nu^c A_\rho^d A_\la^e} 
		\; \gbar_{\mu\nu} \cr
&=0\,\hbar^1 + O(\hbar^2),                         &\numeq\cr
}
$$

\medskip

In principle, due to the presence of $\gam5$, there should be also
evanescent normal products in the expansion of the anomalous
insertion ${\rm N}[\hat\Delta]\cdot\Gamma_{\rm ren}$. Indeed,
the coefficient of $\int\!\N[\om^a\,\hat{\square}\bar\pr_\mu A^{b\mu}]$ is
$$
\eqalignno{
&-(-i)^3 \coefZero{\om^a(p_2\equiv -p_1)A_\mu^b(p_1)}
			\hat p_1{}^2 \;\bar p_{1\mu}     \cr
=&-{1\over8(d-4)} 
	{(-i)^3\pr^3\over\pr\hat p_1^\nu\pr\hat p_{1\nu}\pr\bar p_{1\mu}}
\,\rsp\BonneauGraphZero{\om^a(-p_1)A_\mu^b(p_1)}\Big|_{p_i\equiv0}  \cr
=&-\ucpc {\TL+\TR\over3}\delta^{ab}\, \hbar^1 + O(\hbar^2),   \cr
}
$$
However, a symplification occur, for the coefficient of 
$\int\!\N[\om^a\,\bar{\square}\hat\pr_\mu A^{b\mu}]$ is
$$
\eqalignno{
&-(-i)^3 \coefZero{\om^a(p_2\equiv -p_1)A_\mu^b(p_1)}
			\bar p_1{}^2 \;\hat p_{1\mu}     \cr
&=-{1\over8(d-4)} 
	{(-i)^3\pr^3\over\pr\bar p_1^\nu\pr\bar p_{1\nu}\pr\hat p_{1\mu}}
\,\rsp\BonneauGraphZero{\om^a(-p_1)A_\mu^b(p_1)}\Big|_{p_i\equiv0}  \cr
&=0\, \hbar^1 + O(\hbar^2),   \cr
}
$$
and so on. Due to our expansion in
barred and hatted objects rather than in standard and hatted ones, and because
of the form of the regularized interaction it turns out that {\it the rest of
anomalous coefficients are 0 in the one-loop approximation}. Nevertheless, 
as we know from subsection {\it 2.4.}, 
these anomalous coefficients only matter at the next perturbative 
order, so this vanishing of the coefficiens is only
a simplification relevant to higher order computations.

\medskip
\midinsert

{\settabs 3\columns \def\graphwidth{1.5in}   
\eightpoint
\+\hfil\epsfxsize=\graphwidth\epsffile{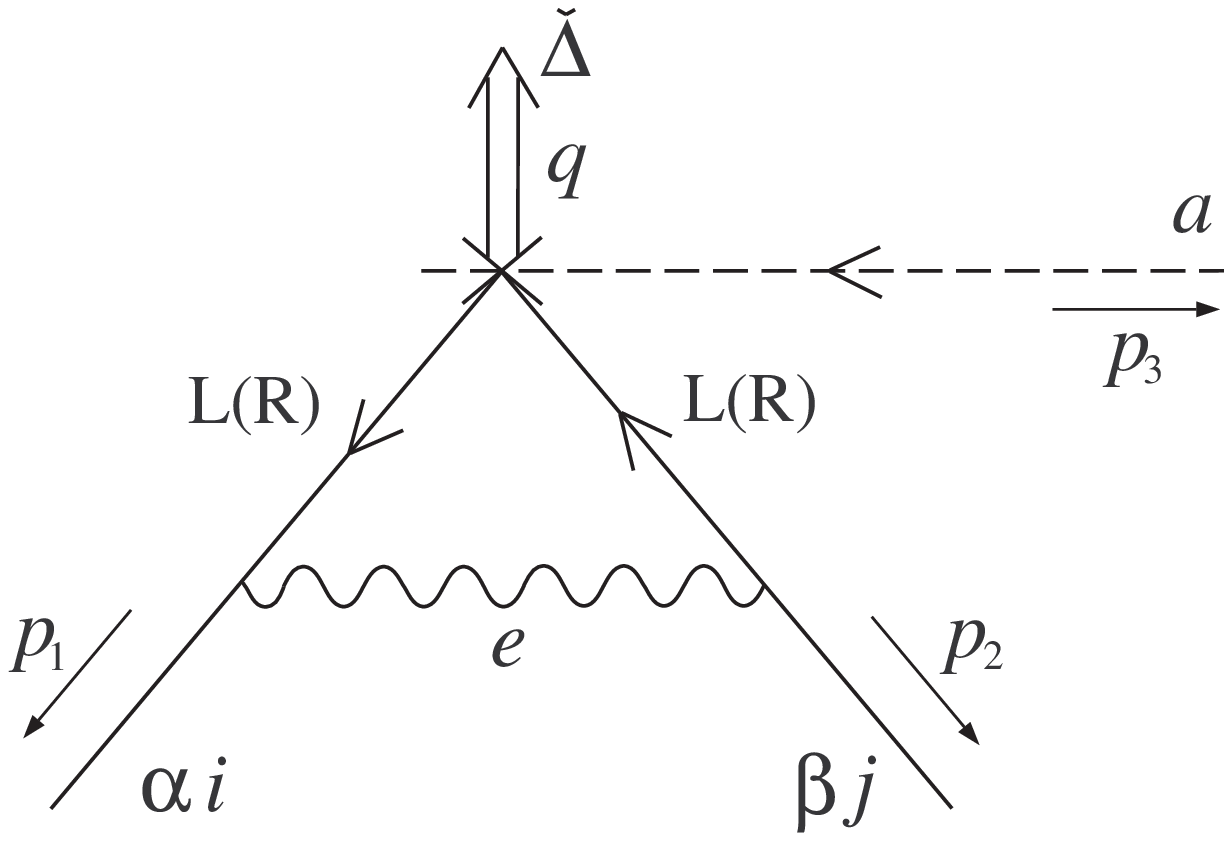}\hfil
	&\hfil\epsfxsize=\graphwidth\epsffile{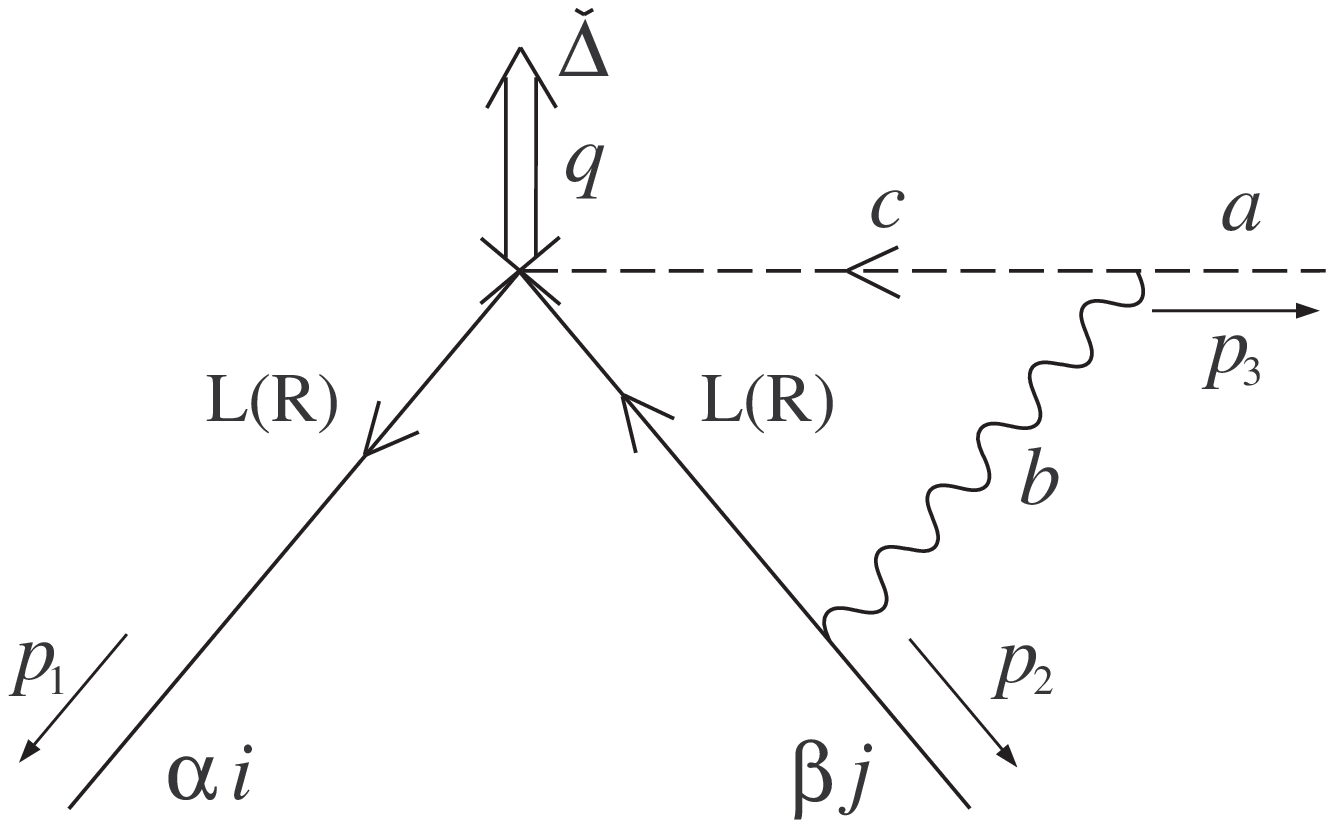}\hfil
	&\hfil\epsfxsize=\graphwidth\epsffile{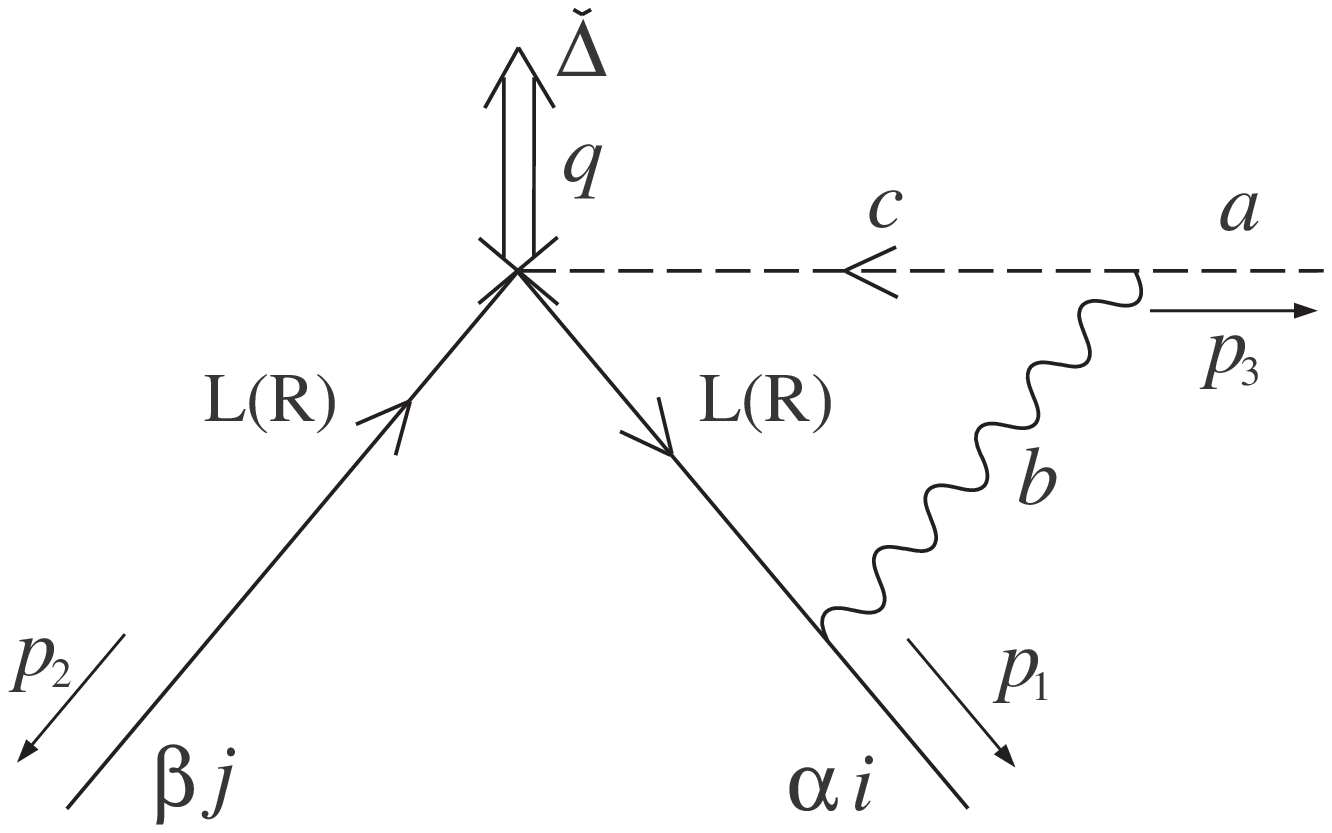}\hfil \cr
\+\hfil$(\TLpR^e \TLpR^a \TLpR^e)_{ij}=$\hfil
	&\hfil $c_{bac}(\TLpR^{[c}\TLpR^{b]})_{ij}=$\hfil
	&\hfil $c_{bac}(\TLpR^{[b}\TLpR^{c]})_{ij}=$\hfil&         \cr
\+\hfil$=(\CLpR-{\CA \over2})\,(\TLpR^a)_{ij}$\hfil
	&\hfil $={i\over2}c_{acb}c_{cbe}(\TLpR^e)_{ij}=$\hfil
	&\hfil $=-{i\over2}c_{abc}c_{bce}(\TLpR^e)_{ij}=$\hfil&         \cr
\+ 
	&\hfil $={i\over2}\,\CA (\TLpR^a)_{ij}$\hfil
	&\hfil $=-{i\over2}\,\CA(\TLpR^a)_{ij}$\hfil&         \cr
}
\vskip 12pt
\narrower\noindent {\bf Figure 9:}
{\eightrm 1PI Feynman diagrams needed to compute eqs.~\relativeq1\
and~\relativeq2.
(Notice that the gauge group structure is explicitly shown 
under these and subsequent diagrams) }
\vskip 0.4cm
\endinsert
\noindent $*$ e) {\sl From 1PI functions with one $\om$ and 
 one $\psi,\bar\psi$ ($\psi',\bar\psi'$) pair} (fig.~9)

$$
\eqalignno{
\b_{10\LpR}^{a,ij}
&=-(-i) 
	\coefZero{\om^a(p_3\equiv -p_1-p_2)%
		\psi_{\b j}^{\pp}(p_2)\bar\psi_{\a i}^{\pp}(p_1)}   \cr
&\qquad  (\bar\pslash_2\PLpR)_{\a\b}               \cr
&=-{1\over4\cdot2} 
	{(-i)\pr\over\pr\bar p_2^\mu}
	\rsp\BonneauGraphZero{\om^a(-p_1-p_2)\psi_{\b j}^{\pp}(p_2)
			\bar\psi_{\a i}^{\pp}(p_1)}
\Big|_{p_i\!\!\equiv0}\!\!\!\!\!\!\cdot (\PLpR\gambar^\mu)_{\!\b\a} \cr
&={i\over4\cdot2} \Tr\,\Bigl[ 
	{\pr\over\pr\bar p_2^\mu}
	\,\rsp\BonneauGraphZero{\om^a(-p_1-p_2)%
		\psi_j^{\pp}(p_2)\bar\psi_i^{\pp}(p_1)}
	\Big|_{p_i\equiv0} \PLpR\gambar^\mu\,\Bigr] \cr
&=-\ucpc g^2 \bigl\{ (\CLpR-{\CA\over4}) [\,\TLpR^a\,]^{ij}+ 
	(\aprime-1) \bigl({\CLpR\over6}-{\CA\over4}\bigr) 
	[\,\TLpR^a\,]^{ij}\bigr\}\;\hbar^1 \cr
&\quad+  O(\hbar^2).                 &\numeq\cr
\b_{11\LpR}^{a,ij}&=-(-i) \coefZero{\om^a(p_3\equiv-p_1-p_2)%
\psi_{\b j}^{\pp}(p_2)\bar\psi_{\a i}^{\pp}(p_1)}          \cr
&\qquad  (\bar\pslash_1\PLpR)_{\a\b}               \cr
&={i\over4\cdot2} \Tr\,\Bigl[ 
	{\pr\over\pr\bar p_1^\mu}
	\,\rsp\BonneauGraphZero{\om^a(-p_1-p_2)%
		\psi_j^{\pp}(p_2)\bar\psi_i^{\pp}(p_1)}
	\Big|_{p_i\equiv0} \PLpR\gambar^\mu\,\Bigr] \cr
&=-\ucpc g^2 \bigl\{ (\CLpR-{\CA\over4}) [\,\TLpR^a\,]^{ij}+ 
	(\aprime-1) \bigl({\CLpR\over6}-{\CA\over4}\bigr) 
	[\,\TLpR^a\,]^{ij}\bigr\}\;\hbar^1 \cr
&\quad +  O(\hbar^2).                &\numeq\cr
}
$$

\vfill\eject
\midinsert

{\settabs 3\columns \def\graphwidth{1.2in} 
\eightpoint
\+\hfil\epsfxsize=\graphwidth\epsffile{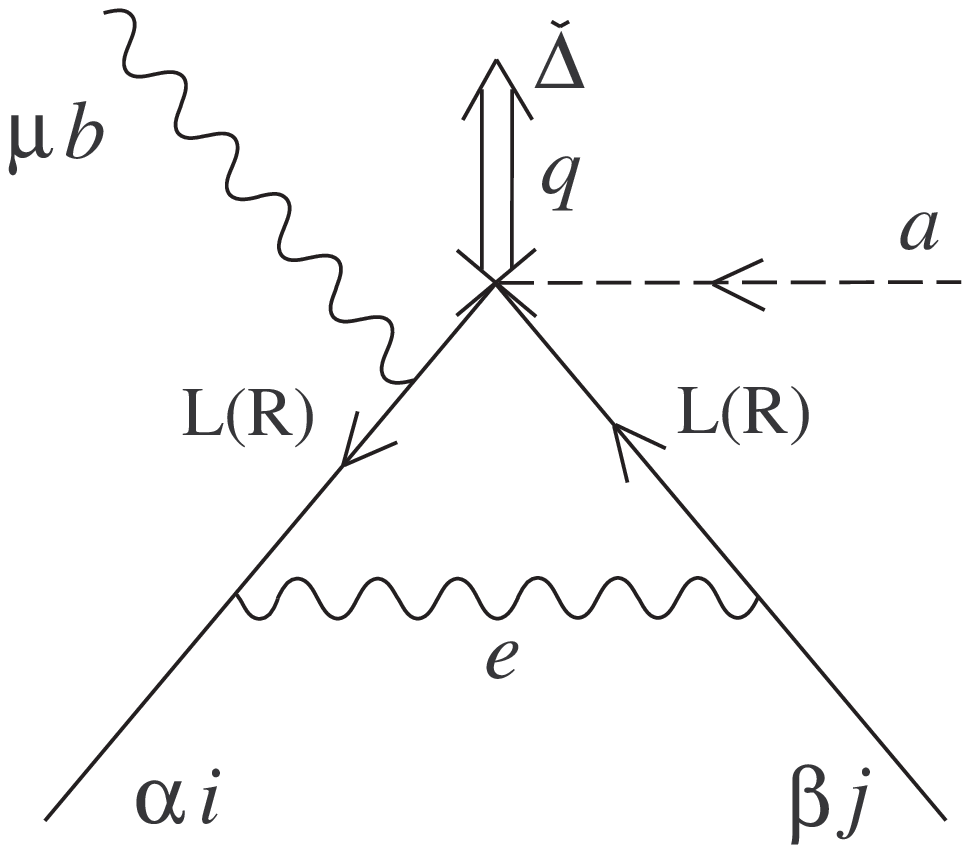}\hfil
	&\hfil\epsfxsize=\graphwidth\epsffile{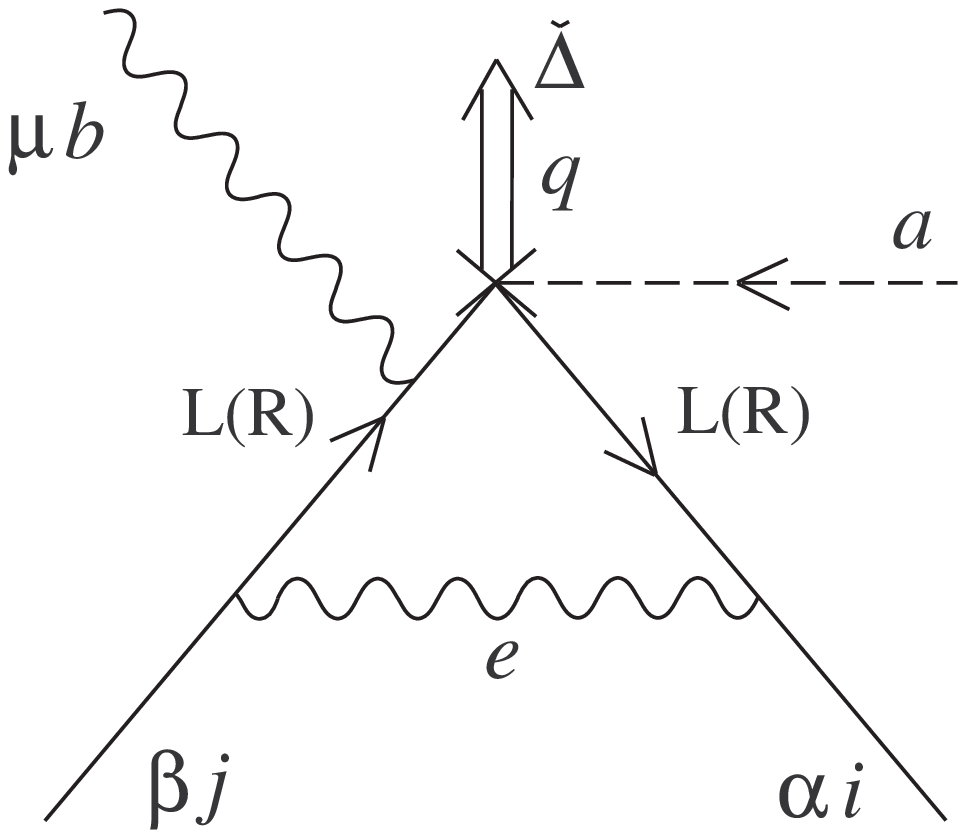}\hfil
	&\hfil\epsfxsize=\graphwidth\epsffile{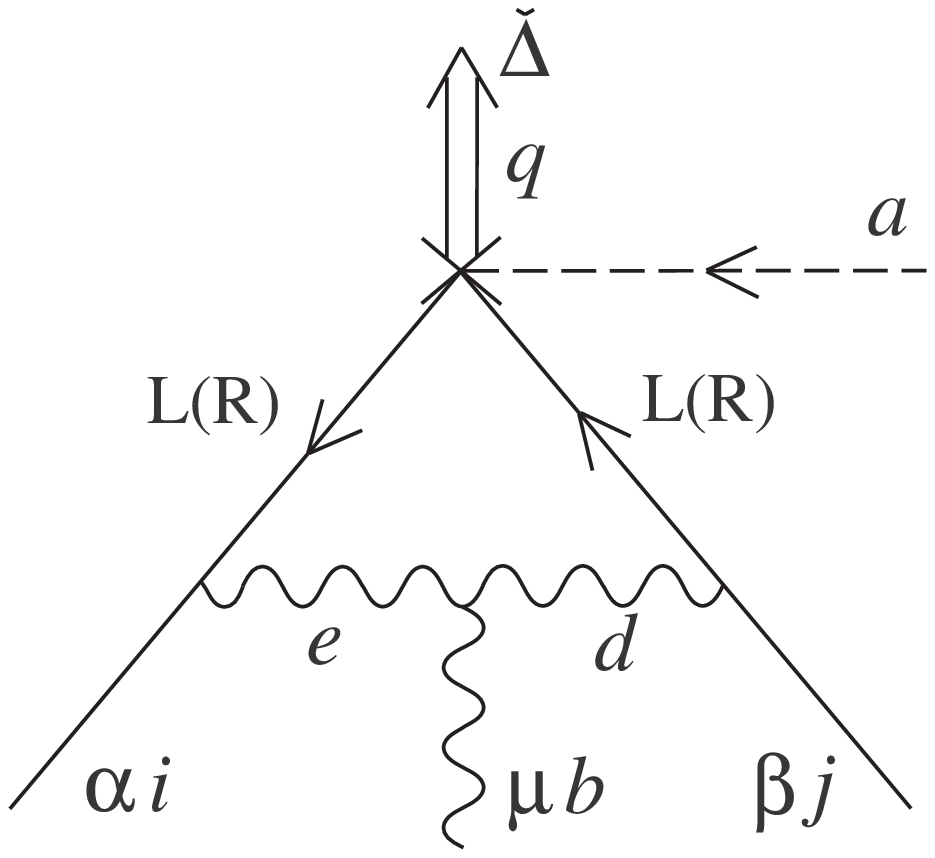}\hfil \cr
\+\hfil$(\TLpR^e \TLpR^b \TLpR^a \TLpR^e)_{ij}$\hfil
	&\hfil $(\TLpR^e \TLpR^a \TLpR^b \TLpR^e)_{ij}$\hfil
	&\hfil $c_{edb}(\TLpR^e \TLpR^a \TLpR^d)_{ij}$\hfil   \cr
}

{\settabs 2\columns \def\graphwidth{1.4in} 
\eightpoint
\+\hfil\epsfxsize=\graphwidth\epsffile{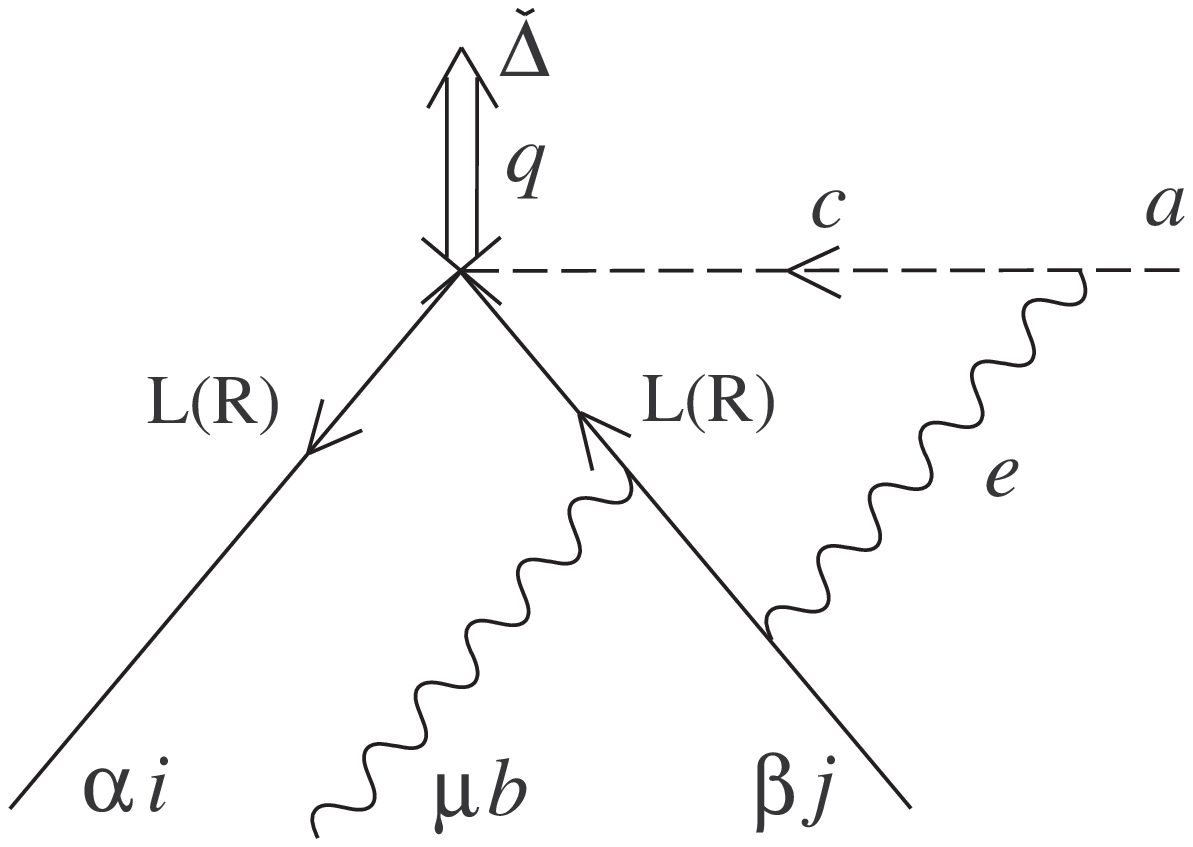}\hfil
	&\hfil\epsfxsize=\graphwidth\epsffile{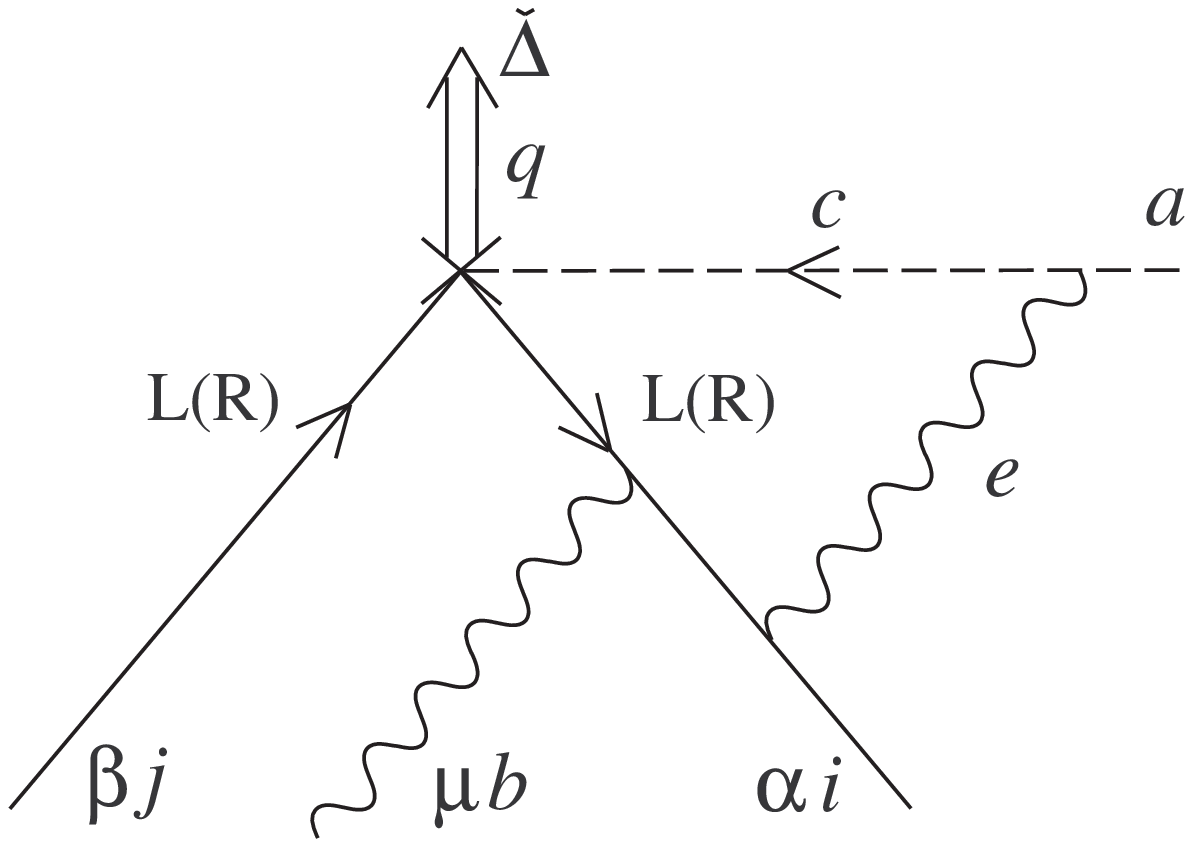}\hfil\cr
\+\hfil $c_{eac}(\TLpR^c \TLpR^b \TLpR^e)_{ij}$\hfil   
	&\hfil $c_{eac}(\TLpR^e \TLpR^b \TLpR^c)_{ij}$\hfil \cr
}

{\settabs 4\columns \def\graphwidth{1.2in}   
\eightpoint
\+\hfil\epsfxsize=\graphwidth\epsffile{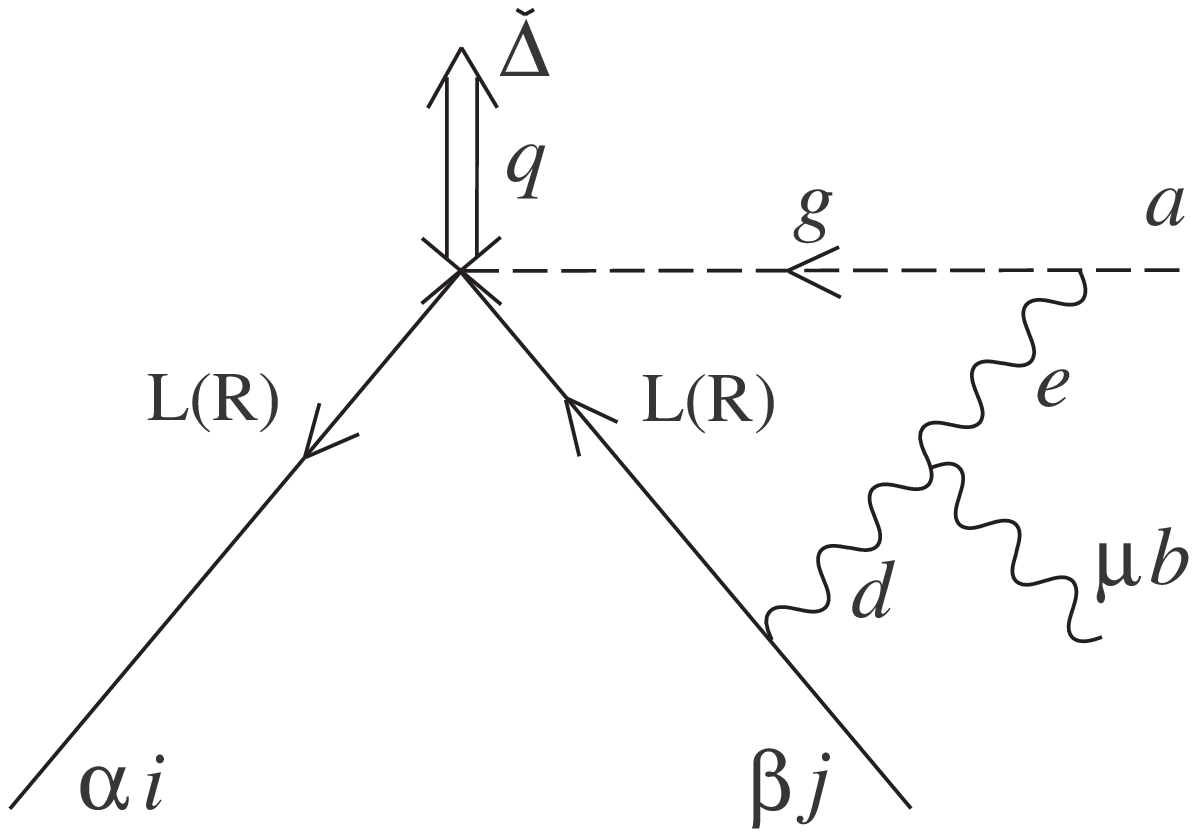}\hfil
	&\hfil\epsfxsize=\graphwidth\epsffile{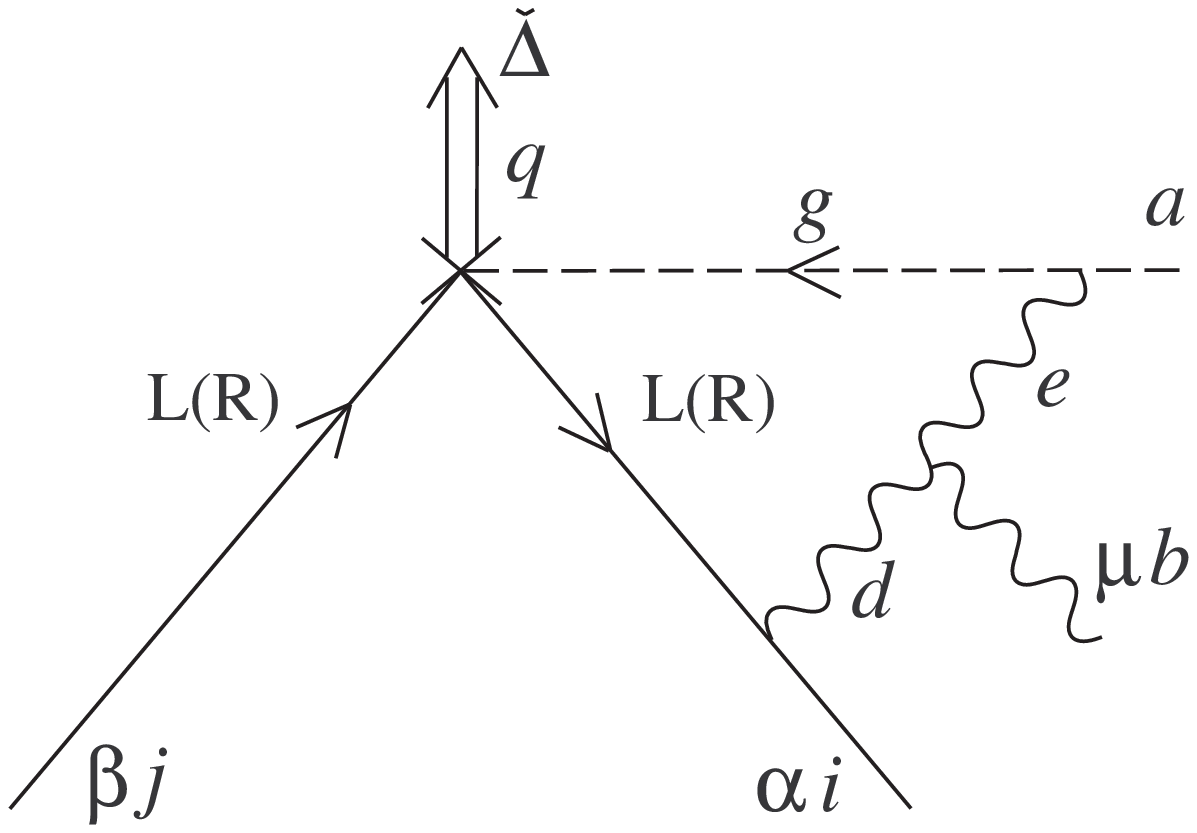}\hfil
	&\hfil\epsfxsize=\graphwidth\epsffile{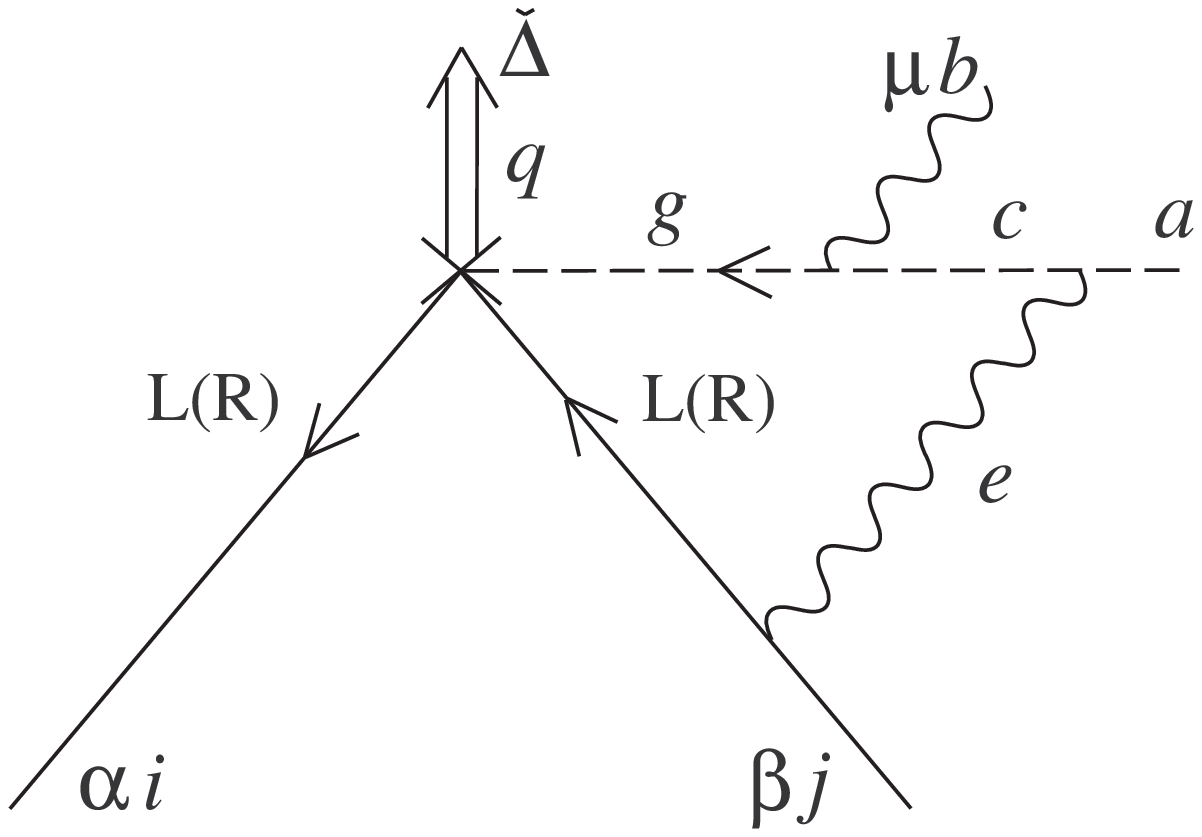}\hfil 
	&\hfil\epsfxsize=\graphwidth\epsffile{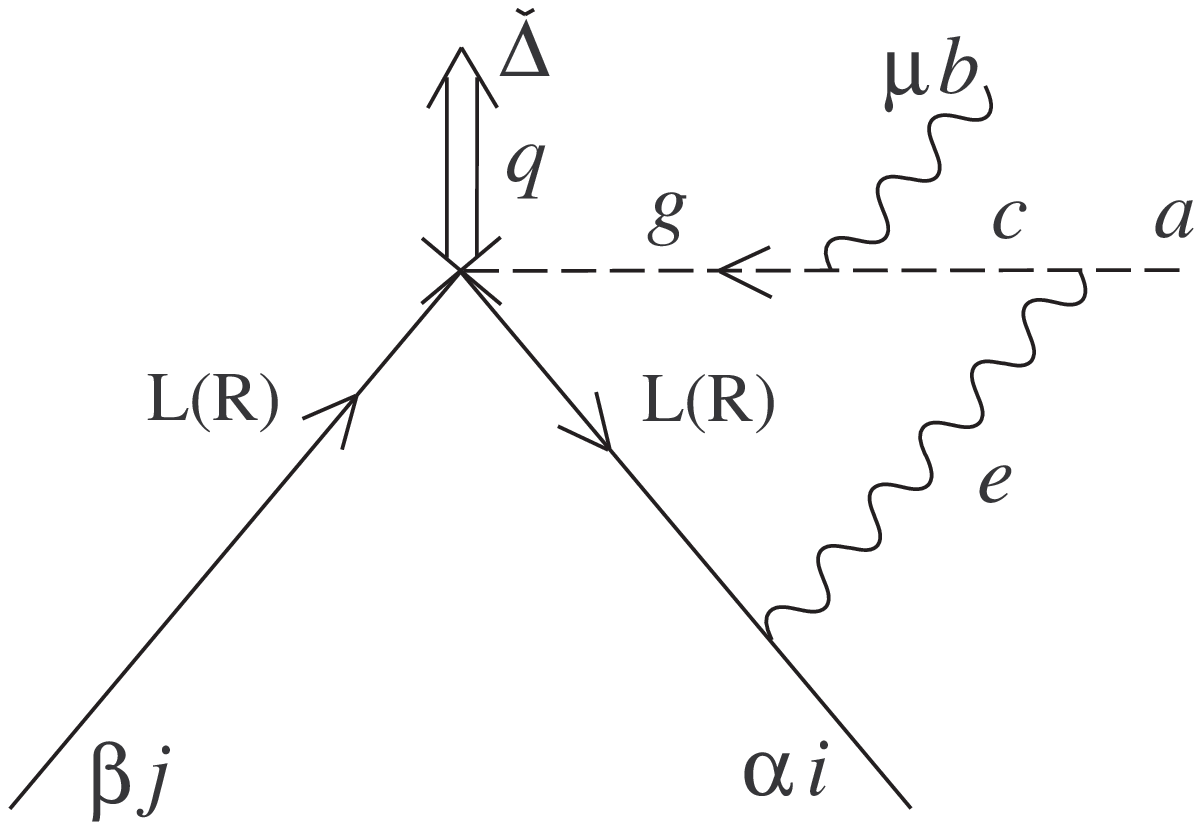}\hfil\cr
\+\hfil$c_{bed}c_{eag}(\TLpR^g\! \TLpR^d)_{ij}$\hfil
	&\hfil $c_{bed}c_{eag}(\TLpR^d \!\TLpR^g)_{ij}$\hfil
	&\hfil $c_{eac}c_{bcg}(\TLpR^g\! \TLpR^e)_{ij}$\hfil
	&\hfil $c_{eac}c_{bcg}(\TLpR^e \TLpR^g)_{ij}$\hfil \cr
}

\vskip 12pt
\narrower\noindent {\bf Figure 10:}
{\eightrm 1PI Feynman diagrams needed to compute eq.~\relativeq1}
\vskip 0.4cm
\endinsert

\noindent $*$ f) {\sl From 1PI functions with one $\om$, one $A$ and one 
   $\psi,\bar\psi$ ($\psi',\bar\psi'$)  pair} (fig.~10)

$$
\eqalignno{
\b_{12\LpR}^{ab,ij}&=-1
	\coefZero{\om^a A_\mu^b \psi_{\b j}^{\pp}\bar\psi_{\a i}^{\pp}} \cr
&\qquad  (\gambar^\mu\PLpR)_{\a\b}               \cr
&=-{1\over4\cdot2} \Tr\,\Bigl[ 
	\,\rsp\BonneauGraphZero{\om^a A_\mu^b
			\psi_j^{\pp}\bar\psi_i^{\pp}}
	\Big|_{p_i\equiv0} \PLpR\gambar^\mu\,\Bigr] \cr
&=-\ucpc g^2 {1+(\aprime-1)\over4}\,\CA\;i
	\bigl[\,\TLpR^a,\TLpR^b\,\bigr]_{ij}\;\hbar^1 + O(\hbar^2). &\numeq\cr
}
$$

Notice that the gauge group structure of the first five diagrams in fig.~10 
involves three or more gauge group generator matrices. The gauge 
group structure of each diagram is rather involved but it turns out 
that the r.s.p.~of each  of the  first three diagrams vanishes, 
that the r.s.p.~of the fourth cancels exactly against the fith,
and, thanks to the Jacobi identity, that the result of the last four 
diagrams  combine in pairs to give eq.~\lasteq,
where the gauge group structure is simply a commutator.This is remarkable, 
because we shall see in Appendix C that the difference between the
one-loop  1PI fermionic vertex computed {\it \`a la} Breitenlhoner 
and Maison and the same 1PI function evaluated with the ``naive'' presciption 
involves only a single  gauge group generator and, hence, it is sensible 
to expect that the finite counterterm that is needed to restore at
the one-loop  level the BRS symmetry in the Breitenlhoner and Maison 
formalism involved only a generator.  The BRS variation, $b$, of this
counterterm yields a contribution linear in $A_\mu^b$ which will  match
the gauge group structure of the symmetry breaking diagrams in fig.~10 thanks
to the various cancellations and simplifications just described.

\medskip
\midinsert
\def\graphwidth{1.5in}

{\eightpoint
$$
\eqalign{\matrix{\epsfxsize=\graphwidth\epsffile{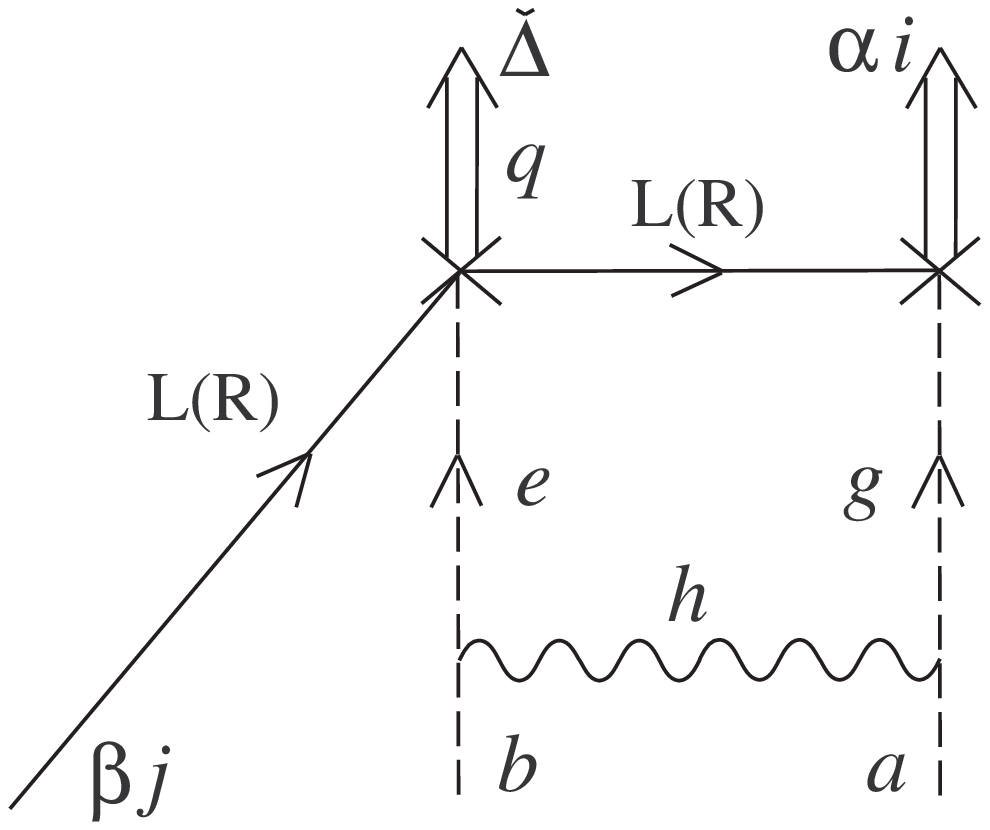}\cr
			c_{hag}c_{hbe}(\TLpR^g \TLpR^e)_{ij}\cr}}
	\quad\eqalign{-}\quad
  \eqalign{a\longleftrightarrow b}  \qquad
\eqalign{\matrix{\epsfxsize=\graphwidth\epsffile{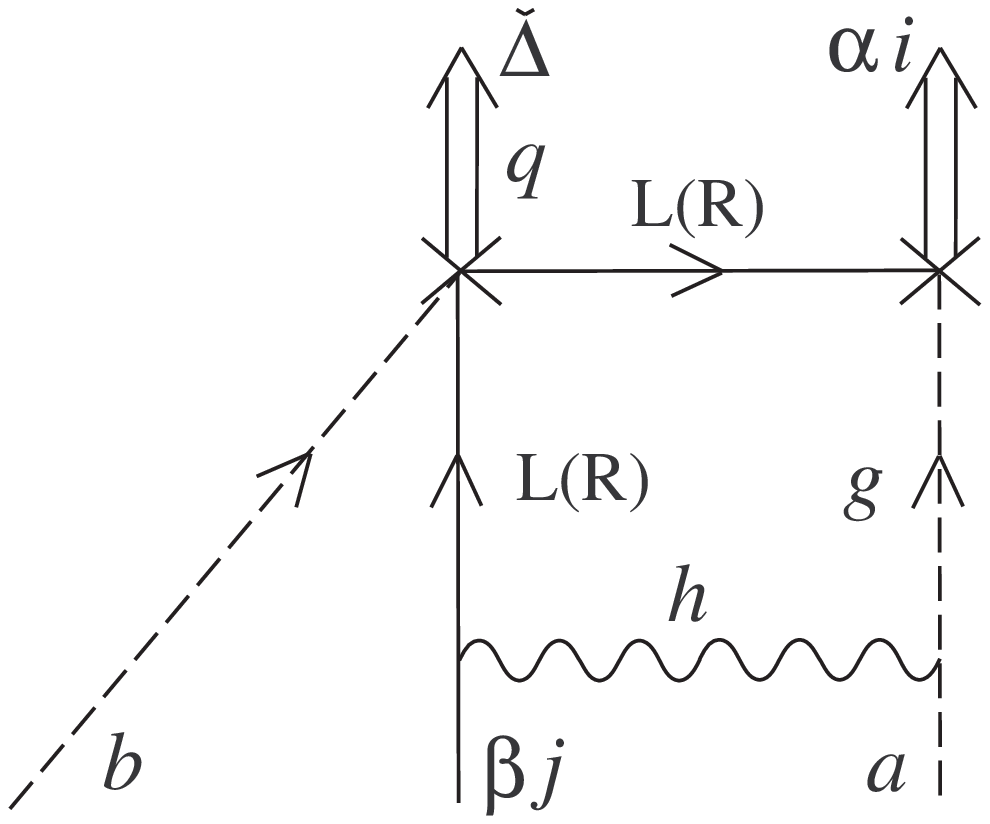}\cr
				c_{hag}(\TLpR^g \TLpR^b \TLpR^h)_{ij}\cr}}
	\quad\eqalign{-}\quad
  \eqalign{a\longleftrightarrow b}
$$
}
\vskip 12pt
\narrower\noindent {\bf Figure 11:}
{\eightrm 1PI Feynman diagrams needed to compute eq.~\relativeq1}
\vskip 0.4cm
\endinsert
\noindent $*$ g) {\sl From 1PI functions with  fermions and external fields}
                     (figs.~11-12)

$$
\eqalignno{
\b_{14\LpR}^{ab,ij}&={i\over2\hbar}
	\coefZero{\om^a \om^b \psi_{\b j}^{\pp}\N[s\psi_{\a i}^{\pp}]} \cr
&\qquad  (\PLpR)_{\a\b}               \cr
&=\ucpc g^2 [1+(\aprime-1)]\,{\CA\over8}\
	\bigl[\,\TLpR^a,\TLpR^b\,\bigr]_{ij}\;\hbar^1 + O(\hbar^2). 
&\numeq\cr
} 
$$

\midinsert
\def\graphwidth{1.5in}
{\eightpoint
$$
\eqalign{\matrix{\epsfxsize=\graphwidth\epsffile{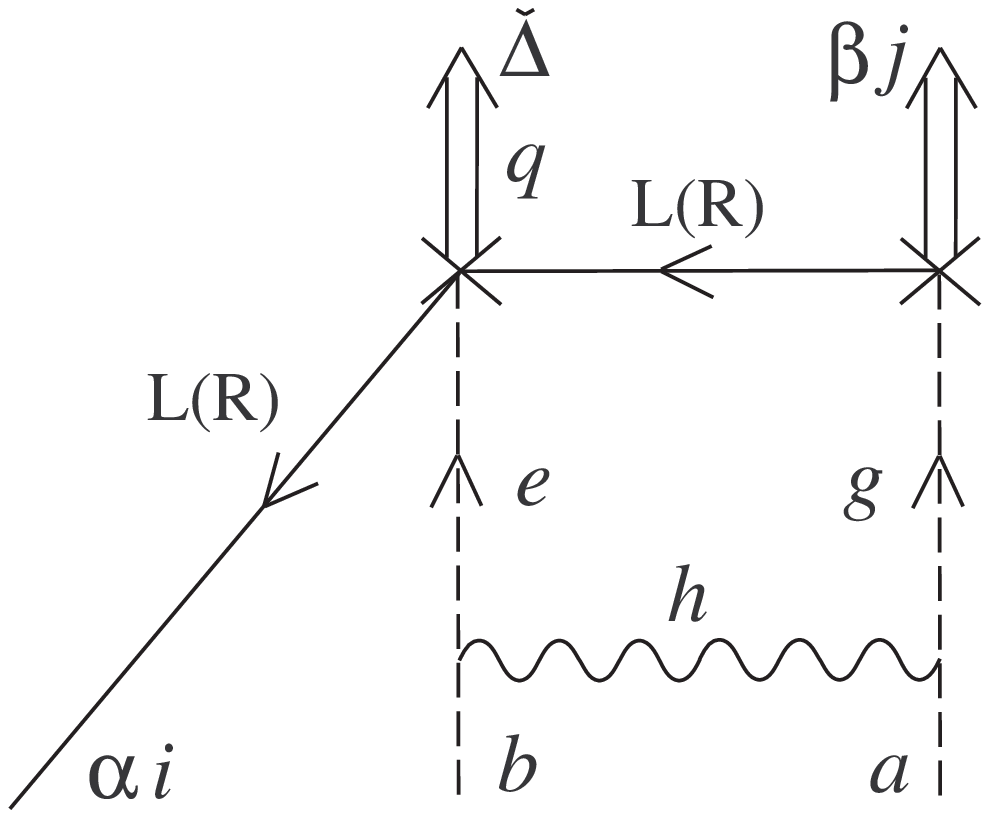}\cr
			c_{hag}c_{hbe}(\TLpR^e \TLpR^g)_{ij}\cr}}
	\quad\eqalign{-}\quad
  \eqalign{a\longleftrightarrow b}  \qquad
\eqalign{\matrix{\epsfxsize=\graphwidth\epsffile{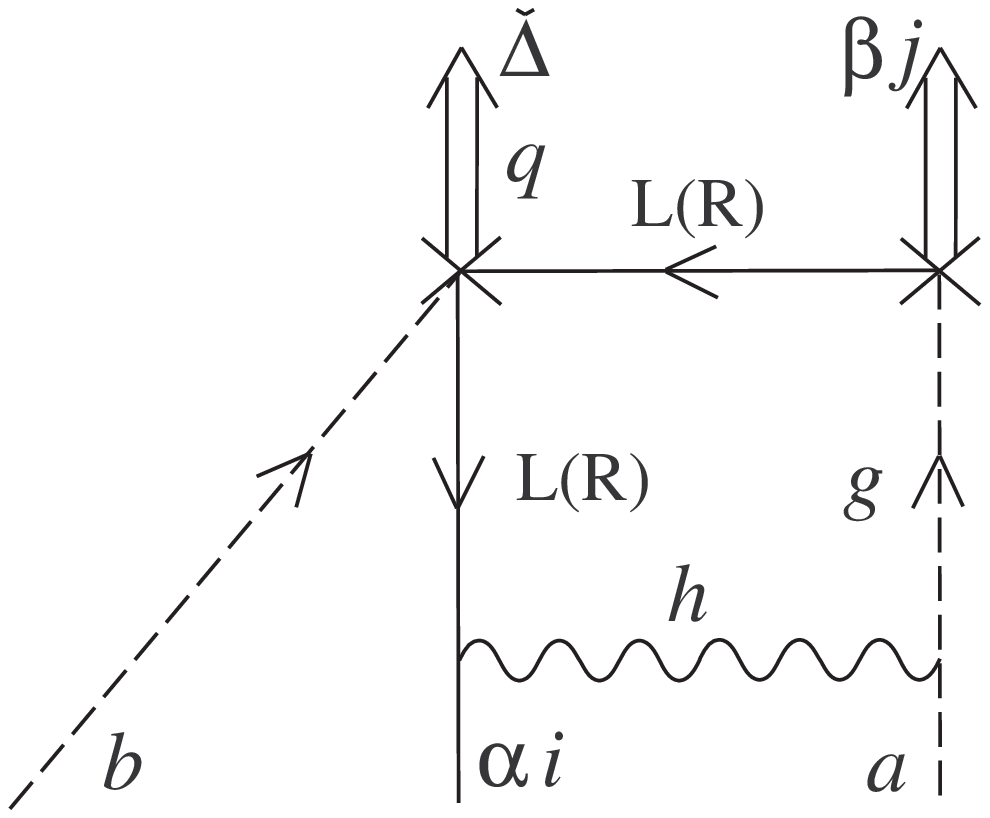}\cr
				c_{hag}(\TLpR^h \TLpR^b \TLpR^g)_{ij}\cr}}
	\quad\eqalign{-}\quad
  \eqalign{a\longleftrightarrow b}
$$
}
\vskip 12pt
\narrower\noindent {\bf Figure 12:}
{\eightrm 1PI Feynman diagrams needed to compute eq.~\relativeq1}
\vskip 0.4cm
\endinsert

$$
\eqalignno{
\b_{15\LpR}^{ab,ij}&={i\over2\hbar}
\coefZero{\om^a \om^b \bar\psi_{\b j}^{\pp}\N[s\bar\psi_{\a i}^{\pp}]} \cr
&\qquad  (\PRpL)_{\a\b}               \cr
&=-\ucpc g^2 [1+(\aprime-1)]\,{\CA\over8}\
	\bigl[\,\TLpR^a,\TLpR^b\,\bigr]_{ij}\;\hbar^1 + O(\hbar^2).
&\numeq\cr
}   
$$

As in  case f), it seems impossible that the gauge group structure of 
the diagrams be matched by gauge group structure of a simple one-loop finite 
counterterm. But the last pair of diagrams of each 1PI function vanishes, and 
the other pair of antisymmetric diagrams conspire with the help of the Jacobi 
identity to give a simple commutator as  result.

\medskip
\noindent $*$ h) {\sl From 1PI functions with $\eps_{\mu\nu\a\b}$}
(figs.~6-8)

These coefficients are very important, because if they are non-zero, give
monomials in the expansion of the breaking that can not be cancelled by
finite counterterms; {\it i.e.} they give the essential non-Abelian 
chiral anomaly.

$$
\eqalignno
{
\b_{50}^{abc}&=\b_{50}^{acb}=\cr
&=-{(-i)^2\over2} \coefZero{\om^a(p_3\equiv -p_1-p_2)%
			A_\mu^b(p_1)A_\nu^c(p_2)} \cr
&\qquad  \eps_{\mu\nu\a\b}\,p_1{}^\a\,p_2{}^\b     \cr
&=-\ucpc {1\over3} \dLR^{abc}\,\hbar^1 + O(\hbar^2),    &\numeq\cr
\b_{51}^{abcd}&=-\b_{51}^{abdc}=\cr     
&=-{(-i)\over2} \coefZero{\om^a(p_4\equiv -\sum_i^3 p_i)%
			A_\mu^b(p_1)A_\nu^c A_\rho^d} \cr
&\qquad  \eps_{\mu\nu\rho\a}\, \bar p_1{}^\a     \cr
&=-\ucpc {1\over6} \DLR^{abcd}\,\hbar^1 + O(\hbar^2),  &\numeq\cr
\b_{52}^{abcde}&=\b_{52}^{a[bcde]}= 
-{1\over4!} \coefZero{\om^a A_\mu^b A_\nu^c A_\rho^d A_\la^e} \cr
&\qquad  \eps_{\mu\nu\rho\la}    &\numeq \cr
&=O(\hbar^2),       \cr      
}
$$
with the definitions
$$
\eqalignno{
\dL^{abc}&=\Tr\,\bigl[\,\TL^a \bigl\{\TL^b,\TL^c\bigr\}\,\bigr]
   =\dL^{\{abc\}}                         \cr
\DL^{abcd}&=-i\,3!\,\Tr\bigl[\,\TL^a\,\TL^{[b}\TL^c\TL^{d]}\,\bigr]=
	{1\over2}\,(\,\dL^{abe} c^{ecd} + \dL^{ace} c^{edb}+
         \dL^{ade} c^{ebc}\,);             \cr
}
$$
same for $\dR^{abc}$ and $\DR^{abcd}$, and $\dLR\equiv\dL-\dR$, 
$\DLR\equiv\DL-\DR$.

Of course, we write $O(\hbar^2)$ because we have computed at first order in
$\hbar$; but due to the Adler-Bardeem theorem for non-Abelian
gauge theories \cite{\PiguetSorella}, all these coefficient should be zero at
higher orders if there would not be  an anomaly  at first order.

 Our one-loop computation of these coefficients give 
the {\it non-Abelian (essential) chiral gauge anomaly}:
$$
\eqalignno{
&\intd^4x \big\{
  \b_{50}^{abc}\,
	\N[\eps_{\mu\nu\a\b}\,\om^a(\pr^\a A^{\mu b})(\pr^\b A^{\nu c})]+
	\b_{51}^{abcd}\,
\N[\eps_{\mu\nu\a\b}\,\om^a(\pr^\a A^{\mu b})A^{\nu c}A^{\rho d}]+ \cr
&\qquad\qquad+\b_{52}^{abcde}\,
	  \N[\eps_{\mu\nu\a\b}\,\om^a A^{\mu b}A^{\nu c}A^{\rho d}A^{\la e}]
	  \big\}\cdot\GR=            \cr
&\quad=-\ucpc {1\over3}\Big\{\,
\eps_{\mu\nu\a\b}\, \dLR^{abc}\,\om^a (\pr^\a A^{\mu b}) (\pr^\b A^{\nu c})+
	{1\over2} \eps_{\mu\nu\rho\a}\, \DLR^{abcd}
		\om^a (\pr^\a A^{\mu b}) A^{\nu c} A^{\rho d}\, \Big\}\cr
&\qquad\qquad\qquad+ O(\hbar^2)=                      \cr
&\quad=\ucpc {1\over3} \eps_{\mu\nu\rho\sigma}\,\om^a\,\pr^\mu\,\left\{
	\dLR^{abc} (\pr^\nu A^{\rho b}) A^{\sigma c} +
{1\over6} \DLR^{abcd}\, A^{\nu b} A^{\rho c} A^{\sigma d} \right\}
		+ O(\hbar^2)=\cr
&\quad=\ucpc {2\over3} \eps_{\mu\nu\rho\sigma}\,\Tr_{\L-\R}\;\left\{\,
	\om\,\pr^\mu\,(\pr^\nu A^\rho A^\sigma -
      {i\over2} A^\nu A^\rho A^\sigma) \right\}+ O(\hbar^2)  \cr
}
$$
Remember that in the formulae for the coefficients, a sum over the
different left or right representations is understood. Therefore, the
matter representation may be chosen such that the coefficients of the 
essential anomaly above vanish.

\medskip

Finally, as explained at the end of  section {\it 2.4.}, the 
coefficients $k_i$ of the true expansion in the quantum (standard) 
basis of insertions are equal to the coefficients $\b_i$ in the one-loop 
approximation. Moreover $\bar k_{i}^{(1)}=k_{i}^{(1)}$.

\medskip
\noindent $*$ {\sl The expression of the breaking at order $\hbar$}
\smallskip
 Putting it all together we get the complete form of the integrated breaking,
$\Bigl[{\rm N}[\hat\Delta]\cdot\Gamma_{\rm ren}\Bigr]^{(1)}$, at order $\hbar$ 
(this formula is a 4 dimensional one !), which reads
$$
\eqalignno{
\Bigl[{\rm N}[\hat\Delta]\cdot\Gamma_{\rm ren}\Bigr]^{(1)}=
	&-\ucpc {\TL+\TR\over3}\,\delta^{ab}\,\M_1^{ab}\;\;\hbar^1+ \cr
	&-\ucpc {\TL+\TR\over3}\,c^{abc}\,\M_2^{abc}\;\;\hbar^1 
		+\ucpc (\TL+\TR){2\over3}\,c^{abc}\,\M_4^{abc}\;\;\hbar^1+ \cr
        &+\ucpc {1\over6} \big[\,\TLR^{acbd}+\TLR^{adbc}-
           (\TL+\TR)\,(c^{ebc}c^{eda}+c^{ebd}c^{eca})\,\big]\,
          \M_7^{abcd}\;\;\hbar^1\cr
         &+\ucpc {1\over3} \big[\,\TLR^{abcd}+\TLR^{adcb}+
          (\TL+\TR)\,(c^{ebc}c^{eda}+c^{eba}c^{edc})\,\big]\,
          \M_8^{abcd}\;\;\hbar^1\cr
	&-\ucpc g^2\,\Big[\,\CL -{\CA\over4} + 
			(\aprime-1)\big({\CL\over6}-{\CA\over4}\big)\Big]\,\,
			\Big(\M_{10\L}+\M_{11\L}\Big)\;\;\hbar^1 \cr
	&-\ucpc g^2\,\Big[\,\CR -{\CA\over4} + 
			(\aprime-1)\big({\CR\over6}-{\CA\over4}\big)\Big]\,
			\Big(\M_{10\R}+\M_{11\R}\Big)\;\;\hbar^1 \cr
	&-\icpc g^2\, {1+(\aprime-1)\over4}\,\CA\,
	\big(\,\M_{12\L}-\M_{13\L}+\M_{12\R}-\M_{13\R}\big)\;\;\hbar^1\cr
	&+\icpc g^2\, {1+(\aprime-1)\over4}\,{\CA\over8}\,c^{abc}\,
			\big(\,\M_{14\L}^{abc}-\M_{15\L}^{abc}+
			\M_{14\R}^{abc}-\M_{15\R}^{abc}\big)\;\;\hbar^1\cr
&+\ucpc\,{2\over3} \intd^4x\; \eps_{\mu\nu\rho\sigma}\,\Tr_{\L-\R} \big[
	\om\,\pr^\mu (\pr^\nu A^\rho A^\sigma 
		- {i\over2} A^\nu A^\rho A^\sigma) \,\big]\;\;\hbar^1,
	                                &\numeq\cr
}
$$\namelasteq\resultbreak
where we have defined
$$
\eqalignno{
\M_{10\LpR}&=(\TLpR^a)^{ij}\,\M_{10\LpR}^{a,ij}=
\intd^4x\;\om^a\,\bar\psi^\pp\gambar^\mu\PLpR\TLpR^a\pr_\mu\psi^\pp\cr
\M_{11\LpR}&=(\TLpR^a)^{ij}\,\M_{11\LpR}^{a,ij}=             
\intd^4x\;\om^a\,(\pr_\mu\bar\psi^\pp)\gambar^\mu\PLpR\TLpR^a\psi^\pp\cr
\M_{12\LpR}&=(\TLpR^a\TLpR^b)^{ij}\,\M_{12\LpR}^{ab,ij}=
	\intd^4x\;\om^a\,\bar\psi^\pp\gambar^\mu
            \PLpR\TLpR^a\TLpR^b\psi^\pp A_\mu^b             \cr
\M_{13\LpR}&=(\TLpR^b\TLpR^a)^{ij}\,\M_{12\LpR}^{ab,ij}=
	\intd^4x\;\om^a\,\bar\psi^\pp\gambar^\mu
				\PLpR\TLpR^b\TLpR^a\psi^\pp A_\mu^b\cr
\M_{14\LpR}^{abc}&=(\TLpR^c)^{ij}\,\M_{14\LpR}^{ab,ij}=
   \intd^4x\;\om^a\,\om^b\,\bar L (\bar R)
            \PLpR\TLpR^c\psi^\pp, \cr
\M_{15\LpR}^{abc}&=(\TLpR^c)^{ij}\,\M_{15\LpR}^{ab,ij}=
   \intd^4x\;\om^a\,\om^b\,\bar\psi^\pp
            \PRpL\TLpR^c L(R)\, , \cr
}
$$
and we have used the following equation:
$$
-2\TLR^{abcd}-2\TLR^{adcb}+   \TLR^{abdc}+\TLR^{acbd}+\TLR^{acdb}+
    \TLR^{adbc}= (\TL+\TR)\,(c^{ebc}\,c^{eda}+c^{eba}\,c^{edc}
            )\, .   \eqno\numeq \namelasteq\EqTRelation
$$

Notice that if the classical theory were CP invariant, the combination of 
monomials obtained have the same CP definite value as the breaking, as we 
would have expected from the fact that Dimensional Regularization respects 
discrete symmetries. But we have not assumed  any discrete symmetry 
property when writing the basis of Bonneau
monomials because we work with general representations.

\bigskip
The  result in eq.~\resultbreak\  can be obtained by computing 
the left hand side of eq.~\resultbreak\ to the one-loop order, {\it i.e.}
 calculating the renormalized part of all 
possible one-loop 1PI diagrams with the insertion $\hat\Delta$; their
finite part has to be a local term, because when going to the 4-dimensional
space all $(d-4)$-objects are set to be 0, so the only remaining finite
part of such a diagram must come from a polynomial singular part 
formed by a 4-dimensional object and a contracted $(d-4)$-object which 
cancel the singularity in $d-4$. Therefore, being local this one-loop
renormalized 1PI functions, there are operators (in the 4-dimensional
space) whose Feynman rules attached with a $\hbar^1$ give the same
result. The sum of this operators is precisely the right hand side of 
eq.~\resultbreak.

In fact, the one-loop calculation of the coefficients of the Bonneau 
identities via r.s.p.~of diagrams with the insertion of $\check\Delta$ are 
exactly the same as the finite part of diagrams with the insertion of
$\hat\Delta$, excepting some factors in the Bonneau coefficients which
directly give the operators whose Feynman rules lead to the same result.

But the Bonneau identities  unravel the problem in higher order computations,
and the formulae of the coefficients given above would be the same (although
with new Feynman rules coming from the finite counterterms we would have
added in the previous recursive step).

\subsection{3.6. Restoring the BRS symmetry: the computation of
$S^{(1)}_{\rm fct}$}

 We know from  algebraic renormalization theory and BRS 
cohomology  \cite{\PiguetSorella} that there exists an integrated 
local functional of the fields and their derivatives, let us call 
it $X^{(1)}$, such that the anomalous contribution 
$\Bigl[{\rm N}[\hat\Delta]\cdot\Gamma_{\rm ren}\Bigr]^{(1)}$, given in
eq.~\resultbreak , can be cast into the form
$$
\Bigl[{\rm N}[\hat\Delta]\cdot\Gamma_{\rm ren}\Bigr]^{(1)}=
\ucpc\,{2\over3} \intd^4x\; \eps_{\mu\nu\rho\sigma}\,\Tr_{\L-\R} \big[
		\om\,\pr^\mu (\pr^\nu A^\rho A^\sigma 
					- {i\over2} A^\nu A^\rho A^\sigma) 
\,\big]\;\;\hbar^1
	+b X^{(1)}. 
$$  
 The integrated local functional, $X^{(1)}$, has ghost number zero and 
ultraviolet dimension 4. Hence, by choosing $S^{(1)}_{\rm fct}$ in
eq.~\Dchiralact\
so that it verifies
$$
S^{(1)}_{\rm fct}=-X^{(1)},
$$
and recalling that eqs.~\YMOneloopbreaking\ and \EqRenYMSTBreaking\ hold, we
conclude that the BRS identity is broken at order $\hbar$ by the essential 
non-Abelian gauge anomaly only:
$$
{\cal S}(\Gamma_{\rm ren})= 
\ucpc\,{2\over3} \intd^4x\; \eps_{\mu\nu\rho\sigma}\,\Tr_{\L-\R} \big[
		\om\,\pr^\mu (\pr^\nu A^\rho A^\sigma 
					- {i\over2} A^\nu A^\rho A^\sigma) 
\,\big]\;\;\hbar^1+\;O(\hbar^2).
$$

The previous equation leads in turn to the result we sought for; namely,
that if the fermion representations meet the anomaly cancellation
criterion \cite{\Alwit}, the BRS symmetry can be restored 
(up to order $\hbar$) by an appropriate
choice of finite counterterms $S^{(1)}_{\rm fct}$.

 Our next task will be to compute $S^{(1)}_{\rm fct}=-X^{(1)}$. 
We shall demand that $S^{(1)}_{\rm fct}$ do not depend neither on $B$ nor
$\tilde \rho$. Recall that these fields do not occur in eq.~\resultbreak . 

The more general approach to this computation, similar to our study of 
the expansion of $\Bigl[{\rm N}[\hat\Delta]\cdot\Gamma_{\rm ren}\Bigr]^{(1)}$,
 would be: 
1. Write the complete list of possible monomials 
of ultraviolet dimension 4, ghost number 0 and with no free Lorentz 
indices \dag%
\vfootnote\dag{This list is a list in the four dimensional space-time. 
Of course, some generalization of them to the $d$-dimensional space-time of
Dimensional Regularization has to be chosen while writing them
in the Dimensional Regularization action; but the differences, 
being $(d-4)$-objects, will begin to produce any effect
at the following order in $\hbar$ (because they require
loops to yield a non-zero contribution and there is already an 
explicit $\hbar$ in $\Sfct^{(1)}$)}%
; 2. Write the 
general expression of the finite counterterms as a linear combination of 
those monomials with coefficients depending on gauge group indices.
3. Take their $b$-variation and finally
4. impose that this variation cancels the expansion of the breaking
term $\Bigl[{\rm N}[\hat\Delta]\cdot\Gamma_{\rm ren}\Bigr]^{(1)}$, excepting
the essential non-Abelian anomaly. 
This should lead to a compatible indeterminate system with
the coefficients as variables and with a degree of indeterminacy exactly
equal to the number of $b$-invariant operators in the classical theory.

But, writing such list and taking its $b$ variation is a tedious task:
lots of indices would come in  and the resolution of the final system
would appear a bit messy.

It is plain that the gauge group structure of the finite counterterms which
would be needed in the one-loop computation to restore the BRS symmetry is
not arbitrary.
A natural ansatz for the gauge group structure of these counterterms is to 
take the one-loop gauge group structure 
which can appear while computing the Feynman diagrams. Therefore, 
we write
$$
\eqalignno{
\Sfct^{(1)}=
&a_1\,\int\!\! (\pr_\mu A^\mu)^2 + a_2\, \int\!\! A_\mu \square\,A^\mu
	+c\,\int\!\! i c^{abc} (\pr_\mu A_\nu^a) A^{b\mu} A^{c\nu} \cr
	&+ (\dL\,\dL^{abc}+ \dR\,\dR^{abc})\,\int\!\!
			(\pr_\mu A_\nu^a)\,A^{b\mu} A^{c\nu}+
	 + (f_\L\,\TL^{abcd}+ f_\R\,\TR^{abcd})\,\int\!\!
			A^{a\mu} A_\mu^b A^{c\nu} A_\nu^d \cr
	&+e\, c^{abc}c^{ade}\,\int\!\! A^{b\mu} A^{c\nu} A_\mu^d A_\nu^e  +
	 n_\L \,\int\!\!{i\over2}\bar\psi\ga^\mu\PL\arrowsim\pr_\mu\psi+
  n_\R \,\int\!\!{i\over2}\bar\psi'\ga^\mu\PR\arrowsim\pr_\mu\psi'\cr
	&+p_\L\,\int\!\!\bar\psi\ga^\mu\PL\TL^a\psi A_\mu^a+
	  p_\R\,\int\!\!\bar\psi'\ga^\mu\PR\TR^a\psi' A_\mu^a \cr
	&+\int\!\big[\,
		u_{1\L}\,\bar L s\psi+u_{1\R}\,\bar R s\psi'+
	u_{2\L}\,L s\bar\psi +u_{2\R}\,R s\bar\psi' \,\big]      &\numeq\cr
}
$$
Notice that because of the reason given above we have not written terms
with $\rho$ or $\bar\om$.

Moreover, we have the combinations of terms given in eq.~\EqbInvariants\ 
which are $b$-invariant. Obviously, these combinations are not relevant
to our computation. Therefore, we can reduce the basis of possible finite
counterterms to a minimum number taking into account these invariant
counterterms.

$L_A$ and $L_\om$ contain $\tilde\rho^\mu$, which we supposed already not
to appear in the finite counterterms, so these invariant terms can not
be used to reduce more the number of finite counterterms. But, using 
$L_g$ we can impose $e=0$ and by taking advantange of $L_{\psi^\pp}$
we can set $p_\LpR=0$, without loss of generality.

Then, there should be a {\it unique} solution for the problem of 
cancellation of spurious anomalies in terms of the rest of variables
$a_1$, $a_2$, $c$, $d_\LpR$, $f_\LpR$, $n_\LpR$, $u_{1\LpR}$ and
$u_{2\LpR}$.
 
We recast eq.~\lasteq\ in the following form:
$$
\eqalignno{
\Sfct^{(1)}=
&a_1\, \tilde\Delta_1+ a_2\, \tilde\Delta_2
		+c\, \tilde\Delta_3 
		+ \dL\,\tilde\Delta_{4\L}+ \dR\,\tilde\Delta_{4\R}
	 + f_\L\,\tilde\Delta_{5\L} + f_\R\,\tilde\Delta_{5\R} \cr
&+n_\L \,\tilde\Delta_{6\L}+n_\R \,\tilde\Delta_{6\R}
	+u_{1\L}\,\tilde\Delta_{7\L} +u_{1\R}\,\tilde\Delta_{7\R}
	+u_{2\L}\,\tilde\Delta_{8\L} +u_{2\R}\,\tilde\Delta_{8\R}
. &\numeq\cr \namelasteq\EqAnsatz
}
$$
\medskip
We need the $b$ variations of the $\tilde\Delta$s expanded in terms
of the integrated monomials we called $\M$s. Moreover, the coefficients
of these expansions should be symmetrized if the gauge group indices of the
corresponding monomials have some symmetry properties. This is necessary
because, for example, $\M_3^{abc}=\M_3^{acb}$ and then not all
$\M_3^{abc}$ are linearly independent. The result reads:                          :
$$
\eqalignno{
b\,\tilde\Delta_1=&\;2\,\delta^{ab}\,\M_1^{ab}- 2\, c^{abc}\,\M_4^{abc},
	\qquad\quad
b\,\tilde\Delta_2=-2\,\delta^{ab}\,\M_1^{ab}+ 2\, c^{abc}\,\M_2^{abc},\cr
\sepeq
b\,\tilde\Delta_3=&-i c^{abc}\,\M_4^{abc}+i c^{abc}\,\M_2^{abc}\cr
&-{i\over2} [c^{ebc}c^{eda}+c^{ebd}c^{eca}]\,\M_7^{abcd}+
	i [c^{ebc}c^{eda}+c^{eba}c^{edc}]\,\M_8^{abcd}, \cr
\sepeq
b\,\tilde\Delta_{4\LpR}=&-\dLpR^{abc}\,\M_2^{abc}-
	\dLpR^{abc}\,\M_3^{abc}+\dLpR^{abc}\,\M_4^{abc}+
		\dLpR^{abc}\,\M_5^{abc} \cr
&-(\dLpR^{ebc} c^{eda} + \dLpR^{ebd} c^{eca})\,\M_7^{abcd}+
	 \dLpR^{ebd} c^{eca}\,\M_8^{abcd},\cr
\sepeq
b\,\tilde\Delta_{5\LpR}=& -\;\big[\,\TLpR^{abcd}+\TLpR^{acdb} +
	\TLpR^{abdc} +\TLpR^{adcb}\,\big]\;\M_7^{abcd}\cr
&-2\, \big[\,\TLpR^{acbd}+\TLpR^{abdc} +
   \TLpR^{acdb} +\TLpR^{adbc}\,\big]\;\M_8^{abcd}    &\numeq\cr
&-{1\over4}\,\big[\,\;\;
(\TLpR^{mcde}+ \TLpR^{mdec}+ \TLpR^{mced}+ \TLpR^{medc}) c^{mab}\cr
&\qquad\;
+ (\TLpR^{mbde}+ \TLpR^{mdeb}+ \TLpR^{mbed}+ \TLpR^{medb}) c^{mac}\cr
&\qquad\;
+ (\TLpR^{mebc}+ \TLpR^{mbce}+ \TLpR^{mecb}+ \TLpR^{mcbe}) c^{mad} \cr
&\qquad\; 
+ (\TLpR^{mdbc}+ \TLpR^{mbcd}+ \TLpR^{mdcb}+ \TLpR^{mcbd}) c^{mae}\,\big]\, 
 \M_9^{abcde},       \cr
=& -\;\big[\,2\TLpR^{acbd}+2\TLpR^{adbc} -
   (\TL+\TR)\,(c^{ebc}c^{eda}+c^{ebd}c^{eca})\,\big]\;\M_7^{abcd}\cr
&-2\, \big[\,2\TLpR^{abcd}+2\TLpR^{adcb} +
   (\TL+\TR)\,(c^{ebc}c^{eda}+c^{eba}c^{edc})\,\big]\;\M_8^{abcd} \cr
&+\;0\;\, \M_9^{abcde},       \cr
\sepeq
b\,\tilde\Delta_{6\LpR}=&\M_{10\LpR}+\M_{11\LpR}, \cr
b\,\tilde\Delta_{7\LpR}=&\M_{11\LpR}+i\,\M_{13\LpR}+
	{i\over2}\,c^{abc}\,\M_{14\LpR}^{abc},\cr
b\,\tilde\Delta_{8\LpR}=&\M_{10\LpR}-i\,\M_{12\LpR}-
	{i\over2}\,c^{abc}\,\M_{15\LpR}^{abc},\cr
}
$$\namelasteq\bvariation
The coefficient in $b\,\cdot\tilde\Delta_{5\LpR}\,$ of $\,\M_9^{abcde}$ 
turns out to be zero!, which can be proved by expressing the $c^{abc}$'s 
as commutators
of the corresponding matrices and then using the cyclicity of the trace.
If this coefficient would not be zero the system of equations below would
be incompatible! We have also used the relationship given in 
eq.~\EqTRelation .
	  
\medskip
Now, we impose the that the sum of the terms on the right 
hand side of eq.~\bvariation\ matches the right hand side 
of eq.~\resultbreak , after removal, of course, of the  
term carrying the essential non-Abelian anomaly. The following linear 
system of equations is thus obtained:
$$
\eqalignno{
&*\;{\rm from}\; \M_1^{ab}:\quad 
	2 a_1-2a_2=\ucpc {\TL+\TR\over3},\cr
&*\;{\rm from}\; \M_2^{abc}:\quad
-2 a_2-i\,c=-\ucpc {\TL+\TR\over3}\quad {\rm and}\quad
				\dLpR=0,\cr
&*\;{\rm from}\; \M_4^{abc}:\quad
2 a_1+i\,c=\ucpc (\TL+\TR){2\over3}\quad {\rm and}\quad
				\dLpR=0,\cr
&*\;{\rm from}\; \M_5^{abc}:\quad \dLpR=0,
\qquad\hbox{ we will not write this unknown anymore}\cr
&*\;{\rm from}\; \M_7^{abcd}:\quad 
-{i\over2}\,c+\TL\,f_\L+\TR\,f_\R=\ucpc {\TL+\TR\over6}
	\quad{\rm and}\quad -2\,f_\LpR=-\ucpc\,{1\over6}\cr
&*\;{\rm from}\; \M_8^{abcd}:\quad 
i\,c-2\TL\, f_\L-2\TL\,f_\R=-\ucpc {\TL+\TR\over3}\quad{\rm and}\quad
	-4\,f_\LpR=-\ucpc\,{1\over3}\cr
&*\;{\rm from}\; \M_{9\LpR}^{abcd}:\quad {f_\LpR\over2}\times0=0
                     \cr
&*\;{\rm from}\; \M_{10\LpR}:\quad 
 n_\LpR+u_{2\LpR}=\ucpc\,g^2\,
	\big[\CLpR-{\CA\over4} + 
		(\aprime-1)\big({\CLpR\over6}-{\CA\over4}\big)\big]\cr
&*\;{\rm from}\; \M_{11\LpR}:\quad 
 n_\LpR+u_{1\LpR}=\ucpc\,g^2\,
	\big[\CLpR-{\CA\over4} + 
		(\aprime-1)\big({\CLpR\over6}-{\CA\over4}\big)\big]\cr
&*\;{\rm from}\; \M_{12\LpR}:\quad 
 -i\,u_{2\LpR}=\icpc\,g^2\,[1+(\aprime-1)]\,{\CA\over4}\cr
&*\;{\rm from}\; \M_{13\LpR}:\quad 
 i\,u_{1\LpR}=-\icpc\,g^2\,[1+(\aprime-1)]\,{\CA\over4}\cr
&*\;{\rm from}\; \M_{14\LpR}^{abc}:\quad 
 -{i\over2}\,u_{1\LpR}=\icpc\,g^2\,[1+(\aprime-1)]\,{\CA\over8}\cr
&*\;{\rm from}\; \M_{15\LpR}^{abc}:\quad 
 {i\over2}\,u_{1\LpR}=-\icpc\,g^2\,[1+(\aprime-1)]\,{\CA\over8}\cr
} 
$$

Notice that there are more constraints than unknowns. In fact, several
equations appear repeated and the system has 14 different equations and 
13 unknowns, which turns to be {\it compatible} whose {\it unique solution}
is:
$$
\eqalignno{
&a_1=\ucpc\, {5\over12}\, (\TL+\TR),\qquad
  a_2=\ucpc\, {1\over4}\, (\TL+\TR), \cr
&c=\icpc\, {1\over6}\, (\TL+\TR),\qquad
  \dLpR=0,          \cr
&f_\LpR=\ucpc\,{1\over12},\qquad
	n_\LpR=\ucpc\,g^2\,\big[\CLpR +(\aprime-1){\CLpR\over6}\big]\cr
&u_{1\LpR}=u_{2\LpR}=-\ucpc\,g^2\,{\CA\over4}\,\aprime .
}
$$

In summary, the following choice of $S^{(1)}_{\rm fct}$ removes up to
order $\hbar$ the spurious anomaly terms from the BRS equation 
(eq.~\EqRenYMSTBreaking ) 

\smallskip
$$
\eqalignno{
S^{(1)}_{\rm fct} = \intd^dx &\Big\{ \,
  \hcpc  (T_L+T_R) 
	 \left[ {5 \over 12} (\pr_\mu A^\mu)^2 +
	{1 \over 4} A_\mu \square\, A^\mu \right] \cr
 &\quad - \hcpc {T_L +T_R \over 6} c^{abc}\, 
	 (\pr_\mu A_\nu^a) A^{b\mu} A^{c\nu} \cr
 &\quad  + \hcpc {T^{abcd}_L + T^{abcd}_R \over 12}\,
		 A^a_\mu A^{b\mu} A^c_\nu A^{d\nu}\cr
 &\quad + \hcpc g^2 \left[ C_L+(\a'-1)C_L /6\right]\,{i \over 2}\,
	 \bar\psi \gambar^\mu P_L \arrowsim\pr_\mu\psi        &\numeq\cr
 &\quad + \hcpc g^2  \left[ C_R+(\a'-1)C_R /6\right]\,{i \over 2}\,
	 \bar\psi' \gambar^\mu P_R \arrowsim\pr_\mu\psi'\cr
 &\quad - \hcpc g^2 {C_A \over 4} [1+(\a'-1)]
	  \left( \bar L s\psi + L s\bar\psi+\bar R s\psi'+R s\bar\psi' \right)
 \Big\}\cr
 +&\hbar^1\,l_g^{(1)}\,L_g + \hbar^1\,l_{\psi}^{(1)}\,L_{\psi} +
 \hbar^1\,l_{\psi^\prime}^{(1)}\,L_{\psi^\prime} + \hbar^1\,l_A^{(1)}\,L_A +
 \hbar^1\,l_\om^{(1)}\,L_\om, \cr
}
$$\namelasteq\EqSfctUno
\smallskip
\noindent where $l_g^{(1)}$, $l_{\psi}^{(1)}$, $l_{\psi^\prime}^{(1)}$,
$l_A^{(1)}$ and $l_\om^{(1)}$ are arbitrary 
coefficients which will determine the renormalization conditions at
the one-loop order and $L_g$, $L_{\psi}$, $L_{\psi^\prime}$,
$L_A$, $L_\om$ are any of the generalizations to the $d$-dimensional 
space-time of Dimensional Regularization of the corresponding $b$-invariant
four-dimensional operators. We have written $S^{(1)}_{\rm fct}$ in the
$d$-dimensional space of Dimensional Regularization so that diagrammatic
computation can be readily done.

Notice that no terms in $\bar\om$ or $\rho$ are added, excepting the
combination $\tilde\rho$ in the invariant counterterms; therefore
the ghost equation holds up to order $\hbar$.

Notice also that, because of finite counterterms depending of the external
fields $L$, $\bar L$, $R$ and $\bar R$ have been 
added, we have now, {\it e.g.} (set $l_\om^{(1)}=0$),
$$
\eqalignno{
{\delta\GR\over\delta \bar L(x)}\Big|_{L\equiv0}&\ne
      \N[s\psi](x)\cdot\GR                         \cr
{\delta\GR \over\delta \bar L(x)}\Big|_{L\equiv0}&=
		\N[s\psi](x)\cdot\GR+ \hbar^1\,u_{1\L}^{(1)}\, s\psi(x) +
			O(\hbar^2)=
\N[s\psi+\hbar^1\,u_{1\L}^{(1)}\, s\psi](x)\cdot\GR + O(\hbar^2),
}        
$$
which can be interpreted as a ``non-minimal'' renormalization of the
operator insertion (we always apply minimal subtraction to the 
regularized diagrams, but the action have explicit higher order finite 
counterterms). Moreover, with $l_A^{(1)}\ne0$ or $l_\om^{(1)}\ne0$, the
insertion of the non-linear BRS-variations of the fields can be renormalized
in different ways, all of them compatible with the 
BRS identities ({\it i.e.} there are several renormalization conditions 
for that insertions compatible with the identities).

Of course, the finite counterterms given in eq.~\EqSfctUno, 
are strongly dependent on the  choice for the 
$O(\hbar^0)$ action, $S_0$ (the Dimensional Regularization classical 
action in eq.~\Dchiralact ). If we 
would have chosen the other mentioned fermionic vertex, the counterterms
would have certainly been completely different (and more complicated). But
anyway the procedure would have been equally correct.

Notice that if the classical original theory is CP invariant the finite 
counterterms of eq.~\EqSfctUno\ are also CP invariant. This was expected, 
due to the comments in the second paragraph below eq.~\EqBreaking.

\bigskip

\subsection{3.7. The renormalization group equation}

It is plain that the standard textbook techniques usually employed to
derive the renormalization group equation for dimensionally renormalized
vector theories like Q.E.D and Q.C.D are of no use here. The formalism of
bare fields, bare coupling constants and multiplicative renormalization
of the classical action breaks down as shows eq.~\OnePIpole . The
generalization of this formalism to include evanescent operators, 
and the corresponding renormalized coupling constants, becomes a bit 
awkward when the regularized theory does not posses the 
symmetries that should hold in the quantum field theory. Indeed a host of 
bare evanescent non-symmetric operators, each bringing in a new renormalized 
coupling constant,  may enter the game. This is  in addition to the finite 
non-evanescent non-symmetric counterterms  needed to restore the 
symmetry broken in the dimensional renormalization process. These finite
counterterms should also come from bare operators. Hence, the construction 
of an appropriate bare action and the derivation of the true ({\it i.e.}
only involving as many coupling constants and anomalous dimensions as
there are classical couplings and fields) renormalization group equation
is a rather involved task (for related information see ref.~\cite{\Bare}).    
We shall abandon this quest altogether.  Indeed,  thanks to the {\it Quantum
Action Principle}, Bonneau identities and the formalism of algebraic
renormalization there is no need to introduce bare fields and bare couplings
to derive the true renormalization group equation. We only need the
renormalized 1PI functional, $\Gamma_{\rm ren}$,  which is symmetric up to 
the order in $\hbar$ demanded, obtained by the method explained in
this paper.

Let  $\Gamma_{\rm ren}[\varphi, \Phi,K_{\Phi};g,\alpha',\mu]$ be the  
dimensionally renormalized up to order $\hbar^n$  1PI functional 
of our theory, which we will take to be anomaly free all along the 
current subsection. $\Gamma_{\rm ren}$ depends 
explicitly on the fields, collectively denoted by  $\varphi$
and  $\Phi$, the external fields, collectively denoted by $K_{\Phi}$, 
the coupling constant $g$, the gauge fixing parameter $\alpha'$ and 
the Dimensional Regularization scale $\mu$, which is introduced along
the lines laid in {\it v)} of subsection {\it 2.2.} 
We shall assume that finite countertems, $S^{(n)}_{\rm fct}$ 
(see eq.~\Dchiralact ), has been chosen
so that the BRS equation $\SS(\GR)=0$, the gauge-fixing condition 
${\cal B} (\GR)=0$ (eq.~\EqGaugeFixingCondition ), and the ghost equation
${\cal G}\,\GR=0$ (eq.~\EqGhostEquation ) hold up to order $\hbar^n$.

The renormalization group equation
of the theory gives the expasion of the functional $\muGR$ in terms of
a certain basis of quantum insertions of  
ultraviolet dimension 4 and ghost number 0
\cite{\BonneauB,\PiguetSorella}. 
The coefficients of this expansion, which are formal power series in
$\hbar$, are the beta functions and anomalous dimensions of the theory.
Since $\Gamma_{\rm ren}$ satisfies, up to order $\hbar^n$, the equations 
mentioned in the previous paragraph, the elements 
of basis of insertions we are seeking is not only constrained  
by power-counting and ghost number. Indeed, the
following three equations hold up to order $\hbar^n$:
$$
\eqalignno{
\SS_{\GR}\;\muGR&=0,             \cr
{\delta\over\delta B}\; \muGR&=0,  &\numeq\cr
{\cal G}\;\muGR&=0.            \cr
}     \namelasteq\EqSymmetricSpace %
$$    
Notice that the operator $\mu{\partial \over \partial\mu}$ commutes with the 
functional operators ${\cal B}$, ${\cal G}$ and that 
$$
\mu{\partial \over \partial\mu}(\SS (\GR))=\SS_{\GR}\;\muGR.
$$ 
$\SS_{\GR}$ stands for the linearized BRS operator:
$$
\eqalignno{\SS_{\GR}=
\intd^4 x\,\,\Bigl\{
&\Tr\,{\delta\Gamma_{\rm ren}\over\delta\rho^\mu}
 {\delta \over\delta A_\mu}+
\Tr\, {\delta\Gamma_{\rm ren}\over\delta A_\mu}
      {\delta\over\delta\rho^\mu}+
\Tr\,{\delta\Gamma_{\rm ren}\over\delta\zeta}
 {\delta\over\delta\om}+
\Tr\,{\delta\Gamma_{\rm ren}\over\delta\om}
{\delta\over\delta\zeta}+
\Tr\; B {\delta\Gamma_{\rm ren}\over\delta\bar\om} \cr
&+ {\delta\Gamma_{\rm ren}\over\delta\bar L} 
{\delta\over\delta\psi}+
{\delta\Gamma_{\rm ren}\over\delta\psi}
{\delta\over\delta\bar L} +
{\delta\Gamma_{\rm ren}\over\delta\bar R}
 {\delta\over\delta\psi'}+
{\delta\Gamma_{\rm ren}\over\delta\psi'}
{\delta\over\delta\bar R}+
{\delta\Gamma_{\rm ren}\over\delta L}
 {\delta\over\delta\bar\psi}+
{\delta\Gamma_{\rm ren}\over\delta\bar\psi}
{\delta\over\delta L}\cr
&+ {\delta\over\delta R} 
{\delta\Gamma_{\rm ren}\over\delta\bar\psi'}+
{\delta\Gamma_{\rm ren}\over\delta\bar\psi'}
 {\delta\over\delta R} 
\Bigr\}.\cr 
}
$$

The functionals with ultraviolet dimension 4 and ghost number 0 which satisfy  
up to order $\hbar^{n}$ the set of equations in eq.~\EqSymmetricSpace\
form a linear space. Let us construct a basis of this  space.
In the classical approximation ({\it i.e.} at order
$\hbar^0$), a basis of this space is given \cite{\PiguetSorella} by the
$b\equiv\SS_{S_{\rm cl}}$-invariant terms $L_g$, $L^{R}_{\psi}$, 
$L^{L}_{\psi'}$, $L^{L}_{\psi}$, $L^R_{\psi'}$, $L_A$ and $L_\omega$, 
defined as follows:
$$
\eqalignno{
 L_g= &g{\pr S_{\rm cl}\over\pr g} =   
   -2\, (S_{{\rm cl}\,AA} + S_{{\rm cl}\,AAA} + S_{{\rm cl}\,AAA}),  \cr
L^{\rm R}_{\psi}=&\NN_{\psi}^\R\,S_{\rm cl}=2 S^\R_{{\rm cl}\,\bar\psi\psi},
     \qquad\qquad\qquad\;
L^{\rm L}_{\psi'}= \NN_{\psi'}^\L\,S_{\rm cl}=
2 S^\L_{{\rm cl}\,\bar\psi'\psi'}, \cr
L^{\rm L}_{\psi}=&\NN_{\psi}^\L\,S_{\rm cl}=2 S^\L_{{\rm cl}\,\bar\psi\psi} +
   2 S_{{\rm cl}\,\bar\psi\psi A},
\qquad
L^{\rm R}_{\psi'}=\NN_{\psi'}^\R\,S_{\rm cl}=
2 S^\R_{{\rm cl}\,\bar\psi'\psi'} + 
	2 S_{{\rm cl}\,\bar\psi'\psi' A},                      &\numeq\cr
L_A=&\NN_A\,S_{\rm cl}=
2 S_{{\rm cl}\,AA} + 3 S_{{\rm cl}\,AAA} + 4 S_{{\rm cl}\,AAAA} +
S_{{\rm cl}\,\bar\psi\psi A} + 
S_{{\rm cl}\,\bar\psi'\psi' A} - S_{{\rm cl}\,\rho\om} -
		S_{{\rm cl}\,\bar\om\om},         \cr
L_\om=&\NN_\om\,S_{\rm cl}=S_{{\rm cl}\,\rho\om}+
S_{{\rm cl}\,\bar\om\om}+S_{{\rm cl}\,\rho\om A} +
S_{{\rm cl}\,\zeta\om\om} +  S_{{\rm cl}\,\bar L s\psi} 
+ S_{{\rm cl}\,\bar R s\psi'} +
	S_{{\rm cl}\,L s\bar\psi} + S_{{\rm cl}\,R s\bar\psi'},         \cr
}
$$ \namelasteq{\EqInvariantTerms}%
where the $S_{{\rm cl}\,\cdots}$ terms are the corresponding 
$B$-independent terms  of the classical action $S_{\rm cl}$ given in 
eq.~\Classicalact . The symmetric differential operators
\cite{\PiguetSorella}
$\NN_{\psi}^\R$, $\NN_{\psi}^\L$, $\NN_{\psi'}^\R$, $\NN_{\psi'}^\L$, 
$\NN_A$ and $\NN_\om$ in eq.~\EqInvariantTerms\ are given by the
following identities:
$$
\eqalignno{
&\NN_{\psi}^{\L}\equiv N_{\psi}^{\L}+N_{\bar\psi}^{\L}
-N_{L}-N_{\bar L},\quad\NN_{\psi}^{\R}\equiv N_{\psi}^{\R}+
N_{\bar\psi}^{\R},\cr
&\NN_{\psi'}^{\L}\equiv N_{\psi'}^{\L}+N_{\bar\psi'}^{\L},\quad
\NN_{\psi'}^{\R}\equiv N_{\psi'}^{\R}+N_{\bar\psi'}^{\R}-N_R-N_{\bar R}\cr
&\NN_A\equiv {\rm Tr}(N_A-N_{\rho}-N_{B}-N_{\bar\omega})+2\alpha{\partial\over
\partial \alpha},&\numeq\cr
&\NN_{\omega}\equiv{\rm Tr}(N_{\omega}-N_{\zeta})
. \cr \namelasteq{\Diffoperators} 
}
$$
The field-counting operators $N_{\psi}^{\L(\R)}$, $N_{\psi'}^{\L(\R)}$,
$N_L$, $N_{\bar L}$,  $N_R$, $N_{\bar R}$, $N_A$, $N_{\omega}$, $N_{\bar\om}$
and $N_{\zeta}$ are the operators in eq.~\Fieldcounting\ for $d=4$. 

We shall work out now a basis (up to order $\hbar^n$, of course) of the 
space of joint solutions of eqs.~\EqSymmetricSpace\ with the help of the 
statement made in the second paragraph of subsection {\it 2.4.} 
(see ref.~\cite{\PiguetSorella} for further details). Up to order $\hbar^n$,
a quantum extension of the classical basis in eq.~\EqInvariantTerms\ is
furnished \cite{\PiguetSorella} by the action on $\GR$ of the 
symmetric differential operators defined above along with 
symmetric differential  operator $g\pr_g$.
Indeed, the  functionals
$$ 
g{\pr\GR\over\pr g},\;\; \NN_{\psi}^\R\,\GR,\;\;  
\NN_{\psi'}^\L\,\GR,\;\; 
 \NN_{\psi}^\L\,\GR,\;\; \NN_{\psi'}^\R\,\GR,\;\;
 \NN_A\,\GR,\;\; \NN_\om\,\GR,
$$
which are to be understood as formal series expansions in $\hbar$, match
the corresponding functionals in eq.~\EqInvariantTerms\ at order $\hbar^0$ 
and the symmetric differential operators 
$g\pr_g$, $\NN_{\psi}^\R$, $\NN_{\psi}^\L$, 
$\NN_{\psi'}^\R$, $\NN_{\psi'}^\L$, $\NN_A$ and $\NN_\om$
are compatible (this is why they are  symmetric), up to order
$\hbar^n$, with the BRS equation, the gauge-fixing condition and the ghost 
equation, in the sense we  spell out next.
 
Let $\cal F$ be the linear space of functionals with ultraviolet dimension 4
and ghost number 0 which satisfy (up to order 
$\hbar^n$) the BRS equation, the gauge-fixing condition and the ghost equation:
$\SS(\Gamma)=0$, ${\cal B}(\Gamma)=0$, ${\cal G}(\Gamma)=0$, 
$\forall \Gamma\in{\cal F}$.   An operator $\cal D$ 
acting on ${\cal F}$ is said to be compatible with the BRS equation, 
the gauge-fixing condition and the ghost equation if, by definition, 
the following set of equations hold: 
$$
\SS_{\Gamma}({\cal D}\Gamma)=0,\qquad
{\delta\over \delta B}({\cal D}\Gamma)=0,\qquad
{\cal G}({\cal D}\Gamma)=0.
$$

In summary, up to order $\hbar^n$, any joint solution 
of eqs.~\EqSymmetricSpace\ is a linear combination with $\hbar$-dependent
coefficients of the following functionals:
$$
g{\pr\GR\over\pr g},\;\; \NN_{\psi}^\R\,\GR,\;\;  \NN_{\psi'}^\L\,\GR,\;\; 
 \NN_{\psi}^\L\,\GR,\;\; \NN_{\psi'}^\R\,\GR,\;\; \NN_A\,\GR,\;\; \NN_\om\,\GR.
\eqno\numeq
$$\namelasteq\Quantumbasis

In particular, if $\mu$ is the arbitrary ``scale'' introduced by hand
for each loop integration in minimal Dimensional Renormalization 
(see {\it v)} in subsection {\it 2.2.}) , then, 
$\muGR$ has an expansion in the quantum basis given above:
$$
\left[\,\mu\,{\pr\over\pr\mu} + \b g\,{\pr\over\pr g} 
	- \ga_\psi^\L \NN_\psi^\L- \ga_{\psi'}^\R \NN_{\psi'}^\R 
	- \ga_\psi^\R \NN_\psi^\R- \ga_{\psi'}^\L \NN_{\psi'}^\L 
	-\ga_A \NN_A -\ga_\om \NN_\om\, \right]\; \GR=0.  \eqno\numeq
$$ \namelasteq\EqRGExacta
This equation holds up to order $\hbar^n$ and it is the renormalization 
group equation of our theory. Notice that the expansion $\muGR$ in terms of
the basis in eq.~\Quantumbasis\ is only possible 
if  the renormalized functional satisfies order by order the BRS identity 
(hence, there is anomaly cancellation, in particular) and the gauge-fixing 
condition. It is important to notice that the regularized 1PI 
functional does not  generally satisfy an equation such as eq.~\EqRGExacta . 

Let us notice that the coefficients of the expansions of 
$\mu\,{\pr\GR\over\pr\mu}$ in the quantum basis of insertions in 
eq.~\Quantumbasis\ are the beta functions and anomalous dimensions of 
the theory. These coefficients are to be understood as formal expansions 
in $\hbar$. The first non-trivial contribution to these expansions is always
of order $\hbar$, since the lowest order contribution to $\GR$ is the
classical action, $S_{\rm cl}$, and $\mu\,{\pr S_{\rm cl}\over\pr\mu}=0$.

We next proceed to the computation of the beta function and anomalous
dimensions of our theory. This is done \cite{\BonneauA} by expressing first 
$\mu\,{\pr\GR\over\pr\mu}$, $g\,{\pr\GR \over\pr g}$ and $\NN_\phi\GR$,
where $\NN_\phi$ denotes any of the differential operators in 
eq.~\Diffoperators , as  insertions in $\GR$ of linearly independent local 
normal products of the fields and their derivatives; then, 
these insertions are substituted into 
the left hand side of the renormalization  
group equation (eq.~\EqRGExacta) and, finally, one solves 
for $\b$, $\ga_\psi^\L$,  $\ga_{\psi'}^\R$, $\ga_\psi^\R$, $\ga_{\psi'}^\L$, 
$\ga_A$ and $\ga_\om$, the
set of equations involving only numbers  which results from the linear 
independence of the  aforementioned local normal products.

That $g\,{\pr\GR \over\pr g}$ and $\NN_\phi\GR$, where $\NN_\phi$ denotes any 
of the differential operators in eq.~\Diffoperators , can be expressed
as insertions in $\GR$ of normal products is a mere consequence of
the {\it Quantum Action Principle}: QAP1 and QAP3, given by eqs.~\QAPone\
and \QAPthree ,  lead to 
$$
{\pr\GR\over\pr g}= \N[{\pr (S_0+S^{(n)}_{\rm fct})\over\pr g}]
\cdot \GR, \eqno\numeq
$$ \namelasteq{\EqPartialg}
and
$$
\NN_\phi\,\GR= \N[(\NN_\phi\, (S_0+S^{(n)}_{\rm fct}))]
\cdot \GR,           \eqno\numeq
$$ \namelasteq{\EqCountingOperator}
respectively. Hence, every element of the symmetric quantum basis in 
eq.~\Quantumbasis\ is a local combination of quantum insertions of integrated 
standard ({\it i.e.} non-evanescent)  monomials with ultraviolet dimension 4
and ghost number 0 (see subsection {\it 2.4.}).
It should not be  overlooked the fact $S_0+S^{(n)}_{\rm fct}$ in 
eqs.~\EqPartialg\ and \EqCountingOperator\ is the non-singular contribution
to the Dimensional Regularization action, $S^{(n)}_{DReg}$  
(see subsection {\it 3.2.}), employed to 
obtain the minimal dimensionally renormalized 1PI functional  up 
to order $\hbar^n$, $\GR$, by following the algorithm
spell out in subsection {\it 2.2.} 
The formal functional $S_0+S^{(n)}_{\rm fct}$ is therefore an object in the
$d$-dimensional space-time of Dimensional Regularization. If $n>2$, it would 
not do to use  some other functional, $S'$, such that 
$(S_0+S^{(n)}_{\rm fct})-S'={\hat S}'$ is an integrated evanescent 
operator whose $\hbar$-expansion has contributions of order $\hbar^m$, with
$m\le n-2$. Recall that the contribution $O(\hbar^{(n-1)})$ in ${\hat S}'$
is not be  relevant here since, being an evanescent object, the lowest order 
contribution coming from its insertion in $\GR$ is $O(\hbar^n)$ and local,  
so that it does not contribute to the renormalization group equation 
at order $\hbar^n$; a similar argument holds for the term of order
$\hbar^n$ in ${\hat S}'$. 

Now, by substituting eqs.~\EqPartialg\ and \EqCountingOperator\ in 
eq.~\EqRGExacta\ one obtains the following equation
$$
\mu\,{\pr\GR\over \pr\mu} = 
-\beta\,g\; \N[{\pr (S_0+S^{(n)}_{\rm fct})\over\pr g}]\cdot \GR+
\sum_{\phi}\gamma_{\phi}\;\N[(\NN_\phi\, (S_0+S^{(n)}_{\rm fct}))]
\cdot \GR, \eqno\numeq  \namelasteq{\Firstexpansion}
$$
where $\phi$ labels the symmetric differential operators in
eq.~\Diffoperators.
Hence, to compute the beta functions and anomalous dimensions of the theory
we only need an independent way of computing $\mu\,{\pr\GR\over \pr\mu}$ in
terms of the insertions on the right hand side of 
eq.~\Firstexpansion . We shall
set up this independent way of computation next. 
    
Unlike for the subtraction algorithm employed in ref.~\cite{\BMa}, in
 minimal dimensional renormalization, one  can not  
use (see ref.~\cite{\BonneauA}) the {\it Quantum Action Principle} to 
obtain an equation for  $\mu\,{\pr\GR\over\pr\mu}$ analogous to 
eqs.~\EqPartialg , \EqCountingOperator. Indeed, in minimal dimensional 
renormalization, $\mu$ is not a parameter of the action 
$(S_0+S^{(n)}_{\rm fct})$, but a parameter which is introduced by hand 
in each loop-momentum integration (as a factor $\mu^{4-d}$) before computing 
the renormalized value of the corresponding Feynman diagram. 

The problem of expressing $\mu\,{\pr\GR\over \pr\mu} $ as an insertion  in 
$\GR$ of a linear combination normal product operators
was also solved by Bonneau \cite{\BonneauA} in the 
framework of rigorous minimal dimensional renormalization. He obtain a formula
similar to eq.~\EqBonneauIdentities, {\it i.e.} a Zimmerman-like identity, 
for a scalar case  with no added finite counterterms in the action. 
Here we need the generalization to the presence of
several type of fields, including external fields, the presence
of hatted objects and also the presence of finite counterterms up to order
$\hbar^n$, in the action $(S_0+S^{(n)}_{\rm fct})$.  The needed generalization 
reads
$$
\eqalignno{
&\mu\,{\pr\GR\over\pr\mu} =
	\sum_{n=0}^{4}\sum_{\{j_1,\cdots, j_n\}} \sum_{N_l>0}\,N_l\;  
	\bigg[\cr
&\sum_{r=0}^{\om(J)}\sum_{\scriptscriptstyle\{i_1\,\ldots\, i_r\} \atop
					\scriptscriptstyle 1\le i_j\le n-1 } 
	\Big\{ \rsp {(-i)^r\over r!} 
		{\pr^r\over\pr p_{i_1}^{\mu_1}\cdots\pr p_{i_r}^{\mu_r} }
(-i\hbar)\overline        
{\bigl\langle                              
	{\tilde\phi_{j_1}(p_1)\ldots\tilde\phi_{j_n}
		(p_n\!=\!\!-\!{\textstyle\sum} p_i)
		\bigr\rangle}}^{\irr, (N_l)}_{K=0} 
	\Big|_{p_i=0}  \Big\}                        \cr
&\qquad\quad\times\;{1\over n!}\; \intd^4x\; 
\N\Bigl[\phi_{j_n}\!\!\!\!\prod_{k=n-1}^{1} 
   \Bigl\{\Bigl(\prod_{\{\a / i_\a=k\}}\!\!\pr_{\mu_\a}\Bigr)
\phi_{j_k}\Bigr\}\Bigr] (x)
		\cdot\GR+                                            \cr
&\sum_{\Phi}\;\sum_{r=0}^{\omega(J;\Phi)}
	\sum_{\scriptscriptstyle\{i_1\,\ldots\, i_r\} \atop
			\scriptscriptstyle 1\le i_j\le n  } \;
	\Big\{\rsp {(-i)^{r}\over (r)!} 
	{\pr^r\over\pr p_{i_1}^{\mu_1}\cdots\pr p_{i_r}^{\mu_r}}     \cr
&\qquad\qquad\qquad\qquad\quad
\overline       
 {\bigl\langle 
	{\tilde\phi_{j_1}(p_1)\ldots\tilde\phi_{j_n}(p_n)
\N[s\Phi](p_{n+1}\!=\!\!-\!{\textstyle\sum} p_i)
\bigr\rangle}}^{\irr, (N_l)}_{K=0} 
	\Big|_{p_i=0}  \Big\}                        \cr
&\qquad\quad\times\;{1\over n!} \;\intd^4x\;( K_{\Phi}(x)\;
	\N\Bigl[ \prod_{k=n}^1 
   \Bigl\{\Bigl(\prod_{\{\a / i_\a=k\}}\!\!\pr_{\mu_\a}\Bigr)
          \phi_{j_k}\Bigr\}\Bigr]
    (x) \cdot\GR \quad  \bigg]\;
      , &\numeq\cr    \namelasteq{\EqPartialmuGeneralization}
} 
$$ 
where $J=\{j_1,\cdots,j_n\}$, and,  $\omega(J)$ and 
$\omega(J;\Phi)$ stand, respectively, for the overall ultraviolet 
degree of divergence of 
${\bigl\langle{\tilde\phi_{j_1}(p_1)\cdot\!\cdot\,\tilde\phi_{j_n}(p_n)
\bigr\rangle}}^{\irr}_{K=0}$ and 
${\bigl\langle {\tilde\phi_{j_1}(p_1)\cdot\!\cdot\,\tilde\phi_{j_n}(p_n)
\N[s\Phi](p_{n+1})\bigr\rangle}}^{\irr}_{K=0}$. The bar upon the 1PI
Green functions means that all subdivergences has been subtracted from the
Feynman diagrams which contribute to them. This is as in
eq.~\EqBonneauGeneralization.  The novel feature here is the presence of
the superscript $N_l$ on the upper right of the Green function, which
indicates that only Feynman diagrams  with precisely $N_l$ loops are to
be considered. Notice that one then sums over all number loops upon 
multiplication by $N_l$ of the $N_l$-contribution.

The r.s.p.~of the subtracted (subdivergences only) 1PI Feynman diagrams 
contributing to the right hand side of  
eq.~\EqPartialmuGeneralization\ is a local polynomial in the $d$-dimensional 
space of Dimensional Regularization of external momenta
associated with the fields of the corresponding 1PI function; so, in general, 
it will contain hatted and barred objects (metrics, momenta, gamma matrices). 
Therefore, the formula will generate automatically also evanescent 
(also called anomalous)  insertions. The meaning of $\times$ in
last formula is the same as in eq.~\EqBonneauIdentities. Hence,
the expansion in eq.~\EqPartialmuGeneralization\ can be recast into the form
$$
\mu {\pr \GR\over\pr\mu} =\sum_i \bar r_{i}\; \N[\bW_i]\cdot\GR+
			\sum_j \hat r_{j}\; \N[\hW_j  ]\cdot\GR 
      \eqno\numeq  \namelasteq{\EqPartialmuAnomala}
$$ 
where  $\bW_i$ and $\hW_j$ denote, 
respectively, the barred (non-evanescent) and hatted (evanescent) 
elements of a basis of  integrated monomials with ghost number 0 and 
ultraviolet dimension 4. 
The symbols  $\bar r_{i}$ and $\hat r_{j}$ in eq.~\EqPartialmuAnomala\ are 
coefficients defined as formal expansions in $\hbar$ (starting at $\hbar$, 
since their computation involves that of the r.s.p.~of the divergent part 
of the appropriate 1PI function). The order $\hbar^m$, $0<m\le n$, 
contribution to these coefficients is obtained as follows:

$*$ Take a divergent, by power-counting, 1PI function with some number of 
	fields of definite type ( maybe including some external field coupled
	to a BRS variation). 
	
$*$ Compute the residue of the simple pole (r.s.p.~) of every 
	$O(\hbar^m)$-graph constructed with the Feynman rules derived from the 
 action $(S_0+S^{(n)}_{\rm fct})$ and contributing to that 
1PI function, with all their subdivergences subtracted. 
	
$*$ Multiply the last quantity by the number of the loops of the graph 

$*$ Sum over all graphs contributing to the 1PI function

$*$ Being the sum a local expression, their different terms can be 
	interpreted as Feynman rules of tree-level integrated insertions of 
   Lorentz invariant (normal or anomalous) composite operators, formed by 
   just the fields of the original 1PI function and some different 
	combination of metrics, derivatives and gamma matrices.

$*$ Do the same steps for every divergent 1PI function. Then you will
obtain the expansion eq.~\lasteq\ with the $O(\hbar^m)$ coefficients. 

Explicit expressions for each $\bar r_i$ and $\hat r_j$ similar to
that ones for the expansion of the anomalous breaking can be
given. For example, if 
$$
\eqalignno{
{\bW}_2^{ab}   &=\int\,\bar{\square}\, A_{\bar\mu}^a\, A^{b \bar\mu},
\qquad
{\hW}_{21}^{ab}=\int\,\bar{\square}\, A_{\hat\mu}^a\, A^{b \hat\mu}, 
\cr
{\hW}_{22}^{ab}&=\int\,\hat{\square}\, A_{\bar\mu}^a\, A^{b \bar\mu}, 
\qquad
{\hW}_{23}^{ab}=\int\,\hat{\square}\, A_{\hat\mu}^a\, A^{b \hat\mu}, 
   \cr
}
$$
then
$$
\eqalignno{
\bar r_{2}^{ab}&={-(-i)^2\over2} \;N_l \;\times   
	\;\hbox{coef. in r.s.p. 
      $(-i\hbar)\,\overline{\bigl\langle 
		A_\mu^a(p_1)A_\nu^b(p_2\!\equiv\!-p_1) \bigr\rangle}^\irr$ 
		of $\,$} 
   \bar p_1{}^2 \;\gbar^{\mu\nu}    \qquad\cr
\hat r_{21}^{ab}&={-(-i)^2\over2} \;N_l \;\times   
	\;\hbox{coef. in r.s.p. 
      $(-i\hbar)\,\overline{\bigl\langle 
		A_\mu^a(p_1)A_\nu^b(p_2\!\equiv\!-p_1) \bigr\rangle}^\irr$ 
		of $\,$} 
   \bar p_1{}^2 \;\ghat^{\mu\nu},         \qquad \cr
\hat r_{22}^{ab}&={-(-i)^2\over2} \;N_l\;\times   
	\;\hbox{coef. in r.s.p. 
      $(-i\hbar)\,\overline{\bigl\langle 
	A_\mu^a(p_1)A_\nu^b(p_2\!\equiv\!-p_1) \bigr\rangle}^\irr$ 
		of $\,$} 
   \hat p_1{}^2 \;\gbar^{\mu\nu},    \qquad\cr
\hat r_{23}^{ab}&={-(-i)^2\over2} \;N_l\;\times   
	\;\hbox{coef. in r.s.p. 
      $(-i\hbar)\,\overline{\bigl\langle 
	A_\mu^a(p_1)A_\nu^b(p_2\!\equiv\!-p_1) \bigr\rangle}^\irr$ 
		of $\,$} 
   \hat p_1{}^2 \;\ghat^{\mu\nu};    \qquad\cr
}
$$
and so on.

Next, by using the Bonneau identities  technique
we shall  express (see subsection {\it 2.4.}) every evanescent insertion
$\N[\hW_j]\cdot\GR$ on the right hand side of eq.~\EqPartialmuAnomala\ as
a linear combination of standard ({\it i.e.} non-evanescent) insertions 
$\N[\bW_i]\cdot\GR$:
$$
\N[\hW_j]\cdot\GR=\sum_i\,c_{ji}\;\N[\bW_i]
    \cdot\GR. \eqno\numeq \namelasteq{\Bonneauhelps}
$$
The coefficients $c_{ji}$ are formal expasions in powers of $\hbar$, having
no order $\hbar^0$ contribution.
The use of the previous equation turns  eq.~\EqPartialmuAnomala\ into the
following equation:
$$
\mu {\pr \GR\over\pr\mu} =\sum_i  r_{i}\; \N[\bW_i]\cdot\GR,
                \eqno\numeq \namelasteq{\EqPartialmu}
$$ 
where $r_i={\bar r}_i+\sum_{j}\,{\hat r}_j\,c_{ji}$, so that $r_i={\bar r}_i$
at order $\hbar$.
Notice that, because the evanescent insertions are to be expanded according
to the Bonneau identities in eq.~\EqBonneauIdentities , 
the computation of $r_i$ up to order $\hbar^n$ only involves the 
contributions to ${\hat r}_j$ up to order $\hbar^{(n-1)}$, whereas the 
contributions to ${\bar r}_i$ are needed  up to order $\hbar^n$.

Now, since $g\,{\pr\over\pr g} [\bar{S}_0+\bar{S}^{(n)}_{\rm fct}]$, 
$\NN_\phi\, [\bar{S}_0+\bar{S}^{(n)}_{\rm fct}]$,
$g\,{\pr\over\pr g}\,[ (S_0- \bar{S}_0)+
(S^{(n)}_{\rm fct}-\bar{S}^{(n)}_{\rm fct})]$ and
$\NN_\phi\,[ (S_0-\bar{S}_0)+
(S^{(n)}_{\rm fct}-\bar{S}^{(n)}_{\rm fct})]$ are linear combinations of
monomials of ultraviolet dimension 4 and ghost number 0, it is advisable 
to choose the basis of monomials $\{\bW_i,\hW_j\}$, 
which occurs in eq.~\EqPartialmuAnomala, in such a way that it contains 
the former set of monomials.  
In other words, the basis $\{\bW_i,\hW_j\}$ so chosen  
makes it possible to obtain the coefficients on the right hand side of the following 
equations
$$
\eqalign{&g\,{\pr\over\pr g} \big[\bar{S}_0+\bar{S}^{(n)}_{\rm fct}\big]
=\sum_i {\bar{\it w}}_{g i}\;{\bW}_i,\quad
\NN_\phi\, \big[\bar{S}_0+\bar{S}^{(n)}_{\rm fct}\big]=
\sum_i {\bar{\it w}}_{\phi i}\;{\bW}_i\cr
&g\,{\pr\over\pr g}\,\big[ (S_0- \bar{S}_0)+
(S^{(n)}_{\rm fct}-\bar{S}^{(n)}_{\rm fct})\big]=\sum_j {\hat{\it w}}_{g j}\;
{\hW}_j,\cr
&\NN_\phi\,\big[ (S_0-\bar{S}_0)+
(S^{(n)}_{\rm fct}-\bar{S}^{(n)}_{\rm fct})\big]=
\sum_j {\hat{\it w}}_{\phi j}\;
{\hW}_j,\cr
}
$$
by reading them from the left hand side of the corresponding equation. 
The coefficients
${\bar{\it w}}_{g i}$, ${\hat{\it w}}_{g i}$, 
${\bar{\it w}}_{\phi i}$ and ${\hat{\it w}}_{\phi i}$ have  expansions 
in powers of $\hbar$; these expansions, though, have now contributions of 
order $\hbar^0$.
 
It is a trivial exercise to  check that eq.~\Firstexpansion\ can be recast 
into the form
$$
\mu {\pr \GR\over\pr\mu} =\sum_i \big(-\beta\,{\bar{\it w}}_{g i}+\sum_{\phi} 
\,\gamma_{\phi}\,{\bar{\it w}}_{\phi i}\big)\;\N[\bW_i]\cdot\GR+
\sum_j (-\beta\,{\hat{\it w}}_{g i}+\sum_{\phi} 
\,\gamma_{\phi}\,{\hat{\it w}}_{\phi i}\big)\; \N[\hW_j  ]\cdot\GR 
$$
 The substitution of eq.~\Bonneauhelps\ into the last equation gives
the following result 
$$
\mu {\pr \GR\over\pr\mu} =\sum_i \big(-\beta\,{\it w}_{g i}+\sum_{\phi} 
\,\gamma_{\phi}\,{\it w}_{\phi i}\big)\;\N[\bW_i]\cdot\GR,
\eqno\numeq
$$\namelasteq{\equationusef}
where
$$
{\it w}_{g i}={\bar{\it w}}_{g i}+\sum_j\,{\hat{\it w}}_{g j}\,c_{ji}
\quad\hbox{and}\quad
{\it w}_{\phi i}={\bar{\it w}}_{\phi i}+\sum_j\,{\hat{\it w}}_{\phi j}\,c_{ji},
$$
where the index $j$ labels the elements of the ``evanescent'' basis 
$\{{\hW_j}\}$.

Finally, by comparing eqs.~\EqPartialmu\ and \equationusef, one obtains an
overdeterminate system of linear equations whose unknowns are the
coefficients $\beta$ and $\gamma_{\phi}$:
$$ 
r_i=-\beta\,{\it w}_{g i}+\sum_{\phi} \,\gamma_{\phi}\,{\it w}_{\phi i},
\eqno\numeq
$$\namelasteq{\linearsystem}
where the index $i$ runs over  the elements of the standard 
({\it i.e.} with no evanescent operator) basis $\{{\bW_i}\}$.
 
Several remarks about eq.~\EqPartialmuGeneralization\ are now in order.

$*$ This equation is only correct if the finite counterterms in 
$S_{\rm fct}^{(n)}$ are independent of $\mu$.
This poses no problem, because the finite counterterms
we need to add to the action to restore the BRS identities  do not
depend on $\mu$, thanks to the fact that in the Bonneau identities we 
have to compute only r.s.p.s of 1PI functions with all subdivergences
minimally subtracted and these r.s.p.~are always mass and $\mu$ 
independent.

$*$ The counterpart of eq.~\EqPartialmuGeneralization\ in 
ref.~\cite{\BonneauA} replaces the factor $N_l$, which counts the number
of loops, with the operator $\hbar\,{\pr\over\pr\hbar}$. This is because, 
{\it if no terms of order} $\hbar^m$, $m\! >\! 0$ are present 
in the action, then the powers in $\hbar$ of a graph 
counts its number of loops. But now we have added 
higher order terms in $\hbar$ to the action and so we cannot use 
$\hbar$ as a loop meter. This can be superseded by introducing 
a new parameter $l$, which divides the whole action, including the 
finite counterterms. Then $l^{N_l-1}$ would be the factor attached 
to a $N_l$-loop Feynman diagram. 
Note that only in the limit $l\rightarrow1$ the finite counterterms restore 
the BRS identities. Therefore, the factor $N_l$ in 
eq.~\EqPartialmuGeneralization\
is replaced by $\lim_{l\rightarrow1}\;\left\{1+l{\pr\over\pr l}\right\}$.
Thanks to the action principles, the action of the operator 
$\lim_{l\rightarrow1}\;\left\{1+l{\pr\over\pr l}\right\}$ 
on a 1PI Green function computed for arbitrary $l$ 
can be replaced with the insertion  of the
integrated operator $1-\N[i(S_0+S^{(n)}_{\rm fct})]$ into that 1PI function
at $l=1$. It should be stressed that this is true only if there is no explicit 
$\mu$-dependence in the finite counterterms.

\medskip

Let us move on and compute the beta function and anomalous dimensions of
our theory at order $\hbar$. At $\hbar$-order, eq.~\linearsystem\ reads
$$
{\bar r}_i^{(1)}=-\beta^{(1)}\,{\bar{\it w}}_{g i}^{(0)}+
\sum_{\phi} \,\gamma_{\phi}^{(1)}\,{\bar {\it w}}_{\phi i}^{(0)},
\eqno\numeq
$$\namelasteq{\orderhbar}
for $r_i={\bar r}_i^{(1)}\hbar+O(\hbar^2)$, 
$\beta=\beta^{(1)}\hbar+O(\hbar^2)$,  
${\it w}_{g i}={\bar{\it w}}_{g i}^{(0)}+O(\hbar)$, 
$\gamma_{\phi}^{(1)}=\gamma_{\phi}^{(1)}\hbar+O(\hbar^2)$ and 
${\it w}_{\phi i}={\bar{\it w}}_{\phi i}^{(0)}+O(\hbar)$. 
To obtain  $\{{\bar r}_i^{(1)}\}$, we just need a suitable basis of integrated
barred monomials ${\{\bW_i}\}$ with ultraviolet
dimension $4$ and ghost number $0$: the evanescent monomials 
${\{\hW_j}\}$ are not needed at this order in $\hbar$.
The basis  ${\{\bW_i}\}$ is furnished by the following list of 
monomials with ghost number 0, dimension 4, free gauge indices and fully
contracted Lorentz indices:
\bigskip

\noindent $*$ a) {\sl Monomials with only $A$s}

{\settabs 2 \columns
\openup1\jot
\+${\bW}_{1}^{ab}\equiv\int\!(\pr_\bmu\pr_\bnu\,A^{a\bmu})A^{b\bnu}$,
   &${\bW}_{2}^{ab}\equiv\int\!(\bar{\square} A_{\bmu}^a)A^{b\bmu}$,  \cr
\+${\bW}_{3}^{abc}\equiv\int\!(\pr_\bmu A_{\bnu}^a) A^{b\bmu} A^{c\bnu}$,
 &${\bW}_{4}^{abc}\equiv\int\!(\pr_\bmu A^{a\bmu}) A_\bnu^b A^{c\bnu}$,\cr
\+${\bW}_{5}^{abc}\equiv\int\!
   \eps_{\mu\nu\rho\a}\,(\pr^\a A^{b\bmu}) A^{a\bnu}A^{c\brho}$,\cr 
\+${\bW}_{6}^{a_1a_2a_3a_4}\equiv\int\!
     A_\bmu^{a_1} A^{a_2\bmu}A_\bnu^{a_3} A^{a_4\bnu}=
{\bW}_{6}^{\{a_1a_2\}\{a_3a_4\}}$,\cr
\+${\bW}_{7}^{a_1a_2a_3a_4}\equiv\int\!
      \eps_{\mu\nu\rho\delta}\,A^{a_1\mu} A^{a_2\nu}A^{a_3\rho}
        A^{a_4\delta}$.    \cr
}
\bigskip

\noindent $*$ b) {\sl Monomials involving fermions and no external fields}

{\settabs 2 \columns
\openup1\jot
\+${\bW}_8^{ij}\equiv\int\!{i\over2}
   \bar\psi_i\,\gambar^\mu\PL\arrowsim\pr_\bmu\psi_j$,
   &${\bW}_9^{ij\,a}\equiv\int\!
    \bar\psi_i\,\gambar^\mu\PL\psi_j\,A_\bmu^a$,     \cr
\+${\bW}_{10}^{ij}\equiv\int\!{i\over2}
   \bar\psi_i^{'}\,\gambar^\mu\PR\arrowsim\pr_\bmu\psi_j^{'}$,
   &${\bW}_{11}^{ij\,a}\equiv\int\!
    \bar\psi_i^{'}\,\gambar^\mu\PR\psi_j^{'}\,A_\bmu^a$,     \cr
\+${\bW}_{12}^{ij}\equiv\int\!{i\over2}
   \bar\psi_i\,\gambar^\mu\PR\arrowsim\pr_\bmu\psi_j$,
   &${\bW}_{13}^{ij\,a}\equiv\int\!
    \bar\psi_i\,\gambar^\mu\PR\psi_j\,A_\bmu^a$,\cr
\+${\bW}_{14}^{ij}\equiv\int\!{i\over2}
   \bar\psi_i^{'}\,\gambar^\mu\PL\arrowsim\pr_\bmu\psi_j^{'}$,
   &${\bW}_{15}^{ij\,a}\equiv\int\!
    \bar\psi_i^{'}\,\gambar^\mu\PL\psi_j^{'}\,A_\bmu^a$.\cr 
}
\bigskip

\noindent $*$ c) {\sl Monomials involving ghost and no external fields}
	
{\settabs 2 \columns
\openup1\jot
\+${\bW}_{16}^{ab}\equiv\int\!-\bar\om^a\,\bar{\square}\,\om^b=
   \int\!(\pr_\bmu\bar\om^a)\,(\pr^\bmu\om^b)$,  \cr
\+${\bW}_{17}^{abc}\equiv\int\!\bar\om^a\,\pr^\bmu
                  (\om^c A_\bmu^b)$,  
 &${\bW}_{18}^{abc}\equiv\int\!\bar\om^a\,
                (\pr^\bmu A^{b\bmu})\,\om^c$,     \cr
\+${\bW}_{19}^{abcd}\equiv\int\!\bar\om^a\om^b
                  A^{c\bmu} A_\bmu^d$,  
 &${\bW}_{20}^{abcd}\equiv\int\!\bar\om^a\bar\om^b
                \om^c\om^d$.     \cr
}
\bigskip

\noindent $*$ d) {\sl Monomials involving external fields}
	
{\settabs 2 \columns
\openup1\jot
\+${\bW}_{21}^{ij\,a}\equiv\int\!
       \om^a\,\bar L_i\PL\psi_j$,
   &${\bW}_{22}^{ij\,a}\equiv\int\!
      \bar\psi_i\PR\om^a L_j$,     \cr
\+${\bW}_{23}^{ij\,a}\equiv\int\!
       \om^a\,\bar R_i\PR\psi^{'}_j$,
   &${\bW}_{24}^{ij\,a}\equiv\int\!
      \bar\psi_i^{'}\PL\om^a R_j$,     \cr
\+${\bW}_{25}^{ij\,a}\equiv\int\!
       \om^a\,\bar L_i \PR\psi_j$,
   &${\bW}_{26}^{ij\,a}\equiv\int\!
      \bar\psi_i\PL\om^a L_j$,     \cr
\+${\bW}_{27}^{ij\,a}\equiv\int\!
       \om^a\,\bar R_i \PL\psi_j^{'}$,
   &${\bW}_{28}^{ij\,a}\equiv\int\!
      \bar\psi_i^{'}\PR\om^a R_j$,     \cr
\+${\bW}_{29}^{ab}\equiv\int\!
       \rho^a_\bmu\,\pr^\bmu\om^b$,
  &${\bW}_{30}^{abc}\equiv\int\!
      \rho^a_\bmu \om^b A^{c\bmu}$,     \cr
\+${\bW}_{31}^{abc}\equiv\int\!\zeta^a\om^b\om^c$. \cr
}
\namelasteq\EqWList
\bigskip

Next, eq.~\EqPartialmuGeneralization\ leads to 
$$
\mu {\pr \GR\over\pr\mu}=
\sum_{i=1}^{31}\sum_{A_i}\;{\bar r}^{(1),A_i}_i\;
\W^{A_i}_i\;+\;O(\hbar^2),
\eqno\numeq \namelasteq{\COEFF}
$$
where $A_i$ denotes the set of gauge indices, $\W^{A_i}_i$ denotes henceforth  
the counterpart of the operator ${\bW}^{A_i}_i$ in four dimensional space-time 
and  the coefficients ${\bar r}^{(1),A_i}_i$ are defined as follows:

$$
\eqalignno{
\bar r_{1}^{\ouno\,ab}=&{(-i)^2\over2}  
   \;\hbox{coef. in r.s.p.  $(-i)$
		$\overline{\bigl\langle 
         A_\mu^a(p_1)A_\nu^b(p_2\!\equiv\!-p_1) \bigr\rangle}^{\irr\ouno}\!$ 
      of $\,$}         
   \bar p_1^\mu \, \bar p_1^\nu=     \cr
  =&\ucpc\,\delta^{ab}\left[\TLR {2\over3}-\CA({5\over3}+{1-\aprime\over2})
                        \right]\,, \cr
\bar r_{2}^{\ouno\,ab}=&{(-i)^2\over2}
   \;\hbox{coef. in r.s.p. $(-i)$
      $\overline{\bigl\langle 
         A_\mu^a(p_1)A_\nu^b(p_2\!\equiv\!-p_1) \bigr\rangle}^{\irr\ouno}\!$ 
      of $\,$} 
   \bar p_1{}^2 \;\gbar^{\mu\nu}=         \cr
   =&\ucpc\,\delta^{ab}\,\left[\TLR{2\over3}-
      \CA({5\over3}+{1-\aprime\over2})\right]\,,  \cr
\bar r_{3}^{\ouno\,abc}=&{(-i)\over3}
   \,\hbox{coef. in r.s.p. $\!(-i)$
      $\overline{\bigl\langle 
         A_\mu^a(p_1)A_\nu^b(p_2)A_\rho^c(p_3\!\equiv\!-p_1\!-\!p_2)
          \bigr\rangle}^{\irr\ouno}\!\!$
      of $\,$} 
   \bar p_1^\nu \,\gbar^{\mu\rho}+         \cr
      &+{(-i)\over3}
   \,\hbox{coef. in r.s.p. $\!\!(-i)$
      $\overline{\bigl\langle 
         A_\mu^a(p_1)A_\nu^c(p_2)A_\rho^b(p_3\!\equiv\!-p_1\!-\!p_2)
          \bigr\rangle}^{\irr\ouno}\!\!\!\!\!$
      of $\,$} 
   \bar p_1^\rho \,\gbar^{\mu\nu}\!\!=         \cr
   =&\ucpc\,{2\over3}\,c^{abc}\,\left[{2\over3}\TLR-
      \CA({2\over3}+{3(1-\aprime)\over4}\right]\,,  \cr
\bar r_{4}^{\ouno\,abc}=&{(-i)\over3}
   \,\hbox{coef. in r.s.p.  $\!(-i)$
      $\overline{\bigl\langle 
         A_\mu^a(p_1)A_\nu^b(p_2)A_\rho^c(p_3\!\equiv\!-p_1\!-\!p_2)
          \bigr\rangle}^{\irr\ouno}\!\!\!$
      of $\,$} 
   \bar p_1^\mu \,\gbar^{\nu\rho}\!=         \cr
   =&\ucpc\,{2\over3}\,c^{abc}\,\left[{2\over3}\TLR-
      \CA({2\over3}+{3(1-\aprime)\over4}\right]\,,  \cr
\bar r_{5}^{\ouno\,abc}=&{(-i)\over3}
   \,\hbox{coef. in r.s.p.  $\!\!(-i)\!$
      $\overline{\bigl\langle 
         A_\mu^a(p_1)A_\nu^b(p_2)A_\rho^c(p_3\!\equiv\!-p_1\!-\!p_2)
          \bigr\rangle}^{\irr\ouno}\!\!\!\!$
      of $\,$} 
   \bar p_1^\a \,\eps_{\mu\nu\!\rho\a}\!= \cr
   =& \;0,        \cr
\bar r_{6}^{\ouno\!\{a_1\!a_2\}\!\{a_3\!a_4\}}\!\!\!=
  &\,r_{6}^{\ouno\{a_3\!a_4\}\!\{a_1\!a_2\}}\!=         \cr
  =&{1\over8}\;\hbox{coef. in r.s.p. $(-i)$
      $\overline{\bigl\langle 
         A_{\mu_1}^{a_1}A_{\mu_2}^{a_2}A_{\mu_3}^{a_3}A_{\mu_4}^{a_4}
          \bigr\rangle}^{\irr\ouno}\!$
      of $\,$} 
   \gbar^{\mu_1\mu_2}\,\gbar^{\mu_3\mu_4}=        \cr
  =&\ucpc\,{1\over8}\,\left[ {4\over3}\,\TLR + 2\, \CA\,
      \left({1\over3}-(1\!-\!\aprime\!)\right)\right] \cr
   &\qquad\quad\left(c^{ea_1a_3}c^{ea_4a_2}- c^{ea_1a_4}c^{ea_2a_3}
              \right)\!,                 &\numeq\cr
\bar r_{7}^{\ouno\,a_1a_2a_3a_4}\!=
  &r_{7}^{\ouno\,\{a_1a_2a_3a_4\}}=           \cr
   =&{1\over4!}\;\hbox{coef. in r.s.p.  $(-i)$
      $\overline{\bigl\langle 
         A_{\mu_1}^{a_1}A_{\mu_2}^{a_2}A_{\mu_3}^{a_3}A_{\mu_4}^{a_4}
          \bigr\rangle}^{\irr\ouno}\!$
      of $\,$} 
   \eps_{\mu_1\mu_2\mu_2\mu_4}=\;0,       \cr
\bar r_{8}^{\ouno\,ij}
  =&i\;\hbox{coef. in r.s.p. 
      $\overline{\bigl\langle 
         \psi_{\b j}(-p)\bar\psi_{\a i}(p)
          \bigr\rangle}^{\irr\ouno\!}$
      of $\,$}
 (\bar\pslash\,\PL)_{\a\b}=\icpc \, 2 g^2 \aprime \CL,  \cr           
\bar r_{9}^{\ouno\,ij\,a}
  =&i\;\hbox{coef. in r.s.p. $(-i)$
      $\overline{\bigl\langle 
         \psi_{\b j}\bar\psi_{\a i}A_a^\mu
          \bigr\rangle}^{\irr\ouno}\!$
      of $\,$}
      (\gambar^\mu\,\PL)_{\a\b}=\cr
  =&\ucpc \, 2 g^2 \CL\,\left[(1-{1-\aprime\over4})\CA +
     \aprime\CL\right]  (\TL^a)_{ij},\cr
\bar r_{10}^{\ouno\,ij}
  =&i\;\hbox{coef. in r.s.p. 
      $\overline{\bigl\langle 
         \psi_{\b j}'(-p)\bar\psi'_{\a i}(p)
          \bigr\rangle}^{\irr\ouno\!}$
      of $\,$}
      (\bar\pslash\,\PR)_{\a\b}= \icpc \, 2 g^2 \aprime \CR, \cr            
\bar r_{11}^{\ouno\,ij\,a}
  =&i\;\hbox{coef. in r.s.p. $(-i)$
      $\overline{\bigl\langle 
         \psi_{\b j}'\bar\psi'_{\a i}A_a^\mu
          \bigr\rangle}^{\irr\ouno}\!$
      of $\,$}
      (\gambar^\mu\,\PR)_{\a\b}=\cr
  =&\ucpc \, 2 g^2 \CR\,\left[(1-{1-\aprime\over4})\CA +
     \aprime\CR\right]  (\TR^a)_{ij},\cr
\bar r_{12}^{\ouno\,ij}
  =&i\;\hbox{coef. in r.s.p. 
      $\overline{\bigl\langle 
         \psi_{\b j}(-p)\bar\psi_{\a i}(p)
          \bigr\rangle}^{\irr\ouno\!}$
      of $\,$}
      (\bar\pslash\,\PR)_{\a\b}=0,\cr          
\bar r_{13}^{\ouno\,ij\,a}
  =&i\;\hbox{coef. in r.s.p. $(-i)$
      $\overline{\bigl\langle 
         \psi_{\b j}\bar\psi_{\a i}A_a^\mu
          \bigr\rangle}^{\irr\ouno}\!$
      of $\,$}
      (\gambar^\mu\,\PR)_{\a\b}=0,\cr
\bar r_{14}^{\ouno\,ij}
  =&i\;\hbox{coef. in r.s.p. 
      $\overline{\bigl\langle 
         \psi_{\b j}'(-p)\bar\psi_{\a i}'(p)
          \bigr\rangle}^{\irr\ouno\!}$
      of $\,$}
      (\bar\pslash\,\PL)_{\a\b}=0,\cr 
\bar r_{15}^{\ouno\,ij\,a}
  =&i\;\hbox{coef. in r.s.p. $(-i)$
      $\overline{\bigl\langle 
         \psi_{\b j}'\bar\psi_{\a i}'A_a^\mu
          \bigr\rangle}^{\irr\ouno}\!$
      of $\,$}
      (\gambar^\mu\,\PL)_{\a\b}=0,\cr
 \bar r_{16}^{\ouno\,ab}
  =&i\;\hbox{coef. in r.s.p. $(-i)$
      $\overline{\bigl\langle 
         \om^b(-p)\bar\om^a(-p)
          \bigr\rangle}^{\irr\ouno}\!$
      of $\,$}
      {\bar p}^2=-\ucpc \,\CA g^2 (1+{1-\aprime\over2}), \cr
\bar r_{17}^{\ouno\,abc}
  =&(-i)\;\hbox{coef. in r.s.p.  $\!(-i)$
      $\overline{\bigl\langle 
         \om^c(-p_1-p_2)\bar\om^a(p_1)A_\mu^b(p_2)
          \bigr\rangle}^{\irr\ouno}\!$
      of $\,$}
  {\bar p_1}^{\mu}=\cr
       =&\ucpc \,\CA g^2 \aprime\,c^{abc},\cr
\bar r_{18}^{\ouno\,abc}
  =&(-i)\;\hbox{coef. in r.s.p. $\!(-i)$
      $\overline{\bigl\langle 
         \om^c(-p_1-p_2)\bar\om^a(p_1)A_\mu^b(p_2)
          \bigr\rangle}^{\irr\ouno}\!$
      of $\,$}
      {\bar p_2}^{\mu}=0,\cr
\bar r_{19}^{\ouno\,abcd}
  =&{1\over2}\;\hbox{coef. in r.s.p. $(-i)$
      $\overline{\bigl\langle 
         \om^b\bar\om^a A_\mu^b A_\nu^d
          \bigr\rangle}^{\irr\ouno}\!$
      of $\,$}
      \gbar^{\mu\nu}=0,\cr
\bar r_{20}^{\ouno\,abcd}
  =&{1\over4}\;\hbox{r.s.p. $(-i)$
      $\overline{\bigl\langle 
         \om^d\om^c\bar\om^b\bar\om^a \bigr\rangle}^{\irr\ouno}\!$}=0,\cr
\bar r_{21}^{\ouno\,ij\,a}
 =&\,\hbox{coef. in r.s.p.
      $\overline{\bigl\langle 
         \psi_{j\b}\om^a{\rm ;}\,s\psi_{i\a}
          \bigr\rangle}^{\irr\ouno}\!$
      of $\,$} (\PL)_{\a\b}\,=\icpc \, g^2 \aprime \CA\,(\TL^a)_{ij},\cr
\bar r_{22}^{\ouno\,ij\,a}
  =&\,\hbox{coef. in r.s.p.
      $\overline{\bigl\langle 
         \om^a\bar\psi_{i\a} {\rm ;}\, s\bar\psi_{j\b}
         \bigr\rangle}^{\irr\ouno}\!$
     of $\,$}
      (\PR)_{\a\b}\,=\icpc \, g^2 \aprime \CA\,(\TL^a)_{ij},\cr
\bar r_{23}^{\ouno\,ij\,a}
 =&\,\hbox{coef. in r.s.p.
      $\overline{\bigl\langle 
         \psi'_{j\b}\om^a{\rm ;}\,s\psi'_{i\a}
          \bigr\rangle}^{\irr\ouno}\!$
      of $\,$} (\PR)_{\a\b}\,=\icpc \, g^2 \aprime \CA\,(\TR^a)_{ij},\cr
\bar r_{24}^{\ouno\,ij\,a}
  =&\,\hbox{coef. in r.s.p.
      $\overline{\bigl\langle 
         \om^a\bar\psi'_{i\a} {\rm ;}\, s\bar\psi'_{j\b}
         \bigr\rangle}^{\irr\ouno}\!$
     of $\,$}
      (\PL)_{\a\b}\,=\icpc \, g^2 \aprime \CA\,(\TR^a)_{ij},\cr
\bar r_{25}^{\ouno\,ij\,a}
 =&\,\hbox{coef. in r.s.p.
      $\overline{\bigl\langle 
         \psi_{j\b}\om^a{\rm ;}\,s\psi_{i\a}
          \bigr\rangle}^{\irr\ouno}\!$
      of $\,$} (\PR)_{\a\b}\,= 0\cr
\bar r_{26}^{\ouno\,ij\,a}
  =&\,\hbox{coef. in r.s.p.
      $\overline{\bigl\langle 
         \om^a\bar\psi_{i\a} {\rm ;}\, s\bar\psi_{j\b}
         \bigr\rangle}^{\irr\ouno}\!$
     of $\,$}
      (\PL)_{\a\b}\,=0\cr
\bar r_{27}^{\ouno\,ij\,a}
 =&\,\hbox{coef. in r.s.p.
      $\overline{\bigl\langle 
         \psi'_{j\b}\om^a{\rm ;}\,s\psi'_{i\a}
          \bigr\rangle}^{\irr\ouno}\!$
      of $\,$} (\PL)_{\a\b}\,=0 \cr
\bar r_{28}^{\ouno\,ij\,a}
  =&\,\hbox{coef. in r.s.p.
      $\overline{\bigl\langle 
         \om^a\bar\psi'_{i\a} {\rm ;}\, s\bar\psi'_{j\b}
         \bigr\rangle}^{\irr\ouno}\!$
     of $\,$}
      (\PR)_{\a\b}\,=0,\cr
\bar r_{29}^{\ouno\,ab}
  =&(-i)\,\hbox{coef. in r.s.p.
      $\overline{\bigl\langle 
         \om^b(p_1) {\rm ;}\,sA_\mu^a(p_2\!=\!-\!p_1)
          \bigr\rangle}^{\irr\ouno}\!$
      of $\,$}
      \bar p_1{}^\mu=\cr
=&-\ucpc \,\delta^{ab}\,\CA g^2 (1+{1-\aprime\over2}), \cr
\bar r_{30}^{\ouno\,ab c}
  =&(-i)\,\hbox{coef. in r.s.p.
      $\overline{\bigl\langle 
         \om^b A_\mu^c {\rm ;}\, sA_\nu^a
          \bigr\rangle}^{\irr\ouno}\!$
      of $\,$}
      \gbar_{\mu\nu}\,=
     -\ucpc \,c^{abc}\,\CA g^2 \,, \cr
r_{31}^{\ouno\,abc}
  =&\, {1\over2}\,\hbox{r.s.p.
      $\overline{\bigl\langle 
         \om^c\om^b {\rm ;}\, s\om^a
          \bigr\rangle}^{\irr\ouno}$
                     }\!\!=\,
     -{1\over2}\ucpc\, c^{abc}\,\CA g^2\,. \cr
}
$$\namelasteq{\listcoeff}

\medskip
To remove much of the redundancy in the linear system in eq.~\orderhbar\ 
we shall rather express $\mu{\pr \GR\over\pr\mu}$ in terms of the 
functionals in eq.~\EqInvariantTerms . By using the following
equations
  
{\settabs 2 \columns
\openup1\jot
\+$S_{{\rm cl}\,AA}=
{\delta^{ab}\over 2g^2}\,({\W}_1^{ab}-{\W}_2^{ab})$,
 &$S_{{\rm cl}\,AAA}=-{c^{abc}\over g^2}\,{\W}_3^{abc}$,     \cr
\+$S_{{\rm cl}\,AAA}=-{1\over 8g^2}\,
      (c_{ace}c_{bde}+c_{bce}c_{ade})\,{\W}_6^{\{ab\}\{cd\}}$,
       \cr
\+$S_{{\rm cl}\,\bar\psi\psi}^{\L}=
 \delta_{ij}\,{\W}_{8}^{ij}$,
 &$S_{{\rm cl}\,\bar\psi\psi\,A}=
 (\TL^a)_{ij}\,{\W}_{9}^{ij}$,     \cr
\+$S_{{\rm cl}\,\bar\psi'\psi'}^{\R}=
 \delta_{ij}\,{\W}_{10}^{ij}$,
 &$S_{{\rm cl}\,\bar\psi'\psi'\,A}=
 (\TR^a)_{ij}\,{\W}_{11}^{ij}$,     \cr
\+$S_{{\rm cl}\,\bar\om\om}=\delta^{ab}\,{\W}_{16}^{ab}$,
 &$S_{{\rm cl}\,\bar\om\om A}=-c^{abc}\,{\W}_{17}^{abc}$, \cr
\+$S_{{\rm cl}\,{\bar L} s\psi}=
    i\,(\TL^a)_{ij}\,{\W}_{21}^{ij\,a}$,
 &$S_{{\rm cl}\, L s\bar\psi}=
    i\,(\TL^a)_{ij}\,{\W}_{22}^{ij\,a}$, \cr
\+$S_{{\rm cl}\,{\bar R} s \psi'}=
    i\,(\TR^a)_{ij}\,{\W}_{23}^{ij\,a}$,
 &$S_{{\rm cl}\, R s \bar\psi'}=
    i\,(\TLpR^a)_{ij}\,{\W}_{24}^{ij\,a}$, \cr
\+$S_{{\rm cl}\,\rho\om}=\delta^{ab}\,{\W}_{29}^{ab}$,
 &$S_{{\rm cl}\,\rho\om A}=-c^{abc}\,{\W}_{30}^{abc}$,     \cr
\+$S_{{\rm cl}\,\zeta\om\om}=-{1\over2}c^{abc}\,{\W}_{31}^{abc}$
,  \cr
}
\bigskip
\noindent one easily obtains the contribution to
$\mu{\pr\GR\over\pr\mu}$ at first order in $\hbar$:

$$
\eqalignno{
\mu\,{\pr\GR\over\pr\mu}=&
\ucpc\,g^2\left[{8\over3}{\TLR\over2}-C_A \left({10\over3} + (1-\a')\right)
				\right]\;S_{{\rm cl}\,AA}  \cr
+&\ucpc\,g^2\left[{8\over3}{\TLR\over2}-C_A \left({4\over3} + 
	{3\over2}(1-\a')\right)\right]\;S_{{\rm cl}\,AAA} \cr
+&\ucpc\,g^2\left[{8\over3}{\TLR\over2}-C_A \left(-{2\over3} + 
	2(1-\a')\right)\right]\;S_{{\rm cl}\,AAAA}     \cr
+&\ucpc\,g^2 \CL 2\,\a'\;S_{{\rm cl}\,\bar\psi\psi}^\L
      +\ucpc\,g^2 \CR 2\,\a'\;S_{{\rm cl}\,\bar\psi'\psi'}^\R \cr
+&\ucpc\,g^2 \left[\left(2-{(1-\a')\over2}\right)C_A + 2\,\a'\CL\right]
			\;S_{{\rm cl}\,\bar\psi\psi A}    &\numeq\cr
+&\ucpc\,g^2 \left[\left(2-{(1-\a')\over2}\right)C_A + 2\,\a'\CR\right]
			\;S_{{\rm cl}\,\bar\psi'\psi' A} \cr
-&\ucpc\,g^2 \left(1+{(1-\a')\over2}\right)C_A\;S_{{\rm cl}\,\bar\om\om}
	+\ucpc\,g^2 \,\a'\,C_A\;S_{{\rm cl}\,\bar\om\om A} \cr    
+&\ucpc\,g^2 \,\a'\,C_A\;
	\left[S_{{\rm cl}\, \bar L s\psi}+S_{{\rm cl}\, \bar R s\psi'}+
	S_{{\rm cl}\,L s\bar\psi}+ S_{{\rm cl}\,R s\bar\psi'}\right]\cr    
-&\ucpc\,g^2 \left(1+{(1-\a')\over2}\right)C_A\;S_{{\rm cl}\,\rho\om} 
      +\ucpc\,g^2 \,\a'\,C_A\; S_{{\rm cl}\,\rho\om A} \cr
+&\ucpc\,g^2 \,\a'\,C_A\; S_{{\rm cl}\,\zeta\om\om}\;+\;O(\hbar^2) \cr
} \namelasteq{\semifinal}
$$
where, $\TLR$ means sum over all left and right representations ({\it e.g.} 
number of ``flavours'')  in the theory.

Now, it is plain that eq.~\Firstexpansion\ reads 
$$
\eqalignno{
\mu\,{\pr\GR\over\pr\mu}
 =&-\beta^{(1)}\,g\,{\pr S_{\rm cl}\over \pr g}+
   \sum_{\phi}\;\gamma_{\phi}^{(1)}\NN_{\phi}S_{\rm cl}\; +\;O(\hbar^2)=\cr
 =&-\beta^{(1)}\,L_g+
   \sum_{\phi}\;\gamma_{\phi}^{(1)}\,L_\phi\; +\;O(\hbar^2)\,, &\numeq\cr
} \namelasteq{\final}
$$
which in turn can be expressed in terms of the functionals 
$S_{{\rm cl}\cdots}$ given in eq.~\EqInvariantTerms. 

Finally, the fact that the left hand sides of eqs.~\semifinal\ 
and \final\ should match gives rise to the following system of equations: 
\medskip
{\settabs 7 \columns
\openup1\jot
\+$\quad *$ From $\quad$ $S_{\clr\,AA}$&$\qquad\qquad\to$ &$
	2\, \b\Ouno +2\, \ga_A\Ouno=
\ucpc\,g^2\left[{8\over3}{\TLR\over2}-C_A \left({10\over3} 
				+ (1-\a')\right)\right]$,      \cr          
\+$\quad *$ From $\quad$ $S_{\clr\,AAA}$&$\qquad\qquad\to$ &$
	2\, \b\Ouno +3\, \ga_A\Ouno=            
\ucpc\,g^2\left[{8\over3}{\TLR\over2}-C_A \left({4\over3} + 
				{3\over2}(1-\a')\right)\right]$,     \cr
\+$\quad*$ From $\quad$ $S_{\clr\,A\!A\!A\!A}$&$\qquad\qquad\to$ &$
	2\, \b\Ouno +4\, \ga_A\Ouno=            
\ucpc\,g^2\left[{8\over3}{\TLR\over2}-C_A \left(-{2\over3} + 
			2(1-\a')\right)\right]$,     \cr
\+$\quad *$ From $\quad$ $S_{\clr\,\bar\psi\psi}^\L$&$\qquad\qquad\to$ &$
	2\,\ga_\psi^\L\Ouno =\ucpc\,g^2 \CL 2\,\a'$, \cr
\+$\quad *$ From $\quad$ $S_{\clr\,\bar\psi'\psi'}^\R$&$\qquad\qquad\to$ &$
	2\,\ga_{\psi'}^\R\Ouno =\ucpc\,g^2 \CR 2\,\a'$,       \cr
\+$\quad *$ From $\quad$ $S_{\clr\,\bar\psi\psi A}^\L$&$\qquad\qquad\to$ &$
	2\,\ga_\psi^\L\Ouno+\ga_A\Ouno=  \ucpc\,g^2 \left[\left(2-
			{(1-\a')\over2}\right)C_A + 2\,\a'\CL\right]$,\cr
\+$\quad *$ From $\quad$ $S_{\clr\,\bar\psi'\psi' A}^\R$&$\qquad\qquad\to$ &$
	2\,\ga_{\psi'}^\R\Ouno+\ga_A\Ouno=\ucpc\,g^2 \left[\left(2
		-{(1-\a')\over2}\right)C_A + 2\,\a'\CR\right]$,\cr
\+$\quad *$ From $\quad$ $S_{\clr\,\bar\om\om}$&$\qquad\qquad\to$ &$
	-\ga_A\Ouno+\ga_\om\Ouno=
         -\ucpc\,g^2 \left(1+{(1-\a')\over2}\right)C_A$,
						&&&&\hfill\numeq&\cr
\+$\quad *$ From $\quad$ $S_{\clr\,\bar\om\om A}$&$\qquad\qquad\to$ &$
	 \ga_\om\Ouno=
				\ucpc\,g^2 \,\a'\,C_A$, \cr    
\+$\quad *$ From $\quad$ $S_{\clr\,\bar L s\psi}$&$\qquad\qquad\to$ &$
	 \ga_\om\Ouno=
				\ucpc\,g^2 \,\a'\,C_A$, \cr    
\+$\quad *$ From $\quad$ $S_{\clr\,\bar R s\psi'}$&$\qquad\qquad\to$ &$
	 \ga_\om\Ouno=
				\ucpc\,g^2 \,\a'\,C_A$, \cr    
\+$\quad *$ From $\quad$ $S_{\clr\,L s\bar\psi}$&$\qquad\qquad\to$ &$
	 \ga_\om\Ouno=
						\ucpc\,g^2 \,\a'\,C_A$, \cr    
\+$\quad *$ From $\quad$ $S_{\clr\,R s\bar\psi'}$&$\qquad\qquad\to$ &$
	 \ga_\om\Ouno=
			\ucpc\,g^2 \,\a'\,C_A$, \cr    
\+$\quad *$ From $\quad$ $S_{\clr\,\rho\om }$&$\qquad\qquad\to$ &$
	 -\ga_A\Ouno+\ga_\om\Ouno=
			-\ucpc\,g^2 \left(1+{(1-\a')\over2}\right)C_A$,  \cr
\+$\quad *$ From $\quad$ $S_{\clr\,\rho\om A}$&$\qquad\qquad\to$ &$
	 \ga_\om\Ouno=
		\ucpc\,g^2 \,\a'\,C_A$,                   \cr
\+$\quad *$ From $\quad$ $S_{\clr\,\zeta\om\om}$&$\qquad\qquad\to$ &$
	 \ga_\om\Ouno=
		\ucpc\,g^2 \,\a'\,C_A$;                   \cr
}\namelasteq{\EqLinearSystem}
\medskip
\noindent which is a simplified version of eq.~\orderhbar. This 
simplification takes place since the 
non-symmetrical finite counterterms, $S_{\rm fct}^{(n)}$, 
needed to restore the BRS symmetry and the evanescent monomials, $\hW_j$, 
only begin to contribute to the renormalization group equation at order 
$\hbar^2$. At order $\hbar^n$, $n>1$, it is the linear system in
eq.~\linearsystem\ which is to be dealt with. 
The  system  in eq.~\EqLinearSystem\ is compatible and 
overdeterminate and its solution reads:
$$
\eqalignno{
\b\Ouno&=\ucpc\,g^2\,\left({4\over3}\,{\TLR\over2} -{11\over3}\,\CA\right),\cr
\ga_A\Ouno&=\ucpc\,g^2\, \left(2-{1-\a'\over2}\right)\,\CA,    \cr
\ga_\om\Ouno&=\ucpc\,g^2\,\a'\,\CA,    &\numeq\cr
\ga_\psi^\L\Ouno&=\ucpc\,g^2\,\a'\,\CL,    \qquad
\ga_{\psi'}^\R\Ouno=\ucpc\,g^2\,\a'\,\CR\,.    \cr
}  \namelasteq{\EqRGSolution}
$$

The reader should notice that to actually compute the beta function and
anomalous dimensions of the theory one does not need to evalute the 
thirtyone coefficients in eq.~\COEFF\ (computed in eq.~\listcoeff)
but just a few of them, if appropriately chosen. Indeed, the computation
of, say, ${\bar r}^{(1)\,ab}_1$, ${\bar r}^{(1)\, ij}_8$, 
${\bar r}^{(1)\, ij}_{10}$, ${\bar r}^{(1)\, ab}_{16}$ and 
$ {\bar r}^{(1)\, abc}_{17}$ would do the job. However, a thorough check
of the formalism demands the computation of the thirtyone coefficients
in question. And this we have done.

We have finished  the computation of the renormalization group equation
at order $\hbar$ for the theory with classical action in eq.~\Classicalact .
Our expasions were expasions in $\hbar$, as suits the cohomology problems
which arise in connection with the BRS symmetry, rather that in the coupling
constant $g$. Our parametrizations of the wave functions were such that 
$g$ only occurs,  in the classical action, in the Yang-Mills term, so that 
$g\pr_g S_{\rm cl}$ is the only element 
(modulo multiplication by a constant) of the local cohomology of the 
BRS operator $b$ (see eq.~\EqOperatorb) over the space of local 
integrated functionals of ghost number 0 and ultraviolet dimension 
4 (Lorentz invariance and rigid gauge symmetry are also assumed). 
The other contributions to the renormalization group equation are 
$b$-exact and have to do with the anomalous dimensions of the theory.

It is often  the case that one choses a parametrization of the wave function
such that, at the tree-level,  the fermionic  and the three-boson 
vertices carry the coupling factor $g$, and the four-boson vertex 
is  proportional to $g^2$. The renormalization group equation for this
parametrization of the wave function is easily retrieved from our results.
Let us denote by 
$\GR^{\star}[A^{\star},\om^{\star},\bar\om^{\star},B^{\star},\rho^{\star},
\zeta^{\star},\psi,\bar\psi,\psi',\bar\psi',
\a^{\star},g]$ the  1PI functional for this new parametrization. Then,
$$
\eqalignno{
&\GR^{\star}[A^{\star},\om^{\star},\bar\om^{\star},B^{\star},\rho^{\star},
\zeta^{\star},\psi,\bar\psi,\psi',\bar\psi',
\a^{\star},g]\equiv      \cr
&\GR[g A^{\star},g\om^{\star},g^{-1}\bar\om^{\star}g^{-1},
g^{-1}B^{\star},g^{-1}\rho^{\star},
g^{-1}\zeta^{\star},\psi,\bar\psi,\psi',\bar\psi',g^2\a^{\star},g]\equiv \cr
&\GR[A,\om,\bar\om,B,\rho,\zeta,\psi,\bar\psi,\psi',
	\bar\psi',\a,g],
}
$$
where $\GR[A,\om,\bar\om,B,\rho,\zeta,\psi,\bar\psi,\psi, \bar\psi',\a,g]$
is the  1PI functional for the parametrization of 
eq.~\Classicalact. Now, since
$$
\eqalignno{
g {\pr\GR\over\pr g} = {\pr\GR^{\star}\over\pr g} 
   &- {1\over g}\, N_{A^{\star}}\,\GR^{\star}- N_{\om^{\star}}\,\GR^{\star}
   + N_{\bar\om^{\star}}\GR^{\star}+ N_{B^{\star}}\,\GR^{\star}     \cr
   &+ N_{\rho^{\star}}\,\GR^{\star} +  N_{\zeta^{\star}}\,\GR^{\star}-
    2\, \a^{\star} {\pr\GR^{\star}\over\pr\a^{\star}},           \cr
}
$$
the renormalization group equation eq.~\EqRGExacta\ becomes 
(in terms of the usual combination $\aS\equiv g^2/(4\pi)$):
$$
\eqalignno{
\Big[\,\mu\,{\pr\over\pr\mu}\, +\, &\bS\,\aS{\pr\over\pr\aS} 
	+\delta_{\a^{\star}}\,\a^{\star}{\pr\over\pr\a^{\star}}
	- \ga_\psi^\L \NN_\psi^\L- \ga_{\psi'}^\R \NN_{\psi'}^\R 
-\ga_{A^{\star}} N_{A^{\star}} -\ga_{\om^{\star}} N_{\om^{\star}}  \cr
&-\ga_{\bar\om^{\star}} N_{\bar\om^{\star}} -\ga_{B^{\star}} N_{B^{\star}}
-\ga_{\rho^{\star}} N_{\rho^{\star}} -\ga_{\zeta^{\star}} N_{\zeta^{\star}}
						\, \Big]\; \GR^{\star}=0, 
}
$$ 
with
$$
\eqalignno{                            
\bS&=\,2\, \b,     \cr
\ga_{A^{\star}}&=-\ga_{\rho^{\star}}=-\ga_{\bar\om^{\star}}=
  -\ga_{B^{\star}}=\ga_A+\b,              \cr
\ga_{\om^{\star}}&=-\ga_{\zeta^{\star}}=\ga_\om+\b, \cr
\delta_{\a^{\star}}&=-2\,\ga_{A^{\star}}=-2\,(\ga_A +\b)\, .            \cr
}     
$$

finally, by using the results in eq.~\EqRGSolution, one obtains
$$
\eqalignno{
\bS\Ouno&={\aS\over\pi}\,\left[{2\over3}\,{\TLR\over2} 
	-{11\over6}\,\CA\right],                                   \cr
\ga_{A^{\star}}\Ouno&={\aS\over\pi}\, 
	\left[{1\over3} \, {\TLR\over2}-  {\CA\over4}
         \left( {13\over6}-{\a^{\star}\over2}\right)\right],        \cr
\ga_{\om^{\star}}\Ouno&={\aS\over\pi}\, 
	\left[{1\over3}\, {\TLR\over2}-{\CA\over4}
	\left({31\over6}+{\a^{\star}\over2}\right)\right] .         \cr
}
$$

\bigskip

\section{4. Chiral Yang-Mills theories: non-simple gauge groups}

\subsection{4.1. The tree level action}

In this section, the gauge group will be a  compact Lie group which is the
direct product of $\NS$ simple groups and  $\NA$ Abelian factors. 
Its (real) Lie algebra is a direct sum: $\bG=\bG_1\oplus\cdots\oplus\bG_\NA$
$\oplus\bG_{\NA+1}\oplus\cdots\oplus\bG_{\NA+\NS}$ of dimension
$d_\bG=\NA +$  $d_{\bG_{\NA+1}}+\cdots+ d_{\bG_{\NA+\NS}}$.
The index $G$, which labels the group factors, will run from 1 to
$\NA+\NS$, the index $\Ac$, which labels the Abelian factors, from $1$ to $\NA$
and the index $\SS$, which labels the simple factors, from $1$ to
$\NS$.

Therefore, there exists a basis for the Lie algebra which can be
split and enumerated as follows:
$$
\eqalignno{
\big\{X_a\big\}_{a=1}^{d_\bG}\,=&\,\big\{X_1,\ldots,X_\NA,
         X_{\NA+1}, \ldots, X_{\NA+d_{\bG_1}},\ldots \big\}           \cr
   \equiv&\, \{\,X_1,\ldots,X_\NA\}
   \bigcup
\big(\cup_{\SS=1}^{\NS}\; \{X^{(\SS)}_{a_\SS}\}_{a_\SS=1}^{d_{\bG_{\NA\SS}}}
\big);      \cr
}
$$
and such that the Killing form of the semisimple part is diagonal and
positive definite; so, we choose $\delta_a^b$ to lower and rise indices.
For any such basis, the structure constants, given by
$[X_a,X_b]=i\,c_{ab}{}^c\,X_c$, are completely antisymmetric,
 and
$c_{a_Gde}\,c_{b_{G^\star}de}=0$ if $G\ne G^\star$,
$c_{a_Gde}\,c_{b_Gde}=$
$c_{a_Gd_Ge_G}\,c_{b_Gd_Ge_G}=\CA^{(G)}\,\delta_{ab}$, with
$\CA^{(\Ac)}\equiv0$ (clearly, $c_{a_{\!\Ac}bc}=0$).

Let us display the field content of the theory and introduce the Dimensional
Regularization classical action, $S_0$. We have first the 
Lie-algebra-valued fields. The pure Yang-Mills part of the tree-level 
action in the $d$-dimensional space of Dimensional Regularization  reads:
$$
S_{\rm YM}=\intd^dx\;-{1\over 4} F_{\mu\nu}^a C_{ab}
   F^{b\,\mu\nu} \eqno\numeq    \namelasteq\EqYMGeneric
$$
where $C$ is a diagonal matrix with 
$$\{g_1^{-2},\ldots, g_\NA^{-2},
 \underbrace{g_{\NA+1}^{-2},\ldots, g_{\NA+1}^{-2}}_{d_{\bG_{\NA+1}}
  {\rm times}}, \ldots,
 \underbrace{g_{\NA+\NS}^{-2},\ldots, g_{\NA+\NS}^{-2}}_{d_{\bG_{\NA+\NS}}
   {\rm times}}\}
$$
\vskip-4pt
\noindent in the diagonal. Hence,
$$
S_{\rm YM}=\intd^dx\;\sum_G\,-{1\over 4g_G^2}  F^{(G)}_{\mu\nu\,a_G}
   F^{(G)}_{a_G\,\mu\nu},    
$$
where $F^{(G)\,a_G}_{\mu\nu}=
   \pr_\mu A_\nu^{(G)\, a_G}-\pr_\nu A_\mu^{(G)\,a_G} +
    c^{a_Gb_Gc_G} A_\mu^{(G)\,b_G} A_\nu^{(G)\,c_G}$. Obviously, all the 
remaining Lie algebra-valued  fields are indexed in the same way yielding:
$$
\eqalignno{
S_{\rm gf} =& \intd^dx \; {1\over2}\,B_a \Lambda^{ab} B_b+
 B_a(\pr^\mu A_\mu^a)- \bar\om_a \,\pr^\mu \nabla_\mu\om^a, \cr
 =& \intd^dx \;\sum_G\; {\a_G\over2}\,B^{(G)\,a_G} B^{(G)}_{a_G}+
 B^{(G)}_{a_G}(\pr^\mu A^{(G)\;a_G}_\mu)-
 \bar\om^{(G)}_{a_G} \,\pr^\mu \nabla_\mu\om^{(G)\,a_G}.   \cr
}
$$

The free boson propagator reads now:
$$
\left(
C^{-1}\,{-i\over k^2 + i\epsilon}\, 
	\Bigl[ \bigl(g^{\mu\nu}-{k^\mu k^\nu \over k^2}\bigr)+
      C\,\Lambda\,{k^\mu k^\nu \over k^2}\Bigr]
       \right)^{ab};   
$$
{\it i.e.} diagonal but no longer multiple of the identity.

Notice that we have also introduced ghosts to control the Abelian
variations. Since ghosts in Abelian theories do not interact with
the rest of fields ($\nabla_\mu\om^{\Ac}=\pr_\mu\om^{\Ac}$), they can
occur in a non-tree-level 1PI diagram only as a part of a vertex insertion
(BRS-variations and   breakings);
so that, by differenciating with respect to $\om^{\Ac}$ the Slavnov-Taylor 
identities, the usual Abelian Ward identities are easily recovered. But we 
insist on keeping the Abelian ghost in order to have an unified notation.

The  matter content of the theory  will be  only non-Majorana fermions.
The corresponding fermion fields, $\psi(\psi')$, carrying a
left-handed (right-handed) fully reducible representation of the gauge group,
so that it can be descomposed in a sum of irreducible representations.
From now on,  until otherwise stated, we will take  both the left-handed
and right-handed fermionic to be irreducible, the general case just involve a 
sum over all the irreducible representations.

We shall index  the fermion fields carrying the irreducible representation of 
the gauge Lie algebra as follows
$\psi^{i_1,\ldots,i_{\NS}}$
$(\psi^{\prime\,i_1,\ldots,i_{\NS}})$,
with $i_\SS$ running from 1 to $d_\SS(d_\SS^\prime)$; 
so that the generators of the Lie algebra are given by
$$
\big\{\TLpR^{\Ac}\big\}_{\Ac=1}^{N_{\Ac}}\,=\,
 \big\{\,\tLpR^{\Ac}\,\big\}_{\Ac=1}^{N_{\Ac}}\,
   \otimes\,1_{d_1^\pp}\cdots\otimes1_{d_\NS^\pp}\, ,
$$
where each $\tLpR^{\Ac}$ is a real number and

$$
\big\{\TLpR^a\big\}_{a=\NA+1}^{\NA+d_\bG}\,=\,
 1\otimes\big\{\,1_{d_1^\pp}\otimes\cdots\otimes\,
  \{\TLpR^{{(\SS)}\,a_\SS}\}_{a_\SS=1}^{d_{\bG_{\NA+\SS}}}\,
  \otimes\cdots 1_{d_\NS^\pp}\,\big\}_{\SS=1}^{\NS}, 
$$
where  $\TLpR^{(\SS)\,a_\SS}\, \in\,
{\rm M}_{d_\SS^\pp}(\Cno )$ are  complex matrices of
dimension $d_\SS^\pp$ which furnish an irreducible representation of 
dimension $d_\SS^\pp$ of the Lie algebra $\bG_{N_{\Ac}+{\SS}}$, 
$\SS=1\cdots N_{\SS}$.
 
The BRS variations of the fermion fields in the $d$-dimensional space
of Dimensional Regularization are defined to be
$$
\eqalignno{
s\psi^\pp=&i\,\om^a\TLpR^a\PLpR\psi^\pp=          \cr
   =&\sum_{\Ac=1}^{\NA}\,i\,\om^{\Ac}\,\tLpR^{\Ac}\,
    1_{d_1^\pp}\otimes\cdots
    \otimes 1_{d_\NS^\pp}\, \PLpR\,\psi^\pp\,+         \cr
   &+ \,\sum_{\SS=1}^\NS\,i
    \sum_{a_\SS=1}^{d_{\bG_{\!\NA\!+\SS}}} \om^{(\SS)\,a_\SS}\,
    1_{d_1^\pp}\otimes\cdots\otimes\TLpR^{(\SS)\,a_\SS}\otimes\cdots
    \otimes 1_{d_\NS^\pp}\, \PLpR\,\psi^\pp,         \cr
}
$$
and the regularized interaction is introduced in the same way.

The following results will be useful
$$
\eqalignno{
&\Tr_{{\rm M}_{d^\pp_\SS}(\Cno)}\,[\TLpR^{(\SS)\,a_\SS}\TLpR^{(\SS)\,b_\SS}]
   \equiv\TLpR{\,}_\SS\,\delta^{a_\SS b_\SS},
\qquad\Tr_{{\rm M}_{d^\pp_\SS}(\Cno)}\,[\TLpR^{(\SS)\,a_\SS}]=0,\cr
&\Tr\,[\TLpR^{a}\TLpR^{b}]
   =\cases{
           \left(\prod\limits_{K\ne \SS}^{\NS} d^\pp_K\right)
           \,\TLpR{}_\SS\,\delta^{a b},\quad
          &if $a=a_{\SS}$ and $b=b_{\SS}$, with $\SS=1,\cdots, N_{\SS}$\cr
        \left(\prod\limits_{\SS=1}^{\NS} d^\pp_\SS\right)\,
          \tLpR^{a}\,\tLpR^{b}, \quad &if $a\le\NA$ and
                $b\le\NA$  \cr
         \qquad 0 &in the remainder \cr
          }                                        \cr
&\sum_{e_\SS}^{d^\pp_\SS}\,\TLpR^{(\SS)\,e_\SS}\,\TLpR^{(\SS)\,e_\SS}
   \equiv\CLpR^{(\SS)}\;
1_{d_\SS^\pp},\qquad(\tLpR^{\Ac})^2\equiv\CLpR^{(\Ac)}\, \cr
&\sum_a\,\TLpR^a\,\TLpR^a = \sum_{G=1}^{\NA\!+\!\NS}\,\CLpR^{(G)}\,
    1_{d_1^\pp}\otimes\cdots\otimes 1_{d_\NS^\pp}\equiv
 \sum_{G}\,\CLpR^{(G)}\,\cr
&\sum_{a,b}\,\TLpR^a\,(C^{-1})^a_b\TLpR^b = \sum_{G=1}^{\NA\!+\!\NS}\,g_G^2
         \CLpR^{(G)}\,\equiv\,\CLpR^g,    \cr
&\sum_{a,b}\,\TLpR^a\,\Lambda_{ab}\TLpR^b = \sum_{G=1}^{\NA\!+\!\NS}\,
    \a_G\CLpR^{(G)}\,\equiv\, \CLpR^\a.    \cr
}
$$

\vfill\eject

\subsection{4.2. The one-loop singular counterterms}

To obtain  the one-loop singular counterterms, it is necessary to 
make in the non-hatted terms in eq.~\OnePIpole\ the following substitutions
$$
\eqalignno{
g^2\,\CA\,S_{0\,X\cdots X\,Y\cdots Y}\,&\longrightarrow\,
   \sum_{\SS}\, g^2_\SS\,\CA^{(\SS)}\,
   S_{0\,X\cdots X\,Y_\SS\cdots Y_\SS}\, , \cr
g^2\,\a' \CA\,S_{0\,X\cdots X\,Y\cdots Y}\,&\longrightarrow\,
   \sum_{\SS}\, g^2_\SS\,\a^{\prime\,S}\CA^{(\SS)}\,
   S_{0\,X\cdots X\,Y_\SS\cdots Y_\SS}\, , \cr
g^2\,\CLpR\, &\longrightarrow\,\CLpR^g ,        \cr
g^2\,\CLpR\, \a'\,&\longrightarrow\,\CLpR^\a ,         \cr
g^2\,\TLpR\,S_{0\,A\cdots A}\,&\longrightarrow\,
\sum_{\Ac=1}^{\NA}\,g^2_A\,
           \left(\prod\limits_{\SS=1}^{\NS} d^\pp_\SS\right)
           \,(\tLpR{}^{\Ac})^2\, S_{0\,A_A\cdots A_\Ac}\,       \cr
&\qquad+\quad\sum_{\SS=1}^{\NS}\,g^2_{\NA+\SS}\,
           \left(\prod\limits_{K\ne \SS}^{\NS} d^\pp_K\right)
           \,\TLpR{}_\SS\, S_{0\,A_\SS\cdots A_\SS}\,.      \cr
}
$$
Above, the symbol $Y$ stands for any Lie-algebra-valued field.
A new type of term should also be added. This term is the Abelian mixing term,
 which is given by
$$
{\hbar\over\eps}\,\ucpc\,\sum_{\Ac_1,\Ac_2\atop\Ac_1\ne \Ac_2}^\NA\,
          \left(\prod\limits_{\SS=1}^{\NS}d^\pp_\SS \right)\,
          \tLpR^{\Ac_1}\,\tLpR^{\Ac_2}\;\intd^dx\; -{1\over4}\,
    \left(\pr_\mu A_\nu^{\Ac_1}-\pr_\nu A_\mu^{\Ac_1}\right)\,
    \left(\pr_\mu A_\nu^{\Ac_2}-\pr_\nu A_\mu^{\Ac_2}\right)\,. 
$$
The last line of eq.~\OnePIpole , {\it i.e.}  the hatted contribution,
has to be modified in the same manner.

The reader should bear in mind that  
$g^2\CLpR\a'\,S_{0\,\bar\psi^\pp\psi^\pp A}$ $\rightarrow$
      $\CLpR^\a\,\sum_G S_{0\,\bar\psi^\pp\psi^\pp A_G}$.

\bigskip

\subsection{4.3. Expression of the breaking}

To obtain the generalization of eq.~\resultbreak\ to current case, one
applies first the same kind of modifications as in the previous subsection.
In addition, the following replacements should be made 
$$
\TLpR\,c^{abc}\,\M^{abc}_4\,\longrightarrow\,
\sum_{\SS=1}^\NS\,
           \left(\prod\limits_{K\ne \SS}^{\NS} d^\pp_K\right)
           \,\TLpR{}_{\SS}\, c^{a_\SS b_\SS c_\SS}
\,\M^{a_\SS b_\SS c_\SS}_4\,, 
$$                   
and similar substitutions for the terms of the same type as 
$\TLpR\,c^{ebc}c^{eda}\M^{abcd}$.

The traces of the product of three and four generators, which
appear for example in the expression of anomaly, keep its form, but 
with the understanding that $\Tr$ is a matrix trace over the full fermion 
representation of the Gauge Lie algebra $\bG$.

\bigskip

\subsection{4.4. Restoration of Slavnov-Taylor Identities}

Apart from modifications of the same type as describe above, we have
to be careful with the resolution of the equations involving  $\M_1^{ab}$
(and also $\M_7$ and $\M_8$, but they actually do not give rise to
any new result).
Now, we substitute in eq.~\EqAnsatz\ $a_1\tilde\Delta_i$ $\rightarrow$
$a_1^G \tilde\Delta_i^G$ $+a_1^{\Ac_1\Ac_2}\tilde\Delta_i^{\Ac_1\Ac_2}$,
$i=1,2$, $\Ac_1\ne \Ac_2$, with the obvious meaning.

Apart from the equation in $\M_1^{a_Gb_G}$, which will give
together with the rest of the system an unique solution for
$$
\eqalignno{
a_1^{\Ac}&=\ucpc{5\over12}
        \big[(\prod_{\SS=1}^\NS d_\SS)(t_L^{\Ac})^2\,+\,
             (\prod_{\SS=1}^\NS d_\SS^\prime)(t_R^{\Ac})^2\big], \cr
a_1^\SS&=\ucpc{5\over12}
        \big[(\prod_{K\ne \SS}d_K)\TL{}_\SS\,+\,
             (\prod_{K\ne \SS}d'_K)\TR{}_\SS\big], \cr
a_2^{\Ac}&=\ucpc{1\over4}
        \big[(\prod_{\SS=1}^\NS d_\SS)(t_L^{\Ac})^2\,+\,
             (\prod_{\SS=1}^\NS d'_\SS)(t_R^{\Ac})^2\big], \cr
a_2^\SS&=\ucpc{1\over4}
        \big[(\prod_{K\ne \SS}d_K)\TL{}_\SS\,+\,
        (\prod_{K\ne \SS}d'_K)\TR{}_\SS\big],          \cr
}
$$
we have new equations in $\M_1^{\Ac_1\Ac_2}$ and
$\M_1^{\Ac_2\Ac_1}$, $\Ac_1\ne \Ac_2$:
$$
\eqalignno{
a_1^{\Ac_1\Ac_2}-a_2^{\Ac_1\Ac_2}\,&=\,\ucpc\,{1\over3}\,
     \left[\,\left(\prod\limits_{\SS=1}^{\NS} d_\SS \right)\,
          \tL^{\Ac_1}\,\tL^{\Ac_2}\;+\;
            \left(\prod\limits_{\SS=1}^{\NS}d_\SS^\prime\right)\,
          \tR^{\Ac_1}\,\tR^{\Ac_2}\;\right]                \cr
a_1^{\Ac_2\Ac_1}-a_2^{\Ac_2\Ac_1}\,&=\,\ucpc\,{1\over3}\,
     \left[\,\left(\prod\limits_{\SS=1}^{\NS}d_\SS \right)\,
          \tL^{\Ac_1}\,\tL^{\Ac_2}\;+\;
            \left(\prod\limits_{\SS=1}^{\NS}d_\SS^\prime\right)\,
          \tR^{\Ac_1}\,\tR^{\Ac_2}\;\right]       \cr
}
$$
whose solution is not unique. This was expected because we have now a new
invariant: $F_{\mu\nu}^{\Ac_1}\,F^{\Ac_2\,\mu\nu}$,
$\Ac_1 \ne \Ac_2$, mixing Abelian 
bosons. We can impose, for example, $a_2^{\Ac_1\Ac_2}\equiv0$.

Therefore, from eq.~\EqSfctUno, with the help of  
the modifications given above (do not forget the new terms), one obtains 
following  result:
$$
\eqalignno{
 S^{(1)}_{\!\rm fct}\!= &\hcpc\!\int\!
  \sum_{\Ac=1}^{\NA}
  \!\left[\!\left(\prod\limits_{\SS=1}^{\NS}\! d_{\!\SS}\!\right)
          \!\! (\tL^{\Ac})^2
        \!+ \!\!
          \left(\prod\limits_{\SS=1}^{\NS}\! d'_{\!\SS}\!\right)
           \!\!(\tR^{\Ac})^2
   \right]
    \left[ {5 \over 12} (\pr^\mu A^{\Ac}_\mu)^2 \!+
   {1 \over 4} A_\mu^\Ac \square\, A^{\Ac\,\mu} \right] \cr
&\qquad\quad+\!\!\!\sum_{\SS\!=\!1}^{\NS} \!
  \left[\!\left(\!\prod\limits_{K\!\ne\! \SS}^{\NS}\!\! d_{\!K}\!\!\!\right)
          \!\! \TL{}_{\SS}
        \!+ \!\!
          \left(\!\prod\limits_{K\!\ne\! \SS}^{\NS}\!\! d'_{\!K}\!\!\!\right)
          \!\! \TR{}_{\SS}
   \right]\!   \!            
    \left[ {5 \over 12}
         (\pr^\mu \!\! A^{(\SS\!)a_{\!\SS}}_\mu\!)(\pr^\nu\!\!
                         A^{(\SS\!)a_{\!\SS}}_\nu\!)\!+\!
   {1 \over 4} A_\mu^{(\SS\!)a_{\!\SS}} \square
               A^{(\SS\!)a_{\!\SS}\mu} \!\right] \cr
 &\qquad\quad  - \!\!\!\sum_{\SS\!=\!1}^\NS \!\!
   \left[\!\left(\prod\limits_{K\!\ne\!\SS}^{\NS} d_K\!\right)
           \!{T_L{}_\SS\over 6}
    \!+\!\! \left(\prod\limits_{K\!\ne\!\SS}^{\NS} d'_K\!\right)
            \!{T_R{}_\SS\over 6}
    \right]\,\!  c^{a_{\!\SS} b_{\!\SS} c_{\!\SS}}
   (\pr_{\!\mu} A_\nu^{(\SS\!)a_{\!\SS\!}}\!)\,
                A^{(\SS\!) b_{\!\SS\!}\mu} A^{(\SS\!)c_{\!\SS\!}\nu}\cr
 &\qquad\quad  +
    {\Tr_{\rm full}[\TL^a\TL^b\TL^c\TL^d] +
     \Tr_{\rm full}[\TR^a\TR^b\TR^c\TR^d]\over 12}\,
		 A^a_\mu A^{b\mu} A^c_\nu A^{d\nu}\cr
 &\qquad\quad + \sum_{G=1}^{\NA\!+\!\NS}\,g_G^2\,\CL^{(G)}\,
     \left[1 + (\a^{\prime\,(G)}-1)/6\right]\,{i \over 2}\,
	 \bar\psi \gambar^\mu P_L \arrowsim\pr_\mu\psi        &\numeq\cr
 &\qquad\quad + \sum_{G=1}^{\NA\!+\!\NS}\,g_G^2\,\CR^{(G)}\,
     \left[1 + (\a^{\prime\,(G)}-1)/6\right]\,{i \over 2}\,
    \bar\psi' \gambar^\mu P_R \arrowsim\pr_\mu\psi'       \cr
 &\qquad\quad -\sum_{\SS=1}^\NS\, g^2_\SS \,{C_A^{(\SS)}\, \over 4}
  [1+(\a^{\prime\,(\SS)}-1)]
  \left( \bar L s\psi + L s\bar\psi+\bar R s\psi'+R s\bar\psi' \right)\cr
 &+\hbar^1\sum_{G=1}^{\NA\!+\!\NS}\!\left(\!
     l_{g_G}^{(1)}\,L_g^{(G)}
     + l_{A^{(G)}}^{(1)}\,L_A^{(G)} +
     l_{\om^{(G)}}^{(1)}\,L_\om^{(G)}\!\right)
 +\hbar^1\,l_{\psi}^{(1)}\,L_{\psi}^{\rm L} +
        \hbar^1\,l_{\psi^\prime}^{(1)}\,L_{\psi^\prime}^{\rm R}\cr
 &+\hcpc\,{1\over3}\!\!\sum_{\Ac_1,\Ac_2\atop\Ac_1\ne \Ac_2}^\NA\!
     \!\!\left[\!\left(\prod\limits_{\SS=1}^{\NS}\!d_\SS\! \right)
          \tL^{\Ac_1}\tL^{\Ac_2}\!\!+\!
            \left(\prod\limits_{\SS=1}^{\NS}\!d_\SS^\prime\!\right)
          \tR^{\Ac_1}\tR^{\Ac_2}\!\right]\,
     \intd^dx\, (\pr_{\!\mu}\! A^{\Ac_1\,\mu})
         \,  (\pr_{\!\mu}\! A^{\Ac_2\,\mu}) \cr
 &+\hbar^1\,\sum_{\Ac_1,\Ac_2\atop\Ac_1\ne \Ac_2}^\NA\,
     l_{\Ac_1\Ac_2}^\ouno\,\intd^dx\;
     {1\over2g_{\Ac_1} g_{\Ac_2}}\,F_{\mu\nu}^{\Ac_1} F^{\Ac_2\,\mu\nu}, \cr
}   \namelasteq\EqSfctUnoNonSimple
$$
\smallskip
\noindent where $l_{g_{G}}^{(1)}$, $l_{A^{(G)}}^{(1)}$,
$l_{\om^{(G)}}^{(1)}$, $l_{\psi}^{(1)}$, $l_{\psi^\prime}^{(1)}$,
and each $l_{\Ac_1\Ac_2}\ouno$ for $\Ac_1 < \Ac_2 \le \NA$  are arbitrary
coefficients which will determine the renormalization conditions at
the one-loop order and $L_g^{(G)}$, $L_A^{(G)}$, $L_\om^{(G)}$
$L_{\psi}^{\rm L}$, $L_{\psi^\prime}^{\rm R}$, are any of the
generalizations to the
$d$-dimensional space-time of Dimensional Regularization of the
corresponding $b$-invariant
four-dimensional operators which are written down explicitly in next
subsection.

\bigskip

\subsection{4.5. Renormalization group equation}

The $O(\hbar^0)$ Lorentz invariant functionals $L_i$ with ultraviolet
dimension 4 and ghost number 0 which satisfy the following restrictions
$$
b\, L_i =0\,\qquad   {\delta\over\delta B_a} L_i=0\,,\qquad
 {\cal G}\, L_i =0\, ,        
$$
form a linear space. It is very convenient to have a basis for such
space constitued by elements which are derived by the action of
a compatible diferential operator over the classical action, because,
as explained in section {\it 3.7.}, then, a quantum basis can be properly
defined. The expansion of $\mu\pr_\mu\Gamma_{\rm ren}$ in such a quantum basis,
is just the renormalization group equation. Let us construct a basis 
for that linear space.

\smallskip
\noindent $*$ a) {\sl Cohomologically trivial terms}

We define for each field
$$
N_\phi{}^a{}_b\equiv\intd^4x\;\phi^a{\delta\over\delta\phi^b},
$$
$a, b$ being group indices.

These are terms which are $b$-exact:
$$
\eqalignno{
L_A{}^a{}_b &\equiv b \cdot\intd^4x\;A_\mu^a\,\tilde\rho_b^\mu
  =\left[N_A{}^a{}_b-\!N_\rho{}_b{}^a -\!N_{\bar\om}{}_b{}^a-
   \!N_B{}_b{}^a + 2
   \Lambda^{ae}\!{\pr\over\pr\Lambda^{eb}}\right]\,S_{\rm cl}\equiv
   \NN_A{}^a{}_b\,S_{\rm cl},    \cr
L_\om{}^a{}_b &\equiv -\,b \cdot\intd^4x\;\om^a\,\zeta_b
  =\left[N_\om{}^a{}_b-N_\zeta{}_b{}^a\right]\,S_{\rm cl}\equiv
   \NN_\om{}^a{}_b\,S_{\rm cl},    \cr
L_{\psi^\pp}^{\LpR}{}^i{}_j &\equiv -\,b \cdot\intd^4x\;\bar L_j(\bar R_j)\,
  \PLpR\,\psi^{\pp\,i}
  =\left[N_{\psi^\pp}^\LpR{}^i{}_j-N_{\bar L(\bar R)}
{}_j{}^i\right]\,S_{\rm cl},    \cr
L_{\bar\psi^\pp}^{\RpL}{}_i{}^j &\equiv -\,b \cdot\intd^4x\;\bar\psi_i^\pp\,
   \PRpL\, L^j(R^j)
  =\left[N_{\bar\psi^\pp}^\RpL{}_i{}^j-N_{L(R)}{}^j{}_ i\right]\,S_{\rm cl},
          \cr
L_{\psi^\pp}^{\RpL}{}^i{}_j &\equiv -b\, \cdot\intd^4x\;\bar L_j(\bar R_j)\,
  \PRpL\,\psi^{\pp\,i}
  =N_{\psi^\pp}^\RpL{}^i{}_j\,S_{\rm cl},    \cr
L_{\bar\psi^\pp}^{\LpR}{}_i{}^j &\equiv -b\, \cdot\intd^4x\;\bar\psi_i\,
  \PLpR\,L^j(R^j)
  =N_{\bar\psi^\pp}^\RpL{}_i{}^j\,S_{\rm cl}, \cr
}
$$
but not all are linearly independent:
$$
\delta^j_i \, L_{\psi^\pp}^{\LpR}{}^i{}_j  =
  \delta^i_j\, L_{\bar\psi^\pp}^{\RpL}{}_i{}^j, \qquad
\delta^j_i \, L_{\psi^\pp}^{\RpL}{}^i{}_j  =
  \delta^i_j\, L_{\bar\psi^\pp}^{\LpR}{}_i{}^j\,. 
$$

In the expression of $L_A$ in terms of field counting operators, it is
used the fact that $S_{\rm cl}$ satisfies the gauge condition and the ghost
equation.

Usually, not only the BRS simmetry is imposed to the 1PI functional, but
also other symmetry of the classical action, the rigid simmetry:
$$
\delta_e^{\rm rig} \phi^a = i [T^{(\phi)}_e]^a_b\, \phi^b,
$$
where  $T^{(\phi)}_e$ is the adjoint representation for
$A_\mu, \om, \bar\om, \rho_\mu, \zeta$ and $B$;
$\TLpR{}_e$ for $\psi^\pp, L(R)$; and
$-\TLpR{}_e$ for $\bar\psi^\pp, \bar L(\bar R)$.

Then $\delta_e^{\rm rig} S_{\rm cl}=0$, which has the consequence
$[\,\delta_e^{\rm rig}, b\,]=0$.

Also, our 1PI renormalized functional constructed from eq.~\EqSfctUnoNonSimple\
with Breitenlohner and Maison  mininal dimensional renormalization realizes 
this symmetry at one-loop level.

If the 1PI functional satisfies this new requirement, the set of admissible
$b$-invariants terms is obviously smaller. Thanks to proposition 
5.9 of ref.~\cite{\PiguetSorella}, the trivial subset formed 
by terms which are $b$-exact  and rigidly symmetric is fully determined 
by studying the restricted cohomology, {\it i.e.} by studing the 
set of terms which are $b$-variations of rigidly symmetric terms.

Some of them are obvious generalizations of the trivial elements of
eq.~\EqbInvariants:
$$
\eqalignno{
L_A^{(\SS)}&=\,b\,\cdot\intd^4x\;
 \tilde\rho^{\mu\,(\SS)}_{a_\SS}\,A^{(\SS)\,a_\SS}_\mu,
\qquad
L_\om^{(\SS)}=\,-b\,\cdot\intd^4x\;\zeta^{(\SS)}_{a_\SS}\,
 \om^{(\SS)\,a_\SS} \cr
L_{\psi}^{\rm L}&=- b\,\cdot\intd^4x\;\bar L\PL\psi + \bar\psi\PR L=
   2\,\intd^4x\;{i\over2}\,\bar\psi\arrowsim\prslash\PL\psi+
	\bar\psi\ga^\mu\PL\TL^a\psi A_\mu^a,    \cr
L_{\psi^\prime}^{\rm R} &=
      - b\,\cdot\intd^4x\;\bar R\PR\psi' + \bar\psi'\PL R=
   2\,\!\intd^4x \; {i\over2}\,\bar\psi'\arrowsim\prslash\PR\psi'+
	\bar\psi'\ga^\mu\PR\TR^a\psi' A_\mu^a,    \cr
L_{\psi}^{\rm R}&=- b\,\cdot\intd^4x\;\bar L\PR\psi + \bar\psi\PL L=
   2\,\intd^4x\;{i\over2}\,\bar\psi\arrowsim\prslash\PR\psi,    \cr
L_{\psi^\prime}^{\rm L} &=
      - b\,\cdot\intd^4x\;\bar R\PL\psi' + \bar\psi'\PR R=
   2\,\intd^4x \; {i\over2}\,\bar\psi'\arrowsim\prslash\PL\psi'\,;    \cr
}
$$
where $\SS$ runs from 1 to $\NS$.

The rest are terms which are $b$-exact but involve posssibly different
Abelian components:
$$
\eqalignno{
L_A^{\Ac_1\Ac_2}\!\!&=\,b\,.\int\!\! \tilde\rho^{\Ac_1\,\mu}\,A_\mu^{\Ac_2}=\cr
  &=\!\! \intd^4x\; 2\,
      {(-1)\over4g^2_{\Ac_1}} F_{\mu\nu}^{\Ac_2} F^{\Ac_1\,\mu\nu}\!\! +
      \bar\psi\gambar^\mu \tL^{\Ac_1}\PL\psi A_\mu^{\Ac_2} \!+
      \bar\psi'\gambar^\mu \tR^{\Ac_1}\PR\psi' A_\mu^{\Ac_2}\!-
       \tilde\rho_\mu^{\Ac_1} \pr^\mu\om^{\Ac_2}, \cr
L_\om^{\Ac_1\Ac_2}\!\!&=\,-\,b\,.\int\!\! \zeta^{\Ac_1}\,\om^{\Ac_2}=\cr
   &=\!\!\intd^4x\,\!\bigg\{\!\!\tilde\rho_\mu^{\Ac_1} \pr^\mu\!\om^{\Ac_2}
    \!\bar L i \om^{\Ac_2}\tL^{\Ac_1}\PL\psi +\!
      \bar\psi\PR\tL^{\Ac_1} i \om^{\Ac_2}\! L +\!\!
      \bar R i \om^{\Ac_2}\tR^{\Ac_1}\PR\psi' +\!
      \bar\psi'\PL\tR^{\Ac_1} i \om^{\Ac_2} R \!\bigg\}.\cr
}
$$

Notice that similar non-diagonal terms are not allowed for
non-Abelian components because of rigid invariance
($L_A^{\Ac_1\Ac_2}$ and $L_\om^{\Ac_1\Ac_2}$ are
trivially rigid invariant even if $\Ac_1\ne\Ac_2$).

$L_A^{\Ac_1\Ac_2}$ and $L_\om^{\Ac_1\Ac_2}$ are related with the following
multiplicative field renormalizations:
\medskip
{\settabs 2 \columns
\openup1\jot
\+$\qquad A_\mu^{\Ac_1}=(Z_A)^{\Ac_1}_{\Ac_2}\,A_\mu^{\star\,\Ac_2} $, 
   &$B_A=(Z_A^{-1})^{\Ac_2}_A\,B^\star_{\Ac_2} $, \cr
\+$\qquad \om^{\Ac_1}=(Z_\om)^{\Ac_1}_{\Ac_2}\,\om^{\star\,\Ac_2} $, 
   &$\bar\om_{\Ac_1}=(Z_A^{-1})_{\Ac_1}^{\Ac_2}\,
            {\bar\om}^\star_{\Ac_2}$, & \cr
\+$\qquad \rho_{\Ac_1}^\mu=(Z_A^{-1})^{\Ac_2}_{\Ac_1}\,
          \rho^{\star\,\mu}_{\Ac_2}$,
 &$\zeta_{\Ac_1}=(Z_\om^{-1})^{\Ac_2}_{\Ac_1}\,\zeta_{\Ac_2}^{\star}$,   \cr 
\+$\qquad \Lambda^{A_1A_2}=(Z_A)^{\Ac_1}_{\Ac'_1}(Z_A)^{\Ac_2}_{\Ac'_2}
             \,\Lambda^{\star\,A'_1A'_2} $ & \cr
}
\medskip

But, they will not be needed in renormalization group computations, 
because in minimal substraction of loop diagrams 
this kind of terms can not appear due to the fact that they 
involve Abelian ghosts.

\medskip
\noindent $*$ b) {\sl Elements of the cohomology}

Imposing the rigid invariance, we have the generalization of the
unique element of the cohomology of the simple group case:
$$
L_g^{(G)}=g_G{\pr S_{\rm cl}\over \pr g_G}=
     {1\over2 g^2_G}\,\intd^4x\;
     F_{\mu\nu}^{(G)\,a_G} F^{(G)\,\mu\nu}_{a_G}\, ,
$$
and the new elements (for $\Ac_1\ne\Ac_2$):
$$
\eqalignno{
L_{\Ac_1\Ac_2}&=
    \intd^4x\, {1\over2\,g_{\Ac1}g_{\Ac_2}}\,
           F_{\mu\nu}^{\Ac_1} F^{\Ac_2\,\mu\nu}  \cr
}
$$
with $\Ac_1, \Ac_2 \le \NA$.

Note that $L_{\Ac\Ac}=L_{g^{\Ac}}=g^{\Ac} \pr_{g^{\Ac}} \, S_{\rm cl}$

\medskip
These terms come from  the renormalization of the Abelian
components of the matrix $C$ of eq.~\EqYMGeneric.
At tree level, it is a diagonal matrix, but, more
generally, it can be changed to a non- diagonal matrix in the
Abelian part while the right hand side of eq.~\EqYMGeneric\ continue
being a rigidly invariant term:
\medskip
{\settabs 2 \columns
\openup1\jot
\+$\qquad C_{\Ac_1\Ac_2}=(Z_C)^{\Ac_1'\Ac_2'}_{\Ac_1A_2}\,
        C^{\star}_{\Ac_1'\Ac_2'}$.
                              &  &\cr 
}

\medskip

We define:
$$
L_C^{\Ac_1\Ac_2}={\pr\over\pr C_{\Ac_1\Ac_2}}
      \,S_{\rm cl}=-{L^{\Ac_1\Ac_2}\over2} 
$$

\medskip

Therefore, eq.~\EqRGExacta\ becomes
$$
\eqalignno{
\Big[\,\mu\,{\pr\over\pr\mu} &+\! \sum_G^{\NA+\NS}\b^G g_G\,{\pr\over\pr g_G}
  - \ga_\psi^\L \NN_\psi^\L- \ga_{\psi'}^\R \NN_{\psi'}^\R 
   - \ga_\psi^\R \NN_\psi^\R- \ga_{\psi'}^\L \NN_{\psi'}^\L  \cr
  &-\!\sum_{\SS=1}^\NS\ga_A^{\,\SS} \NN_A^{(\SS)}
   -\!\sum_{\SS=1}^{\NS}\ga_\om^{\,\SS} \NN_\om^{(\SS)}
   -\!\sum_{\Ac_1,\Ac_2\atop\Ac_1\ne \Ac_2}^\NA
      \ga_C^{\Ac_1\Ac_2}{\pr\over\pr C_{\Ac_1\Ac_2}}\, \Big]
   \, \GR=0,  \cr
}
$$
which at order $\hbar$ reads
$$
\eqalignno{
\mu\,{\pr\GR\over\pr\mu}=&-\! \sum_{G=1}^{\NA+\NS}\b^G{}\Ouno\,L_g^{(G)} 
    + \ga_\psi^\L\Ouno L_{\psi}^{\rm L}+
    \ga_{\psi'}^\R\Ouno L_{\psi'}^{\rm R} \cr
  & +\!\sum_{\SS=1}^{\NS}\ga_A^{\,\SS}{}\Ouno L_A^{(\SS)} +
   \! \sum_{\SS=1}^{\NS}\ga_\om^{\,\SS}{}\Ouno L_\om^{(\SS)}
  + \!\sum_{\Ac_1,\Ac_2\atop\Ac_1\ne \Ac_2}^\NA
         \ga_C^{\Ac_1\Ac_2\,\ouno} L_C^{\Ac_1\Ac_2}\,
             +O(\hbar^2).            \cr
}
$$
Eq.~\final\ is thus generalized.  

Eq.~\semifinal\ should  be modified by applying the rules stated
above and adding the new term:
$$
\eqalignno{
+&\ucpc\,{8\over3}\;
\sum_{\Ac_1,\Ac_2\atop\Ac_1\ne \Ac_2}^\NA\,
          \left(\prod\limits_{\SS=1}^{\NS}d^\pp_\SS \right)\,
          \tLpR^{\Ac_1}\,\tLpR^{\Ac_2}\;\intd^4x\; -{1\over4}\,
    F_{\mu\nu}^{\Ac_1}\,F^{\Ac_2\,\mu\nu}       \;S_{{\rm cl}\,AA} \cr
}
$$

Therefore,
$$
\eqalignno{
\b^{\Ac\,\ouno}&=\ucpc\,g^2_\Ac\,
{2\over3}\,
  \left[\!\left(\prod\limits_{\SS=1}^{\NS}\! d_{\!\SS}\!\right)
          \! (\tL^{\Ac})^2
        \!+ \!
          \left(\prod\limits_{\SS=1}^{\NS}\! d'_{\!\SS}\!\right)
           \!(\tR^{\Ac})^2
   \right]\,,\cr
\b^{\SS\,\ouno}&=\ucpc\,g^2_\SS\,
\left(
{2\over3}\,
  \left[\!\left(\prod\limits_{K\!\ne\! \SS}^{\NS}\! d_{\!K}\!\!\!\right)
          \!\! \TL{}_{\SS}
        \!+
          \left(\prod\limits_{K\!\ne\! \SS}^{\NS}\! d'_{\!K}\!\!\!\right)
          \!\! \TR{}_{\SS}
   \right]\!
 -\!{11\over3}\,\CA^{(\SS)}
\right)\,,\cr
\ga_A^{\SS\ouno}&=\ucpc\,g^2_\SS\, \left(2-{1-\a^{\prime\,\SS}\over2}
 \right)\,\CA^{(\SS)},    \cr
\ga_\om^{\SS\ouno}&=\ucpc\,g^2_\SS\,\a^{\prime\,\SS}\,\CA^{(\SS)}, \cr
\ga_\psi^\L\Ouno&=\ucpc\,
      \sum_{G=1}^{\NA\!+\!\NS} g^2_G\,\a^{\prime\,G}\, \CL^{(G)},  \qquad
\ga_{\psi'}^\R\Ouno=\ucpc\,
      \sum_{G=1}^{\NA\!+\!\NS} g^2_G\,\a^{\prime\,G}\, \CR^{(G)},    \cr
\ga_C^{\Ac_1\Ac_2\,\ouno}&=\ucpc \,{4\over3}\,
   \big( \tL^{\Ac_1}\tL^{\Ac_2}+\tR^{\Ac_1}\tR^{\Ac_2} \big)\,. \cr
}
$$
where $\a^{\prime\,G}\equiv\a^G/g^2_G$, $1\le\Ac, \Ac_1, \Ac_2 \le \NA$,
$1\le\SS\le\NS$.

\bigskip

\section{6. Conclusions}

We have shown by performing  explicit thorough one-loop computations that 
Dimensional Regularization {\it \'a la} Breitehloner and Maison 
can be used blindly to carry out calculations in chiral gauge theories. 
Up to best of our knowledge it is the first time that such a complete one-loop
study has been carried out. It shows that the renormalization
method at hand has all the properties that on general grounds were expected.  
Notice that here the cancellation of anomalies is not used 
to try to hide any inconsistency. Even when there is not cancellation 
of anomalies, the renormalized theory is finite and unambigous: in particular,
the BRS anomaly is  obtained without doing any tinkering. 
This is in stark contrast to the ``naive'' approach, were cancellation of 
anomalies is invoked to dangerously try to hide some incosistencies.

We would like to stress the fact that it would not have been feasible to
carry out the computations made in this paper without the help of the
action principles and the BRS cohomolgy theory. This makes manifest 
the importance of these two theoretical tools in practical computations 
in chiral gauge theories.

\bigskip

\section{Appendix A: Slavnov-Taylor Identities}

This appendix is devoted to establishing eq.~\EqRegSTBreakingct\ by using
the {\it Regularized Action Principle} as stated in eqs.~\Varfield\ and
\Varexternal. 

Let $S^{(n)}_{\rm DReg}$ be given by 
$$
S_{\rm free}[\varphi,\Phi]+S_{\rm int} [\varphi,\Phi;K_{\Phi}],
$$
where 
$$
\eqalign{&S_{\rm free}[\varphi;\Phi]=S_{0\rm free}[\varphi;\Phi],\cr
&S_{\rm int} [\varphi,\Phi;K_{\Phi};\lambda]=S_{0 \rm int}[\varphi,\Phi]+
\intd^d x\, K_{\Phi}(x) s_d\Phi(x)+ 
S^{(n)}_{\rm ct}[\varphi,\Phi;K_{\Phi}].\cr}
$$
The symbols $\varphi$ and $\Phi$ stand, respectively, for  sets of fields  
which transforms linearly and non-linearly under arbitrary BRS
transformations. The generalization to ``$d$-dimensions'' of the
BRS transformations are denoted by $s_d\phi(x)$ and $s_d\Phi(x)$.
We define $s_d K_\Phi(x)=0$

We next define 
$$
S_{\rm INT}[\varphi,\Phi;J_{\varphi},J_{\Phi},K_{\Phi}]=
	S_{\rm int}[\varphi,\Phi;K_\Phi]+ 
\int d^d x \bigl(J_\varphi(x)\varphi(x)+ J_\Phi(x)\Phi(x)\bigr)
$$
Following eq.~\EqZDefinition , the generating functional $Z_{\rm DReg}$ 
is introduced next
$$
Z_{\rm DReg}[J_\varphi,J_\Phi,K_\Phi]=
\bigl\langle \exp\Bigl\{ {i\over\hbar}
	S_{\rm INT}[\varphi,\Phi;J_\varphi,J_\Phi,K_\Phi]\Bigr\}\bigr\rangle_0,
$$
where the symbol $\langle \cdots \rangle_0$ is defined as in the paragraph
below eq.~\EqZDefinition .
 
Let $\Delta$ and $\Delta_{\rm ct}$ de defined as follows:
$$
\Delta=s_d S_0,\qquad\Delta_{\rm ct}=s_d S^{(n)}_{\rm ct},
$$
where $S_0=S_{0\rm free}[\varphi;\Phi]+S_{0 \rm int}[\varphi,\Phi]+
\intd^d x\, K_{\Phi}(x) s_d\Phi(x)$.
 
We now introduce $S^{(\Upsilon)}_{\rm INT}$ for reasons that will become clear 
later:
$$
S^{(\Upsilon)}_{\rm INT}=S_{\rm INT}[\varphi,\Phi;J_\varphi,J_\Phi,K_\Phi]+
S_{\rm aux},
$$
where
$$ 
S_{\rm aux}=
\intd^d x \bigl(\Upsilon_1(x)\Delta (x)+\Upsilon_2(x)\Delta_{\rm ct}(x)
+ \Upsilon_3^\Phi(x){\delta S^{(n)}_{\rm ct}\over\delta K_\Phi(x)}\bigr),
$$
and $\Delta=\intd^d x\,\Delta(x)$ and 
$\Delta_{\rm ct}=\int d^d x\,\Delta_{\rm ct}(x)$. $\Upsilon_{i}$, 
$i=1,2$ and $3$ are external fields. We define $s_d \Upsilon_{i}(x)$ to be
zero for any exteranl field $\Upsilon_{i}$. 
Then, we introduce the generating
functional $Z_{\rm DReg}^{\Upsilon}$ as follows
$$
Z_{\rm DReg}^{\Upsilon}[J_\varphi,J_\Phi,K_\Phi,\Upsilon_1,
\Upsilon_2,\Upsilon_3^\Phi]
=
\bigl\langle \exp\Bigl\{ {i\over\hbar}
S_{\rm INT}^{\Upsilon}[\varphi,\Phi;J_\varphi,J_\Phi,K_\Phi,
\Upsilon_i]\Bigr\}\bigr\rangle_0,
$$
where $\langle \cdots \rangle_0$ is defined as for $Z_{\rm DReg}$.

It is obvious that the following equation holds in Dimensional Regularization:
$$ 
Z_{\rm DReg}[J_\varphi,J_\Phi,K_\Phi]=
Z_{\rm DReg}^{\Upsilon}[J_\varphi,J_\Phi,K_\Phi,
\Upsilon_1 =0,\Upsilon_2 = 0,\Upsilon_3^\Phi =0]
$$
The {\it Regularized Action Principle} (see eq.~\Varfield ) implies that 
the following equation holds in Dimensional Regularization:
$$
\eqalignno{&s_d Z_{\rm DReg}^{\Upsilon}\equiv
\intd^dx\;\Big\langle\,\Big( \Delta(x)+\Delta_{\rm ct}(x) +
(-1)^\varphi J_\varphi(x) s\varphi(x)+ (-1)^\Phi J_\Phi (x) s\Phi(x)\cr 
	&+\hbox{ terms proportional to $\Upsilon_i$} \Big)
	\, \exp\Bigl\{ {i\over\hbar}
S^{\Upsilon}_{\rm INT}[\varphi,\Phi,J_\varphi,J_\Phi,K_\Phi,\Upsilon_i\Bigr\} 
   \, \Bigr\rangle_0=0.            &\eq{{\rm A}.1}\cr
}
$$  
Next, by using eq.~\Varexternal , one easily obtains that
$$
\Big\langle\, J_\varphi(x) s\varphi(x) \, 
	\exp\Bigl\{ {i\over\hbar}S^{\Upsilon}_{\rm INT} \Bigr\} 
	\, \Bigr\rangle_0= {\hbar\over i}\,J_\varphi(x)
\Big[a_{\varphi\varphi} {\delta Z^{\Upsilon}_{\rm DReg}
\over\delta J_\varphi(x)}+
a_{\varphi\Phi} {\delta Z^{\Upsilon}_{\rm DReg}\over\delta J_\Phi(x)}\Big]
\eqno{({\rm A}.2)}
$$
and 
$$
\eqalignno{
&\Big\langle\, J_\Phi(x) s\Phi(x) \, \exp\Bigl\{ 
 {i\over\hbar}S^{\Upsilon}_{\rm INT} \Bigr\}
	\, \Bigr\rangle_0=
J_\Phi(x)\Big\langle\, 
{\delta(S^{\Upsilon}-S_{\rm ct}-S_{\rm aux})\over\delta K_\Phi(x)}
		\,\Big\rangle_0=              \cr
&\qquad {\hbar\over i}\, J_\Phi(x)
{\delta Z^{\Upsilon}_{\rm DReg}\over\delta K_\Phi(x)}-
{\hbar\over i}\,J_\Phi(x) 
{\delta Z^{\Upsilon}_{\rm DReg}\over\delta\Upsilon_3^\Phi(x)} +
   \hbox{ terms proportional to $\Upsilon_i$}.          &\eq{{\rm A}.3}\cr
}
$$
In eq.~(\A.2) the linear BRS transformations $s_d \varphi$ are given by
$s_d \varphi= a_{\varphi\varphi}\varphi+a_{\varphi\Phi}\Phi$.

Now, by substituting eqs.~(\A.2) and (\A.3) in eq.~(\A.1), one arrives at
$$
\eqalignno{
\intd^dx\; &\Bigg[ 
	(-1)^\varphi {\hbar\over i} J_\varphi (x)
\Big( a_{\varphi\varphi} 
{\delta Z^{\Upsilon}_{\rm DReg}\over\delta J_\varphi(x)} +
a_{\varphi\Phi} {\delta Z^{\Upsilon}_{\rm DReg}\over\delta J_\Phi(x)}\Big) +
(-1)^\Phi {\hbar\over i} J_\Phi(x) \Big({\delta Z^{\Upsilon}_{\rm DReg}
\over K_\Phi (x))}-
 {\delta Z^{\Upsilon}_{\rm DReg}\over\delta \Upsilon_3^\Phi(x)}\Big)\cr
&\qquad+ 
{\hbar\over i} {\delta Z^{\Upsilon}_{\rm DReg}\over\delta\Upsilon_1(x)} +
{\hbar\over i} {\delta Z^{\Upsilon}_{\rm DReg}\over\delta\Upsilon_2(x)} +
	\hbox{ terms proportional  to $\Upsilon_i$}  \Bigg]
   =0.       &\eq{{\rm A}.4}\cr
}
$$

Let us introduce as usual the generating functionals 
$Z^{\Upsilon}_{c\,{\rm DReg}}[J_\varphi,J_\Phi;K_\Phi,\Upsilon_i]$
and \break\hfill
$\Gamma^{\Upsilon}_{\rm DReg}[\varphi,\Phi;K_\Phi,\Upsilon_i]$:
$$
\eqalignno{
&Z^{\Upsilon}_{\rm DReg}[J_\varphi,J_\Phi;K_\Phi,\Upsilon_i]=
\hbox{exp}\Bigl(\;{\imath\over\hbar}
Z^{\Upsilon}_{c\,{\rm DReg}}[J_\varphi,J_\Phi;K_\Phi,\Upsilon_i]\Bigr)\cr
&\Gamma^{\Upsilon}_{\rm DReg}[\varphi,\Phi;K_\Phi,\Upsilon_i]=
Z^{\Upsilon}_{c\,{\rm DReg}}[J_\varphi,J_\Phi;K_\Phi,\Upsilon_i]-
\intd^d x \bigl(J_\varphi(x)\varphi(x)+ J_\Phi(x)\Phi(x)\bigr).
         \qquad &\eq{{\rm A}.5}\cr
}
$$
where the functionals in the previous equations are to be understood as
formal power series in $\hbar$ and the fields (both external and quantum).
By  taking advantage of eq.~(\A.5) one turns eq.~(\A.4) into an
equation for the 1PI functional 
$\Gamma^{\Upsilon}_{\rm DReg}[\varphi,\Phi;K_\Phi,\Upsilon_i]$: 
$$
\eqalignno{
\intd^dx\; &\Bigg[ 
	-{\delta\Gamma^{\Upsilon}_{\rm DReg}\over\delta\varphi(x)}
		(a_{\varphi\varphi} \varphi(x)+a_{\varphi\Phi} \Phi(x) ) 
	-{\delta\Gamma^{\Upsilon}_{\rm DReg}\over\delta\Phi(x)} 
	\Big({\delta\Gamma^{\Upsilon}_{\rm DReg}\over\delta K_\Phi(x)}-
{\delta\Gamma^{\Upsilon}_{\rm DReg}\over\delta \Upsilon_3^\Phi(x)}\Big)\cr
&\qquad+{\delta\Gamma^{\Upsilon}_{\rm DReg}\over\delta\Upsilon_1(x)} +
   {\delta\Gamma^{\Upsilon}_{\rm DReg}\over\delta\Upsilon_3^\Phi(x)} +
   \hbox{ term prop. to $\Upsilon_i$}  \Bigg] =0   &\eq{{\rm A}.6}\cr
}
$$
Finally, by setting $\Upsilon_i(x)=0$, $\forall i$, in eq.~(\A.6)
and taking into account that 
$$
\Gamma_{\rm DReg}[\varphi,\Phi;K_\Phi]=
\Gamma^{\Upsilon}_{\rm DReg}[\varphi,\Phi;K_\Phi, \Upsilon_i=0]
$$
one obtains eq.~\EqRegSTBreakingct\   from eq.~(\A.6):
$$
\eqalignno{
\intd^dx\;&  
(s_d\varphi){\delta\Gamma_{\rm DReg}\over\delta\varphi(x)}
  +{\delta\Gamma_{\rm DReg}\over\delta K_\Phi(x)}
 {\delta\Gamma_{\rm DReg}\over\delta\Phi(x)}=   &\eq{{\rm A}.7}\cr
&\intd^dx\; \Bigg[\Delta(x)\cdot\Gamma_{\rm DReg} 
+\Delta_{\rm ct}(x)\cdot\Gamma_{\rm DReg}+
\Big( {\delta S_{\rm ct}\over\delta K_\Phi(x)}\cdot\Gamma_{\rm DReg}\Big)\,
{\delta\Gamma_{\rm DReg}\over\delta\Phi(x)}\Bigg].  \cr
}
$$
Notice that 
$$
\eqalign{
&\Bigg[{\delta\Gamma^{\Upsilon}_{\rm DReg}\over\delta K_\Phi(x)}-
{\delta\Gamma^{\Upsilon}_{\rm DReg}\over\delta \Upsilon_3^\Phi(x)}
\Bigg]_{\Upsilon_i=0}=
{\delta\Gamma_{\rm DReg}\over\delta K_\Phi(x)}-
{\delta S_{\rm ct}\over\delta K_\Phi(x)}\cdot\Gamma_{\rm DReg}\,,\cr
&\Bigg[{\delta\Gamma^{\Upsilon}_{\rm DReg}\over\delta\Upsilon_1(x)}
\Bigg]_{\Upsilon_i=0}\quad=\qquad
\Delta(x)\cdot\Gamma_{\rm DReg}\qquad\hbox{and}\cr
&\Bigg[{\delta\Gamma^{\Upsilon}_{\rm DReg}\over\delta\Upsilon_2(x)}
\Bigg]_{\Upsilon_i=0}\quad =\qquad
\Delta_{\rm ct}(x)\cdot\Gamma_{\rm DReg},\cr
}
$$
for eq.~\Varexternal\ holds.
\bigskip

\section{  Appendix B. Slavnov-Taylor Identities for $\eps$-dependent BRS}

This appendix is devoted to the study of the contributions to 
the right hand side of eq.~\Breaking\ coming from BRS variations 
which depend explicitly on $\eps=4-d$.
We shall see that, unlike  for the BRS transformations in eq.~\DBRStrans,
the variation under these new BRS transformations of the singular 
counterterms yield a finite contribution to 
the right hand side of eq.~\Breaking.  

To introduce an explicit $\eps$ dependence on the Dimensional Regularization
BRS transformations, we shall  replace the BRS transformation of $A_\mu$ 
given in eq.~\DBRStrans\ with  
$s'_d A_\mu = \nabla_\mu\om  + C (d-4) \nabla_\mu\om$.  We shall assume,
however, that $s'_d$ acts on the rest of the fields
as $s_d$ does. Hence, $s'_d$ defines BRS transformations with explicit 
$\eps$ dependence. 

Although we have changed the Dimensional Reguarization BRS transformations,
we shall not take a new Dimensional Regularization classical action, which
still will be $S_0$ in eq.~\DRegclass. Thus, we will not modify the
minimally renormalized 1PI functional $\Gamma_{\rm minren}$, so that we will
have  the effect of the new Dimensional Regularization BRS 
transformations isolated.
 
The  $s'_d$-variation of $S_0$ is given by:
$$
\eqalignno{
s'_d S_0 &= \hat\Delta + \Delta_{\rm f}^\eps + \Delta_{\rm gf}^\eps +
      \Delta_\rho^\eps.
}
$$
Here, $\hat\Delta=s_d S_0$ is the breaking given in eq.~\EqBreaking\
and 
$$
\eqalign{
\Delta_{\rm f}^\eps &= -\eps\,C\,\intd^dx\;
   (\bar\psi\gambar^\mu\PL\TL^a\psi +\bar\psi'\gambar^\mu\TR^a\psi')
   \, (\nabla_\mu\om)^a,\cr
\Delta_{\rm gf}^\eps &=
   -\eps\,C\,\intd^dx\; B^a \pr^\mu(\nabla_\mu\om)^a\,-\,
   c^{abc}\pr^\mu\bar\om^a(\nabla_\mu\om)^b\om^c,\cr
\Delta_\rho^\eps &=
  +\eps\,C\,\intd^dx\; c^{abc}\, \rho^{a\mu}
   (\nabla_\mu\om)^b\om^c.
}
$$
are the new breakings (which are also evanescent operators).
Notice that ${s'_d}^2\, A_\mu^a \ne0$, whereas ${s_d}^2\, A_\mu^a = 0$.

Now, let $\Gamma_0$ denote dimensionally regularized  1PI functional 
obtained from $S_0$, then,  regularized Slavnov-Taylor equation for the 
$s'_d$-transformations reads:
$$
\SS'_d(\Gamma_0) = \left[
\hat\Delta + \Delta_{\rm f}^\eps + \Delta_{\rm gf}^\eps +
   \Delta_\rho^\eps \right]\cdot\Gamma_0, \eqno{({\rm B}.1)}
$$
where $\SS'_d$ is the Slavnov-Taylor operator for $s'_d$. Notice that
$$
\eqalign{\SS'_d(\Gamma_0)\equiv 
&\intd^d x\,\,\Bigl\{
(1-C\eps)\,\Tr\,{\delta\Gamma_0\over\delta\rho^\mu}
 {\delta\Gamma_0\over\delta A_\mu}+
\Tr\,{\delta\Gamma_0\over\delta\zeta}
 {\delta\Gamma_0\over\delta\om}+
\Tr\; B {\delta\Gamma_0\over\delta\bar\om} \cr
&+ {\delta\Gamma_0\over\delta\bar L} 
{\delta\Gamma_0\over\delta\psi}+
{\delta\Gamma_0\over\delta\bar R}
 {\delta\Gamma_0\over\delta\psi'}+
{\delta\Gamma_0\over\delta L}
 {\delta\Gamma_0\over\delta\bar\psi}+
{\delta\Gamma_0\over\delta R} 
{\delta\Gamma_0\over\delta\bar\psi'}\Bigr\}.
}
$$
The next issue to address is the derivation of the 
(anomalous) Slavnov-Taylor identity satisfied by the minimal dimensional 
renormalization 1PI functional, $\Gamma_{\rm minren}$. If we, naively, 
just replace in eq.~(B.1) every  regularized object with their 
minimally renormalized counterpart, we will obtain the following equation
$$
\SS(\Gamma_{\rm minren} )= \N\left[
\hat\Delta + \Delta_{\rm f}^\eps + \Delta_{\rm gf}^\eps +
   \Delta_\rho^\eps \right]\cdot\Gamma_{\rm minren}.\eqno{({\rm B}.2)}
$$
This equation does not hold, though, since the coefficients of the regularized
Slavnov-Taylor equation, eq.~(B.1), do have an explicit $\eps$ dependence. 
Indeed, at  the one-loop level the following identity holds
$$
\SS(\Gamma_{\rm minren} )= 
\Bigl[\N\left[\hat\Delta\right]\cdot\Gamma_{\rm minren}\Bigr]^{(1)}+
O(\hbar^2).\eqno{({\rm B}.3)}
$$
This has been shown in subsection {\it 3.4.} Now, let us substitute the
previous equation in eq.~(B.2). We thus come to the 
conclusion that  $\N\left[\Delta_{\rm f}^\eps + \Delta_{\rm gf}^\eps +
   \Delta_\rho^\eps \right]\cdot\Gamma_{\rm minren}$ should vanish at the
one-loop level. Which is nonsense, for a  diagramatic  computation shows that
$$
\eqalignno{
\Bigl[\N\left[\Delta_{\rm f}^\eps +\Delta_{\rm gf}^\eps +
   \Delta_\rho^\eps \right]&\cdot\Gamma_{\rm minren}\Bigr]^{(1)}=\cr
\qquad{-\,\hbar\over (4\pi)^2}\,C\,\,g^2\,\bigg\{
&\left( \a'\CA+2\,\a'\CL\right)\;
\intd^4x\;{\delta S_{{\rm cl}\,\bar\psi\psi A} \over \delta A^a_{\mu}}\,
 \pr_\mu\om^a \cr
&+\left[\left(3-{3\over2}(1-\a')\right)\CA+2\,\a'\CL\right]\;
\intd^4x\;{\delta S_{{\rm cl}\,\bar\psi\psi A} \over \delta A^a_{\mu}}\,
 c^{abc}\, A_\mu^b \om ^c \cr
&+\left( \a'\CA+2\,\a'\CR\right)\;
\intd^4x\;{\delta S_{{\rm cl}\,\bar\psi'\psi' A} \over \delta A^a_{\mu}}\,
 \pr_\mu\om^a            \cr
&+\left[\left(3-{3\over2}(1-\a')\right)\CA+2\,\a'\CR\right]\;
\intd^4x\;{\delta S_{{\rm cl}\,\bar\psi'\psi' A} \over \delta A^a_{\mu}}\,
 c^{abc}\, A_\mu^b \om ^c \cr
&-{3\over2}\, (1-\a')\CA\;\intd^4x\;
\left({\delta S_{{\rm cl}\,\bar\om\om A}\over \delta A^a_{\mu}}
  +   {\delta S_{{\rm cl}\,\rho\om A}\over \delta A^a_{\mu}} \right)
\,\pr_\mu\om^a                     &\eq{{\rm B}.4}\cr
&+2\,\a'\,\CA\;\intd^4x\;
\left({\delta S_{{\rm cl}\,\bar\om\om A}\over \delta A^a_{\mu}}
  +   {\delta S_{{\rm cl}\,\rho\om A}\over \delta A^a_{\mu}}\right)
\,c^{abc}\,A_\mu^b\om^c \cr
&+\left(1+{1\over2}(1-\a')\right)\,\CA\;\intd^4x\;
(\pr_\mu B^a) \,(\pr^\mu\om^a)\cr 
&-\,\a'\,\CA\;\intd^4x\;
(\pr_\mu B^a)
  \,c^{abc}\,A_\mu^b\om^c \,\bigg\}. \cr
}
$$
Hence, a contribution to the right hand side of eq.~(B.2) is missing. 
This missing contribution should apparently be furnished by the second
and third terms on the right hand side of eq.~\Breaking, which is a consequence
of eq.~(A.7). But, it is essential to remember the 
assumption made in Appendix A in order to properly deduce eq.~(A.7): 
the dependence in $K_\Phi$ in the
starting dimensional regularized action should be
$\int K_\Phi\, s'_d\Phi$, if $s'_d$ is the variation under study.

Therefore, due to the fact that the dependence in
$\rho_\mu$ of the chosen action is just $\int \rho_\mu\, s_d A^\mu$,
an extra factor of $1 -C\eps$ is needed in eq.~(A.3), when
studying $s'_d$ variations:
$$
\eqalignno{
&\Big\langle\, J_A{}_a^\mu(x) s'_d A^a_\mu(x) \, \exp\Bigl\{ 
 {i\over\hbar}S^{\Upsilon}_{\rm INT} \Bigr\}
	\, \Bigr\rangle_0=
(1-C\,\eps)\,\Big\langle\, J_A{}_a^\mu(x) s'_d A^a_\mu(x) \, \exp\Bigl\{ 
 {i\over\hbar}S^{\Upsilon}_{\rm INT} \Bigr\}\, \Bigr\rangle_0=\cr
&\qquad (1-C\,\eps){\hbar\over i}\, J_A{}_a^\mu(x) \left(
{\delta Z^{\Upsilon}_{\rm DReg}\over\delta\rho_\mu^a(x)}-
{\delta Z^{\Upsilon}_{\rm DReg}\over\delta\Upsilon_3^A{}_A^\mu(x)}\right) +
   \hbox{ terms proportional to $\Upsilon_i$}.      \cr
}
$$

Then, the equation (A.7) becomes
$$
\eqalignno{
\SS'_d&(\Gamma_{\rm DReg})=                                                   
s'_d(S_0 +S_{\rm ct})\cdot\Gamma_{\rm DReg} 
+  \cr
&\intd^dx\;\left[
\sum_{\Phi\ne A}\,\Big(
{\delta S_{\rm ct}\over\delta K_\Phi(x)}\cdot\Gamma_{\rm DReg}\Big)\,
{\delta\Gamma_{\rm DReg}\over\delta\Phi(x)}
+ (1-C\eps) \Big(
{\delta S_{\rm ct}\over\delta\rho_\mu^a(x)}\cdot\Gamma_{\rm DReg}\Big)\,
{\delta\Gamma_{\rm DReg}\over\delta A_\mu^a(x)}\right], \cr
}
$$
and the correct modification of eq.~(B.2) is:
$$
\eqalignno{
\SS(\Gamma_{\rm minren})=&\Bigl[\N\left[
\hat\Delta + \Delta_{\rm f}^\eps + \Delta_{\rm gf}^\eps + 
\Delta_\rho^\eps \right]\cdot\Gamma_{\rm minren}\Bigr]^{\ouno}+\cr
& C
\intd^4x\;\left[(\nabla_\mu\om)^a\,{\delta {\cal F}\over \delta A^a_{\mu}}\, +
{\delta S_{\rm cl}\over\delta A_\mu^a}{\delta {\cal F}\over\delta\rho_a^\mu}
\right]\,+ O(\hbar^2).  &\eq{{\rm B}.5}\cr
}
$$
Here, ${\cal F}= {\rm LIM}_{d\rightarrow 4}\Big\lbrace
\eps\Gamma^{\ouno}_{0\,{\rm sing}}\Big\rbrace$, with 
$\Gamma^{\ouno}_{0\,{\rm sing}}$ as  given by eqs.~\Singcount\ and  \OnePIpole.
Notice that the second term on the right hand side of 
(B.5) cancels the contribution
displayed in eq.~(B.4). Eq.~(B.3) is thus recovered.

Of course, if we want to use eqs.~\EqRegSTBreakingct, \Breaking\ and
\Oneloopbreaking ---with $s'_d$ instead of $s_d$---, we have to
choose the following new Dimensional Regularization classical action:
$$
S'_0=\left(S_0 - \intd^dx\, \rho_\mu^a s_dA^{a\mu}\right) +
          \intd^dx\, \rho_\mu^a s'_d A^{a\mu}\,,
$$
which cause the breaking
$$
\eqalignno{
s'_d S'_0 &= \hat\Delta + \Delta_{\rm f}^\eps + \Delta_{\rm gf}^\eps +
     (1-C\eps) \Delta_\rho^\eps,
}
$$
and a new perturbative renormalized 1PI functional
$\Gamma'_{\rm minren}$.

Therefore, in this case, from eq.~\EqRegSTBreakingct, \Breaking\ and
\Oneloopbreaking, we conclude that the (anomalous) one-loop renormalized 
Slavnov-Taylor identity reads 
is
$$
\eqalignno{
\SS(\Gamma'_{\rm minren})=
&\Bigl[\N\left[s'_d S'_0\right]\cdot\Gamma'_{\rm minren}\Bigr]^{\ouno}+
C
\intd^4x\;\left[(\nabla_\mu\om)^a\,{\delta {\cal F}\over \delta A^a_{\mu}}\, +
{\delta S_{\rm cl}\over\delta A_\mu^a}{\delta {\cal F}\over\delta\rho_a^\mu}
\right] +  \cr
&\intd^4x\;
{\delta (S_{{\rm cl}\,\rho\om} + S_{{\rm cl})\,\rho\om A}\over\delta\om^a}
{\delta {\cal F}\over\delta\zeta^a} \,+ O(\hbar^2)\,. \cr
}
$$

\medskip

\doublesection{Appendix C. Comparison between  ``naive'' and
Breitenlohner and Maison}{minimal dimensional renormalization schemes}

In this appendix  we shall make a thorough comparison of the minimally 
renormalized one-loop 1PI functionals obtained within the ``naive'' 
dimensional renormalization prescription, on the one hand, and 
the Breitenlohner and Maison Dimensional Regularization scheme, on the other.
The quantum field theory under scrutiny will be the theory already studied
in section {\it 3}. Obviously, at the one-loop level, differences only 
arise in diagrams involving fermions. We will denote these contributions 
with the subscript ${}_{\F}$. All along this appendix, 
the subscripts BM and ``naive'' will indicate 
that the corresponding quantity has been evaluated by using  the 
Breitenlohner and Maison scheme and the ``naive'' prescription, respectively.

Let us begin with the self-energy of the gauge field. Here, only the fermionic
loops may give rise to differences between the two calculational techniques. 
There are terms with one $\gam5$ and four $\ga$s  in the trace, but their
contribution explicitly vanish,  the reason being  the contraction of
two $\ga$s of the antisymmetic trace with the same momentum $q$. The  
``naive'' prescription does not yield ambiguous the results in this 
instance. These results read
$$
\eqalign{
\Gapuno_{\F\,AA}{\,}^{\mu\nu,ab}_\n (p)=&{1\over4\pi^2} (\TL+\TR) 
   \left[{1\over3\eps}-{\ga\over6} +
		{5\over18} - {1\over6}\ln{-p^2+i\epsilon\over4\pi\mu^2}
      + O(\eps)\right]\cr
&\qquad \delta^{ab}\;(p^\mu p^\nu- p^2 g^{\mu\nu}),      
}
$$
for the regularized  theory and 
$$
\GRpuno_{\F\,AA}{\,}^{\mu\nu,ab}_\n (p)={1\over4\pi^2}(\TL+\TR) 
	\left[-{\ga\over6} +
		{5\over18} - {1\over6}\ln{-p^2+i\epsilon\over4\pi\mu^2}\right]\,
		\delta^{ab}\,(p^\mu p^\nu- p^2 g^{\mu\nu}),     
$$
for the renormalized Green function. It  should be understood a sum over 
the different fermionic multi\-plets: 
$\sum_{k=1}^{n_\L} \TL^k+\sum_{r=1}^{n_\R} \TR^r$. Notice that both results 
are transverse, so that the corresponding Slavnov-Taylor identity holds.

The calculation within the Breitenlohner and Maison framework is also 
straightforward and it yields the following regularized  expression 
$$
\eqalignno{
\Gapuno_{\F\,AA}{\,}^{\mu\nu,ab}_\BM (p)=
&\;\delta^{ab}\,\bigg\{{1\over4\pi^2} (\TL+\TR) 
	\left[{1\over3\eps}-{\ga\over6} +
		{5\over18} - {1\over6}\ln{-p^2+i\epsilon\over4\pi\mu^2}\right]
(\bar p^\mu \bar p^\nu- \bar p^2 \gbar^{\mu\nu})                   \cr
&-{1\over4\pi^2} (\TL+\TR) {1\over12} 
\gbar^{\mu\nu} \bar p^2&\eq{{\rm C}.1}\cr
&-{1\over4\pi^2} {(\TL+\TR) \over2}
	\left[{1\over3\eps}-{\ga\over6} +
{5\over18} - {1\over6}\ln{-p^2+i\epsilon\over4\pi\mu^2}+{1\over6}\right]
    \gbar^{\mu\nu} \hat p^2 \,\bigg\}+\, O(\eps).     \cr
}
$$

Notice that there is a singular contribution at $\eps =0 $ whose residue is a
an ``evanescent'' polynomial of the momentum. Minimal subtraction has to
be applied also to this contribution; in the language of ultraviolet divergent 
counterterms: counterterms with hatted objects are also needed! This is 
important for the consistency of calculations in the next orders \cite{\Bare}.

The renormalized 1PI fermionic contribution to the gauge field two-point
function defined by minimal substraction {\it \'a la} Breitenlhoner 
and Maison reads 
$$
\GRpuno_{\F\,AA}{\,}^{\mu\nu,ab}_\BM(p)=
	\GRpuno_{\F\,AA}{\,}^{\mu\nu,ab}_\n (p)
		-\delta^{ab}\,{1\over4\pi^2}{(\TL+\TR)\over12} 
			p^2 g^{\mu\nu}, 
$$
which differs from the ``naive'' result in a non-transversal quantity, 
so that it is clear that this result do not  satisfy the Slavnov-Taylor 
identity.

But, we learned from eq.~\EqRegYMSTBreaking\ that the Slavnov-Taylor 
identities for the regularized  theory {\it \'a la} Breitenlohner and 
Maison   have a breaking, so instead of transversality the following 
equation should hold
$$
-i p_\mu \Gapuno_{AA}{\,}^{\mu\nu,ab}_\BM(p)=
		\Gapuno_{A\om;\hat\Delta}{\,}^{\nu,ab}_\BM (p;0)  
\eqno{({\rm C}.2)}
$$
(because the insertion of the breaking is integrated, it is an insertion 
at zero momentum).

The right hand side of eq.~(C.2) is just the sum of the 
1PI diagrams in fig.~5 with $\hat\Delta$ instead of 
$\check\Delta$ and $p$ instead of $p_1$. Their value is:
$$
\eqalignno{
	\Gapuno_{A\om;\hat\Delta}{\,}^{\mu,ab}_\BM(p;q)=\delta^{ab}\,\bigg\{
	&\;{i\over4\pi^2} {(\TL+\TR)\over2}
		\left[{1\over3\eps}-{\ga\over6} +
	{5\over18} - {1\over6}\ln{-p^2+i\epsilon\over4\pi\mu^2}\right]
	\bar p^\mu \hat p^2                                         \cr
	&+{i\over4\pi^2} {(\TL+\TR)\over12} \bar p^\mu p^2          
         \,\bigg\}\,+\, O(\eps).                       &\eq{{\rm C}.3}\cr
}
$$
 
As it should be. Indeed, eqs.~(C.1) and (C.3) satisfy ($O(\eps)$) eq.~(C.2).
The contribution from the sum of diagrams with  bosonic and ghost loops
is obviously transversal and thus it is of no bearing here.  

From eq.~\EqRenYMSTBreaking\ we get the renormalized counterpart to eq.~(C.2):
$$
-i p_\mu \GRpuno_{AA}{\,}^{\mu\nu,ab}_\BM (p)=
		\GRpuno_{A\om;N[\hat\Delta]}{\,}^{\nu,ab}_\BM (p;0), 
$$
which is also satisfied because minimal subtraction applied to eq.~(C.3) 
yields
$$
\GRpuno_{A\om;N[\hat\Delta]}{\,}^{\mu,ab}_\BM(p;q)=\delta^{ab}\,
 {i\over4\pi^2} {(\TL+\TR)\over12}  p^\mu p^2.                 
$$
We have checked thus eq.~\EqRenYMSTBreaking\ by explicit computation of both 
its sides.
 
\medskip

The following results will be needed as well
$$
\eqalign{
&\GRpuno_{\bar\psi\psi}{\,}^{\a i,\b j}_\BM(p)-
\GRpuno_{\bar\psi\psi}{\,}^{\a i,\b j}_\n (p)=     \cr
&\qquad\qquad=-\ucpc g^2 \left[1 + (\aprime-1) {2\over3}\right] 
		\CL \left[\pslash \PL \right]_{\a\b}\,\delta_{ij},\cr
&\GRpuno_{A\bar\psi\psi}{\,}^{\nu,\a i,\b j}_\BM(q,p)-
\GRpuno_{A\bar\psi\psi}{\,}^{\nu, \a i,\b j}_\n (q,p)   = \cr
&\qquad\qquad=-\ucpc g^2 (\CL -{\CA\over2}) \,
	\left[2 + (\aprime-1) {5\over6}\right] \left[\ga^\mu\PL\right]_{\a\b}
	\left[\TL^a\right]_{ij} +  \cr
&\qquad\qquad\;\;-\ucpc g^2 \CA \,
	\left[1 + (\aprime-1) {5\over12}\right] \left[\ga^\mu\PL\right]_{\a\b}
			\left[\TL^a\right]_{ij} \cr
&\qquad\qquad=-\ucpc g^2 \CL \,
	\left[2 + (\aprime-1) {5\over6}\right] \left[\ga^\mu\PL\right]_{\a\b}
		\left[\TL^a\right]_{ij}.            \cr
}
$$
The first term of the sum corresponds to the QED-like diagram and the
second term to the diagram with the three boson vertex. 

\bigskip

Let us move on to the computation of the most interesting one-loop 
1PI functions to compare, namely, the three and four point 1PI  functions 
for the gauge field. In both these cases the ``naive'' prescription leads 
to ambiguities.

The fermionic diagrams which contribute to the one-loop three boson vertex 
are shown in fig.~13. The contribution from the diagram drawn there  
will be denoted with $i\L_{abc}^{\mu\nu\rho}(k_1,k_2,k_3)$. The full
contribution will thus read: 
$$
\eqalign{
i\Gapuno_{\F\,AAA}{}_{abc}^{\mu\nu\rho}(k_1,k_2,k_3)\equiv
  &i\L_{abc}^{\mu\nu\rho}(k_1,k_2,k_3)+i\L_{bac}^{\nu\mu\rho}(k_2,k_1,k_3)\cr
 +&i\R_{abc}^{\mu\nu\rho}(k_1,k_2,k_3)+i\R_{bac}^{\nu\mu\rho}(k_2,k_1,k_3).\cr
}
$$ 

\midinsert
\def\graphwidth{1.5in}
$$
\eqalign{\epsfxsize=\graphwidth\epsffile{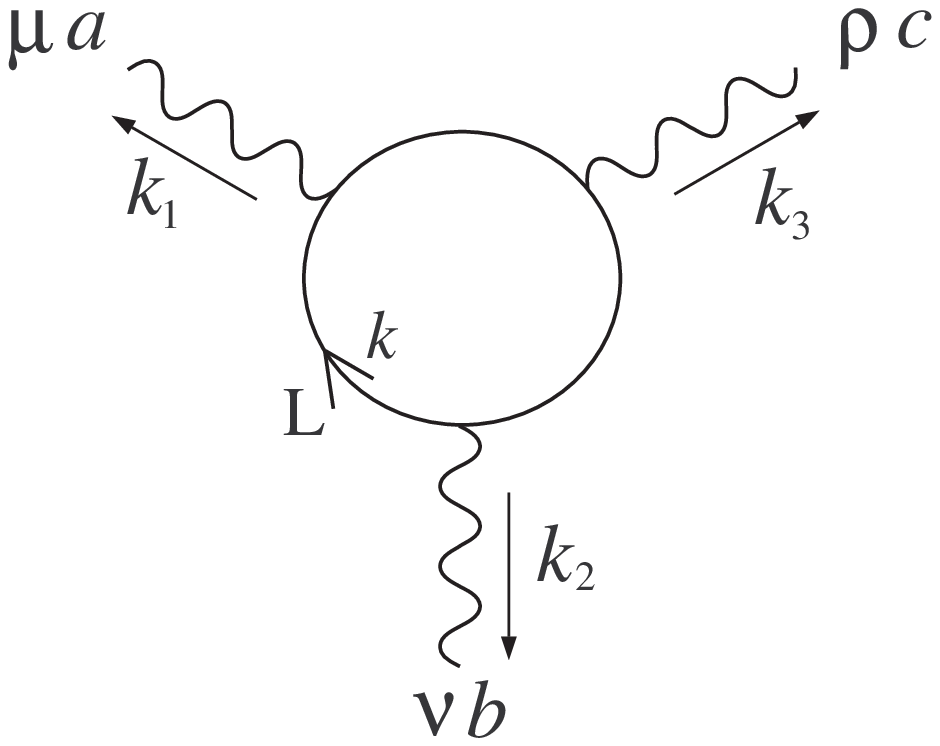}}
  \quad\eqalign{+}\quad
		\eqalign{\matrix{\hbox{permutation} \cr
				\hbox{of}\cr
			\hbox{legs 1 and 2}\cr
								}
					}
	\quad\eqalign{+}\quad
  \eqalign{\hbox{idem with R fermions}}
$$
\vskip -0.3cm
\narrower\noindent {\bf Figure 13:}
{\eightrm The one-loop 1PI Feynman diagrams with three bosonic legs}
\vskip 0.4cm
\endinsert

We shall focus first on the calculation of $i\L_{abc}^{\mu\nu\rho}$ 
$+i\R_{abc}^{\mu\nu\rho}$. The other diagrams are obtained simply
by exchanging $(k_1,\mu,a)$ with $(k_2,\nu,b)$. Of course, the three
momenta are not independent: for example, $k_3=-k_2-k_1$.

The ``naive'' and Breitenlohner and Maison dimensionally regularized  
integrals take the form:
$$
\eqalign{
&i \L_{abc}^{\mu\nu\rho}(k_1,k_2,k_3)+i\R_{abc}^{\mu\nu\rho}(k_1,k_2,k_3)=\cr 
&\qquad\qquad\qquad\;\intdk {N_{abc}^{\mu\nu\rho} \over  
	 (k-k_1+i\epsilon)^2 (k+i\epsilon)^2 
	 (k+k_2+i\epsilon)^2}\; ,\cr
}
$$
where the numerator depends on the regularization algorithm:
$$
\eqalign{
N_{abc}^{\mu\nu\rho}\,{}_{\n}
=&\TL^{abc}\; \Tr\,[\,(\kslash-\kslash_1)\ga^\mu\PL\kslash\ga^\nu\PL
	(\kslash+\kslash_2)\ga^\rho\PL\,] \; +       \cr
 &\TR^{abc}\; \Tr\,[\,(\kslash-\kslash_1)\ga^\mu\PR\kslash\ga^\nu\PR
	(\kslash+\kslash_2)\ga^\rho\PR\,] \, ,\cr        
N_{abc}^{\mu\nu\rho}\,{}_{\BM}
=&\TL^{abc}\; \Tr\,[\,(\kslash-\kslash_1)\gambar^\mu\PL\kslash\gambar^\nu\PL
	(\kslash+\kslash_2)\gambar^\rho\PL\,] \; +       \cr
 &\TR^{abc}\; \Tr\,[\,(\kslash-\kslash_1)\gambar^\mu\PR\kslash\gambar^\nu\PR
	(\kslash+\kslash_2)\gambar^\rho\PR\,] \, .    \cr
}
$$

Now, we anticommute the $\gam5$s of the projectors in the ``naive'' expression 
towards the right to get
$$
\eqalignno{
N_{abc}^{\mu\nu\rho}\,{}^{\n}_1
=&{\TR^{abc}+\TL^{abc} \over 2}\; 
	\Tr\,[\,(\kslash-\kslash_1)\ga^\mu\kslash\ga^\nu
		(\kslash+\kslash_2)\ga^\rho\,] \; +       \cr
 &{\TR^{abc}-\TL^{abc} \over 2}\; 
	\Tr\,[\,(\kslash-\kslash_1)\ga^\mu\kslash\ga^\nu
		(\kslash+\kslash_2)\ga^\rho\,\gam5\,].     &\eq{{\rm C}.4}\cr
}
$$

Notice that when the integration over the $k$s is done a trace over 
six gammas, two of them contracted, and a $\gam5$ occurs. This is 
precisely the source of the ambiguity displayed by  eq.~\EqNaiveAmbiguity\ ! 
Indeed, if we had decided to leave the matrix $\gam5$ in another place, 
the minimally renormalized results should differ in a local quantity:  there 
is a divergent loop present, and the ambiguity of eq.~\EqNaiveAmbiguity\ is 
proportional to $\eps=4-d$. This is the reason why we have labeled the 
numerator with the subscript $1$.

Therefore, we can write
$$
\eqalignno{
&i\L_{abc}^{\mu\nu\rho}(k_1,k_2,k_3)+i\R_{abc}^{\mu\nu\rho}(k_1,k_2,k_3)=\cr
&\qquad{\TR^{abc}+\TL^{abc} \over 2}\;\T_{\V\V\V}^{\mu\nu\rho}(k_1,k_2,k_3)+
	{\TR^{abc}-\TL^{abc} \over 2}\;
	\T_{\V\V\A}^{\mu\nu\rho}(k_1,k_2,k_3), &\eq{{\rm C}.5}\cr 
} 
$$
where $\T_{\V\V\V}$ stands for the Abelian Green function  
({\it i.e.} without gauge group matrices)
and with three vector-like vertices and $\T_{\V\V\A}^{\mu\nu\rho}(k_1,k_2,k_3)$
denotes the Abelian diagram with an axial vertex $\ga^\rho\gam5$ in 
the leg carrying momentum $k_3$.

The Breitenlohner and Maison counterpart to eq.~(C.4) is 
(we have used eq.~\EqRegProjectors) the following
$$
\eqalign{
N_{abc}^{\mu\nu\rho}\,{}_{\BM}
=&{\TR^{abc}+\TL^{abc} \over 2}\; 
	\Tr\,[\,(\bar\kslash-\bar\kslash_1)\gambar^\mu\bar\kslash\gambar^\nu
		(\bar\kslash+\bar\kslash_2)\gambar^\rho\,] \; +       \cr
 &{\TR^{abc}-\TL^{abc} \over 2}\; 
	\Tr\,[\,(\bar\kslash-\bar\kslash_1)\gambar^\mu\bar\kslash\gambar^\nu
(\bar\kslash+\bar\kslash_2)\gambar^\rho\,\gam5\,].    \cr
}
$$

It is easy to calculate the difference between the renormalized result 
obtained with the help of eq.~(C.4) ({\it do not move around the $\gamma_5$ 
``matrix''}) and the renormalized result obtained from eq.~(C.5):  
the only contribution to this difference comes
from the renormalization of the dimensionally regularized integral whose
numerator is given by
$$
\eqalignno{
{\TR^{abc}+\TL^{abc} \over 2}\;\big\{\, 
	&\Tr\,[\,\hat\kslash\gambar^\mu\hat\kslash\gambar^\nu
		(\bar\kslash+\bar\kslash_2)\gambar^\rho\,] \; +  
  \Tr\,[\,\hat\kslash\gambar^\mu\bar\kslash\gambar^\nu
		\hat\kslash\gambar^\rho\,] \; +       
	\Tr\,[\,(\bar\kslash-\bar\kslash_1)\gambar^\mu\hat\kslash\gambar^\nu
		\hat\kslash\gambar^\rho\,] \,\big\} \; +       \cr
{\TR^{abc}-\TL^{abc} \over 2}\;\big\{\, 
	&\Tr\,[\,\hat\kslash\gambar^\mu\hat\kslash\gambar^\nu
		(\bar\kslash+\bar\kslash_2)\gambar^\rho\,\gam5\,] \; +  
\Tr\,[\,\hat\kslash\gambar^\mu\bar\kslash\gambar^\nu
		\hat\kslash\gambar^\rho\,\gam5\,] \; +       \cr     
&\qquad\qquad\qquad\qquad\qquad\qquad
	\Tr\,[\,(\bar\kslash-\bar\kslash_1)\gambar^\mu\hat\kslash\gambar^\nu
		\hat\kslash\gambar^\rho\,\gam5\,]\,\big\}\; .\cr
}
$$

The value of that difference is thus the following
$$
\eqalignno{
&i\L_{abc}^{\mu\nu\rho}{}^\R_\BM(k_1,k_2,k_3)+
	i\R_{abc}^{\mu\nu\rho}{}^\R_\BM(k_1,k_2,k_3)       \cr
&-i\L_{abc}^{\mu\nu\rho}{}^\R_{\n_1}(k_1,k_2,k_3)-
	i\R_{abc}^{\mu\nu\rho}{}^\R_{\n_1}(k_1,k_2,k_3)=   \cr
&\quad=\icpc {\TR^{abc}+\TL^{abc} \over 2}\,4\, 
	\big[ g^{\mu\nu} {(k_1-k_2)^\rho\over3} +
			g^{\nu\rho} {(k_1+2k_2)^\mu\over3}-
			g^{\mu\rho} {(2k_1+k_2)^\nu\over3} \big]     \cr
&\qquad+\icpc {\TR^{abc}-\TL^{abc} \over 2}\; 
	\Tr\;\big[\,{(\kslash_1-\kslash_2)\over3}\,
		\ga^\mu\ga^\nu\ga^\rho\gam5\,\big]\, .
		&\eq{{\rm C}.6}\cr
}
$$

Notice that the second term of the previous equation is not cyclic, 
{\it i.e.} if we make the change $(k_1,\mu,a)\rightarrow(k_2,\nu,b)$,
$(k_2,\nu,b)\rightarrow(k_3,\rho,c)$,
$(k_3,\rho,c)\rightarrow(k_1,\mu,a)$ the term changes. This seems to
be incompatible with having a trace which is  
cyclic (or with the fact that we can start to read a closed fermion loop
wherever we want). Of course, if a prescription for Feynman rules is 
consistent, the renormalized result should be also cyclic. This is the
case for the Breitenlohner amd Maison result, as can be checked explicitly. 
But in the ``naive'' prescription case, if we insist on an 
anticommuting $\gam5$, the cyclicity of the trace is to be abandoned, 
lest the trace of four gammas and one $\gam5$  vanishes. The lack of 
cyclicity of eq.~(C.6) is due, of course, to the inconsistency of the 
``naive'' prescription with the ciclicity of the trace.

One could insist on keeping an anticommuting $\gam5$ and
cyclicity of the trace. But, then, as we
know from eq.~\EqNaiveInconsistency, the trace of four gammas and a $\gam5$
should unavoidably vanish. This would not do since it would set to zero any
dimensionally regularized Feynman integral that is finite by power-counting 
at $d=4$ and whose Dirac algebra yields a contribution proportional to
the trace of four gammas and one $\gamma_5$.  

The cyclicity of the diagram is also needed for obtaining a Bose symmetric
1PI function after summing the crossed diagram. Taking into account this
crossed diagram the difference between the complete 1PI functions in 
both schemes is:
$$
\eqalignno{
&\Gapuno{}_{abc}^{\mu\nu\rho}{}^\R_\BM(k_1,k_2,k_3)
-\Gapuno{}_{abc}^{\mu\nu\rho}{}^\R_{\n_1}(k_1,k_2,k_3)=       \cr
&\quad=\ucpc\,{2\over3}\, i\, c^{abc}\, (\TL+\TR)\,
	\big[ g^{\mu\nu} (k_1-k_2)^\rho +
			g^{\nu\rho} (k_2-k_3)^\mu-
			g^{\mu\rho} (k_3-k_1)^\nu \big]                    \cr
&\qquad+\ucpc\, {1\over6}\,\big[
	\Tr\{\TR^a,\TR^b\}\TR^c\,-\Tr\{\TL^a,\TL^b\}\TL^c\,\big] \; 
	\Tr\;\big[\,(\kslash_2-\kslash_1)\,
\ga^\mu\ga^\nu\ga^\rho\gam5\,\big]\, .  &\eq{{\rm C}.7}\cr
}
$$
The second term on the right hand side of the previuous equation is not Bose 
symmetric unless it is zero. 

Notice that the problem of the inconsistency of the ``naive'' prescription
can be swept under the carpet by chosing the matter
content  so that $\Tr\{\TR^a,\TR^b\}\TR^c$ 
$-\Tr\{\TL^a,\TL^b\}\TL^c$ ($\equiv\dR^{abc}-\dL^{abc}$ $\equiv\dRL$) 
vanish, {\it i.e.} if there is cancellation of 
anomalies. Moreover, we have compared the unambiguous Breitenlohner and  
Maison computation with the ``naive'' result obtained by putting 
the $\gam5$ after the $\ga^\rho$ (see eq.~(C.4)). If we had 
put it after, say, the $\ga^\mu$, the new difference with the 
 result {\it \'a la} Beitenlohner and Maison  has exactly the same first
term as the difference in eq.~(C.7) but its second term is 
the second contribution in the right hand side of 
eq.~(C.7), upon having changed 
cyclically the indices and momenta of the latter. That is, {\it the one-loop 
difference between two possible ``naive'' calculations is a term involving a 
product of an epsilon tensor and the coefficient\/} $\dRL^{bca}=$ 
$\dRL^{abc}=$ $\dRL^{\{abc\}}$\dag.%
\vfootnote\dag{$\,\{\cdots\}\,$ enclosing indices  means symmetrization
and $\,[\cdots]\,$ antisymmetrization}
This justifies the statement  that
{\it if there is cancellation of anomalies there is no ambiguity in
the one-loop ``naive'' computation of the three boson vertex\/}.
In ref.~\cite{\Gottlieb}, it was correctly claimed that the one-loop essential 
anomaly could be also obtained with the ``naive'' prescription for the
Abelian case from a VVA triangle diagram by simply not moving around 
the {\it unique} $\gam5$ which appear in the traces, 
computing explicitly the dimensional integrals and contracting, 
at the end of the day, the final expression with the momentum in the
axial vertex. In fact, with our notation
$$
k_3^\rho\, \T_{\V\V\A}^{\mu\nu\rho}(k_1,k_2,k_3) =
	-\icpc\,\Tr[\kslash_1\kslash_2\ga^\mu\ga^\nu\gam5], \eqno{({\rm C}.8)}
$$ 
which, with the crossed diagram, gives correctly the Abelian chiral anomaly.
Of course, also is checked explicitly that
$$
k_1^\mu\, \T_{\V\V\A}^{\mu\nu\rho}(k_1,k_2,k_3) =
k_2^\nu\, \T_{\V\V\A}^{\mu\nu\rho}(k_1,k_2,k_3) =0.         
$$

It is widely known that a zero (so, incorrect) result can be 
obtained by performing shifts in the integration variables and using
the cyclicity of the trace and the {\it anticommutativity of $\gam5$}.
Therefore, the correctness of the results above seems to be due to the fact 
that in the explicit calculations no use of the anticommutativity of $\gam5$ 
have been made. Unfortunately, for this restricted -the $\gamma_5$ is not moved
around- ``naive'' prescription  action principles cannot be 
usefully applied (the action is not invariant
if the $\gam5$ is not anticommuted) and, more importantly, {\it in local gauge
chiral theories}, such as the model studied here, {\it triangle diagrams\/ 
{\rm VAA} and\/ {\rm AAA} also appear}. If one decides not to anticommute 
the $\gam5$s before the subtraction is made, one cannot tell before hand 
whether the result will satisfy the Slavnov-Taylor identities with the correct 
essential breaking, unless explicit check  is made order by order. 
And if one naively anticommutes the $\gam5$s, then one will 
obtain ambiguous results (which turns out to be safe in 
one-loop calculations if there is cancellation of anomalies, but it is not 
clear at all what happens at higher orders)

Of course, the Breitenlohner and Maison scheme is a consistent method and 
has none of these problems. From eq.~\EqRenYMSTBreaking\ we
get the explicit identity  which involves $\GRpuno_{AAA}$:
$$
\eqalign{
&-i (k+p)^\mu \GRpuno_{AAA}{\,}_{\nu\rho\mu}^{bca} (k,p)\cr
&\qquad- i\, c^{bce}\, [\,(2k+p)_\rho\, g_{\mu\nu}-(k+2p)_\nu\, g_{\rho\mu}+
	(p-k)_\mu\, g_{\nu\rho}\,]\; \GRpuno_{w;\N[sA]}{\,}_\mu^{ae} (;k+p)\cr
&\qquad+c^{eba}\,\GRpuno_{AA}{\,}_{\rho\nu}^{ce} (p)+
		c^{eca}\,\GRpuno_{AA}{\,}_{\nu\rho}^{be} (k)-
1/g^2\, p^2\,\PT^{\rho\mu}(p)\,\GRpuno_{Aw;\N[sA]}{\,}_{\nu\mu}^{bac}(k;p)\cr 
&\qquad\qquad\qquad-
 1/g^2\, k^2\,\PT^{\rho\mu}(k)\,\GRpuno_{Aw;\N[sA]}{\,}_{\rho\mu}^{cab}(p;k)
 \;=\;\GRpuno_{AAw;\N[\hat\Delta]}{\,}_{\nu\rho}^{bca}(k,p;0),\cr
}
$$
where $\PT^{\mu\nu}(k)$ is the transversal Lorentz projector
$g^{\mu\nu}-{k^\mu k^\nu\over k^2}$.

The diagrams without fermions satisfy the Slavnov-Taylor identity
without breaking; therefore,
$$
\eqalignno{
-i (k+p)_\mu \GRpuno_{\F AAA}{\,}_{\nu\rho\mu}^{bca} (k,p)
+c^{eba}\,\GRpuno_{\F AA}{\,}_{\rho\nu}^{ce} (p)+
 &c^{eca}\,\GRpuno_{\F AA}{\,}_{\nu\rho}^{be} (k)=    \cr
 &\qquad\quad\GRpuno_{AAw;\N[\hat\Delta]}
      {\,}_{\nu\rho}^{bca}(k,p;0).&\eq{{\rm C}.9} \cr
} 
$$

The explicit calculation of the right hand side 
of the last equation gives a result
which contains spurious anomalies, {\it i.e.} which can be canceled by
adding finite counterterms to the action, this we have already seen 
in subsection {\it 3.6.}, and a part which is the correct essential anomaly:
$$
\eqalign{
\GRpuno_{AAw;\N[\hat\Delta]}{\,}_{\nu\rho}^{bca}{\,}^\BM(k,p;0)=
&-\ucpc\,{1\over3}(p^2-k^2)\,c^{bca}\,g_{\nu\rho}\, (\TR+\TL)  \cr
&-\ucpc\,{2\over3}(k_\nu k_\rho-p_\nu p_\rho)\,c^{bca}\, (\TR+\TL)\cr
&+\ucpc\,{1\over3}\,{\dRL^{bca}\over2}\,
	\Tr\,[\kslash\pslash\,\ga_\nu\ga_\rho\,].             \cr
}
$$
Notice that even if there $\dRL=0$ there are spurious anomalies in the
calculation.

For the ``naive'' prescription, by using eq.~(C.5) and {\it anticommuting
the\/} $\gam5$s, it is possible to write
$$
\eqalign{
\GRpuno_{\F AAA}{\,}_{\nu\rho\mu}^{bca}{\,}^\n (k,p)=&
\, i\,c^{bca}\,{\TR+\TL\over2}\, \T_{\V\V\V}^{\nu\rho\mu}(k,p,-k-p)\cr
&+{\dRL^{bca}\over2}\, \T_{\V\V\A}^{\nu\rho\mu} (k,p,-k-p).  \cr
}
$$

Therefore, the left hand side of eq.~(C.9) in the ``naive'' prescription 
(anticommuting $\gam5$) reads:
$$
\eqalign{
&c^{bca}\,{\TR+\TL\over2}\,
   \big\{(k+p)_\mu\,\T_{\V\V\V}^{\nu\rho\mu}(k,p,-k-p)-
		\T_{\V\V}^{\nu\rho} (p)+ \T_{\V\V}^{\nu\rho} (k)\,\big\}\cr
&-{\dRL^{bca}\over2}\,i\,(k+p)_\mu\,\T_{\V\V\A}^{\nu\rho\mu}(k,p,-k-p).  \cr
}
$$

But the first term between brackets vanish in the ``naive'' prescription 
without incurring into any inconsistency -no $\gam5$ is involved. Hence,  
we arrive at the conclusion that 
$$
\hbox{l.h.s.~of eq.~(C.9) }=
{\dRL^{bca}\over2}\,i\,(-k-p)^\mu\,\T_{\V\V\A}^{\nu\rho\mu}(k,p,-k-p),
						\eqno{({\rm C}.10)}
$$
in the ``naive'' dimensional regularization prescription 
(anticommuting $\gam5$).
Next, one could think of applying eq.~(C.8) to obtain an ``anomaly''
$\ucpc\,{\dRL^{bca}\over2}$ 
$\Tr\,[\kslash\pslash\,\ga_\nu\ga_\rho\,]$, which is different
from the correct Breitenlohner and Maison result. Moreover, 
eq.~(C.10) was obtained by putting the one $\gam5$ that is left after 
the anticommutation process is done in the position $\mu$. If we had put it
in the position $\nu$, we would not have obtained any 
anomaly at all. (The ``anomaly'' obtained with the first
choice for a place for the $\gam5$ is three 
times the correct Breitenlohner and Maison result and 
three is the number of choices). This shows clearly that 
the ``naive'' prescription cannot give consistently
the one-loop anomaly in chiral local gauge theories. Therefore, the
claim that  ``we do not need discuss the graphs containing the AAA triangle
since these behave in the same manner as the AVV triangle'' of \cite{\Ovrut}
seems to be not well justified.

Let us move on to computation of the one-loop four gauge field vertex.
This vertex also contributes to the anomaly. We have already shown in
subsection {\it 3.5.} that the Breitenlohner and Maison scheme gives  
the correct relative factors between the anomaly from the three gauge 
field vertex and from four gauge field vertex. 

The one-loop 1PI four gauge field function is given by the sum of the diagram
depicted in fig.~14 and all the diagrams which are obtained
from the first by performing all the possible permutations of three legs. 

\midinsert
\def\graphwidth{1.5in}
$$
\eqalign{\epsfxsize=\graphwidth\epsffile{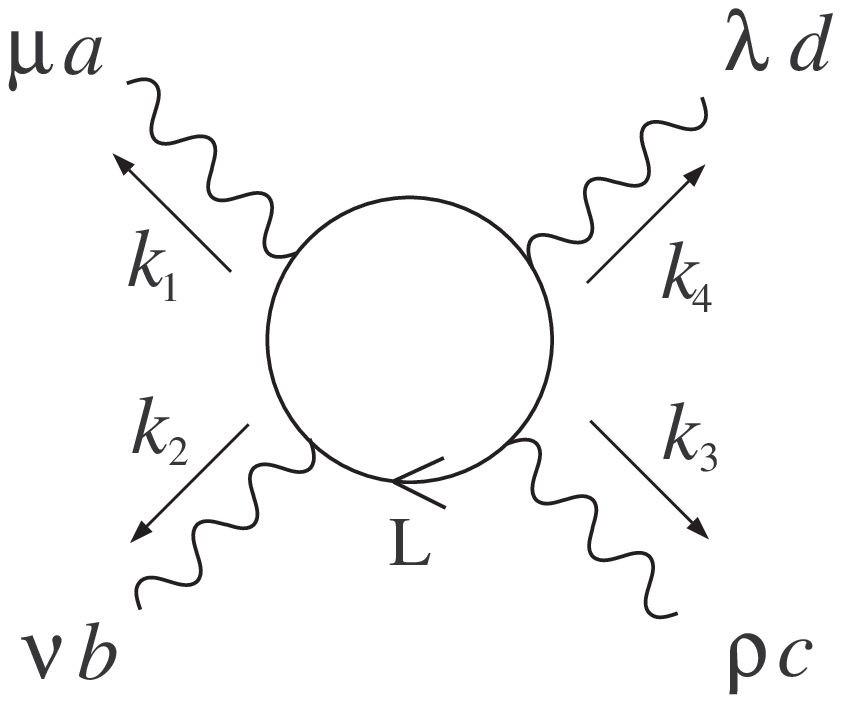}}
	\quad\eqalign{+}\quad
  \eqalign{\hbox{idem with R fermions}}
$$
\vskip -0.3cm
\narrower\noindent {\bf Figure 14:}
{\eightrm one-loop 1PI Feynman diagrams with four bosonic legs}
\vskip 0.4cm
\endinsert

Proceeding as for the three boson vertex case, the difference between 
the renormalized minimal Breitenloner and Maison result for the diagrams 
of fig.~14  and the ``naive'' result defined with the 
by putting the $\gam5$ after the $\ga^\la$ -the result depends on the
possition of the $\gam5$- 
$$
\eqalign{
-\icpc\,{2\over3}\,
	\Big\{ &g^{\mu\nu}g^{\rho\la}\,
		\big[\,3 {\TLR^{abcd}\over2}+3{\TLR^{abdc}\over2}+
				3{\TLR^{acdb}\over2}+3{\TLR^{adcb}\over2}-
			5{\TLR^{acbd}\over2}-5{\TLR^{adbc}\over2}\,\big]+\cr
	&g^{\mu\la}g^{\nu\rho}\,
		\big[\,3 {\TLR^{abcd}\over2}+3{\TLR^{acbd}\over2}+
				3{\TLR^{adbc}\over2}+3{\TLR^{adcb}\over2}-
			5{\TLR^{abdc}\over2}-5{\TLR^{acdb}\over2}\,\big]+\cr
	&g^{\mu\rho}g^{\nu\la}\,
		\big[\,3 {\TLR^{acbd}\over2}+3{\TLR^{acdb}\over2}+
				3{\TLR^{adbc}\over2}+3{\TLR^{abdc}\over2}-
		5{\TLR^{abcd}\over2}-5{\TLR^{adcb}\over2}\,\big] \Big\} \cr 
-\icpc {1\over2}\,\Tr\,[\,&\ga^\mu\ga^\nu\ga^\rho\ga^\la\,\gam5\,]
	\big\{ {\TRmL^{abcd}\over2} + {\TRmL^{acdb}\over2}+{\TRmL^{adbc}\over2}
	-{\TRmL^{acbd}\over2}-{\TRmL^{abdc}\over2}-{\TRmL^{adcb}\over2} \big\}.
							\cr
}
$$

Notice that the first term between brackets is cyclic but that the second
one is, in fact, $3\,\TRmL^{a[bcd]}$, {\it i.e.} antisymmetric in the 
last three indices but not in the four indices, so, again, is not cyclic. 

Then, the complete difference, taking into account all diagrams, is:
$$
\eqalign{
-\icpc\,{2\over3}\,
	\Big\{ &g^{\mu\nu}g^{\rho\la}\,
		\big[\,3 {\TLR^{\{a,b\}\,\{c,d\}}\over2}-5{\TLR^{acbd}\over2}-
				5{\TLR^{adbc}\over2} \,\big] + \cr
	&g^{\mu\la}g^{\nu\rho}\,
		\big[\,3 {\TLR^{\{a,d\}\,\{c,b\}}\over2}-5{\TLR^{abdc}\over2}-
				5{\TLR^{acdb}\over2} \,\big] + \cr
	&g^{\mu\rho}g^{\nu\la}\,
		\big[\,3 {\TLR^{\{a,c\}\,\{b,d\}}\over2}-5{\TLR^{abcd}\over2}-
				5{\TLR^{adcb}\over2} \,\big]\,\Big\} \cr 
+\icpc {1\over2}\,\Tr\,[\,&\ga^\mu\ga^\nu\ga^\rho\ga^\la\,\gam5\,]
	\big\{ \,\DR^{abcd}-\DL^{abcd}\,\big\},  \cr
}
$$
where we have defined
$$
\eqalign{
\TLR^{\{a,b\}\,\{c,d\}}&=\Tr\,\{\TL^a,\TL^b\}\,\{\TL^c,\TL^d\}+
			 \Tr\,\{\TR^a,\TR^b\}\,\{\TR^c,\TR^d\}\, , \cr
\DL^{abcd}&=-i\,3!\,\TL^a\,\TL^{[b}\TL^c\TL^{d]}=                    
	{1\over2}\,(\,\dL^{abe} c^{ecd} + \dL^{ace} c^{edb}+
			\dL^{ade} c^{ebc}\,)=\DL^{a[bcd]},       \cr
}
$$
and similarly $\DR^{abcd}$.

Again, the first term between brackets is a local term which can be
checked to be just proportional to the Feynman rules of 
$\Tr_{\R+\L}\, A_\mu A^\mu A_\nu A^\nu$ and 
$\Tr_{\R+\L}\, A_\mu A_\nu A^\mu A^\nu$; whereas the second term
is not Bose symmetric and reflects the inconsistency of the ``naive'' 
prescription. Also, we can use these result to show that the
{\it ambiguities in the ``naive'' calculation of the one-loop four boson
vertex\/} is proportional to the epsilon tensor and to the coefficient
$\DR^{abcd}-\DL^{abcd}$, which vanish if there is cancellation of anomalies.
For example, the one-loop
 renormalized difference between the ``naive'' 
calculation with the $\gam5$ after $\ga^\la\PLpR$ and 
the ``naive'' calculation 
with the $\gam5$ after $\ga^\mu\PLpR$ is $i/{8\pi^2}$ $\eps_{\mu\nu\rho\la}$
($\DR^{abcd}-\DL^{abcd}$ $+\DR^{bcda}-\DL^{bcda})$.

We have exhibited by explicit computation the many inconsistencies that plague 
the computations leading to a renormalized one-loop chiral gauge theory if the
``naive'' Dimensional Regularization presciption is used. The classical 
action of the theory was given in subsection {\it 3.1.} To close this
appendix we shall show that in spite of these inconsistencies, and if 
the chiral theory is anomaly free- a compulsory constraint-, the renormalized 
BRS invariant chiral theories obtained by means of the ``naive'' dimensional 
regularization presciption, on the one hand, and Breitenlohner and Maison 
Dimensional Regularization procedure, on the other, are equivalent at the
one-loop level, in the sense that they are related by finite BRS symmetric
renormalizations of the fields and couplings  of the theory.

As we have previously compared the  Breitenlohner and Maison 
renormalized theory with the ``naive'' renormalized theory when no finite 
BRS-asymmetric counterterms have been added to the former to restore 
the BRS symmetry, the 
comparison between the two renormalized theories, when the BRS-asymmetric 
counterterms have been added to the minimal Breitelhoner and Masion 
renormalized theory, is very easy. 

Due to the fact that, in the situation of cancellation of anomalies, 
the ``naive'' minimal dimensionally renormalized theory  satisfy the 
Slavnov-Taylor identities at order $\hbar^1$,
the finite counterterms relating the two renormalized theories should appear 
in $b$-invariant combinations. And this is what nicely happen, 
upon inclusion of finite BRS asymmetric counterterms, the  Breitenlohner and 
Maison renormalized theory equals the ``naive'' renormalized theory  if
the coefficients -$l_g^{(1)}$, $l_{\psi}^{(1)}$, $l_{\psi^\prime}^{(1)}$,
$l_A^{(1)}$ and $l_\om^{(1)}$- of the BRS-symmetric  counterterms in
eq.~\EqSfctUno\ are chosen to be given by   
$$
l_g^{(1)}=-\ucpc\,{5\over24}\, g^2,\quad 
l_{\psi^\pp}^{(1)}=\ucpc\,g^2\,\CLpR\,\big[1+(\aprime-1){5\over12}\big],
\quad l_A^{(1)}=l_\om^{(1)}=0.       
$$
This corresponds, of course, to a mass-independent multiplicative 
finite renormalization of the fields and parameters of the effective action.
Hence, the beta function and anomalous dimensions of both renormalized 
theories are the same at the one-loop level. The finite renormalizations
we have just mentioned read
\medskip
{\settabs 4 \columns
\openup1\jot
\+$\qquad A=A^\star $, &$\qquad B=B^\star $, &$\qquad \om=\om^\star$,
		&$\qquad\bar\om=\bar\om^\star$,      \cr
\+$\qquad \rho=\rho^\star$, &$\qquad\zeta=\zeta^\star$,      
		&$\qquad\a=\a^\star$,                \cr
\+$\qquad g=(1 + \ucpc\,{5\over48}\,{g^\star}^2\,\hbar^1)\,g^\star $, \cr 
\+$\qquad \psi=(1 + 
	\ucpc\,{g^\star}^2\,\CL\,\left[1+(\a'-1){5\over12}\right]
		\,\hbar^1\,\PL)\,\psi^\star $,       \cr
\+$\qquad\bar\psi=\bar\psi^\star\,(1 +
	\ucpc\,{g^\star}^2\,\CL\,\left[1+(\a'-1){5\over12}\right]    
		 \,\hbar^1\,\PR)$, \cr
\+$\qquad \psi'=(1 + 
	\ucpc\,{g^\star}^2\,\CR\,\left[1+(\a'-1){5\over12}\right]
	\,\hbar^1\,\PR)\,\psi'{}^\star $,    \cr
\+$\qquad\bar\psi'=\bar\psi'{}^\star\,(1 + 
	\ucpc\,{g^\star}^2\,\CR\,\left[1+(\a'-1){5\over12}\right]
	\,\hbar^1\,\PL)$, \cr
\+$\qquad\bar L=\bar L^\star\, (1 + 
	\ucpc\,{g^\star}^2\,\CL\,\left[1+(\a'-1){5\over12}\right]
	\,\hbar^1\,\PL)$,     \cr
\+$\qquad L=(1 + 
	\ucpc\,{g^\star}^2\,\CL\,\left[1+(\a'-1){5\over12}\right]
	\,\hbar^1\,\PR)\,L^\star$,                   \cr 
\+$\qquad \bar R=\bar R^\star\, (1 + 
	\ucpc\,{g^\star}^2\,\CR\,\left[1+(\a'-1){5\over12}\right]
	\,\hbar^1\,\PR)$,       \cr
\+$\qquad R=(1 +
	\ucpc\,{g^\star}^2\,\CR\,\left[1+(\a'-1){5\over12}\right]
	\,\hbar^1\,\PL)\,R^\star$.                 \cr 
}
\medskip
Where the fields and constants with a star are the renormalized fields and
constants in the ``naive'' minimal Dimensional Regularization prescrition and
the fields and constant without stars stand for the corresponding 
renormalized quantities in renormalized theory obtained by means of 
the Breitenlohner and Maison scheme (minimal subtraction plus addition of the
finite counterterms in eq.~\EqSfctUno, with 
 $l_g^{(1)}$, $l_{\psi}^{(1)}$, $l_{\psi^\prime}^{(1)}$,
$l_A^{(1)}$ and $l_\om^{(1)}$ set to zero).

\section{Acknowledgments}
The financial support from the Universidad Complutense de Madrid under grant 
{\tt PR156/97-7164}  and the Consejer{\'\i}a de Educaci\'on y
Cultura of the Junta de Comunidades de Castilla-La Mancha are acknowledged.

\section{References}

\frenchspacing
\refno\Ash.
J. Ashmore, Lett. Nuovo Cimento 4 (1972) 289; C.G.Bollini and J.J. Giambiagi,
Phys. Lett. B40 (1972) 566.

\refno\HV.
G. t'Hooft and M. Veltman, Nucl. Phys. B44 (1972) 189-213.

\refno\BMa.
P. Breitenlohner and D. Maison, Comm. Math. Phys 52 (1977) 11-38.

\refno\BMb.
P. Breitenlohner and D. Maison, Comm. Math. Phys 52 (1977) 39-54.

\refno\BMc.
P. Breitenlohner and D. Maison, Comm. Math. Phys 52 (1977) 55-75.

\refno\BonneauA.
G. Bonneau, Nucl. Phys. B167 (1980) 261-284.

\refno\BonneauB.
G. Bonneau, Nucl. Phys. B180 (1980) 477-508.

\refno\BonneauC.
G. Bonneau, Nucl. Phys. B177 (1980) 523-527.

\refno\Leib.
G. Leibbrandt, Rev. Mod. Phys. 47 4 (1975) 849.

\refno\Collins.
J. Collins, Renormalization
(Cambridge University Press, Cambridge, 1984).

\refno\Min.
G. `t Hooft, Nucl. Phy. B61 (1973) 455.

\refno\Tech.
G. `t Hooft, Nucl. Phys. B62 (1973) 444; O.V. Tarasov, A.A. Vladimirov and
A.N. Zharkov, Phys. Lett. B93 (1980); F.V. Tachkov, Phys. Lett. B100 (1981) 65;
D.I. Kazakov, Phys. Lett. B133 6 (1983) 406; J.M. Campbell, E.W.N. Glover
and D.J. Miller, Nucl. Phys. B498 (1997) 397; L. Br\"ucher, J. Franzkowski
and D. Kreimer, {\it Oneloop 2.0} {\tt hep-th 9709209}.

\refno\Multi.
M.E. Machacek and M.T. Vaughn, Nucl. Phys. B222 (1983) 83;
{\it  ibid.} B236 (1984) 221; I. Jack and D.R.T. Jones, Nucl. Phys. 
B249 (1985) 472; C. Ford, I. Jack
and D.R.T. Jones, Nucl. Phys. B387 (1992) 373; C. Ford, D.R.T. Jones, 
P.W. Stephenson and M.B. Einhorn, Nucl. Phys. B395 (1993) 17; T. Van Ritbergen
and J.A.M. Vermaseren, Phys. Lett. B400 (1997) 379; B.A. Kniehl, Int. J. Mod.
Phys. A10 4 (1995) 443 {\it and references therein.}

\refno\BonneauRemarks.
G. Bonneau, Phys. Lett. B96 (1980) 148-150.

\refno\Aky.
D.A. Akyeampong and R. Delbourgo, Nuovo Cimento 17A (1973) 578; {\it ibid.}
18A (1973) 94; {\it ibid.} 19A (1974) 219.

\refno\Kreimer.
D. Kreimer, Phys. Lett. B237 (1990) 59, ``The role of $\gam5$ in Dimensional
Regularization'', {\tt hep-ph/9401354}.

\refno\Colnor.
J.C. Collins, Nucl. Phys. B92 (1975) 477.

\refno\PiguetSorella.
O. Piguet and S. Sorella, Algebraic renormalization (Springer-Verlag, Berlin,
1995) {\it and references therein.}

\refno\Brandt.
F. Brandt, Commun. Math. Phys. 190 (1997) 459; E. Krauss, Renormalization of
the Electroweak Standard Model to all orders, {\tt hep-th/9709154}, {\it
to appear in Ann. Phys.} 

\refno\Hepp.
K. Hepp, Renormalization theory in statistical mechanics and quantum field
theory, in ``Les Houches XX 1970'' (Gordon and Breach, New York, 1971).

\refno\BonneauReview.
G. Bonneau, Int. Journ. Mod. Phys. A5 (1990) 3831-3859.

\refno\naive.
W.A.  Bardeen, R. Gastmans and B. Lautrup, Nucl. Phys. B46 (1972) 319;
M. Chanowitz, M. Furman and I. Hinchliffe, Nucl. Phys. B159 (1979) 225.

\refno\Alwit.
L. Alvarez-Gaum\'e and E. Witten, Nucl. Phys. B234 (1984) 269.

\refno\Wilson.
K. Wilson, Phys. Rev. D7 (1973), 2924-2926.

\refno\PiguetRouet.
O. Piguet and A. Rouet, Phys. Rep. 76 (1981) 1-77.

\refno\Korner.
J.G. K\"orner, N. Nasrallah and K. Schilcher, Phys. Rev. D41 (1990) 888-890.

\refno\Ferrari.
R. Ferrari, A. Le Yaouanc, L. Oliver and J.C. Raynal, Phys. Rev. D52 (1995)
3036-3047.

\refno\Barroso.
A. Barroso, M.A. Doncheski, H. Grotch, J.G. K\"orner and K. Schilcher, 
Phys. Lett. B261 (1991) 123-126.

\refno\Gottlieb.
S. Gottlieb and J.T. Donohue, Phys. Rev. D20 (1979) 3378-3389.

\refno\Ovrut.
B.A. Ovrut, Nuc. Phys. B213 (1983) 241-265.

\refno\IZ.
C. Itzikson and J.B. Zuber, Quantum Field Theory (McGraw-Hill, 1980)

\refno\Bare.
M. Bos, Ann. Phys. 181 (1988) 197; H. Osborn, Ann. Phys. 200 (1990) 1;
I. Jack, D.R.T. Jones and K.L. Roberts, Z. Phys. C 63 (1994) 151.

\bye